\documentclass[11pt, a4paper, twoside, openright]{book}

\usepackage{amsmath,amssymb,amscd,amsfonts,amsthm}     
\usepackage{psboxit,pstricks,psfrag}
\usepackage{graphics,subfigure,epsfig,graphicx}        
\usepackage{lettrine}   
\usepackage{titlesec}   
\usepackage{setspace}   
\usepackage{fancyhdr}   
\usepackage{cite}       
\usepackage{slashed}    
\usepackage{pifont}     

\usepackage{axodraw}    
\usepackage{color}
\usepackage{colortbl}    

\usepackage[T1]{fontenc} 
\usepackage{emerald}    

\newcommand{\cmarkfont}{\usefont{T1}{pag}{b}{sc}}
\newcommand{\smarkfont}{\usefont{T1}{pag}{b}{n}}
\newcommand{\dropcapfont}{\ECFAugie \bfseries \fontsize{60}{70}\selectfont}
\newcommand{\dropcaptext}{\ECFAugie \bfseries}
\newcommand{\dropcapfontC}{\ECFAugie \bfseries \fontsize{60}{70}\color[rgb]{0.3,0.3,0.3}
                          \selectfont}
\newcommand{\dropcaptextC}{\ECFAugie \bfseries \color[rgb]{0.3,0.3,0.3}\selectfont}                          
                          
\newcommand{\ECap}[4]{
 \lettrine[lines=3, findent=#3pt, nindent=#4pt]{\dropcapfont #1}{\dropcaptext #2}
} 

\newcommand{\ECapC}[4]{
 \lettrine[lines=3, findent=#3pt, nindent=#4pt]{\dropcapfontC #1}{\dropcaptextC #2}
}

\makeatletter
\def\@makechapterhead#1{
  {\parindent \z@ \raggedleft \reset@font
            \scshape {\Large\bfseries \color[rgb]{0.5,0.5,0.5} \@chapapp{}} \par
            \vspace{25\p@} 
       {
      \usefont{T1}{ppl}{b}{n}       
       \fontsize{100}{130}\selectfont \color[rgb]{0.5,0.5,0.5}     
            \thechapter} 
     \par\vspace{25\p@} \nobreak
     \interlinepenalty\@M
    \Huge 
    {\usefont{T1}{pag}{b}{n} #1}
    \par\nobreak
    \rule{\textwidth}{1pt}
     \par\nobreak
    \vskip 50\p@
  }  
}

\def\@makeschapterhead#1{
  {\parindent \z@ \raggedleft \reset@font
            \scshape \vphantom{\@chapapp{} }
        \par\nobreak
        \interlinepenalty\@M
    \Huge
    {\usefont{T1}{pag}{b}{n} #1}\par\nobreak
    \rule{\textwidth}{1pt}
    \par\nobreak
    \vskip 30\p@
  }
}

\def\cleardoublepage{\clearpage\if@twoside \ifodd\c@page\else
	\hbox{}
	\vspace*{\fill}
	\thispagestyle{empty}
	\newpage
	\if@twocolumn\hbox{}\newpage\fi\fi\fi}
\makeatother

\titleformat{\section}{\Large\smarkfont}{\thesection}{1em}{}

\renewcommand{\chaptermark}[1]{\markboth{\cmarkfont \chaptername~\thechapter. #1}{}}

\fancyheadoffset[RO,LE]{\marginparsep} 

\fancypagestyle{plain}{
 \fancyhf{}     
 \fancyfoot[RO]{\smarkfont \thepage}

}

\def\ththMat#1#2#3#4#5#6#7#8#9{\ensuremath{\begin{pmatrix}#1&#2&#3\\
                                                          #4&#5&#6\\
                                                          #7&#8&#9\end{pmatrix}}}
\def\twtwMat#1#2#3#4{\ensuremath{\begin{pmatrix}#1&#2\\
                                                  #3&#4\\
                                                  \end{pmatrix}}}
\def\thtwMat#1#2#3#4#5#6{\ensuremath{\begin{pmatrix}#1&#2\\
                                                    #3&#4\\
                                                    #5&#6\end{pmatrix}}}
                                                    
\def\fivecMat#1#2#3#4#5{\ensuremath{\begin{pmatrix}#1\\#2\\#3\\#4\\#5\end{pmatrix}}}
\def\threecMat#1#2#3{\ensuremath{\begin{pmatrix}#1\\#2\\#3\end{pmatrix}}}

\def\threerMat#1#2#3{\ensuremath{\begin{pmatrix}#1&#2&#3\end{pmatrix}}}
\def\tworMat#1#2{\ensuremath{\begin{pmatrix}#1&#2\end{pmatrix}}}
\def\twocMat#1#2{\ensuremath{\begin{pmatrix}#1\\#2\end{pmatrix}}}
\def\Matdiag#1#2#3{\ensuremath{\begin{pmatrix}#1&0&0\\
                                              0&#2&0\\
                                              0&0&#3\end{pmatrix}}}
\def\MatCase#1#2{\ensuremath{\begin{matrix}#1\\
                                           #2\end{matrix}}}
\def\MatU{\ensuremath{\frac{1}{\sqrt{3}}
                         \begin{pmatrix}1 &1 &1\\
                                     1 &\omega &\omega^2\\
                                     1 &\omega^2 &\omega\end{pmatrix}}}                                              

\def\msol{\ensuremath{m_\textrm{sol}}}
\def\matm{\ensuremath{m_\textrm{atm}}}
\def\order#1{\ensuremath{\mathcal{O}\left(#1\right)}}
\def\vev#1{\ensuremath{\langle #1\rangle}}
\def\ket#1{\ensuremath{|#1\rangle}}
\def\bra#1{\ensuremath{\langle #1}} 
\def\rept{\ensuremath{\underline{3}}}
\def\rep{\ensuremath{\underline{1}}}
\def\repp{\ensuremath{\underline{1}'}}
\def\reppp{\ensuremath{\underline{1}''}}
\def\Upmns{\ensuremath{U_\text{PMNS}}}
\def\Uckm{\ensuremath{U_\text{CKM}}}

\def\Dvev{\ensuremath{\langle\Phi\rangle}}
\def\Svev{\ensuremath{\langle\chi\rangle}}
\def\curlN{\ensuremath{\mathcal{N}}}
\def\Mpl{\ensuremath{M_\textrm{Pl}}}
\def\odifone#1#2{\ensuremath{\frac{d #1}{d #2}}}
\def\Ztwo{\ensuremath{\mathbb{Z}_2}}
\def\Zfour{\ensuremath{\mathbb{Z}_4}}
\def\Mbare{\ensuremath{M_\textrm{bare}}}

\def\mnu{\ensuremath{m_\nu}}
\def\mnud{\ensuremath{m_\nu^D}}
\def\mnuhat{\ensuremath{\widehat{m}_\nu}}
\def\Uw{\ensuremath{U_\omega}}

\def\dslash{\ensuremath{\slashed{\partial}}}
\def\pslash{\ensuremath{\slashed{p}}}
\def\ppslash{\ensuremath{\slashed{p}'}}
\def\qslash{\ensuremath{\slashed{q}}}
\def\qoneslash{\ensuremath{\slashed{q}_1}}
\def\qtwoslash{\ensuremath{\slashed{q}_2}}
\def\qthreeslash{\ensuremath{\slashed{q}_3}}

\def\mmat{\ensuremath{\mathcal{M}}}
\def\ubarj{\ensuremath{\overline{u}_j}}
\def\ubark{\ensuremath{\overline{u}_k}}
\def\vbarj{\ensuremath{\overline{v}_j}}

\def\mNm{\ensuremath{M_{m}}}
\def\mNk{\ensuremath{M_{k}}}
\def\costh{\ensuremath{\cos\theta}}
\def\halfmNk{\ensuremath{\frac{M_k}{2}}}
\def\Eq1{\ensuremath{E_{q1}}}
\def\modq1{\ensuremath{|\vec q_1|}}
\def\gammazero{\ensuremath{\gamma^0}}
\def\gammamu{\ensuremath{\gamma^\mu}}
\def\LCtensor{\ensuremath{\epsilon^{\mu\nu\rho\sigma}}}

\begin{document}



%
\begin{titlepage}
~\vspace{-30pt}
 \begin{center}
 {\usefont{T1}{pag}{m}{n}\fontsize{38}{40}\selectfont 
    Neutrino Models\\
     \vspace{-10pt} and\\ \vspace{-10pt}  Leptogenesis\\
 }\vspace{30pt} 
 Submitted in total fulfilment of the requirements of the degree of\par
  {\em Doctor of Philosophy} \par\vspace{90pt}
   \par\vspace{20pt}
  {\LARGE\bfseries  Sandy Sheung Che Law}\par\vspace{40pt}
    School of Physics\par
    The University of Melbourne\par
    AUSTRALIA\par\vspace{20pt}
    December, 2008
    \vfill\phantom{\fbox{\small Produced on archival quality paper}}
  \end{center}
\end{titlepage}

%
\begin{titlepage}
 ~\vskip 6cm
 \begin{center}
 \end{center}
\end{titlepage}

\cleardoublepage 
 \pagenumbering{roman}
  \setcounter{page}{1}


%
%
%
%
%


{\singlespacing
\chapter*{Abstract\markboth{\cmarkfont Abstract}{\cmarkfont Abstract}}
}

\ECap{N}{e~u~t~r~i~n~o}{7}{0}
properties can play a crucial role in determining the matter-antimatter asymmetry of the universe if thermal leptogenesis is the correct solution to the baryogenesis problem. Owing to this, the study of neutrino models goes beyond the mere purpose of generating tiny neutrino masses, and it is natural to incorporate the puzzle of the cosmic baryon asymmetry.
To this end, we have investigated several different extensions of the neutrino model based on the type~I seesaw mechanism with particular emphasis on their leptogenesis implications. 

In the first part of our work, we present a thorough analysis, in the leptogenesis context, of  constrained neutrino models which have an abelian family symmetry and one extra Higgs singlet. The general result is that although these models contain less free parameters than the default type~I seesaw setup, they do not hinder the possibility of successful leptogenesis in both the one-flavor approximation and when flavor effects are included. In fact, we have discovered that they do not modify or provide significant constraints on the typical leptogenesis scenario with hierarchical heavy right-handed neutrinos.

We then explore how the seesaw sector in neutrino mass models may be constrained through symmetries to be completely determined in terms of low-energy mass, mixing angle and $CP$ violating phase observables, with the subsequent aim to study their leptogenesis implications. The key ingredient in achieving this connection between the two distinct energy sectors is to simultaneously employ intra- and inter-flavor symmetries so that the neutrino Dirac mass matrix is only given in terms of the charged-lepton or quark mass matrices while all relevant diagonalization matrices can be fully predicted by the theory. We have built illustrative models to demonstrate this and shown that leptogenesis can succeed in small regions of parameter space for the case where the neutrino Dirac and up-quark mass matrix are identical. Also, it has been found that when the neutrino Dirac mass matrix is equal to its charged-lepton counterpart, TeV scale heavy neutrinos are possible although it is unlikely that they can be detected in colliders.

Finally, we present a new leptogenesis scenario, where the lepton asymmetry is generated by $CP$ violating decays of heavy right-handed neutrinos via electromagnetic dipole moment couplings to the ordinary light neutrinos. Akin to the usual scenario where the decays are mediated through Yukawa interactions, we have shown, by explicit calculations, that the desired asymmetry can be produced through the interference of the corresponding tree-level and one-loop decay amplitudes involving the effective dipole moment operators. Furthermore, we find that the relationship of the leptogenesis scale to the light neutrino masses is similar to that for the standard Yukawa-mediated mechanism.



{\singlespacing
\chapter*{Declaration}}
\vspace{30pt}
\noindent
{\bf This is to certify that}\\

\begin{itemize}
\item[(i)] the thesis comprises only my original work towards the PhD except where indicated in the \textbf{Preface},
\item[(ii)] due acknowledgement has been made in the text to all other material used,
\item[(iii)] the thesis is less than 100,000 words in length,
exclusive of table, maps, bibliographies, appendices and footnotes.
\end{itemize}\vspace{5em}
\begin{flushright}
  \makebox[6.0cm]{\dotfill}\\
  \makebox[6.0cm]{\small \em Sandy S. C. Law}\\
  \vspace{30pt}
  {\em Date:}\makebox[6.0cm]{\hspace{1.5cm}/\hspace{1.5cm}/\hspace{1.5cm}}\\
\end{flushright}

{\singlespacing
\chapter*{Preface\markboth{\cmarkfont Preface}{\cmarkfont Preface}}}

\noindent
The work presented here comprises of five main chapters (excluding auxiliary materials such as bibliography and appendices):\\
{\bf Chapter~\ref{ch_intro}} is an original literature review on the various topics relating to neutrino properties, baryon asymmetry of the universe and thermal leptogenesis;\\
{\bf Chapter~\ref{ch_work_LV}}, 
{\bf \ref{ch_work_HLV}} and 
{\bf \ref{ch_work_BKL}} 
are based on the work presented in Publications~1, 3 and 2 (see page~\pageref{listofpub}) respectively. They were done in collaboration with N.~F.~Bell (Chapter~\ref{ch_work_BKL}), X.~G.~He (Chapter~\ref{ch_work_HLV}), B.~J.~Kayser (Chapter~\ref{ch_work_BKL}) and R.~R.~Volkas (Chapter~\ref{ch_work_LV} and \ref{ch_work_HLV}). 
Most of the initial ideas for these investigations are due to N.~F.~Bell (Chapter~\ref{ch_work_BKL}) and R.~R.~Volkas (Chapter~\ref{ch_work_LV} and \ref{ch_work_HLV}).
But, all subsequent analyses and methods of presentation are my own work with the exceptions of Sec.~\ref{subsec:HLV_collider} where X.~G.~He has made a significant contribution, and the proof in Sec.~\ref{subsec:BKL_CP_needs_loops} which is due to B.~J.~Kayser;\\
{\bf Chapter~\ref{ch_conc}} is the conclusion.\\


%
%
%
%
%


{\singlespacing
\chapter*{Acknowledgements\markboth{\cmarkfont Acknowledgements}{\cmarkfont Acknowledgements}}
}

\begin{center}
\begin{quote}
\textit{``I do physics not because it, as a subject, is simple, but because the spirit of physics,
that turns the complicated and seemingly unrelated into something that can be described by a common set of rules, makes everything else looks simple.''}
\end{quote}
\end{center}
\vspace{15pt}

\ECap{O}{f}{7}{0}
all the people I wish to thank, perhaps the most important on the list are my parents. Indeed, this journey did not begin four years ago when I entered graduate school; nor ten years ago when I became a freshman at the University of Melbourne; nor fifteen years ago when I touched down in Australia, but twenty-eight years ago when I was given a name which carries the meaning: ``to ascertain knowledge and wisdom''. Although as this journey is coming to a close, it is unclear whether I have achieved what these words may entail, there is no doubt that my parents' unyielding support has been present every bit of the way. So it is fitting that I thank them here first and foremost.

Needless to say, however, there are many others who have played a role (big or small) in helping me get over the line. Although I shall not be able to list them all here due to the limited space available, their contribution is very much appreciated and noted. But for those I shall acknowledge here, let me now mention them below.

It is very rare that a student can succeed without the guidance and endless source of ideas of his/her advisors. Thus, to get this far, I must thank my two advisors, Ray Volkas and Nicole Bell. Ray has been particularly influential not only because he is my primary thesis supervisor, but also because he taught us at the undergraduate level. The latter allowed me to find out what theoretical physics is all about, to know what a role model theorist is like, and above all, to make the easy decision to switch from mathematics to physics. The rest is history. Nicole entered the picture a bit later but the project she suggested formed a pivotal part of my candidature. It must be emphasized that because of them, I had the privilege to collaborate with outstanding scientists, Xiao-Gang He (NTU, Taiwan) and Boris Kayser (Fermilab, USA), to whom I thank very much also.

Next, I would like to thank some of my fellow graduate students in physics (past and present). First of all, Damien George for his numerous ``lessons'' in physics, mathematics and computing over the years (even before we embarked on this PhD path together under Ray), and for the countless discussions we had on miscellaneous topics that have made me a smarter/better person~\footnote{It was during one such discussion when we accidentally discovered that there is a secret message contained in this thesis.}. Secondly, 
Andrew Coulthurst, Catherine Low and Alison Demaria for being great officemates and mentors; and Kristian McDonald, Benedict Carson, Jason Doukas and Jared Cole (quantum computing) for being great mentors, as well as for all the interesting chats. Thirdly, special mentions are due to Paul Fraser (Nuclear/Condense matter theory), Thomas Jacques, Melissa Makin (and Melissa Wals, quantum computing), Swati Singh (astrophysics), Nadine Pesor, Jayne Thompson, Rhys Davies (now at Oxford) and  David Curtin (now at Cornell).

Furthermore, it is a pleasure to thank Bruce~H.~J.~McKellar (emeritus professor) for his expert help and advices; Norm Frankel and Girish Joshi for many interesting conversations; Geoffrey Taylor for being one of my supervisory panel member; Kamala Lekamge, the physics librarian, for countless assistances with the books and journals; Colin Entwisle (tutoring and lab admin) for making 1st year teaching manageable for graduate students like me; Geoff Shute (emeritus and 1st year lab guru) for the numerous conversations, innovative ideas and answering my dumb questions on undergraduate physics; and Wing Hin Lai, my uncle in Singapore, for his ongoing support.

Finally, I would like to thank some of my good friends who have played a part over the years in making who I am: Jeffrey~K.~C.~Chan (grad student in computer science, Melb), Alex~S.~C.~Leong (PhD in Elec.~Eng., Melb), Siegfried~S.~F.~Leung (grad student in chemistry and long time friend, Yale), Roy Whitby (teacher, mentor and friend who has been pivotal to many of my achievements), and last but not least Fido, the black Labrador, who reminded me during my 68 days ``isolation'' from the rest of the world in producing this that life is more than just a thesis.\\

\begin{flushright}
{\small
Sandy~S.~C.~Law\\
December 2008}
\end{flushright}



{\singlespacing  

  \tableofcontents                
\renewcommand{\rightmark}{\cmarkfont \contentsname}
\renewcommand{\leftmark}{\cmarkfont \contentsname}

  \listoffigures                  
   \addcontentsline{toc}{section}{
   List of Figures} 
   
\renewcommand{\rightmark}{\cmarkfont \listfigurename}
\renewcommand{\leftmark}{\cmarkfont \listfigurename}

  \listoftables                  
   \addcontentsline{toc}{section}{
   List of Tables} 
\renewcommand{\rightmark}{\cmarkfont \listtablename}
\renewcommand{\leftmark}{\cmarkfont \listtablename}

  \chapter*{List of Acronyms\markboth{\cmarkfont List of Acronyms}{\cmarkfont List of Acronyms}}
   \addcontentsline{toc}{section}{
   List of Acronyms} 
   
} 



\begin{tabbing}
\hspace{40pt}\=BBN\hspace{80pt}\=Big Bang Nucleosynthesis\\
\>BooNE  \>Booster Neutrino Experiment\\
\>CKM  \>Cabibbo-Kobayashi-Maskawa\\
\>CMB  \>Cosmic Microwave Background\\
\>CUORE  \>Cryogenic Underground Observatory for Rare Events\\
\>CUORICINO  \>``small'' CUORE (test bench for CUORE)\\
\>EMDM  \>ElectroMagnetic Dipole Moment\\
\>GERDA  \>GERmanium Detector Array\\
\>GNO  \>Gallium Neutrino Observatory  \\
\>GUTs \>Grand Unified Theories\\
\>IGEX \>International Germanium EXperiment\\
\>ILC  \>International Linear Collider\\
\>IMB  \>Irvine-Michigan-Brookhaven\\
\>K2K  \>KEK to Super-K (Long Baseline Experiment)\\
\>Kamiokande  \>Kamioka Nucleon Decay Experiment\\
\>KamLAND  \>Kamioka Liquid scintillator AntiNeutrino Detector\\
\>KATRIN  \>KArlsruhe TRItium Neutrino experiment\\
\>KEK  \>{\it Ko Enerugi Kasokuki Kenkyu Kiko} \\
\>    \>({\small High Energy Accelerator Research Organization, Tsukuba})\\ 
\>LH  \>Left-Handed\\
\>LHC \>Large Hadron Collider\\
\>LHS \>Left Hand Side\\
\>LMA \>Large Mixing Angle (MSW solution)\\
\>LSND \>Liquid Scintillator Neutrino Detector\\
\>MiniBooNE  \>(first phase of) BooNE\\
\>MINOS  \>Main Injector Neutrino Oscillation Search\\
\>MSW  \>Mikheyev-Smirnov-Wolfenstein\\
\>PMNS \>Pontecorvo-Maki-Nakagawa-Sakata\\
\>RH \>Right-Handed\\
\>RHS \>Right Hand Side\\
\>SAGE  \>Soviet-American Gallium Experiment\\
\>SDSS  \>Sloan Digital Sky Survey\\
\>SM \>Standard Model (of particle physics)\\
\>SNO  \>Sudbury Neutrino Observatory\\
\>SUSY \>SUperSYmmetry\\
\>VEV  \>Vacuum Expectation Value\\
\>WMAP \>Wilkinson Microwave Anisotropy Probe
\end{tabbing}

\cleardoublepage   

\pagenumbering{arabic}
\setcounter{page}{1}


\renewcommand{\topfraction}{0.9}	
\renewcommand{\bottomfraction}{0.8}	


%
%
%
%
%



\chapter{Introduction}\label{ch_intro}

\ECapC{O}{u~r}{7}{0}
quest to understand how the universe works at the most fundamental levels has been the impetus behind the constant experimental and theoretical efforts in various fields of physics. Over time, there have been many important breakthroughs, but just as many (if not more) puzzling questions remain. Amongst the most intriguing of these is the origin of matter---~the substance which almost everything we see today is made of. Indeed, the standard hot Big Bang model does not discriminate between matter and antimatter as far as their primordial evolution is concerned. Therefore, in order to produce the abundance of matter we observe today, other mechanisms must have been at play in the early universe such that the expected catastrophic annihilation between matter and antimatter was avoided.

In the pursuit of a satisfactory answer for this phenomenon, it is immediately clear that high energy physics will play an essential role since the primordial environment created soon after the Big Bang was very hot and dense. But similar to its cosmological counterpart, the Standard Model of Particle Physics (denoted SM throughout) is incomplete despite the enormous success it has enjoyed over the past decades. The first 
evidence of physics beyond the SM is the observation of neutrino oscillations. Its utmost importance is certainly not restricted to within the realm of precision particle physics, and as we shall explain in the sections to come, far-reaching implications for the cosmological evolution of matter and antimatter can result in certain scenarios.

As a consequence of this potential connection, the studies of neutrino properties and interactions contain added significance. From the theoretical standpoint, this interplay will inevitably lead to constraints in neutrino model building. It is therefore the aim of this work to explore several classes of neutrino models and their possible implications for the origin of matter.

To elucidate all these ideas and build the background for the chapters to come, we begin by briefly reviewing the relevant topics on the theory of neutrinos and cosmic baryon asymmetry generation.

\section{The phenomenon of neutrino oscillations}\label{sec_nu_osc}

\subsection{Experimental evidence}\label{subsec_osc_ex}

The first real indication of the neutrino oscillation phenomenon (i.e. neutrinos can change flavor as they propagate) emerged as early as 1967 when Davis and his collaborators conducted experiments to measure the electron neutrino flux produced from the Sun using a chlorine detector ($\nu_e +{}^{37}$Cl $\rightarrow$ $e^- +{}^{37}$Ar) in the Homestake mine (South Dakota) \cite{RDavis} and found that there was a noticeable deficit in the observed value compared to that predicted by the standard solar model \cite{bahcall}. But at that time the oscillation hypothesis was not taken seriously for it required a very large mixing angle to work, which seemed at odds with the tiny quark mixing. It was not until many years later (in the period 1989 to 1996) that Kamiokande (Kamioka Nucleon Decay Experiment, in Japan)  provided new experimental data to strengthen the case for solar $\nu_e$ disappearance. They employed water $\check{\rm C}$erenkov techniques to investigate elastic scattering events $(\nu e^- \rightarrow \nu e^-)$ and discovered that the measured flux was only about 55\% of the expected size \cite{kam89, kam96}. Later, in 2001, these findings were consolidated in the much larger 50,000 ton experiment of Super-Kamiokande (about 15 times bigger than Kamiokande) \cite{superK01}.

It should be noted that the Kamiokande experiments were only sensitive to the higher energy $^8\text{B}$ solar neutrinos (above 5~MeV), in contrast to those targeted by Davis et al. which included lower energy $^7\text{Be}$ neutrinos. Hence, complementary experiments using radiochemical method with gallium ($\nu_e + {}^{71}$Ga $\rightarrow$ $e^- + {}^{71}$Ge) to target the lowest energy neutrinos from $p+p$ nuclear reaction (as well as $^7\text{Be}$ neutrinos) were conducted by the Gallex/GNO \cite{Gallex} and later the SAGE collaborations \cite{sage}. Both experiments detected a deficit in the solar neutrino flux, and importantly, their results were also consistent with the idea of matter enhanced neutrino oscillations \cite{MSW}.

Further improvement to the solar neutrino analysis was achieved when the SNO (Sudbury Neutrino Observatory, Canada) experiment managed to make reliable measurements of the \emph{total} solar neutrino flux using a heavy water ($\text{D}_2\text{O}$) detector to look for both charged-current induced ($\nu_e + d \rightarrow e^- + p + p$) and neutral-current induced ($\nu + d \rightarrow \nu + p + n$) reactions \cite{SNO}, where $d$ denotes deuteron. This result provided conclusive evidence for solar neutrino oscillation and confirmed the hypothesis at a significant 3.3$\sigma$ level. Other experiments such as KamLAND (in Japan), a long-baseline (source-detector distance $\sim 180$~km) reactor experiment \cite{kamland} looking for disappearance of $\overline{\nu}_e$, and Borexino (in Italy), a real time scintillator based $^7\text{Be}$ low energy neutrino experiment \cite{borexino} also played an important role in authenticating the oscillation picture for solar neutrinos.

An independent set of (perhaps more) compelling experimental evidence for neutrino oscillations came from the studies of atmospheric neutrinos. When primary cosmic rays (usually high energy protons) collide with nuclei in the upper atmosphere, a shower of hadrons, mostly pions (and some kaons) are created. These pions then undergo a chain of reactions (e.g. $\pi^- \rightarrow \overline{\nu}_\mu\,\mu^- \rightarrow \overline{\nu}_\mu  \nu_\mu \nu_e\,e^-$) producing muon and electron fluxes that can be measured. As early as the 1980s, two groups, the IMB (Irvine-Michigan-Brookhaven) collaboration \cite{imb86} and Kamiokande \cite{kam_atm} had found a significant deficit in the muon neutrino flux that was too big to be ascribed to statistical errors. This opened up the interpretation that $\nu_\mu \leftrightarrow \nu_\tau$ oscillation might be at play. Motivated by their initial results, the Kamiokande group expanded their investigation, and in 1998, Super-Kamiokande reported a $\nu_\mu$-$\nu_e$ ratio and asymmetry in the zenith angle distribution of $\mu$-type events that were very much consistent with the atmospheric neutrino oscillation picture \cite{kam_atm98}. More recently, experiments with accelerator $\nu_\mu$'s such as K2K in Japan \cite{K2K} and MINOS in USA \cite{MINOS} have further confirmed this interpretation.

\subsection{Theoretical formulation}\label{subsec_osc_th}

The experimental observation that neutrinos can oscillate from one flavor to another as they propagate is the strongest indication (barring some by now strongly disfavored exotic possibilities) for nonzero neutrino masses and mixing. To see this, consider the case of three neutrino species. We define the neutrino weak eigenstate $\nu_\alpha$ with flavor $\alpha$ (where $\alpha = e, \mu, \text{ or }\tau$) such that it is produced in association with the charged antilepton
$\overline{\ell}_\alpha$ in a tree-level interaction with the $W$ boson. These weak eigenstates are in general different from the neutrino mass eigenstates $\nu_i$ (with $i = 1,2,\text{ and } 3$), each having a (rest) mass given by $m_i$. One can relate the weak and mass eigenstates via a unitary transformation\footnote{Unitarity will guarantee that charged-current interactions of $\nu_\alpha$ always produce a charged lepton with the same flavor $\alpha$.}, and write $\nu_\alpha$ as a coherent superposition of the $\nu_i$ fields:
\begin{equation}\label{eq1:osc_th_super}
 \ket{\nu_\alpha} = \sum_i\, U_{\alpha i}^* \,\ket{\nu_i}\;,
\end{equation}
where the nine quantities $U_{\alpha i}$ make up the unitary matrix of basis transformation. The meaning of leptonic mixing can therefore be understood as the fact that in a charged-current interaction, the neutrino mass eigenstate which accompanies the production of (or otherwise involves) $\overline{\ell}_\alpha$ may not always be the same $\nu_i$. 

When we are working in the charged lepton mass eigenbasis, this unitary matrix \emph{is} the leptonic mixing matrix, also known as the Pontecorvo-Maki-Nakagawa-Sakata (PMNS) matrix \cite{pmns}, which is often parametrized as
\begin{equation}\label{eq1:UPMNS_param}
 \Upmns =
 \ththMat{c_{12}c_{13}}{s_{12}c_{13}}{s_{13}\,e^{-i\delta}}
 {-s_{12}c_{23}-c_{12}s_{23}s_{13}\,e^{i\delta}}
 {c_{12}c_{23}-s_{12}s_{23}s_{13}\,e^{i\delta}}
 {s_{23}c_{13}}
 {-s_{12}s_{23}+c_{12}c_{23}s_{13}\,e^{i\delta}}
 {c_{12}s_{23}+s_{12}c_{23}s_{13}\,e^{i\delta}}
 {-c_{23}c_{13}} 
 \hspace{-5pt}
 \Matdiag{e^{i\alpha_1/2}}{e^{i\alpha_2/2}}{1}, 
\end{equation}
where $s_{mn}=\sin\theta_{mn}, c_{mn}=\cos\theta_{mn}$, $\delta$ is the $CP$ violating Dirac phase, while $\alpha_1$ and $\alpha_2$ denote the two Majorana phases \footnote{Neutrinos can be Majorana particles.}.

To quantify the phenomenon of a neutrino changing from flavor-$\alpha$ to flavor-$\beta$ as it propagates in vacuum \footnote{In this discussion, we will ignore neutrino oscillations in matter (the MSW effect) \cite{MSW}.}, we are interested in the probability with which this happens, i.e. $\text{Pr}\left(\nu_\alpha \rightarrow \nu_\beta\right)$, a quantity that depends on how the \ket{\nu_\alpha} state in (\ref{eq1:osc_th_super}) evolves with time.
Because $\nu_\alpha$ is a superposition of the mass eigenfields $\nu_i$ whose quantum mechanical evolution may be different, we must solve each $\ket{\nu_i(t)}$ state individually and then coherently sum them up. Working in the rest frame of $\nu_i$, the state vector at the proper time $\tau_i$ will satisfy the Schr\"odinger equation:
\begin{equation}\label{eq1:osc_SEqn}
 i\,\frac{\partial}{\partial \tau_i}\; \ket{\nu_i(\tau_i)} = m_i\, \ket{\nu_i(\tau_i)}\;.
\end{equation}
The solution to Eq.~(\ref{eq1:osc_SEqn}) is simply given by:
\begin{equation}\label{eq1:osc_SEqn_sol}
 \ket{\nu_i(\tau_i)} = e^{-i\,m_i \tau_i} \ket{\nu_i(0)}\;,
\end{equation}
where \ket{\nu_i(0)} is the original state before propagation. By Lorentz invariance, we can express $m_i \tau_i$ in terms of laboratory variables:
\begin{equation}\label{eq1:osc_lorentz}
 m_i \,\tau_i = E_i \,t - |p_i| \,L\;,
\end{equation}
where $E_i$ and $|p_i|$ are respectively, the energy and momentum of the $\nu_i$ component of the neutrino, while $L$ and $t$ are the source-detector distance and the time it takes to traverse $L$, as measured in the lab-frame \footnote{In practice, $t$ is not measured.}. For neutrino beams used in experiments, it can be shown that only components of the beam with the \emph{same energy} will contribute coherently to the oscillation signal \cite{boris_slac4}. Therefore, we can replace $E_i$ with $E$, and momentum $|p_i|$ is now given by
\begin{equation}\label{eq1:osc_pi}
 |p_i| = \sqrt{E^2-m_i^2}\;\simeq \;E -  \frac{m_i^2}{2E}\;,
\end{equation}
where in the last step, we have assumed $m_i \ll E$. With this approximation for $|p_i|$ and applying relation (\ref{eq1:osc_lorentz}), solution (\ref{eq1:osc_SEqn_sol}) becomes
\begin{align}
 \ket{\nu_i(t)} &\simeq e^{-i\,\left[E(t-L) + m_i^2 L / 2E\right]}\; \ket{\nu_i(0)}\;,
    \nonumber\\
 &=
   e^{-i\,E(t-L)} \,e^{-i\,m_i^2 L / 2E}\; \ket{\nu_i(0)}\;.\label{eq1:osc_sol_t}
\end{align}
Since the phase factor $e^{-i\,E(t-L)}$ is the same for all $i$, it may be neglected in (\ref{eq1:osc_sol_t}). Thus, the amplitude for a neutrino $\nu_\alpha$ of energy $E$ changing to $\nu_\beta$ after propagating a distance $L$ through vacuum is
\begin{equation}\label{eq1:osc_amp}
 \bra{\nu_\beta}\ket{\nu_\alpha}
  \simeq \sum_i\, U_{\alpha i}^*\, e^{-i\,m_i^2 L / 2E}\, U_{\beta i}\;.
\end{equation}
Squaring (\ref{eq1:osc_amp}) and after some algebra, we deduce the corresponding probability as:
\begin{align}
 \text{Pr}(\nu_\alpha\rightarrow\nu_\beta)
 \equiv
 |\bra{\nu_\beta}\ket{\nu_\alpha}|^2
 &=
  \delta_{\alpha\beta}
   -4\,\sum_{i>j}\,\text{Re}\left(U_{\alpha i}^*U_{\beta i}U_{\alpha j}U_{\beta j}^*\right)
   \,\sin^2\left[\frac{\Delta m^2_{ij} \,L}{4E}\right] \nonumber\\
 &\qquad
  +2\,\sum_{i>j}\,\text{Im}\left(U_{\alpha i}^*U_{\beta i}U_{\alpha j}U_{\beta j}^*\right)
   \,\sin \left[\frac{\Delta m^2_{ij} \,L}{2E}\right]
  \;, \label{eq1:osc_pr}
\end{align}
where $\Delta m^2_{ij} \equiv m_i^2-m_j^2$.

From this result, it is quite clear that when all neutrino masses $m_i$'s are zero (or nonzero but degenerate) and hence, the second and third term in Eq.~(\ref{eq1:osc_pr}) disappear, neutrino oscillation is not possible. By the same token, the observation that $\nu_e$ and $\nu_\mu$ do change flavor during propagation implies that (at least two of) $\nu_i$'s must be massive.

\subsection{Oscillation parameters}\label{subsec_osc_param}

One can combine the plethora of neutrino oscillation data gathered to date from the various solar, atmospheric, reactor and accelerator experiments (see Sec.~\ref{subsec_osc_ex}) \footnote{Here, we shall ignore the LSND anomaly \cite{LSND} because an oscillation explanation has been ruled out by MiniBooNE \cite{MiniBooNE}.} and place constraints on the relevant parameters shown on the right-hand side (RHS) of Eq.~(\ref{eq1:osc_pr}). For the three-flavor neutrino oscillation picture and invoking the parametrisation of (\ref{eq1:UPMNS_param}), one can see that variables that are potentially relevant in controlling the effects of oscillation are the three mixing angles ($\theta_{12},\theta_{23}$ and $\theta_{13}$), the $CP$ violating Dirac phase ($\delta$), the two Majorana phases ($\alpha_1$  and $\alpha_2$), and two squared-mass differences (e.g. $\Delta m_{21}^{2}$ and $\Delta m_{31}^{2}$).

It turns out that oscillation probabilities are insensitive to the Majorana phases while the null result from the short-baseline ($L \sim 1$km) reactor experiment CHOOZ (in France) \cite{chooz} \footnote{This experiment looked for the disappearance of $\overline{\nu}_e$ at $E \sim$ 3~GeV, which is sensitive to the large squared-mass gap $\Delta m_{31}^{2}$, thus allowing the properties of $\nu_3$ to be probed.} suggests that $\theta_{13}$ must be quite small, consequently making the observation of the Dirac phase rather difficult. The reason for this latter point is that in (\ref{eq1:UPMNS_param}), the Dirac phase always appears as the combination: $\sin \theta_{13}\,e^{-i\,\delta}$, which means a sufficiently small $\theta_{13}$ can always mask any $CP$ violating effects 
\footnote{It must be emphasized that the choice of associating $\theta_{13}$ with $\delta$ in the parametrization of (\ref{eq1:UPMNS_param}) is not vital to our conclusion here. As long as there is at least one mixing angle which is very small, one can always re-phase the mixing matrix so that $\delta$ is associated with the term containing the small angle. Our choice of parametrization is naturally motivated by the CHOOZ result which indicates a tiny $\theta_{13}$.}.

Overall, neutrino experiments have so far determined that there are two large  ($\theta_{12}, \theta_{23}$) and one small ($\theta_{13}$) mixing angles \footnote{This is in stark contrast to the quark sector where the CKM matrix \cite{CKM} has three small mixing angles.}, as well as, a ``small'' ($\Delta\msol^2$) and a ``large'' ($\Delta\matm^2$) squared-mass differences which drive the solar and atmospheric neutrino oscillations respectively. Because the sign of $\Delta\matm^2$ is not known, two arrangements for the neutrino mass spectrum are possible: with \emph{normal} hierarchy, we identify $\Delta\matm^2 \equiv \Delta m_{31}^2 > 0$, which gives $ m_1 < m_2 < m_3$ with
\begin{equation}
 m_2 = \sqrt{m_1^2 +\Delta \msol^2}\;,\qquad 
 m_3 = \sqrt{m_1^2 +\Delta \matm^2}\;, \label{eq1:NH_m2_m3}
\end{equation}
and for \emph{inverted} hierarchy, we take $\Delta\matm^2 \equiv \Delta m_{32}^2 < 0$ instead, implying $m_3 < m_1 < m_2$ with
\begin{equation}
 m_1 = \sqrt{m_3^2 +\Delta \matm^2-\Delta \msol^2}\;,\qquad 
 m_2 = \sqrt{m_1^2 +\Delta \msol^2}\;. \label{eq1:IH_m1_m2}
\end{equation}
Note that in both cases, we have used $\Delta \msol^2 \equiv \Delta m_{21}^2 > 0$ (as implied by the matter-affected solar neutrino oscillations).  
The best-fit values at $1 \sigma$ error level for these neutrino oscillation parameters in the three-flavor framework are summarised as follows \cite{nu_best-fit}:
\begin{align}
 \sin^2 \theta_{12} = 0.304^{+0.022}_{-0.016}\;,\quad
 \sin^2 \theta_{23} = 0.50^{+0.07}_{-0.06}\;,\quad
 \sin^2 \theta_{13} \leq 0.01^{+0.016}_{-0.011}\;, \label{eq1:best-fit_angle}\\
 \Delta\msol^2 = 7.65^{+0.23}_{-0.20} \times 10^{-5} \text{ eV}^2 \;,\quad
 |\Delta\matm^2| = 2.40^{+0.12}_{-0.11} \times 10^{-3} \text{ eV}^2 \;.
 \label{eq1:best-fit_mass}
\end{align}
In this work, we shall take these values as standard inputs for many of our analyses.

\subsection{Tribimaximal mixing}\label{subsec_osc_TB}

An interesting observation from the best-fit parameter values shown in (\ref{eq1:best-fit_angle}) is that $\theta_{23}$ and $\theta_{13}$ correspond to almost maximal and no mixing respectively. This prompts the question as to whether Nature selects these extreme values at random or there is a more profound reason. If one assumes that the maximal and no mixing is exact (i.e.  $\theta_{23} = \pi/4$ and $\theta_{13}=0$), then up to the Majorana phases, the mixing matrix (\ref{eq1:UPMNS_param}) becomes
\begin{equation}\label{eq1:osc_BiLarge}
 U_\text{BL} = 
 \ththMat
 {\cos\theta_{12}}{\sin\theta_{12}}{0}
 {-\frac{\sin\theta_{12}}{\sqrt{2}}}{\frac{\cos\theta_{12}}{\sqrt{2}}}{\frac{1}{\sqrt{2}}}
 {-\frac{\sin\theta_{12}}{\sqrt{2}}}{\frac{\cos\theta_{12}}{\sqrt{2}}}{-\frac{1}{\sqrt{2}}}
 \;,
\end{equation}
which is often referred to as ``bi-large'' mixing. A particularly beautiful aspect of $U_\text{BL}$ is that any neutrino mass matrix in the flavor (i.e. weak interaction) basis which has a $\mu$-$\tau$ symmetry \footnote{In other words, exchanging the $\mu$ and $\tau$ columns followed by the $\mu$ and $\tau$ rows will leave the mass matrix unchanged.} can be diagonalised by it if one chooses $\theta_{12}$ carefully. Moreover, it can be shown that the specific value needed for $\theta_{12}$ to diagonalise these types of mass matrices is \emph{independent} of the mass eigenvalues. This suggests that for certain neutrino mass matrix structures, mixing angles and masses are not necessarily interconnected even though both of them are related to the same source. In fact, this feature is quite common for mass matrices in models with an underlying \emph{flavor symmetry} (or family symmetry \footnote{We shall use both terms interchangeably through out this work.}). The class of mass matrices which has a diagonalisation matrix that is independent of any of the parameters in the original matrix or its eigenvalues has been dubbed ``form-diagonalisable'' \cite{Low01+thesis}. We shall make use of this idea quite often in our neutrino model building in the chapters to come.

Another important mixing structure which is related to $U_\text{BL}$ is the so-called ``tribimaximal'' mixing scheme \cite{TBmixing}. Noting that the best-fit value for $\theta_{12}$ is very well approximated by the relation: $\sin^2 \theta_{12} = 1/3$, we substitute it into $U_\text{BL}$ and the result is the tribimaximal mixing matrix \footnote{Tri-maximal and bi-maximal mixing in the 2nd and 3rd columns respectively.}:
\begin{equation}\label{eq1:osc_TriBi}
 U_\text{TB} =
  \ththMat
  {\sqrt{\frac{2}{3}}}{\frac{1}{\sqrt{3}}}{0}
  {-\frac{1}{\sqrt{6}}}{\frac{1}{\sqrt{3}}}{\frac{1}{\sqrt{2}}}
  {-\frac{1}{\sqrt{6}}}{\frac{1}{\sqrt{3}}}{-\frac{1}{\sqrt{2}}}
  \;.
\end{equation}
This neutrino mixing structure has attracted ample attention over the years with many model builders attempting to reconstruct it via symmetries and auxiliary fields. Most notable examples are models with discrete (e.g. $A_4,\Delta_{27}$) and continuous (e.g. $SO(3), SU(3)$) family symmetries. Some recent examples can be found in \cite{TB_a4_list, TB_other_list}. It should be noted though that many of these are often quite intricate and require additional constraints or a non-trivial Higgs sector for them to be viable, further highlighting the difficulties in explaining the peculiar mixing pattern of neutrinos. 

Nevertheless, the tribimaximal structure itself presents a relatively simple manifestation of the neutrino mixing matrix that is more or less consistent with current experimental bounds. In the light of this, it is theoretically appealing to take $U_\text{TB}$ (with or without the Majorana phases) as the starting point in any model building or analyses involving the neutrino mass matrix. Our work here in the future chapters will be no exception.

\section{The case of massive neutrinos}\label{sec_case_numass}

\subsection{Absolute scale for neutrino masses}\label{subsec_abs_numass}

Because oscillation experiments have strongly indicated that at least some neutrinos must be massive, it is natural to ask the scale at which these masses are situated. In the discussion of oscillation parameters in Sec.~\ref{subsec_osc_param}, we have learnt that those experiments can help us determine the splitting between the neutrino mass eigenstates, leading to Eqs.~(\ref{eq1:NH_m2_m3}) and (\ref{eq1:IH_m1_m2}). However, it is clear that those relations do not tell us the overall scale of masses, and therefore other methods must be used to probe for a better understanding of the neutrino mass spectrum.
 
One way to obtain meaningful bounds on the absolute scale for the neutrinos is to look for kinematic effects due to their nonzero masses in tritium $\beta$-decay (${}^3$H $\rightarrow {}^3\text{He} + \overline{\nu}_e + e^-$). By studying the resultant electron energy spectrum, one can probe the quantity:
\begin{equation}\label{eq1:numass_beta}
 m_{\beta} \equiv \sqrt{\sum_{i=1}^{3} |U_{ei}|^2\, m_i^2} \;.
\end{equation}
The value of this then gives an upper limit on the absolute neutrino mass scale. Two groups, Mainz \cite{mainz} and Troitsk \cite{troitsk}, have reported bounds of $m_\nu < 2.3$~eV and $m_\nu < 2.5$~eV respectively. An upcoming experiment, KATRIN \cite{katrin}, is expected to have a sensitivity down to about 0.3~eV, which will further narrow down the scale of the neutrino spectrum. It should be remarked that whether neutrinos are of Dirac or Majorana type \footnote{See the next two subsections for more discussions on Dirac and Majorana neutrinos.} has no bearing on the capability of this method of ``directly'' searching for their masses.

Another possible way to probe the neutrino mass scale is via studies of lepton number ($L$) violating neutrinoless double $\beta$-decay 
($^{A}_{Z}\left[\text{Nucl}\right] 
\rightarrow \;^{\;\;\;\;A}_{Z+2}\left[\text{Nucl}'\right] + 2 e^-$),
whose observation would imply that neutrinos are Majorana fermions \cite{bb_Maj_test}. Assuming that there are no new $L$ violating interactions at play, one can conclude that the amplitude for this decay is proportional to the so-called effective Majorana neutrino mass:
\begin{equation}\label{eq1:numass_0vbb}
 m_{\beta\beta} \equiv \left|\sum_{i=1}^{3} U_{ei}^2\, m_i \right| \;.
\end{equation}
Several groups such as the Heidelberg-Moscow \cite{heidelberg-moscow} and IGEX \cite{IGEX} collaborations
conducted experiments with $^{76}\text{Ge}$, while the more recent CUORICINO experiment \cite{CUORICINO} used $^{130}\text{Te}$ to test for this. So far there are no confirmed discoveries of the neutrinoless double $\beta$-decay 
\footnote{There is actually a claim by a subset of the Heidelberg-Moscow group that neutrinoless double $\beta$-decay was observed \cite{HM_bb_claim}. For a more thorough discussion on this controversial finding, see \cite{HM_bb_discuss}.},
but the best upper bounds on the decay lifetimes are presently provided by CUORICINO (which is still running), whose results are translated to~\footnote{The large range is due to the uncertainty in the nuclear matrix elements.}
\begin{equation}\label{eq1:numass_CUOR}
 m_\nu \equiv m_{\beta\beta} < 0.19 -0.68 \text{ eV } \;(90\% \text{ C.L.}) \;,
\end{equation}
for the neutrino mass. Relation (\ref{eq1:numass_CUOR}) is a much tighter bound than the one obtained from tritium $\beta$-decay. Upcoming experiments like CUORE \cite{CUORE}, GERDA \cite{GERDA} and \textsl{Majorana} \cite{MAJORANA} are expected to further improve these results with projected sensitivity of about 0.05~eV.

\subsubsection{Neutrino masses from cosmology}

Finally, it must be mentioned that some of the strongest bounds on the overall scale for neutrino masses come from cosmology. This is one of the important examples that illustrates the intricate connections between neutrino physics and the evolution of the early universe.
During the epoch of structure formation, free-streaming neutrinos with a sizable mass can have significant effects on the growth of structure and hence on the eventual
galaxy power spectrum. Therefore, an accurate measurement of it can help set limits on the neutrino mass scale given the standard theory of structure formation. The studies of the data from the Wilkinson Microwave Anisotropy Probe (WMAP) and the Sloan Digital Sky Survey (SDSS) have deduced that the sum of neutrino masses (three species assumed) is constrained by $\sum_i \,|m_i| \lesssim 0.6$ \cite{wmap_numass} and $1.6$~eV \cite{sdss_numass} respectively at a confidence level of about 95\%. Since the observed squared-mass splittings ($\Delta\msol^2, \Delta\matm^2$) imply that $|m_i-m_j|\ll \order{0.1}$~eV for all $i$ and $j$, taking $\sum_i \,|m_i| \lesssim 0.6$
gives an absolute upper bound for each individual neutrino mass of about
\begin{equation}\label{eq1:numass_cosmo}
 |m_i| \;\lesssim\; 0.2 \text{ eV }\;\; (95\% \text{ C.L.})\quad \text{for all } i\;.
\end{equation}
This estimation is similar to the least upper bound imposed by the CUORICINO experiment, which further confirms that the absolute neutrino mass scale must be in the sub-eV range.

The immediate impact of these findings is that it raises the question as to why the neutrino is so light compared to other SM particles (e.g. $m_\nu/m_e \sim 10^{-6}$), an issue we will return to in Secs.~\ref{subsec:seesaw1} to \ref{subsec:radcorr}.

\subsection{Neutrino mass terms}\label{subsec:massterms}

The minimal Standard Model (SM) does not include neutrino masses. But as we have discussed, the existence of massive neutrinos is the most natural explanation for the   oscillation experiments.
Therefore, it is imperative to modify the SM to include this fact. The only question is: how to do this?

Since only chirally left-handed (LH) fermion fields participate in weak interactions, it makes sense to express the SM fields in terms of their chiral components. To this end, we can define the following two-component Weyl spinors:
\begin{equation}\label{equ1:weyl_def}
 \psi_{L,R} \equiv P_{L,R}\;\psi = \frac{1\mp \gamma^5}{2}\;\psi \;,
\end{equation}
where $\psi$ is a four-component Dirac spinor. Because a standard
mass term for fermions has the form given by $m\overline{\psi}\psi$, when expressed in $\psi_{L,R}$ one finds that only fields of opposite chirality couple together. Consequently, two types of mass term are possible:
\begin{equation}\label{eq1:dirac_term}
 m_\mathcal{D}\, \overline{\psi_L} \,\psi_R + \text{h.c.}\;,
\end{equation}
which is known as the Dirac mass term, and
\begin{equation}\label{eq1:maj_term}
  M_\mathcal{L}\, \overline{\psi_L} \,(\psi_L)^c 
 +M_\mathcal{R}\,\overline{(\psi_R)^c} \,\psi_R
 + \text{h.c.}\;,
\end{equation}
which are called Majorana mass terms. The charge conjugate field is defined by
$\psi^c \equiv C\, \overline{\psi}^T$ where $C$ denotes the charge conjugation matrix. 
These mass terms are so named because (\ref{eq1:dirac_term}) arises from the coupling of  Dirac type fields: $\psi = e^{i\varphi_1}\psi_L +e^{i\varphi_2}\psi_R$, whereas the terms in 
(\ref{eq1:maj_term}) come from coupling of Majorana type fields: 
$\Psi_{a,b}= e^{i\varphi_1}\psi_{L,R} +e^{i\varphi_2}(\psi_{L,R})^c$, which satisfy the Majorana condition:
\begin{equation}\label{eq1:nu_maj_cond}
 (\Psi_{a,b})^c = e^{-i\,(\varphi_1+\varphi_2)}\, \Psi_{a,b}\;,
\end{equation}
where $\varphi_{1,2}$ are constants \footnote{In (\ref{eq1:dirac_term}) and (\ref{eq1:maj_term}), overall phases involving $\varphi_{1,2}$ are absorbed into the definitions of $m_\mathcal{D}$ and $M_{\mathcal{L},\mathcal{R}}$.}. The important implication of (\ref{eq1:nu_maj_cond}) is that $\Psi_{a,b}$ are identical to their antiparticles.

At this point it is worth emphasizing that because of electric charge conservation, the existence of Majorana type masses for all charged fermions in the SM is forbidden. But since neutrinos are electrically neutral, all mass terms in (\ref{eq1:dirac_term}) and (\ref{eq1:maj_term}) may be relevant, and as a result, one can have either massive Dirac or Majorana neutrinos.

For massive Dirac neutrinos, it is obvious from (\ref{eq1:dirac_term}) that two different Weyl fields are required to construct the mass term. So, in addition to the $\nu_L$ as seen in weak interactions, one must introduce a chirally right-handed (RH) neutrino field, $\nu_R$, to the SM. These RH neutrinos are weak isospin singlets and can therefore couple to the $\nu_L$ and $\phi$, the SM Higgs doublet, to form the Yukawa term:
\begin{equation}\label{eq1:nu_Yukawa_term}
 Y_\nu \; \overline{\ell}_L \,\phi \,\nu_R + \text{h.c.}\;,
\end{equation}
where $\ell_L = (e_L, \nu_L)^T$ is a doublet of $SU(2)_L$.
When the neutral component of the Higgs field $\phi$ acquires a vacuum expectation value, (\ref{eq1:nu_Yukawa_term}) will induce a Dirac mass term for the neutrino. This is the same mechanism through which all quarks and charged leptons get their masses. However, in order to explain why neutrino is much lighter than other fermions, one has to impose a hierarchy in the Yukawas: $Y_\nu \ll Y_e$, making it seem rather ad-hoc. We shall discuss some possible resolutions to this later.

Turning to the case of massive Majorana neutrinos, it is clear from (\ref{eq1:maj_term}) that only one type of Weyl field is required in principle. Hence, given $\nu_L$, one can already construct a Majorana mass term for the neutrino. But since $\overline{\ell}_L\,\ell_L^c$ is a weak isospin triplet, within the framework of the SM, the simplest possibility is therefore the dimension-5 mass term of the form \cite{numass_dim5_op}:
\begin{equation}\label{eq1:nu_dim5_term}
 \frac{y^2}{\Lambda}\, \overline{\ell}_L\,\phi\, \phi^T\ell_L^c +\text{h.c.}\;,
\end{equation}
where $y$ denotes some dimensionless coupling constant and $\Lambda$ is the high-energy cutoff scale above which this non-renormalisable interaction is no longer valid. After spontaneous symmetry breaking, the term in (\ref{eq1:nu_dim5_term}) will induce an effective Majorana mass for the neutrino:
\begin{equation}\label{eq1:numass_eff}
 m_\nu = \frac{y^2 \vev{\phi^0}^2}{\Lambda}\;.
\end{equation}
It is interesting to note here that if $\Lambda \gg \vev{\phi^0}$, then expression (\ref{eq1:numass_eff}) suggests that neutrinos can naturally have a much smaller mass than other fermions without the need to fine-tune $y$ with the corresponding Yukawa coupling of the charged leptons. Given this potential benefit in models with Majorana neutrino masses, and the fact that no SM symmetry forbids their inclusion, it seems natural that neutrinos ought to be Majorana particles. Consequently, we shall assume throughout this work that neutrinos are Majorana unless otherwise indicated.

Returning to the initial problem of how to modify the SM to incorporate neutrino masses, we see that both the Dirac mass term of (\ref{eq1:nu_Yukawa_term}) and the non-renormalisable term of (\ref{eq1:nu_dim5_term}) imply that some new physics must be introduced. While (\ref{eq1:nu_Yukawa_term}) demands the addition of RH singlet neutrinos, (\ref{eq1:nu_dim5_term}) can lead to many possibilities. So, it is the subject of the next few subsections to explore several possible scenarios which can achieve this within framework of renormalisable interactions.

\subsection{Seesaw mechanism: type I}\label{subsec:seesaw1}

Within the class of renormalisable models that can give rise to the effective interaction of form (\ref{eq1:nu_dim5_term}), the type~I seesaw mechanism \cite{type1_seesaw} is perhaps the most elegant solution of all. Not only can it provide a way to generate tiny but nonzero neutrino masses, it also contains all the necessary ingredients for explaining the cosmic baryon asymmetry, as will be discussed in more detail later in Sec.~\ref{sec:lepto_main}.

\begin{figure}[tb]
\begin{center}
 \includegraphics[width=0.40\textwidth]{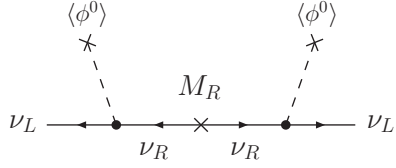}
\end{center} 
 \caption{Diagram representing the type I seesaw realisation of the small Majorana mass for the LH neutrino with $m_\nu \simeq \vev{\phi^0}^2\, Y_\nu\,M_R^{-1}\,Y_\nu^T$ (up to a multiplicative constant).}
 \label{fig1:seesaw1}
\end{figure}

The idea of the type I seesaw model is quite simple. It springs from the inclusion of  heavy neutral singlet fermions, the RH neutrinos (one for each generation of light neutrino) in the SM, just as one would require to have a Dirac neutrino mass term. But since the RH neutrinos are electroweak singlets, one can also include a Majorana mass term for them. As a result of having \emph{both} Dirac and Majorana mass terms, one gets the following  Lagrangian which is SM gauge invariant:
\begin{equation}\label{eq1:L_mass_type1} 
 -\mathcal{L}_\text{type-I} =
  Y_\nu \, \overline{\ell}_L\, \phi\, \nu_R
   + \frac{M_R}{2}\, \overline{(\nu_R)^c}\,\nu_R +\text{h.c.}\;,
\end{equation}
where $M_R$ denotes the bare mass for the RH neutrino. Since the SM does not predict or restrict the size of $M_R$, we may assume that it is arbitrarily large. So, by integrating out the heavy RH neutrino field, one gets an effective Lagrangian that is of the same form as (\ref{eq1:nu_dim5_term}):
\begin{equation}\label{eq1:L_eff_type1}
 \mathcal{L}_\text{eff}^{\rm I} =  \frac{Y_\nu^2}{2 M_R}\, 
 \overline{\ell}_L\, \phi\,\phi^T\ell_L^c\;,
\end{equation}
if we identify $y$ and $\Lambda$ of Eq.~(\ref{eq1:nu_dim5_term}) with  $Y_\nu$ and $M_R$ respectively. The interaction of  (\ref{eq1:L_eff_type1}) will naturally give rise to a very small Majorana neutrino mass: 
$m_\nu \simeq \vev{\phi^0}^2\, Y_\nu\,M_R^{-1}\,Y_\nu^T$
if one assumes that $M_R \gg \vev{\phi}$. This situation is illustrated in Fig.~\ref{fig1:seesaw1}.

Equivalently, one can obtain this by writing (\ref{eq1:L_mass_type1}) in matrix form and subsequently block-diagonalising the resultant $2n\times 2n$ neutrino matrix, where $n$ is the number of generations. To elucidate this, we begin by rewriting the Lagrangian (\ref{eq1:L_mass_type1}) in index form with the generation indices $i,j$ included (ignoring the charged lepton terms):
\begin{equation}\label{eq1:L_nu_index}
   -2\mathcal{L}_{\nu} =
   \sum_{i,j}^n\,
  \overline{\nu}_{iL}\,(m_D)_{ij}\,\nu_{jR}
  + \overline{(\nu_{jR})^c}\,(m_D)_{ji}\,\nu_{iL}^c
   + \overline{(\nu_{Ri})^c}\,(M_R)_{ij}\,\nu_{jR} +\text{h.c.}\;,
\end{equation}
where $m_D \equiv Y_\nu\vev{\phi^0}$ and we have used the fact: $\overline{\nu_L}\nu_R \equiv \overline{(\nu_R)^c} (\nu_L)^c$. Then, (\ref{eq1:L_nu_index}) can be re-packaged into:

\begin{equation}\label{eq1:L_nu_matrix}
 -2\mathcal{L}_{\nu} = 
 \tworMat{\overline{\nu}_L}{\overline{(\nu_R)^c}}
 \underbrace{\twtwMat{0}{m_D}{m_D^T}{M_R}}_{M}
 \twocMat{(\nu_L)^c}{\nu_R} +\text{h.c.}\;,
\end{equation}
where $\nu_L \equiv (\nu_{1L}, \ldots, \nu_{nL})^T$ and 
$\nu_R \equiv (\nu_{1R}, \ldots, \nu_{nR})^T$. Using the key assumption that all eigenvalues of $m_D$ are much less than those of $M_R$, the neutrino mass matrix in (\ref{eq1:L_nu_matrix}) may be block-diagonalised (to first order in $M_R^{-1}m_D^T$) by:
\begin{align}
D_\nu &= V \,M\, V^T\;, \nonumber\\ 
      &\simeq \twtwMat{m_D\, M_R^{-1}\, m_D^T}{0}
                 {0}{M_R}\;, \label{eq1:L_nu_Mdiag} \\
\intertext{where}
V &= \twtwMat{I}{(M_R^{-1}\, m_D^T)^\dagger}{-M_R^{-1}\, m_D^T}{I}
  \twtwMat{i\,I}{0}{0}{I}\;. \label{eq1:L_nu_V}
\end{align}
The second matrix in the definition of (\ref{eq1:L_nu_V}) is included to ensure all mass eigenvalues are positive. If we let $(\nu' , N')^T = (\nu_L , \nu_R^c)^T V^T$ and define a set of Majorana fields:
\begin{equation}\label{eq1:L_nu_Maj_field}
 \twocMat{\nu}{N} \equiv \twocMat{\nu'+(\nu')^c}{N' + (N')^c}\;,
\end{equation}
then Lagrangian (\ref{eq1:L_nu_matrix}) becomes
\begin{align}
 -2\mathcal{L}_{\nu} &= 
 (m_D\,M_R^{-1}\,m_D^T)\,\overline{\nu}\,\nu  
 + M_R\,\overline{N}\,N
 +\text{h.c.}\;,\nonumber\\
 &= 
 m_\nu \,\overline{\nu}\,\nu  
 + M_R\,\overline{N}\,N
 +\text{h.c.}\;.
 \label{eq1:L_nu_Maj_Lag}
\end{align}
So from this, it is easy to see that this model gives rise to two sets of Majorana neutrinos: the light ones ($\nu$) with mass matrix $m_\nu \simeq m_D\,M_R^{-1}\,m_D^T \equiv \vev{\phi^0}^2\, Y_\nu\,M_R^{-1}\,Y_\nu^T$, and the heavy ones ($N$) with $M_N \simeq M_R$. A particularly attractive feature of this is that the smallness of $m_\nu$ is a direct consequence of the large mass scale of $M_R$, which may have its origin from higher unification theories. Furthermore, the addition of RH neutrinos per se seems quite natural from the point of view of making the SM more ``left-right'' (and ``quark-lepton'') symmetric, in the sense that for each LH fermion, there is a RH version. Owing to these benefits, the type I seesaw framework will form the central theme of all of our subsequent investigations in this present work.

\subsection{Other seesaw models}\label{subsec:seesaw23}

Besides type I seesaw, there exist other extensions to the SM  which can lead to the effective Majorana mass term in (\ref{eq1:nu_dim5_term}). These seesaw-like models provide alternative ways to understand the smallness of neutrino mass \cite{Ma_pathway2_small_nu}, and hence are of great interest to many model builders \footnote{Small neutrino masses can also be generated without seesaw in models with extra dimensions \cite{numass_exdim}.}. So, for completeness, we will briefly review them here despite the fact that they play no direct role in our current work.

\subsubsection{Type II seesaw}

\begin{figure}[tbh]
\begin{center}
 \includegraphics[width=0.85\textwidth]{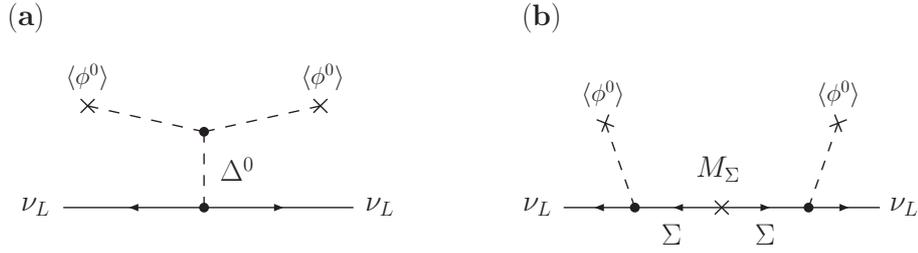}
\end{center} 
 \caption{{\bf (a)} The process induced by the type II seesaw Lagrangian that will give rise to small neutrino Majorana masses. 
 {\bf (b)}~The corresponding process in the type III seesaw case with heavy triplet fermion $\Sigma$ instead.}
 \label{fig1:seesaw23}
\end{figure}

Instead of extending the SM by adding heavy singlet fermions as in Sec.~\ref{subsec:seesaw1}, one can make use of the fact that $\overline{\ell}_L\,\ell_L^c$ is an $SU(2)_L$ triplet and introduce a heavy \emph{triplet scalar} to the Higgs sector \cite{type2_seesaw}, so that a gauge invariant and renormalisable $\overline{\ell}_L\,\ell_L^c$-type mass term can be formed. Specifically, suppose we have a heavy  $SU(2)_L$ triplet scalar field $\Delta$ with hypercharge $Y=-2$ and a convenient $2\times 2$ matrix parametrization given by
\begin{equation}\label{eq1:triplet_delta_2rep}
 \Delta = \twtwMat{\Delta^- /\sqrt{2}}{\Delta^{--}}
                  {\Delta^0}{-\Delta^-/\sqrt{2}}\;,
\end{equation}
then the Lagrangian 
\begin{equation}\label{eq1:L_mass_type2}
 -\mathcal{L}_\text{type-II} =
  \frac{Y_\Delta}{2}\, \overline{\ell}_L i\tau_2\, \Delta \,\ell_L^c
  + \mu_\Delta\, \phi^T \Delta\,\phi + M_\Delta^2\,\Delta^\dagger\Delta
  +\text{h.c.}\;,
\end{equation}
will give rise to the process depicted in Fig.~\ref{fig1:seesaw23}a. This then leads to an effective mass term 
\begin{align}
 m_\text{eff}^{\rm II} &\simeq \,
  \mu_\Delta\, Y_\Delta \frac{\vev{\phi^0}^2}{M_\Delta^2}\;, \nonumber
\\  
  &=\,
  \lambda_\Delta\, Y_\Delta \frac{\vev{\phi^0}^2}{M_\Delta} \;,
  \quad \text{ after setting }\;\; \mu_\Delta \equiv \lambda_\Delta M_\Delta \;.
  \label{eq1:type2_mass_eff}
\end{align}
This expression has the same form as (\ref{eq1:numass_eff}) with the couplings $\lambda_\Delta Y_\Delta$ playing the role of $y^2$. So, when $M_\Delta \gg \vev{\phi^0}$, small neutrino masses can be induced. This mechanism is known as type II seesaw \cite{type2_seesaw}.

\subsubsection{Type III seesaw}

Another possibility is to replace the RH neutrinos with heavy triplet fermions and allow them to interact with the ordinary lepton doublets via Yukawa couplings \cite{type3_seesaw}. In this scenario, the Higgs sector is unmodified, and a set of  self-conjugate $SU(2)_L$ triplets of exotic leptons with hypercharge $Y=0$ are added:
\begin{equation}\label{eq1:triplet_sigma_2rep}
 \Sigma = \twtwMat{\Sigma^-}{\Sigma^0/\sqrt{2}}
                  {\Sigma^0/\sqrt{2}}{\Sigma^+}\;.
\end{equation} 

The corresponding Lagrangian for this model is given by
\begin{equation}\label{eq1:L_mass_type3}
 -\mathcal{L}_\text{type-III} =
  Y_\Sigma\, \overline{\ell}_L\, i\tau_2\, \Sigma \,\phi
  + M_\Sigma\, \text{Tr}\left(\overline{\Sigma^c}\, \Sigma\right)
  +\text{h.c.}\;,
\end{equation} 

This gives rise to the diagram shown in Fig.~\ref{fig1:seesaw23}b, and after integrating out the heavy $\Sigma$ field, one obtains the desired form for the seesaw neutrino mass
\begin{equation}
 m_\text{eff}^{\rm III} \simeq \,
  Y_\Sigma\,\frac{\vev{\phi^0}^2}{M_\Sigma}\,Y_\Sigma^T
  \;.
  \label{eq1:type3_mass_eff}
\end{equation}

Hence by setting $M_\Sigma \gg \vev{\phi^0}$, one can explain the smallness of neutrino masses, and as a result, this is often referred to as the type III seesaw mechanism \cite{type3_seesaw}.

\subsection{Neutrino mass from radiative corrections}\label{subsec:radcorr}

Finally, it should be mentioned that neutrino masses can also be induced from radiative corrections. Typically, this involves the enlargement of the Higgs sector and having the newly introduced charged scalars coupling to the LH lepton doublets, so that interactions of the form of (\ref{eq1:nu_dim5_term}) are created at loop level.

\subsubsection{Zee model}

\begin{figure}[t]
\begin{center}
 \includegraphics[width=0.95\textwidth]{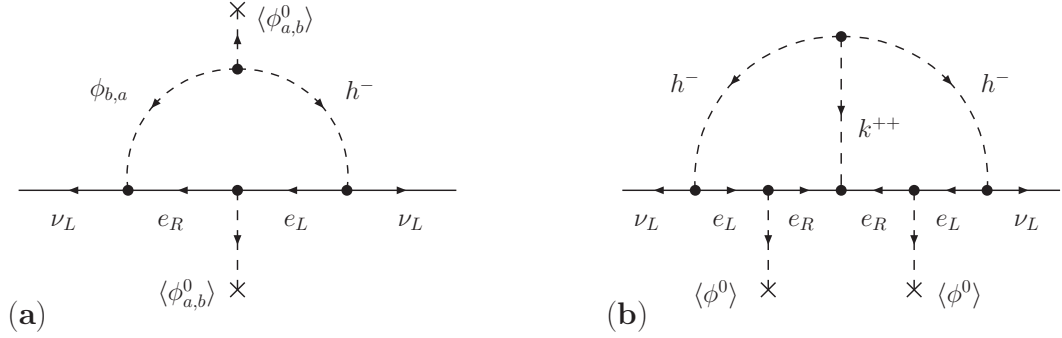}
\end{center} 
 \caption{{\bf (a)} One-loop correction graph of the Zee model that generates a neutrino Majorana mass. 
 {\bf (b)}~Two-loop diagram of the Babu-Zee model that contributes to the neutrino Majorana mass term.}
 \label{fig1:zee_babu}
\end{figure}

One realisation of this concept is the so-called Zee model \cite{zee_model}. In this particular extension to the SM, there are two (or more) $SU(2)_L$ Higgs doublets, all with the same hypercharge ($Y(\phi)=1$) and a charged scalar singlet $h^-$ which has $Y(h^-)=-2$. With these, the scalar field $h^-$ can couple to the LH lepton doublets in the following way:
\begin{equation}\label{eq1:zee_yukawa}
 \mathcal{L}_\text{yuk}^\text{Z} = \kappa\,\epsilon_{ij}\,
 \overline{\ell}_L^i \, (\ell_L^j)^c\, h^- +\text{h.c.}\;,
\end{equation}
where $i, j$ are indices in $SU(2)_L$ and $\epsilon_{ij}$ denotes the Levi-Civita tensor. Furthermore, in the Higgs potential, one gets a cubic coupling term between $h^-$ and other doublets in the model:
\begin{equation}\label{eq1:zee_cubic}
 \mathcal{L}_\text{cubic}^\text{Z} = 
  \mu_{ab}\,\epsilon_{ij}\,\phi_a^i \,\phi_b^j\,h^- +\text{h.c.}\;,
\end{equation}
where $a, b = 1,2,\ldots$ are labels for the different Higgs doublets. When combining the tree-level interactions of (\ref{eq1:zee_yukawa}), (\ref{eq1:zee_cubic}) and the standard Higgs Yukawa ($\overline{\ell}_L \,\phi_{a,b}\, e_R$) together, a Majorana neutrino mass term of the form of (\ref{eq1:nu_dim5_term}) is induced via the one-loop diagram shown in Fig.~\ref{fig1:zee_babu}a. Tiny neutrino masses can then be realised by carefully choosing the scale of the Higgses and assuming a small coupling for $\kappa$.

\subsubsection{Babu-Zee model}

Another example of neutrino masses generated from radiative corrections is the so-called Babu-Zee model \cite{babu_model, zee_model_2}. In contrast to the Zee setup, it contains two charged scalar singlets: one singly charged ($h^-$) and the other doubly charged ($k^{++}$), together with only one Higgs doublet (the SM Higgs). The relevant interaction Lagrangian is then given by:
\begin{equation}\label{eq1:babu_zee_int}
 \mathcal{L}_\text{int}^\text{BZ} = 
  \kappa\,\epsilon_{ij}\,
 \overline{\ell}_L^i \, (\ell_L^j)^c\, h^- 
+ \lambda\,
 \overline{(e_R)^c} \, e_R\, k^{++}
+ \mu\,k^{++}\,h^-\,h^-
+ Y_e\,\overline{\ell}_L\,\phi\,e_R
 +\text{h.c.}\;,
\end{equation}
where $\kappa, \lambda$ and $Y_e$ are dimensionless coupling constants while constant $\mu$ has a dimension of mass. These interactions can give rise to Majorana neutrino masses at the two-loop level as indicated in Fig.~\ref{fig1:zee_babu}b.

\subsection{Neutrino electromagnetic dipole moments}\label{subsec:numass_emdm}

One of the important implications of massive neutrinos is that they can in general possess a nonzero transition magnetic and electric dipole moment (both for Dirac and Majorana neutrinos), regardless of the mechanism by which they gain their mass. If neutrinos are Dirac particles, then they can also have diagonal electromagnetic dipole moments
\cite{general_emdm_bg, general_emdm_calc,SV_nu_dm,
Wolf_nu_dm}, unlike their Majorana counterparts. We can see this by first noting that the electromagnetic current that corresponds to magnetic and electric dipole moment interactions between two neutrinos has the form:
\begin{equation}\label{eq1:mdm_formfac}
 \bra{\nu_j(p')}| J_{\mu}^\text{dm} \ket{\nu_k(p)} = 
   \overline{\nu}_j  \left[ 
   F(q^2) + i \gamma^5 G(q^2)
   \right] \sigma_{\mu\nu}\,q^\nu\nu_k\;,
\end{equation}
where $q = p'-p$, while  $F(q^2)$ and $G(q^2)$ are the magnetic and electric dipole form factors respectively. For a Majorana field $\Psi$  and $j=k$ (the diagonal case), Eq.~(\ref{eq1:mdm_formfac}) becomes
\begin{align}
   \overline{\Psi}  \left[F+ i \gamma^5 G \right] \sigma_{\mu\nu}\,q^\nu \Psi
   &\equiv  
   \left(\overline{\Psi} \left[F+ i \gamma^5 G \right]\sigma_{\mu\nu}\,q^\nu   \Psi
   \right)^T\;, \label{eq1:mdm_formfac_maj_A}\\
   &= - \Psi^T  \left[F+ i \gamma^5 G \right]^T (\sigma_{\mu\nu})^T q^\nu \,
   \overline{\Psi}^T\;, \label{eq1:mdm_formfac_maj_B}
\end{align}
where the negative sign in (\ref{eq1:mdm_formfac_maj_B}) comes from anticommuting fermion fields and we have used the fact that $\gamma^5 \sigma_{\mu\nu} = \sigma_{\mu\nu}\gamma^5$. 
Using the following conventions for charge conjugation
\begin{equation} \label{eq1:C_convention}
 \psi^c = C\overline{\psi}^T\;, \quad
 C^\dagger\gamma^5 C = (\gamma^5)^T\;, \quad 
 C^\dagger\sigma_{\mu\nu} C = (-\sigma_{\mu\nu})^T\;, \quad
\end{equation}
expression (\ref{eq1:mdm_formfac_maj_B}) can be rewritten as
\begin{align}
 &- \Psi^T C^\dagger \left[F+ i \gamma^5 G \right] (-\sigma_{\mu\nu}) q^\nu \,
   C \overline{\Psi}^T\;, \label{eq1:mdm_formfac_maj_C}\\
 =& - \overline{\Psi^c}  \left[F+ i \gamma^5 G \right] \sigma_{\mu\nu} q^\nu \Psi^c\;,
  \label{eq1:mdm_formfac_maj_D}\\
 =& - \overline{\Psi}  \left[F+ i \gamma^5 G \right] \sigma_{\mu\nu} q^\nu
   \Psi\;,
  \label{eq1:mdm_formfac_maj_F}  
\end{align}
where we have used the Majorana condition $\Psi^c = e^{i\varphi}\Psi$. So, combining LHS of  (\ref{eq1:mdm_formfac_maj_A}) and (\ref{eq1:mdm_formfac_maj_F}) implies that $F$ and $G$ must vanish. In other words, Majorana neutrinos cannot have diagonal dipole moments although they may possess transition moments.

\begin{figure}[bht]
\begin{center}
 \includegraphics[width=0.70\textwidth]{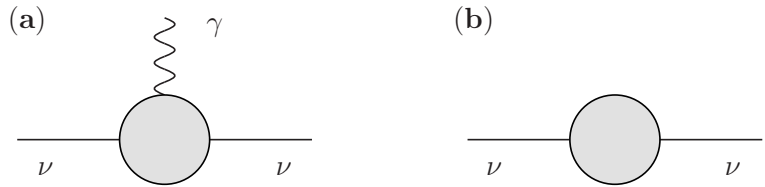}
\end{center} 
 \caption{{\bf (a)} A generic Feynman graph representing the contribution to neutrino electromagnetic dipole moments. 
 {\bf (b)}~The corresponding graph with the photon line removed which gives a radiative contribution to neutrino mass. The shaded blob denotes some new physics beyond the SM which gives rise to the neutrino dipole moments and/or masses.} 
\label{fig1:emdm_fey_graphs}
\end{figure}

To understand the connection between neutrino mass and neutrino dipole moment, one can compare the Feynman graphs which correspond to the generic contributions of these quantities as depicted in Fig.~\ref{fig1:emdm_fey_graphs}. By applying simple dimensional analysis on a generic dipole moment operator:
\begin{equation}\label{eq1:emdm_op_generic}
 \mathcal{L}^\text{dm} =
  \overline{\nu}_j\, (\mu_{jk} +i\gamma^5 d_{jk})\, \sigma_{\alpha\beta}\, \nu_k \,F^{\alpha\beta}
  \;,
\end{equation}
where $F^{\alpha\beta}$ denotes the photon field tensor, we see that the magnetic ($\mu_{jk}$) and electric ($d_{jk}$) dipole moments have dimension of inverse mass. Therefore, the contribution from the dipole moment interaction as shown in Fig.~\ref{fig1:emdm_fey_graphs}a can be expressed as
\begin{equation}\label{eq1:emdm_feyn_a}
  \mu_\nu \sim \frac{e\,\mathcal{G}}{\Lambda}
  \;,
\end{equation}
where $e$ is the coupling from the photon vertex, $\mathcal{G}$ denotes the combination of dimensionless parameters such as coupling constants (except $e$), mass ratios, mixing angles and $1/16\pi^2$ loop factors which may appear in the graph, while $\Lambda$ is some energy scale beyond the SM at which the dipole moment is generated. Similar analysis for Fig.~\ref{fig1:emdm_fey_graphs}b where the external photon line has been removed gives a radiative correction to the neutrino mass of order
\begin{equation}\label{eq1:emdm_feyn_b}
  m_\nu \sim \mathcal{G}\,\Lambda
  \;.
\end{equation}
Therefore, putting (\ref{eq1:emdm_feyn_a}) and (\ref{eq1:emdm_feyn_b}) together, a simple relationship between neutrino mass and dipole moment can be obtained:
\begin{equation}\label{eq1:emdm_mass_dm_rel}
  \mu_\nu \sim \frac{e\,m_\nu}{\Lambda^2}
  \simeq
  10^{-18}\left(\frac{\phantom{I}m_\nu\phantom{I}}{1\text{ eV}}\right)
  \left(\frac{1\text{ TeV}}{\Lambda}\right)^2 \mu_B 
  \;,
\end{equation}
where $\mu_B = e/2m_e$ is the Bohr magneton.

Although, in a more general context, where new symmetries or physics are employed \cite{bell_fest_5_8, Fuk_Yan_book}, zero neutrino masses do not necessarily imply vanishing dipole moments, the case of massive neutrinos, however, naturally allows for dipole moments to exist as indicated by (\ref{eq1:emdm_mass_dm_rel}). 
In fact, for the SM with massive Dirac neutrinos, the diagonal magnetic dipole moment induced by radiative corrections may be calculated for the mass eigenstate $\nu_j$ \cite{general_emdm_calc,general_emdm_other}:
\begin{equation}\label{eq1:emdm_SM_calc}
  \mu_{\nu_j} \simeq \frac{3e\,G_F}{8\pi^2\sqrt{2}}\,m_{\nu_j} \approx
  3\times 10^{-19} \left(\frac{m_{\nu_j}}{1 \text{ eV}}\right) \mu_B 
  \;,
\end{equation}
where $G_F$ is the Fermi constant. Moreover, one can perform a model-independent ``naturalness'' estimate of the contribution to neutrino masses from the dipole moment operators, thus gaining important insights into the size of $\mu_\nu$ in relation to $m_\nu$. In the case of Dirac neutrinos, one finds that the magnetic moment is bounded from above by \cite{Bell_mdm_PRL}  
\begin{equation}\label{eq1:mdm_natural_bound_D}
 \mu_\nu^\text{D} \lesssim 3 \times 10^{-15} 
 \left(\frac{\phantom{I}m_\nu\phantom{I}}{1\text{ eV}}\right)
  \left(\frac{1\text{ TeV}}{\Lambda}\right)^2 \mu_B
  \;,
\end{equation}
while for Majorana neutrinos, the transition moment is restricted by \cite{Bell_mdm_PLB}
\begin{equation}\label{eq1:mdm_natural_bound_M}
 (\mu_\nu^\text{M})_{\alpha\beta} \lesssim 4 \times 10^{-9} 
 \left(\frac{(m_\nu)_{\alpha\beta}}{1\text{ eV}}\right)
   \left(\frac{1\text{ TeV}}{\Lambda}\right)^2 
   \left|\frac{m_\tau^2}{m_\alpha^2-m_\beta^2}\right|
   \mu_B
  \;,
\end{equation}
where $\alpha, \beta$ are flavor indices, $m_{\alpha}$ denotes the mass of the charged lepton of flavor $\alpha$ and $m_\tau$ is the mass of the tauon. In both (\ref{eq1:mdm_natural_bound_D}) and (\ref{eq1:mdm_natural_bound_M}), $\Lambda$ represents the energy scale of any new physics that gives rise to the dipole moments.

Once neutrinos have electromagnetic dipole moments (diagonal or transition), it is clear that new interactions between neutrinos and other fermions are possible. For instance, on top of the usual weak interactions, there can be a new contribution to neutrino-electron scattering due to photon exchange, hence modifying the cross section. Also, the existence of transition moments can lead to neutrino decays. In particular, if the transition moments between the ordinary LH and heavy RH neutrinos (from the minimally extended SM) are non-vanishing, then the radiative decay of the heavy RH neutrinos can have important implications in the cosmological evolution of matter in the early universe. This will be the main topic of our work in Chapter~\ref{ch_work_BKL}.

The current laboratory limits on the magnetic dipole moment are obtained from the low-energy scattering processes mentioned above and they give a bound of about \cite{borexino,mdm_bound_exp1, mdm_bound_exp2}
\begin{equation}\label{eq1:emdm_exp_bonud_earth}
 \mu_\nu \lesssim 0.54 \times 10^{-10} \mu_B \;\;(90\%\text{ C.L.})\;.
\end{equation}
A slightly stronger constraint can be obtained from astrophysical data on plasmon decay in globular-cluster stars which gives $\mu_\nu \lesssim 3 \times 10^{-12} \mu_B$ \cite{mdm_astro_Raffelt}.

Another interesting point to note is that both of these limits lie between the upper bounds indicated by (\ref{eq1:mdm_natural_bound_D}) and (\ref{eq1:mdm_natural_bound_M}). As a consequence, measuring the size of the neutrino electromagnetic moments may provide information on the nature of neutrinos as their detection near the upper limit of (\ref{eq1:emdm_exp_bonud_earth})  will suggest that neutrinos are Majorana particles.\\

\section{Baryon asymmetry of the universe}\label{sec:BAU}

In the next few sections, we turn our attention to another intriguing puzzle, one which has troubled cosmologists for some time, and whose resolution is still rather speculative. This is the puzzle of the predominance of matter over antimatter in the universe.
Conventional wisdom has it that soon after the Big Bang, the universe should contain the same amount of baryons and antibaryons. So assuming that both evolved in an identical way (as the standard hot Big Bang theory suggests), there would be no apparent reason for baryons to exist in such a large amount whereas antibaryons are so rare in the universe today. 
Therefore, the challenge is to find a plausible explanation to this asymmetry 
within the framework of the standard cosmological model, which has been  successful on so many fronts.

\subsection{Evidence for a baryon asymmetry}\label{subsec:evid4BAU}

In everyday life, we know that antimatter is rare. Almost everything that we interact with is made of matter. Terrestrially, noticeable amounts of antimatter only appear in situations like nuclear medicine imaging or accelerator experiments. But even then, the abundance in question is of the order of one trillionth of a gram or less!

Beyond the earth, information about the ambient antimatter density of the Milky Way (and perhaps other nearby galaxies) may be inferred from the studies of antiproton flux in cosmic ray experiments conducted on earth. It has been observed that the ratio of antiprotons to protons is about: $\overline{p}/p \sim 10^{-4}$
\cite{BAU:cosmic_ray_flux}, a result which is consistent with the view that 
these antiprotons are only byproducts produced in cosmic ray collisions with the inter-stellar media, and cannot be originated from a non-negligible antiproton density in the galaxies. In other words, there is strong evidence that antimatter is also rare in our 
local galaxy.

On an even larger scale, the presence of intra-cluster hydrogen gas clouds (as indicated by their x-ray emissions) implies that patches containing large amount of antibaryons cannot exist within galaxy clusters. The reason for this is that the existence of such patches would give rise to strong $\gamma$-ray emission from  baryon-antibaryon annihilations near the interfaces with the gas clouds. But no such $\gamma$-ray flux has been seen. As a result, one can compellingly argue that should there be large regions of space containing antimatter, they must be beyond the observable universe \cite{BAU:cohen_glashow}, and hence on general grounds, one can conclude that the possibility of our universe containing the same amount of matter and antimatter today is empirically excluded.

Along with the puzzle of why there is an excess of matter over antimatter in the universe today, recent measurements of the temperature anisotropy of the Cosmic Microwave Background (CMB) radiation by the WMAP probe \cite{wmap_numass}, together with studies of large scale structure \cite{sdss_numass}, have given us a reliable estimate of the baryon-to-photon ratio at the current epoch of
\begin{equation}\label{eq1:BAU_B-asmy_CMB}
 \eta_B^\text{CMB} \equiv \frac{n_B}{n_\gamma} = (6.1\pm 0.2) \times 10^{-10}\;,
\end{equation}
which prompts the question as to why this amount. In (\ref{eq1:BAU_B-asmy_CMB}), $n_B$ and $n_\gamma$ denote the number density of baryons and photons respectively, with $n_\gamma = \zeta(3)g_*  T^3/\pi^2$, where $g_*=2$ counts the internal degrees of freedom, $\zeta(3) \approx 1.202$ is the Riemann zeta function of order 3 and $T$ denotes the temperature.

To understand why the size of $\eta_B^\text{CMB}$ in (\ref{eq1:BAU_B-asmy_CMB}) is of relevance, we first note that the standard Big Bang Nucleosynthesis (BBN) analysis of the primordial abundances of $^3$He, $^4$He, (D)euterium, $^6$Li and $^7$Li depends crucially on the value of the baryon-to-photon ratio. In fact, astrophysical observations have inferred that \cite{BAU:BBN_intro}
\begin{equation}\label{eq1:BAU_B-asmy_BBN}
 \eta_B^\text{BBN} \approx (4.7-6.5) \times 10^{-10}\;,
\end{equation}
which is very much consistent with (\ref{eq1:BAU_B-asmy_CMB}), hence demonstrating the validity of BBN and standard cosmology.

More importantly though, the amount of \order{10^{-10}} for this ratio signifies that there must have been a \emph{primordial} baryon asymmetry in the early universe. This is because if the universe was baryon-antibaryon symmetric at $T \simeq \order{100}$~MeV, the annihilation process $B+\overline{B} \rightarrow 2\,\gamma$ would significantly reduce both the value of $n_B/n_\gamma$ and $n_{\overline{B}}/n_\gamma$, before they subsequently froze out at $T \sim 22$~MeV when the annihilations became ineffective. By studying the Boltzmann evolution of the number density of the (anti)baryons in this scenario, one can estimate the expected baryon-to-photon ratio for today to be \cite{kolb_turner}
\begin{equation}\label{eq1:BAU_B-sym_est}
 \frac{n_B}{n_\gamma} = \frac{n_{\overline{B}}}{n_\gamma} = \order{10^{-18}}\;.
\end{equation}
Therefore, barring some exotic possibilities, the apparent discrepancy between  (\ref{eq1:BAU_B-asmy_CMB}) and (\ref{eq1:BAU_B-sym_est}) is a clear indication that during  primordial times, the universe must already have been matter-antimatter asymmetric, and the current scarcity of antimatter is just a manifestation of that fact. So in reality,  the observed value of the baryon-to-photon ratio also characterizes the amount of asymmetry between matter and antimatter in the universe:\footnote{From now on, $\eta_B$ will  be taken as a measure of the amount of matter-antimatter asymmetry in the universe.}
\begin{equation}\label{eq1:BAU_B-sym_generalised}
 \eta_B 
  \simeq \frac{n_B -n_{\overline{B}}}{\eta_\gamma}  
  \;.
\end{equation}

In the wake of this necessity for a primordial asymmetry, the key focus becomes understanding its physical origins. Although taking the asymmetry as an initial condition set by the Big Bang is perhaps the simplest solution, in the context of an inflationary universe this approach is strongly disfavored, for any pre-existing asymmetry would be diluted away by the proliferation of entropy during reheating. Thus, the expectation is that this excess of baryons over antibaryons must have been dynamically generated during the evolution of the early universe. But the question of whether there are physical processes (within or beyond the SM) that would allow this to happen will need to be addressed.\\

\subsection{Elements of baryogenesis}\label{subsec:elem_baryog}

The puzzle of creating a baryon asymmetry from an initially symmetric state 
so that the observed baryon-to-photon ratio can be accounted for is known as the \emph{baryogenesis problem}. While the mechanisms that can lead to this asymmetry can be quite different, they must in general satisfy the three basic conditions for baryogenesis as pointed out by Sakharov in 1967 \cite{sakharov}: a dynamical model should contain processes that 
\begin{enumerate}
 \item violate baryon number, $B$,
 \item violate $C$ and $CP$, and
 \item are out of thermal equilibrium.
\end{enumerate}
These are often referred to as the \emph{Sakharov conditions}. It should be noted at this point that, later on, when we discuss the mechanism of baryogenesis via leptogenesis (see Sec.~\ref{sec:lepto_main}), these conditions will be extended to include lepton number ($L$) violation processes. This is because an excess in $L$ must be created for that scenario to work. 

It is customary to assign a positive number $B$ for baryons while their respective antiparticle counterparts are given a negative number $\overline{B} \equiv -B$ for their baryon number. 
The first Sakharov's criterion is obvious as no increase or decrease of baryon number $B$ can happen if all interactions in the model are $B$ conserving. 

Owning to the fact that for every $B$ violating interaction which involves a baryon, $X\rightarrow qq$, there will be a mirror process,  $\overline{X}\rightarrow \overline{q}\,\overline{q}$, for the corresponding antibaryon that can create an exact negative amount of $B$, no net $B$ asymmetry may result if both types of processes are equally likely. Hence, Sakharov's second condition demands that $C$ (charge conjugation) and $CP$ (charge conjugation plus parity flip) violations are necessary as they will lead to different rates for the particle and antiparticle processes, i.e.
$\Gamma(X\rightarrow qq) \neq \Gamma(\overline{X}\rightarrow \overline{q}\,\overline{q})$.

The condition of deviation from thermal equilibrium for these processes is essential because the Hamiltonian of the system under consideration is usually assumed to be $CPT$ invariant (i.e. $\Theta^{-1} \mathcal{H} \Theta = \mathcal{H}$, where $\Theta$ is the $CPT$ operator and $\mathcal{H}$ denotes the Hamiltonian). So, given the fact that baryon number is $CPT$-odd (i.e. $\Theta^{-1} B \Theta = -B$),  it can be easily demonstrated that no baryon asymmetry can be generated in thermal equilibrium. Suppose $B(t)$ denotes the baryon number at time $t$. For a system in thermal equilibrium, we can define the thermal average for $B(t)$ at temperature $T$ by
\begin{equation}\label{eq1:thermal_average}
 \bra{B(t)}\rangle_{T} = \text{Tr}\left[ B(t)\,e^{-\mathcal{H}/T} \right]\;,
\end{equation}
where $e^{-\mathcal{H}/T}$ is the density operator. Given that the time evolution of baryon number is $B(t) = e^{-i \mathcal{H} t} \,B(0)\,e^{i \mathcal{H} t}$, then we have
\begin{align}
 \bra{B(t)}\rangle_{T} 
 &= 
 \text{Tr}\left[ e^{-i \mathcal{H} t} \,B(0)\,e^{i \mathcal{H} t}\,e^{-\mathcal{H}/T} \right]\;,\nonumber\\
 &=
 \text{Tr}\left[ B(0)\,e^{i \mathcal{H} t}\,e^{-\mathcal{H}/T}\, e^{-i \mathcal{H} t} \right]\;,\nonumber \\
 &=
 \text{Tr}\left[ B(0)\,e^{-\mathcal{H}/T}\right]\;,\nonumber \\
 &\equiv \bra{B(0)}\rangle_{T}\;.
  \label{eq1:thermal_average_time}
\end{align}
In other words, a system in thermal equilibrium must be stationary. Furthermore, using the $CPT$ property of $B$ and $\mathcal{H}$, we can see that
\begin{align}
 \bra{B(t)}\rangle_{T} 
 \equiv \text{Tr}\left[ B(t)\,e^{-\mathcal{H}/T} \right] 
   &= \text{Tr}\left[\Theta \Theta^{-1} B(t)\,\Theta \Theta^{-1} e^{-\mathcal{H}/T} \right]\;,
   \nonumber\\
 &= \text{Tr}\left[\Theta^{-1} B(t)\,\Theta\, \Theta^{-1} e^{-\mathcal{H}/T}\,\Theta  \right]\;, 
 \nonumber\\
 &= \text{Tr}\left[-B(t)\, e^{-\mathcal{H}/T}\right]\;,  \nonumber\\
 &= -\bra{B(t)}\rangle_{T}\;,  \label{eq1:thermal_eqbm_CPTa}\\
 \Rightarrow\quad
  \bra{B(t)}\rangle_{T} &=0 \;.\label{eq1:thermal_eqbm_CPTb}
\end{align}
So, the thermal expectation value of $B$ vanishes in equilibrium. Consequently, baryogenesis cannot succeed without the ``arrow of time'' provided by the departure from thermal equilibrium.

It may come as a surprise to some that standard electroweak theory actually contains all of the above ingredients, and so in principle at least, the baryogenesis problem may be solved within the framework of the SM alone. The reason that SM interactions are insufficient to produce the observed baryon asymmetry is because precision experiments on electroweak processes have placed stringent constraints on the allowed parameter space. In particular, it is found that the $CP$ violation observed in the quark sector \cite{KM_CPviolation} (e.g. in $K^0$-$\bar{K}^0$ or $B^0$-$\bar{B}^0$ mesons system) is far too small \cite{CP_in_SM_tooSmall} to give rise to the observed $\eta_B$. Moreover, the present empirical lower limit on the Higgs mass, $m_\text{Higgs} > 114$~GeV \cite{limit_higgs_mass}, implies that the electroweak phase transition cannot be first order \cite{EWPT_first_order, EWPT_shaposhnikov}, making it difficult for the baryon number violating sphaleron processes (see Sec.~\ref{subsec:sphalerons}) in the SM to go out of thermal equilibrium \footnote{See \cite{cohen_kaplan_nelson} for a more detailed discussion on the standard \emph{electroweak baryogenesis} theory.}. 

Although the SM per se cannot solve the baryogenesis problem, the understanding of its behavior at high energies plays a crucial role in finding a suitable extension that can work. In particular, the non-perturbative interactions that can violate baryon and lepton numbers in the early universe are especially important as we shall review in the next subsection.\\

\subsection{$B+L$ anomaly and sphalerons}\label{subsec:sphalerons}

From the particle phenomenology at energy levels reachable by present-day experiments, it seems that baryon ($B$) and lepton ($L$) numbers are conserved in all interactions. Indeed, $B$ and $L$ violations in the SM only enter in a subtle and non-perturbative way, and therefore, to a good approximation, they can usually be ignored. However, at the very high temperatures typical of the early universe, these effects on particle evolution become prominent and must be taken into account. 

To see how $B$ and $L$ violations come about while at the same time reconciling their apparent conservation at low energies, it is instructive to study the electroweak theory at both the classical and quantum mechanical levels. A well known fact of the classical SM Lagrangian is that it has global $U(1)_B$ and $U(1)_L$ symmetries and is therefore invariant under the following transformations of the quark and lepton fields:
\begin{align}
 U(1)_B:\quad &q(x) \rightarrow  q(x)\, e^{i\theta}\;;\quad \ell(x) \rightarrow 
  \ell(x)\, \phantom{e^{i\phi}}\;,\label{eq1:U1_B_trans}\\
 U(1)_L:\quad &q(x) \rightarrow  q(x)\, \phantom{e^{i\theta}}\;;\quad 
 \ell(x) \rightarrow \ell(x)\, e^{i\phi}\;, \label{eq1:U1_L_trans}
\end{align}
where $\theta$ and $\phi$ are constants. Noether's theorem then implies that the classical $J_{\mu}^B$ and $J_{\mu}^L$ currents are conserved:
\begin{align}
 \partial^\mu J_{\mu}^B &= \partial^\mu 
   \mathop{\sum_{\text{\tiny flavors}}}_\text{\tiny colors} 
   \frac{1}{3} \left( \overline{q}_L \gamma_\mu q_L +\overline{u}_R \gamma_\mu u_R 
    +\overline{d}_R \gamma_\mu d_R\right) = 0\;,
    \label{eq1:U1_B_current}\\
 &\partial^\mu J_{\mu}^L = \partial^\mu 
   \sum_\text{\tiny flavors} 
   \left( \overline{\ell}_L \gamma_\mu \ell_L +\overline{e}_R \gamma_\mu e_R 
    \right) = 0\;,
    \label{eq1:U1_L_current}
\end{align}
where we have conveniently defined the baryon and lepton numbers for quarks and leptons as: $B_\text{quark} = 1/3, B_\text{lepton} =0, L_\text{quark} = 0$ and $ L_\text{lepton} = 1$.

But due to the chiral nature of the electroweak sector, quantum corrections through the triangle anomaly \cite{triangle_anomaly} will render the divergences (\ref{eq1:U1_B_current}) and (\ref{eq1:U1_L_current}) nonzero. The situation can be made explicit by studying the anomaly equations for the LH and RH chiral components of the vector currents, which have the general form:
\begin{align}
 \partial^\mu \left(\overline{\psi}_L \gamma_\mu \psi_L \right)
 &= -c_L\, \frac{g_G^2}{32\pi^2} \, F_{\mu\nu}^{a} \widetilde{F}_{a}^{\mu\nu}\;, 
 \label{eq1:chiral_current_L}\\
 \partial^\mu \left(\overline{\psi}_R \gamma_\mu \psi_R \right)
 &= +c_R\, \frac{g_G^2}{32\pi^2} \, F_{\mu\nu}^{a} \widetilde{F}_{a}^{\mu\nu}\;, 
 \label{eq1:chiral_current_R}
\end{align}
where $\psi_{L,R}$ denotes any of the LH or RH quark and lepton fields, $c_L$ and $c_R$ are constants~\footnote{They do however depend on the representations of $\psi_{L,R}$.}, and $g_G$ is the gauge coupling of some gauge group $G$, with $F_{\mu\nu}^{a}$ and $\widetilde{F}_{a}^{\mu\nu}$ being the corresponding field strength and dual tensor respectively. 

Analysing this for the SM gauge group, $SU(3)_c \otimes SU(2)_L \otimes U(1)_Y$, one can deduce, after some algebra, that:
\begin{equation}\label{eq1:B_L_anomaly_current_SM}
 \partial^\mu J_{\mu}^B = \partial^\mu J_{\mu}^L = 
 \frac{N_f}{32\pi^2} \left( -g^2 W_{\mu\nu}^{a}\widetilde{W}_{a}^{\mu\nu}
 + {g'}^2 B_{\mu\nu}\widetilde{B}^{\mu\nu}\right)\;, 
\end{equation}
where $g$ and $g'$ are the gauge couplings of $SU(2)_L$ and $U(1)_Y$ respectively, with $W_{\mu\nu}^{a}$ and $B_{\mu\nu}$ the corresponding field tensors, and $N_f$ denotes the number of generations. It should be mentioned that no $SU(3)_c$ related elements appear in (\ref{eq1:B_L_anomaly_current_SM}) because gluons couple to all quark currents with the same strength regardless of chirality (and leptons do not couple to them at all). As a result, $c_L^\text{strong} = c_R^\text{strong}$ in the $SU(3)_c$ version of (\ref{eq1:chiral_current_L}) and (\ref{eq1:chiral_current_R}), and so the anomaly cancels for this sub-sector. 

Another important observation from (\ref{eq1:B_L_anomaly_current_SM}) is that $\partial^\mu J_{\mu}^B$ and $\partial^\mu J_{\mu}^L$ are identical and hence,
\begin{equation}\label{eq1:B-L_current_zero}
 \partial^\mu \left(J_{\mu}^{B} -J_{\mu}^{L}\right) =0\;.
\end{equation}
In other words, the $B-L$ quantum  number is strictly conserved in the SM. However, it is also clear from (\ref{eq1:B_L_anomaly_current_SM}) that $B+L$ must be violated. To deduce the corresponding change in the $B+L$ quantum number, one must evaluate the Euclidean integral of $\partial^\mu (J_{\mu}^{B}+J_{\mu}^{L})$ over $d^4 x$: 
\begin{align}
 \Delta(B+L) \equiv \int d^4 x\; \partial^\mu J_{\mu}^{B+L}
  &=
  \int d^4 x\;
  \frac{2 N_f}{32\pi^2} \left( -g^2 W_{\mu\nu}^{a}\widetilde{W}_{a}^{\mu\nu}
 + {g'}^2 B_{\mu\nu}\widetilde{B}^{\mu\nu}\right)\;,
 \\
 &= 2\,N_f\,\Delta N_{\rm cs}\;, \label{eq1:B+L_chern-simons}
\end{align}
where $\Delta N_{\rm cs} = \pm 1, \pm 2, \ldots$ is the change in the Chern-Simons number. This is typical of any non-abelian gauge theory (like the SM) where infinitely many degenerate vacua exist. The transitions among these ground states correspond to jumps between the different topological sectors of the gauge theory, which are classified by the Chern-Simons numbers. 

The main consequence of relation (\ref{eq1:B+L_chern-simons}) is that for the SM with $N_f = 3$, the smallest change in $B+L$ must be $\pm 6$ (or more precisely at least $\pm 3$ for $\Delta B$ and $\Delta L$). Also, from (\ref{eq1:B_L_anomaly_current_SM}), we note that $B_i - B_j$ and $L_i - L_j$, where $i$ and $j$ are family indices, must be conserved by sphaleron processes in addition to $B-L$.
Therefore, a quantum transition from one ground state to another which results in $\Delta N_{\rm cs} = \pm 1$ will give rise to the following 12-fermion interaction which involves all three families:
\begin{equation}\label{eq1:12-fermion_interactions}
 \text{vacuum} \;\;\leftrightarrow\;\;
 u_L\, d_L\, d_L\, \ell_{L_e}\; 
 c_L\, s_L\, s_L\, \ell_{L_\mu}\; 
 t_L\, b_L\, b_L\, \ell_{L_\tau} \;,
\end{equation}
where $\ell_{L_\alpha}$ ($\alpha = e,\mu \;\text{or}\; \tau$) is either the LH charged or neutral lepton.
Note that reaction (\ref{eq1:12-fermion_interactions}) conserves color and electric charge.

However, since the degenerate vacua in different topological regions of the field space are separated by a potential barrier \cite{Klink_Manton_sphaleron} with energy, $E_\text{sph}$ (called the \emph{sphaleron energy}), for any ``small'' gauge field quantum fluctuations around the perturbative vacuum, transitions with $\Delta N_{\rm cs} \neq 0$ are negligible. 
It has been shown that quantum tunnellings through such barriers due to instantons are exponentially suppressed at zero temperature \cite{tHooft_instanton}, with transition rate:
$\Gamma^{T=0} \sim e^{-16\pi^2/g^2} \simeq \order{10^{-165}}$. 

For finite temperatures, transition between gauge vacua can happen at a much greater rate because of (non-perturbative) thermal fluctuations \footnote{In this work, we shall refer to all such vacuum transitions as ``sphaleron processes'' or simply ``sphalerons'' although, strictly speaking, these processes induced by thermal fluctuations are unrelated to the sphaleron saddle-point solution that influences the quantum tunnellings.} over the barrier \cite{KRS_mechanism, Arnold_belowEW_rate}. Depending on whether the temperature $T$ is above or below the critical temperature, $T_\text{ew}$, for electroweak symmetry restoration, the transitions will proceed at significantly different rates. It has been shown that for $T<T_\text{ew} \simeq 100$~GeV, the rate is Boltzmann suppressed by $e^{-E_\text{sph}/T}$ \cite{Arnold_belowEW_rate, moore_belowEW_rate}, whereas for $T>T_\text{ew}$, the rate $\Gamma_\text{sph}^{T>T_{\rm ew}}$ is proportional to $T^4$ (at leading order) \cite{Arnold_aboveEW_rate, bodeker_moore_aboveEW}. Therefore, in the early universe where $T \gg T_\text{ew}$, these sphaleron processes  are very potent, while
at low temperatures such as those accessible in conventional experiments, baryon and lepton violations due to quantum corrections are physically irrelevant, and $B$ and $L$ can be regarded as conserved quantities to good approximation.

By comparing the sphaleron rate $\Gamma_\text{sph}^{T>T_{\rm ew}}$ with the Hubble expansion rate at $T$:
\begin{equation}\label{eq1:hubble_rate}
 H \simeq 1.66 \sqrt{g^*_s}\, \frac{T^2}{M_\text{Pl}} \;,
\end{equation}
where $g^*_s$ is the number of relativistic degrees of freedom and $M_\text{Pl} \approx 1.22 \times 10 ^{19}$~GeV is the Planck mass, one can check that for $T$ in the range:
\begin{equation}\label{eq1:sphaleron_temp}
 T_\text{ew}\simeq 100~\text{GeV } < T \lesssim 10^{12}~\text{GeV}\;,
\end{equation}
$B+L$ violating sphaleron interactions are in \emph{thermal equilibrium}. This observation is important because Sakharov's 3rd condition then implies that any baryogenesis mechanism which operates above $T_\text{ew}$ cannot generate an excess of $B$ and $L$ unless they also violate $B-L$. The reason for this is that $B-L$ is conserved by the sphaleron processes, and therefore, any asymmetry in $B-L$ generated from other interactions in the model will not be erased. We will make good use of this fact later on in Sec.~\ref{sec:lepto_main}.\\

\subsection{$CP$ violation in heavy particle decays}\label{subsec:CP-violation}

As mentioned in Sec.~\ref{subsec:elem_baryog}, a key ingredient for $B$ asymmetry generation is the condition of $CP$ violation in processes that can create baryon number. Given that $B+L$ violating sphalerons are in thermal equilibrium during the epoch of interest and hence would not be able to create an excess of $B$ regardless of their $CP$ properties, it is useful to investigate other type of processes that may do the job. However, no SM interactions (besides those induced by sphalerons) violate $B$, and therefore one must invoke physics beyond the SM.

A typical way to do this is to expand the particle content with exotic heavy particles and include new ($B$ violating) interaction terms that couple them to other constituents of the model. Such heavy particles could be a byproduct of the enlargement in the model's symmetry as typical in grand unification theories (GUTs) and supersymmetry (SUSY) models. However, their precise origins are not of concern in the current discussion.

\begin{figure}[tb]
\begin{center}
 \includegraphics[width=0.35\textwidth]{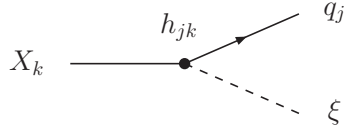} 
\end{center} 
 \caption{Tree-level Feynman diagram for the heavy particle decay of $X_k \rightarrow q_j\,\xi$, where the arrow denotes the flow of baryon number.}
 \label{fig1:CP_tree_Xq}
\end{figure}

To illustrate how $CP$ violation can arise in the decay of such heavy particles, we consider a toy model with a set of exotic particles $X_k$'s which can interact with other fermions $q_j$'s and scalars $\xi$'s through the Yukawa terms:
\begin{equation}\label{eq1:CP_toymodel_Lag}
  \mathcal{L}_\text{int} = h_{jk}\,\overline{q}_j\,\xi\, X_k + \text{h.c.}\;,
\end{equation}
where indices $j,k =1,2,\ldots$ are labels for the different particles within a set and $h_{jk}$ denotes the coupling which is a complex quantity in general. The tree-level Feynman diagram for the decay of $X_k$ induced by this term is shown in Fig.~\ref{fig1:CP_tree_Xq}. Suppose that there are other interactions besides (\ref{eq1:CP_toymodel_Lag}) and that they link $X_k$ to a final state with a \emph{different} baryon number to the state $q_j\,\xi$, then the decay of $X_k$ must violate $B$. Without loss of generality, let us 
assume that the decay: $X_k \rightarrow q_j\, \xi$ gives a change of $\Delta B_{X} = +1$, while the antiparticle decay:  $\bar{X}_k \rightarrow \bar{q}_j\, \bar{\xi}$ has $\Delta B_{\bar{X}} = -1$, then the $CP$ asymmetry in baryon number produced by these decays can be quantified by
\begin{align}
 \varepsilon_{CP} 
 &= \frac{\Delta B_{X}\, \Gamma(X_k \rightarrow q_j\, \xi)}{\Gamma_\text{tot}}
    +\frac{\Delta B_{\bar{X}}\, \Gamma(\bar{X}_k \rightarrow \bar{q}_j\, \bar{\xi})}{\Gamma_\text{tot}}
    \;,\nonumber\\
 &= \frac{(+1)\, \Gamma(X_k \rightarrow q_j\, \xi)
 +(-1)\, \Gamma(\bar{X}_k \rightarrow \bar{q}_j\, \bar{\xi})}{\Gamma_\text{tot}}
    \;,\nonumber\\
  &= \frac{\Gamma -\bar{\Gamma}}{\Gamma_\text{tot}}
    \;,    
    \label{eq1:CP_asym_general}
\end{align}
where $\Gamma_\text{tot} = \Gamma +\bar{\Gamma}$ is the total decay rate with $\Gamma \equiv \Gamma(X_k \rightarrow q_j\, \xi)$ and $\bar{\Gamma}\equiv \Gamma(\bar{X}_k \rightarrow \bar{q}_j\, \bar{\xi})$. As expected, (\ref{eq1:CP_asym_general}) confirms the requirement of unequal rates for particle and antiparticle decays in order to produce an asymmetry in baryon number. Therefore, we seek the general conditions under which $\Gamma$ and $\bar{\Gamma}$ can be different.

\begin{figure}[tb]
\begin{center}
 \includegraphics[width=0.35\textwidth]{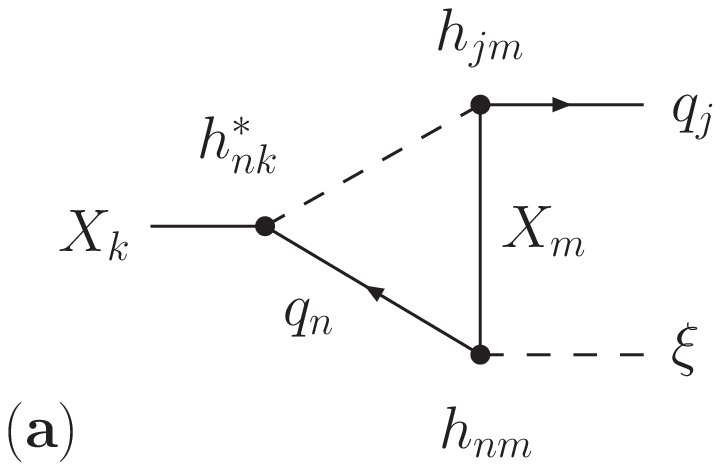}
~~~
 \includegraphics[width=0.45\textwidth]{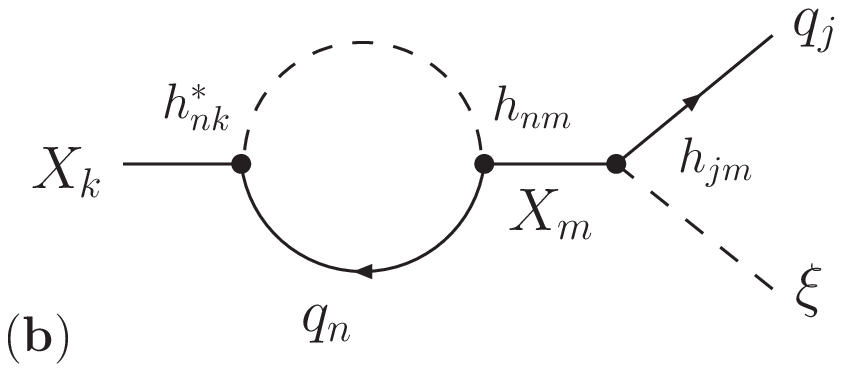}
\end{center} 
 \caption{Feynman graphs of the {\bf (a)} one-loop vertex correction for $X_k \rightarrow q_j\,\xi$ and {\bf (b)} the corresponding one-loop self-energy correction.}
 \label{fig1:CP_loops_Xq}
\end{figure}

It turns out that because of $CPT$ invariance, there can never be a difference between $\Gamma$ and $\bar{\Gamma}$ if one only considers the tree-level process depicted in Fig.~\ref{fig1:CP_tree_Xq} as $\Gamma = |h_{jk}|^2 I_\text{tree} = |h_{jk}|^2 \bar{I}_\text{tree} = \bar{\Gamma}$, where the kinematic factors $I_\text{tree}$ and $\bar{I}_\text{tree}$, which come from integrating over phase space, are necessarily equal.
As a result, one must go beyond the lowest order~\footnote{We will present a general proof of this in Sec.~\ref{subsec:BKL_CP_needs_loops}}. The first nonzero contribution to $\varepsilon_{CP}$ comes from the interference between the tree-level graph and the one-loop corrections shown in Fig.~\ref{fig1:CP_loops_Xq}. Writing out the terms up to \order{h^4} in the couplings, we have:
\begin{align}
 \Gamma
 &= 
\mathop{\int\limits_{\text{phase-}}}_\text{space}
  \left|\rule{0pt}{20pt}\right.
  \SetScale{0.5}
  \begin{picture}(45,0)(40,-73) 
    \SetWidth{0.8}
    \SetColor{Black}
    \Line(90,-141)(125,-141)
    \ArrowLine(125,-141)(155,-115)
    \DashLine(125,-141)(155,-167){4}
    \SetWidth{0.5}
    \Vertex(125,-141){3}
  \end{picture}
  +
\underbrace{\rule[-16pt]{0pt}{16pt}
  \SetScale{0.5}
  \begin{picture}(45,0) (40,-73)
    \SetWidth{0.8}
    \SetColor{Black}
    \Line(90,-141)(125,-141)
    \ArrowLine(134,-134)(155,-115)
    \DashLine(125,-141)(155,-167){4}
    \GOval(125,-141)(13,13)(0){0.882}
  \end{picture}
  }_\text{1-loop graphs}
  + \;\;\cdots\;\;
  \left.\rule{0pt}{20pt}\right|^{\;2} \;, \nonumber\\
  &= 
 \mathop{\int\limits_{\text{phase-}}}_\text{space} 
  \left|\rule{0pt}{20pt}\right.
  \SetScale{0.5}
  \begin{picture}(35,0)(45,-73) 
    \SetWidth{0.8}
    \SetColor{Black}
    \Line(90,-141)(125,-141)
    \ArrowLine(125,-141)(155,-115)
    \DashLine(125,-141)(155,-167){4}
    \SetWidth{0.5}
    \Vertex(125,-141){3}
  \end{picture}
   \left.\rule{0pt}{20pt}\right|^{\;2}
   +
    \left(\rule{0pt}{20pt}\right.
  \SetScale{0.5}
  \begin{picture}(35,0)(45,-73) 
    \SetWidth{0.8}
    \SetColor{Black}
    \Line(90,-141)(125,-141)
    \ArrowLine(125,-141)(155,-115)
    \DashLine(125,-141)(155,-167){4}
    \SetWidth{0.5}
    \Vertex(125,-141){3}
  \end{picture}
   \left.\rule{0pt}{20pt}\right)^{\dagger}
   \left(\rule{0pt}{20pt}\right.  
     \SetScale{0.5}
  \begin{picture}(35,0) (45,-73)
    \SetWidth{0.8}
    \SetColor{Black}
    \Line(90,-141)(125,-141)
    \ArrowLine(134,-134)(155,-115)
    \DashLine(125,-141)(155,-167){4}
    \GOval(125,-141)(13,13)(0){0.882}
  \end{picture}
   \left.\rule{0pt}{20pt}\right)   
  +
   \left(\rule{0pt}{20pt}\right.  
     \SetScale{0.5}
  \begin{picture}(35,0) (45,-73)
    \SetWidth{0.8}
    \SetColor{Black}
    \Line(90,-141)(125,-141)
    \ArrowLine(134,-134)(155,-115)
    \DashLine(125,-141)(155,-167){4}
    \GOval(125,-141)(13,13)(0){0.882}
  \end{picture}
   \left.\rule{0pt}{20pt}\right)^{\dagger}
    \left(\rule{0pt}{20pt}\right.
  \SetScale{0.5}
  \begin{picture}(35,0)(45,-73) 
    \SetWidth{0.8}
    \SetColor{Black}
    \Line(90,-141)(125,-141)
    \ArrowLine(125,-141)(155,-115)
    \DashLine(125,-141)(155,-167){4}
    \SetWidth{0.5}
    \Vertex(125,-141){3}
  \end{picture}
   \left.\rule{0pt}{20pt}\right)
   +\;\;
   \cdots \;,   
   \nonumber\\
 &= |h_{jk}|^2 I_\text{tree} 
   + h_{jk}^* h_{jm}h _{nm} h_{nk}^* \,I_\text{loop}
   + h_{jk} h_{jm}^* h_{nm}^* h_{nk} \,I_\text{loop}^* + \order{h^6}\;,  
  \label{eq1:CP_loop_expand_X}
\end{align}   
where $I_\text{loop}$ denotes the kinematic factor associated with the one-loop diagrams in Fig.~\ref{fig1:CP_loops_Xq} that accounts for integration over the phase space of final states, as well as any internal loop momenta. Repeating this for the antiparticle decay, one obtains
\begin{equation}\label{eq1:CP_loop_expand_Xbar}
 \bar{\Gamma} = 
  |h_{jk}|^2 I_\text{tree} 
   + h_{jk} h_{jm}^* h_{nm}^* h_{nk} \,I_\text{loop}
   +  h_{jk}^* h_{jm}h _{nm} h_{nk}^*\,I_\text{loop}^* + \order{h^6}\;,
\end{equation}
where we have used the fact that $\bar{I}_\text{loop} \equiv I_\text{loop}$. So, putting (\ref{eq1:CP_loop_expand_X}) and (\ref{eq1:CP_loop_expand_Xbar}) into the definition for $\varepsilon_{CP}$ in (\ref{eq1:CP_asym_general}) and ignoring the higher order terms, we have
\begin{align}
 \varepsilon_{CP} &=
 \frac{1}{\Gamma_\text{tot}}\,
 \left(
 A_h \,I_\text{loop}
 + A_h^* \,I_\text{loop}^*
 -A_h^* \,I_\text{loop}
 -A_h\,I_\text{loop}^*
 \right)\;,
 \\
 \intertext{where $A_h \equiv h_{jk}^* h_{jm}h _{nm} h_{nk}^*$ and $\Gamma_\text{tot} \simeq 2|h_{jk}|^2 I_\text{tree}$ to the lowest order. Thus,}
 \varepsilon_{CP}
 &= \frac{1}{\Gamma_\text{tot}}\,
 \left( 
  A_h -A_h^*\right)
  \left(
  I_\text{loop}
 -I_\text{loop}^*
   \right)\;, \nonumber\\
 &= \frac{1}{\Gamma_\text{tot}}\;
 2i\text{Im}\left(A_h\right)
  2i\text{Im}\left(
  I_\text{loop}
   \right)\;, \nonumber\\
 &=
 -\frac{4}{\Gamma_\text{tot}}\,   
  \text{Im}\left(h_{jk}^* h_{jm}h _{nm} h_{nk}^*\right)
  \text{Im}\left(I_\text{loop} \right)\;. \label{eq1:CP_aysm_final_form}
\end{align}
Equation (\ref{eq1:CP_aysm_final_form}) is the main result of this section. It highlights the three essential ingredients required for any model to have a nonzero $CP$ asymmetry. Firstly, the couplings $h$ must be complex so that the imaginary part of their products are in general non-vanishing. Secondly, there must be at least two heavy particles $X_k$ present in the model because if $k=m$ in (\ref{eq1:CP_aysm_final_form}), then $\text{Im}(h_{jk}^* h_{jm}h _{nm} h_{nk}^*) = \text{Im}(|h_{jk}|^2|h_{nk}|^2) = 0$ and no asymmetry can be generated. Thirdly, the $\text{Im}\left(I_\text{loop} \right)$ term demands that the mass of $X_k$ must be greater than the combined mass of the two intermediate state particles: $q_n$ and $\xi$. This is because the imaginary part of the internal loop function corresponds to the \emph{on-shell} contribution of the diagram, and hence kinematical restriction implies that $X_k$ must be heavy enough in order to give $\varepsilon_{CP} \neq 0$. Also, the dependence on $\text{Im}\left(I_\text{loop} \right)$ indicates why self-energy loop graphs of Fig.~\ref{fig1:CP_loops_Xq}b are actually relevant and can contribute to the overall asymmetry~\cite{Liu:1993tg}.

This  is the typical way in which $CP$ violation enters in many of the baryogenesis scenarios, including thermal leptogenesis (see Sec.~\ref{sec:lepto_main}). It will also forms the cornerstone of our discussion in Chapter~\ref{ch_work_BKL}.
\newpage

\section{Thermal leptogenesis}\label{sec:lepto_main}

The failure of the minimal SM to dynamically generate the correct amount of baryon asymmetry together with the fact that SM sphalerons strictly conserve the $B-L$ quantum number have prompted us to look for new physics that can violate lepton number $L$ when tackling the baryogenesis problem. As was hinted in the beginning of Sec.~\ref{subsec:seesaw1}, such new physics could be closely related to neutrino mass models. Indeed, if neutrinos are Majorana, then the induced dim-5 mass term of (\ref{eq1:nu_dim5_term}): $y^2 \overline{\ell}_L\,\phi\, \phi^T\ell_L^c /\Lambda$, will violate $L$ by two units. Therefore, it is natural to ask whether such lepton violating interactions can actually lead to the observed baryon asymmetry. 

In order to answer this, we must first tackle the question of whether, in principle, a lepton asymmetry can be turned into a baryon asymmetry during the primordial evolution, and if so, whether this can be achieved without introducing more new physics. Upon closer inspection, one realises that both are possible, and the mechanism that can implement this conversion is the SM sphalerons. Recall from Sec.~\ref{subsec:sphalerons} that the non-perturbative sphalerons will give rise to the 12-fermion interactions of (\ref{eq1:12-fermion_interactions}) when the ambient environment is hot enough. In addition, for temperatures in the range of $10^2 \lesssim T \lesssim 10^{12}$~GeV, these 12-fermion interactions are in thermal equilibrium, and hence, one can write down a relation among the various chemical potentials $\mu_X$ of particle species $X$ in the primordial plasma for this process as:
\begin{equation}\label{eq1:lepto_sphaleron_chem_pot}
 \sum_{\text{flavor }\alpha} 
 \left(\mu_{u_{\alpha L}}+2 \mu_{d_{\alpha L}} +  \mu_{\nu_{\alpha L}}\right) = 0 \;,
\end{equation}
where subscripts $u_{\alpha L}$, $d_{\alpha L}$ and $\nu_{\alpha L}$ denote LH up-type quark, down-type quark and neutrino species respectively. Moreover, since  electroweak symmetry is restored in this epoch of the early universe, the sum of hypercharge ($Y$) of all particle species must vanish, thus:
\begin{equation}
 Y = 
 2 \mu_{H}
 +\sum_{\text{flavor }\alpha} 
 \left(\mu_{u_{\alpha L}} + \mu_{d_{\alpha L}} + 2\mu_{u_{\alpha R}} 
  -\mu_{d_{\alpha R}} -\mu_{e_{\alpha R}} -\mu_{e_{\alpha L}} -\mu_{\nu_{\alpha L}}
 \right)
  =0\;, \label{eq1:lepto_Y_chem_pot}
\end{equation}
where $H$ is a generic symbol for either one of four degrees of freedom of the complex doublet Higgs, $\phi^+,\phi^0,\phi^-,\phi^{0*}$, while $u_{\alpha R}$, $d_{\alpha R}$ and $e_{\alpha R}$ are the RH up-type quark, down-type quark and charged lepton species respectively, and $e_{\alpha L}$ denotes the LH charged lepton species.
In (\ref{eq1:lepto_Y_chem_pot}), we have implicitly summed over colors for the quarks and set $\mu_{\phi^{+}}=\mu_{\phi^{0}}=\mu_{\phi^{-}}=\mu_{\phi^{0*}}\equiv\mu_{H}$ since they belong to the same doublets. Similarly, we expect $\mu_{u_{\alpha L}} =\mu_{d_{\alpha L}}$ and $\mu_{e_{\alpha L}} =\mu_{\nu_{\alpha L}}$ by the same argument.
Depending on the temperature of the plasma, some or all of the following SM Yukawa interactions can also be in equilibrium:
\begin{align}
 \phi^0\leftrightarrow \bar{d}_{\alpha R}\; d_{\beta L}\;,\qquad &\qquad
 \phi^+\leftrightarrow \bar{d}_{\alpha R}\; u_{\beta L}\;,\nonumber\\
 \phi^{0*}\leftrightarrow \bar{u}_{\alpha R}\; u_{\beta L}\;,\qquad &\qquad
 \phi^{-}\leftrightarrow \bar{u}_{\alpha R}\; d_{\beta L}\;,\nonumber\\
 \phi^0\leftrightarrow \bar{e}_{\alpha R}\; e_{\beta L}\;,\qquad &\qquad
 \phi^+\leftrightarrow \bar{e}_{\alpha R}\; \nu_{\beta L}\;.
  \label{eq1:lepto_yuk_interactions}
\end{align}
When all the processes (with arbitrary combination in flavors $\alpha$ and $\beta$) in (\ref{eq1:lepto_yuk_interactions}) are in thermal equilibrium, all LH or RH up-quark, down-quark, charged lepton and LH neutrino flavor states will be adequately mixed, and one may then assume that $\mu_{f_{\alpha L,R}} = \mu_{f_{\beta L,R}}$ for all $\alpha, \beta$ where $f$ denotes $u,d,e$ or $\nu$. Using this approximation and equating the chemical potentials, one gets three more relations:
\begin{equation}\label{eq1:lepto_yuk_chem_pot}
 \mu_{q_{L}} - \mu_{d_{R}} -\mu_{H}=0\;,\;\;\;\;
 \mu_{q_{L}} - \mu_{u_{R}} +\mu_{H}=0\;,\;\;\;\;
 \mu_{\ell_{L}} - \mu_{e_{R}} -\mu_{H}=0\;,
\end{equation}
where we have dropped the flavor index and set $\mu_{q_{L}} \equiv \mu_{u_{L}}=\mu_{d_{L}}$ and $\mu_{\ell_{L}} \equiv \mu_{\nu_{L}}=\mu_{e_{L}}$. 

If we assume that all particle species involved are approximately massless and behave as a weakly coupled relativistic gas at this temperature, then we can invoke the standard result \cite{kolb_turner, harvey_turner_PRD42}: 
\begin{equation}\label{eq1:lepto_n_densities_chem_pot}
 n_X - \bar{n}_X = \frac{g_i\,T^3}{6}\left[ \frac{\mu_X}{T} + \order{\frac{\mu_X^3}{T^3}}\right]
 \;,
\end{equation}
which relates the asymmetry in the fermion and antifermion number densities to their chemical potentials \footnote{Here, $g_i$ denotes the number of internal degrees of freedom.} and subsequently write down the relations for baryon and lepton number:
\begin{align}
 B &= \sum_{\text{flavor}} 
 \left(
 2 \mu_{q_{L}}+ \mu_{d_{L}} +  \mu_{d_{R}}
 \right)\;, \label{eq1:lepto_B_chem_pot}\\
 L &= \sum_{\text{flavor}} 
 \left(
 2 \mu_{\ell_{L}}+  \mu_{e_{R}}
 \right)\;. \label{eq1:lepto_L_chem_pot}
\end{align}
Solving (\ref{eq1:lepto_sphaleron_chem_pot}), (\ref{eq1:lepto_Y_chem_pot}), (\ref{eq1:lepto_yuk_chem_pot}), (\ref{eq1:lepto_B_chem_pot}) and (\ref{eq1:lepto_L_chem_pot}) simultaneously for $B$ and expressing the answer in terms of $B-L$ or $L$, we arrive at the key result \cite{harvey_turner_PRD42, khle_shaposh_et_al}
\begin{equation}\label{eq1:lepto_B_proj_B-L}
 B = \frac{28}{78}\, (B-L) = -\frac{28}{51}\, L \;,
\end{equation}
from which one can conclude that an initial $B-L$ asymmetry can be partially converted into a $B$ asymmetry by sphalerons and other SM processes.

Given this relationship between $B$ and $L$, we can indeed tackle the problem of baryogenesis by first finding a solution for \emph{leptogenesis}. As mentioned before, models with massive Majorana neutrinos naturally provide a source of lepton violation---~an essential criterion for generating a $L$ asymmetry dynamically, and thus it is particularly fruitful to study them in this context because the new physics introduced to give neutrinos a mass may simultaneously explain the cosmic baryon asymmetry.

In this work, we are especially interested in the leptogenesis scenario involving type~I seesaw models \cite{fukugita_yanagida_86} because, in our opinion, it presents the most ``elegant'' solution to both the smallness of neutrino masses and the observed baryon-to-photon ratio, while it only requires a rather modest extension of the SM. Although we shall not discuss the leptogenesis implications of the other neutrino mass generation methods (see Sec.~\ref{subsec:seesaw23} and \ref{subsec:radcorr}), it should be added in passing that leptogenesis based on type II \cite{eg_type_II_lepto, eg_type_II_lepto_also}, type III \cite{eg_type_III_lepto} seesaw, as well as Babu-Zee type models \cite{eg_radcorr_lepto} are also possible \footnote{Henceforth, we shall often refer to leptogenesis in type I seesaw models as simply ``leptogenesis''.}.

\subsection{Leptogenesis with hierarchical RH neutrinos}\label{subsec:lepto_N1}

The classic leptogenesis scenario of Fukugita and Yanagida \cite{fukugita_yanagida_86}, which we shall often refer to as the ``standard'' case in our subsequent discussions, involves taking the type I seesaw Lagrangian of (\ref{eq1:L_mass_type1}) with (usually) three heavy RH Majorana neutrinos, 
so that the $L$ violating Yukawa interactions between the RH neutrinos and the ordinary LH leptons can generate a $B-L$ asymmetry during the primordial times. Furthermore, it is customary to assume that the spectrum of the RH neutrino masses in this scenario is hierarchical, and therefore the asymmetry created will be dominated by the decays of the lightest RH neutrinos (denoted $N_1$) due to the efficient washout of any $N_{2,3}$-generated asymmetries by $N_1$ mediated $\Delta L\neq 0$ scattering processes in equilibrium.

\begin{figure}[tb]
\begin{center}
 \includegraphics[width=\textwidth]{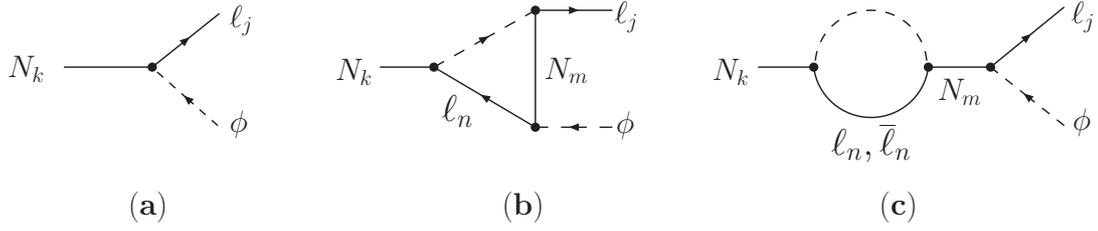}
\end{center} 
 \caption{The {\bf (a)} tree-level,  {\bf (b)} one-loop vertex correction,  and {\bf (c)} one-loop self-energy correction graphs for the decay: $N_k \rightarrow \ell_j\,\overline{\phi}$.}
 \label{fig1:lepto_std_decay_graphs}
\end{figure}

To enunciate these ideas, we begin by rewriting Lagrangian (\ref{eq1:L_mass_type1}) in the mass eigenbasis of the heavy RH neutrinos (and with some notation changes \footnote{For the ease of presentation, we shall disregard the small subtleties involving the precise definitions of $N$ or $\nu$ used before when explaining the seesaw mechanism (see Sec.~\ref{subsec:seesaw1}), and simply denote the heavy RH Majorana neutrinos with $N\equiv {\nu_R}' + ({\nu_R}')^c$ where ${\nu_R}'$ is the
mass eigenstate after the change of basis from $\nu_R$.}):
\begin{equation}\label{eq1:lepto_std_Lag}
 \mathcal{L}_\text{int} =
 - y_{\alpha\beta}\, \overline{\ell}_\alpha\, \widetilde{\phi}\, e_{\beta}
 - h_{jk}\, \overline{\ell}_j\, \phi\, N_{k}
 -\frac{1}{2}\, \overline{N}_k\,M_k\, N_k 
 +\text{h.c.}\;,
\end{equation}
where flavor indices $\alpha, \beta, j$ can be one of $e,\mu$ or $\tau$, and $k=1,2,3$ are labels for the lightest to heaviest RH neutrinos (with mass $M_k$). The $SU(2)_L$ doublets: $\ell_\alpha = (\nu_L, e_L)^T_\alpha$ and $\phi = (\phi^0, \phi^-)^T$ have their usual meanings, with $\widetilde{\phi} = i\sigma_2\phi^*$ being the charge conjugate Higgs. The Yukawa couplings $h_{jk}\, \overline{\ell}_j\, \phi\, N_{k}$ in (\ref{eq1:lepto_std_Lag}) can then induce heavy RH neutrino decays via two channels:
\begin{equation}\label{eq1:lepto_std_decay_channels}
 N_k \rightarrow
 \left\{ 
  \begin{matrix}
  \;\;\; \ell_j + \overline{\phi} \;,\\
  \;\;\; \overline{\ell}_j + \phi \;, 
  \end{matrix} 
 \right.
\end{equation}
which violate lepton number by one unit. All Sakharov's conditions for leptogenesis will be satisfied if these decays also violate $CP$ and go out of equilibrium at some stage during the evolution of the early universe. As shown in Sec.~\ref{subsec:CP-violation}, the requirement for $CP$ violation means that coupling matrix $h$ in (\ref{eq1:lepto_std_Lag}) must be complex and the mass of $N_k$ must be greater than the combined mass of $\ell_j$ and $\phi$, so that interferences between the tree-level process (Fig.~\ref{fig1:lepto_std_decay_graphs}a) and the one-loop corrections (Fig.~\ref{fig1:lepto_std_decay_graphs}b, c) with on-shell intermediate states will be nonzero. Clearly, both of these are possible as type I seesaw mechanism naturally implies a very large $M_k$ in order to induce small LH neutrino masses, while it does not forbid the presence of $CP$ violating phases in the RH neutrino sector. The condition of thermal non-equilibrium is achieved when the expansion rate of the universe exceeds the decay rate of $N_k$. One may quickly check using  (\ref{eq1:hubble_rate}) that this is actually possible for a wide range of mass $M_k$.

So, with all the essential ingredients for leptogenesis positively identified, the remaining question is whether quantitatively this model can generate the correct amount of asymmetry, and if so, what are the constraints (if any) on the parameter space. To this end, using the definition of (\ref{eq1:CP_asym_general}) for $L$, we begin by writing down the $CP$ asymmetry in the lepton number production due to $N_k$ decays: 
\begin{equation}\label{eq1:lepto_std_CP_general_kj}
 \varepsilon_{kj} = 
   \frac{\Gamma(N_k\rightarrow \ell_j\,\overline{\phi})
   -\Gamma(N_k\rightarrow \overline{\ell_j}\,\phi)}
        {\Gamma(N_k\rightarrow \ell_j\,\overline{\phi})
        +\Gamma(N_k\rightarrow \overline{\ell_j}\,\phi)} \;.
\end{equation}
If we assume that the evolution of the $L$ asymmetry is ``flavor blind'' (a condition that we shall relax in the next subsection), then we are only interested in the quantity after summing over lepton flavor $j$. In addition, for hierarchical RH neutrinos, we have the $N_1$-dominated scenario, and therefore, we can set $k=1$. Explicit calculation of the interference terms will then result in \cite{Covi:1996wh, CP_calculations_Flanz, CP_calculations_BP}: \footnote{We will illustrate how to calculate this result explicitly as part of the discussion in Chapter~\ref{ch_work_BKL}.}
\begin{align}
  \varepsilon_{1} 
  &= 
  \frac{1}{8\pi}\sum_{m\neq 1} 
              \frac{\textrm{Im}\left[(h^\dagger h)^2_{1m}\right]}
               {(h^\dagger h)_{11}}
             \left\{f_V\left(\frac{M_m^2}{M_1^2}\right)+f_S\left(\frac{M_m^2}{M_1^2}\right)
             \right\}
   \;,\label{eq1:lepto_std_CP_general_N1}\\
\intertext{where $f_V(x)$ and $f_S(x)$ are given by}
    f_V(x) &= \sqrt{x}\left[1-(1+x)\ln\left(\frac{1+x}{x}\right)\right]\quad\textrm{and}\quad
 f_S(x) = \frac{\sqrt{x}}{1-x}\;
  \label{eq1:lepto_std_CP_general_N1_VS_fn}
\end{align}
which denote the vertex and self-energy contributions respectively. The tree-level $N_1$ decay rate (at $T=0$) used to calculate the denominator of (\ref{eq1:lepto_std_CP_general_kj}) with $j$ summed is given by: 
\begin{equation}\label{eq1:lepto_std_N1_tree}
 \Gamma(N_1\rightarrow \ell\,\overline{\phi})
 \equiv
 \Gamma(N_1\rightarrow \overline{\ell}\,\phi)
 = \frac{(h^\dagger h)_{11}}{16\pi}\,M_1\;.
\end{equation}

Suppose that $|h_{jk}| \leq |h_{33}|$ for all $j$ and $k$, then in the hierarchical limit of $M_1 \ll M_{2,3}$, the seesaw relation gives:
\begin{equation}\label{eq1:lepto_std_eps_est_seesaw}
 m_3 \simeq \frac{|h_{33}|^2\,\vev{\phi}^2}{M_3}\;,
\end{equation}
where $m_3$ is mass of the heaviest LH neutrino. Assuming these conditions, and using the fact that
\begin{equation}\label{eq1:lepto_std_eps_fVS_limit}
 \left|f_V(x)+f_S(x)\right| \simeq \frac{3}{2\sqrt{x}}
 \;,\qquad \text{for } x \gg 1\;, 
\end{equation}
one can estimate the $CP$ asymmetry as
\begin{align}\label{eq1:lepto_std_eps_est_1}
 |\varepsilon_1| &\simeq \frac{3}{16\pi}\; |h_{33}|^2 
   \left(\frac{M_1}{M_3}\right) \sin\,\delta_N \;, \\
   &= \frac{3}{16\pi}\,
   \frac{m_3\,M_1}{\vev{\phi}^2}\; \sin\,\delta_N \;,
   \label{eq1:lepto_std_eps_est_2}
\end{align}
where in the last line we have used (\ref{eq1:lepto_std_eps_est_seesaw}). The quantity: $\sin\,\delta_N$, is a measure of the amount of $CP$ violation in the decay with $\delta_N = \text{arg}\left[(h^\dagger h)_{13}^2\right]$ which is in general different from the $CP$ phase appearing in neutrino oscillations. Relation (\ref{eq1:lepto_std_eps_est_2}) implies that the size of $|\varepsilon_1|$ cannot be arbitrarily large for a given $M_1$. Taking $m_3 \simeq 0.05$~eV and $\vev{\phi} \simeq 174$~GeV, one gets a useful ballpark estimate of the maximum $CP$ asymmetry as
\begin{equation}
 |\varepsilon_1|^\text{max}
  \approx
   10^{-6} \left(\frac{M_1}{10^{10} \text{ GeV}}\right) \;.
   \label{eq1:lepto_std_eps_est_3}
\end{equation}
Within the type I seesaw paradigm, this result actually holds in general as long as the LH neutrinos are strongly hierarchical \cite{davidson_ibarra_eps_bound}.

Taking Eq.~(\ref{eq1:lepto_std_eps_est_3}) at face value, it appears that leptogenesis should be able to produce the correct baryon-to-photon ratio of $\eta_B \simeq 10^{-10}$ quite easily by simply tuning $\delta_N$ and $M_1$. While this freedom to adjust $\varepsilon_1$ (partly) ensures the eventual success of baryogenesis via leptogenesis quantitatively, the solution to the problem is actually quite subtle and involves carefully tracking the evolution of the $N_1$'s abundance in the thermal plasma, as well as the $B-L$ asymmetry they generate.  

It is not difficult to visualise the reasons for these subtleties. Firstly, to have $N_1$ decays happening, the plasma must contain a nonzero amount of them. If inflation or some other mechanism implies that the initial $N_1$ density is zero, then one needs to work out whether the $N_1$-scattering processes involving SM particles would be  strong enough to generate a sufficient amount of $N_1$'s so that their subsequent decays can give rise to the correct asymmetry. Secondly, if the $\Delta L\neq 0$ interactions such as the inverse decay ($\ell\,\phi\rightarrow N_1$) are too strong, they may washout any asymmetry already generated. Hence, depending on when these processes go in (or out) of thermal equilibrium, they may have a substantial effect on the final asymmetry. This interplay between $N_1$ generation and washout effects demands a closer study into the evolution of all particle species involved, and as a result, one must solve the relevant Boltzmann transport equations.

In general, the Boltzmann equation for particle species $a$ has the form:
\begin{equation}
 \widehat{O}_L\, f_a(p)
  = -\frac{1}{2} \,C\left[f_a(p)\right]\;, \label{eq1:lepto_std_boltzmann_form}
\end{equation}
where $\widehat{O}_L$ denotes some Liouville operator, $f_a(p)$ is the phase-space density function of species $a$, and $C\left[f_a(p)\right]$ is the so-called ``collision'' integral which is defined by \footnote{We have made the simplification that all species obey Maxwell-Boltzmann statistics. See e.g. \cite{kolb_turner, kolb_wolfram}.}  
\begin{align}
  C\left[f_a(p)\right]
  \equiv
  \sum_{a X\leftrightarrow Y} 
  & \int dw_X\, dw_Y (2\pi)^4\delta^4(p_a+p_X-p_Y)
   \nonumber\\
 &\times\left[f_a f_X |\mathcal{M}(aX\rightarrow Y)|^2
  -f_Y |\mathcal{M}(Y\rightarrow aX)|^2\right]\;,
  \label{eq1:lepto_std_collision_intg}
\end{align}
where $X$ and $Y$ denote some multiparticle states and all allowed processes $a\,X \leftrightarrow Y$ are summed. The transition amplitude with all internal degrees of freedom in both initial and final states averaged, and with the appropriate symmetry factors accounted for, is represented by $|\mathcal{M}(\ldots)|^2$, whilst
\begin{equation}\label{eq1:lepto_std_BE_abbr}
 p_{X,Y} = \sum_{b\in X,Y}\,p_{b}\;,
 \quad
 f_{X,Y}=\prod_{b\in X,Y}\, f_b\;, \text{ and }
 \quad
 dw_{X,Y} = \prod_{b\in X,Y}\, dw_b
 \;,
\end{equation}
with the measure given by
\begin{equation}\label{eq1:lepto_std_BE_measures}
 dw_b = \frac{g_b\,d^3p_b}{2E_b(2\pi)^3}
 \;,
\end{equation}
where $g_b$ and $E_b$ are the number of internal degrees of freedom and the total energy of species $b$ respectively. In calculations, one is concerned with the number density of a particle species which is related to the phase-space distribution via
\begin{equation}\label{eq1:lepto_std_BE_n_density}
 n_b = \frac{g_b}{(2\pi)^3}\int d^3p_b \; f_b(p) \;.
\end{equation}
So, integrating (\ref{eq1:lepto_std_boltzmann_form}) to put it in terms of number density and using the fact that the Liouville operator representing an isotropic and homogeneous (Robertson-Walker) universe is \cite{kolb_turner, bernstein}
\begin{equation}\label{eq1:lepto_std_BE_RW_Lop}
 \widehat{O}_L\,f_a(p) \equiv E_a\,\frac{\partial f_a}{\partial t}-H |\vec p_a|^2\,
 \frac{\partial f_a}{\partial E_a} \;,
\end{equation}
where $H$ is the Hubble parameter, one can express the LHS of (\ref{eq1:lepto_std_boltzmann_form}) as
\begin{align}
 \frac{1}{2E_a}\frac{g_a}{(2\pi)^3}\int d^3p_a\; \widehat{O}_L\, f_a(p)
 &=\frac{g_a}{2\,(2\pi)^3}\int d^3p_a\;
   \left[\frac{\partial f_a}{\partial t}-H \frac{|\vec p_a|^2}{E_a}
   \frac{\partial f_a}{\partial E_a}\right]\;, \nonumber\\
 &=\frac{1}{2}\left[\frac{d n_a}{d t} +3Hn_a\right]
 \;,\label{eq1:lepto_std_BE_LHS_general}
\end{align}
where we have also divided through by $2E_a$. Similarly, the RHS of (\ref{eq1:lepto_std_boltzmann_form}) becomes
\begin{align}
  -\frac{1}{2}\, C\left[f_a(p)\right]
  &=
   -\frac{1}{2} \sum_{a X\leftrightarrow Y} 
   \int dw_a\,dw_X\, dw_Y (2\pi)^4\delta^4(p_a+p_X-p_Y)
   \nonumber\\
 &\qquad\qquad\;\;\;\;\;\;
 \times\left[f_a f_X |\mathcal{M}(aX\rightarrow Y)|^2
  -f_Y |\mathcal{M}(Y\rightarrow aX)|^2\right]\;,\\
  &=
   -\frac{1}{2} \sum_{a X\leftrightarrow Y} 
   \int dw_a\,dw_X\, dw_Y (2\pi)^4\delta^4(p_a+p_X-p_Y)
   \nonumber\\
 &\qquad\qquad\;\;\;\;\;\;
 \times\left[\frac{n_a}{n_a^\text{eq}} f_a^\text{eq}
  \left(\prod_{b\in X} \frac{n_b}{n_b^\text{eq}} f_b^\text{eq}\right)
   |\mathcal{M}(aX\rightarrow Y)|^2 \right.\nonumber\\
 &\qquad\qquad\;\;\;\;\;\;\qquad\qquad\;\;\;
  \left.
  -\left(\prod_{c\in Y} \frac{n_c}{n_c^\text{eq}} f_c^\text{eq}\right) |\mathcal{M}(Y\rightarrow aX)|^2\right]\;,
  \label{eq1:lepto_std_BE_RHS_general}
\end{align}
where we have assumed that all particles are in kinetic equilibrium so that $f_i \equiv (n_i/n_i^\text{eq}) f_i^\text{eq}$ with $f_i^\text{eq} = e^{-E_i/T}$ being the equilibrium distribution for Maxwell-Boltzmann statistics and $n_i^\text{eq}$ the corresponding number density.

It is convenient to define the thermally averaged reaction density as
\begin{align}
 \gamma(aX\rightarrow Y) &= \int dw_a\, dw_X\, dw_Y\,
   (2\pi)^4\delta^4(p_a+p_X-p_Y)\,
   \, f_a^{eq} f_X^\text{eq}\,  
       |\mathcal{M}(aX\rightarrow Y)|^2\;, \nonumber\\
   &\equiv n_a^{eq} \left(\prod_{b\in X} n_b^{eq}\right)
       \langle\sigma(aX\rightarrow Y)\,|\vec{v}|\rangle\;,
   \label{eq1:lepto_std_BE_averaged_density}
\end{align}
where $\vev{\sigma |\vec v|}$ is the thermally averaged cross section times the relative velocity of the interacting particles. For decays, $\gamma(a\rightarrow Y)$ is related to the more familiar thermally averaged reaction rate $\vev{\Gamma(a\rightarrow Y)}$ via
\begin{align}\label{eq1:lepto_std_BE_averaged_rate}
 \gamma(a\rightarrow Y) &= n^\text{eq}_a\,\vev{\Gamma(a\rightarrow Y)}
 \;,\\
 &= n^\text{eq}_a\, \frac{\mathcal{K}_1(z)}{\mathcal{K}_2(z)}\,
 \Gamma_{0}(a\rightarrow Y)
 \;,\quad z\equiv \frac{m_a}{T}\;,
 \label{eq1:lepto_std_BE_averaged_rate0}
\end{align}
where $\Gamma_{0}(a\rightarrow Y)$ is the decay rate at temperature $T=0$ (usually obtained from quantum field theory calculations), $m_a$ is the mass of the particle $a$ and $\mathcal{K}_n(z)$ denotes the $n$th order modified Bessel function of the second kind. In the case of two-body scattering $\gamma(a b\rightarrow Y)$, one has
\begin{equation}\label{eq1:lepto_std_BE_averaged_2body}
 \gamma(a b \rightarrow Y)
  = \frac{T}{64\pi^4} \int_{(m_a+m_b)^2}^{\infty} ds\, \sqrt{s} \;
  \mathcal{K}_1(\sqrt{s}/T)\; \widehat{\sigma}(s)\;,
\end{equation}
where $s$ is the squared centre-of-mass energy and $\widehat{\sigma}(s)$ is the reduced cross section for the process with $s\, \widehat{\sigma}(s) = 8 \left[(p_a\cdot p_b)^2 - m_a^2\, m_b^2 \right] \sigma(s)$, where $\sigma(s)$ is the usual total cross section.

Using definition (\ref{eq1:lepto_std_BE_averaged_density}) and equating (\ref{eq1:lepto_std_BE_LHS_general}) and (\ref{eq1:lepto_std_BE_RHS_general}), the Boltzmann equation for particle evolution in the expanding universe becomes:
\begin{equation}\label{eq1:lepto_std_BE_final_form}
 \frac{d n_a}{d t} +3Hn_a =
 -\sum_{a X\leftrightarrow Y}\left[
   \frac{n_a}{n_a^\text{eq}} \left(\prod_{b\in X}\frac{n_b}{n_b^\text{eq}}\right)
   \gamma(aX\rightarrow Y)
   -\left(\prod_{c\in Y}\frac{n_c}{n_c^\text{eq}}\right)
   \gamma(Y\rightarrow aX)
 \right]\;,
\end{equation}
with the term $3 H n_a$ signifying the change in $n_a$ due to the expansion.

\begin{figure}[tb]
\begin{center}
 \includegraphics[width=\textwidth]{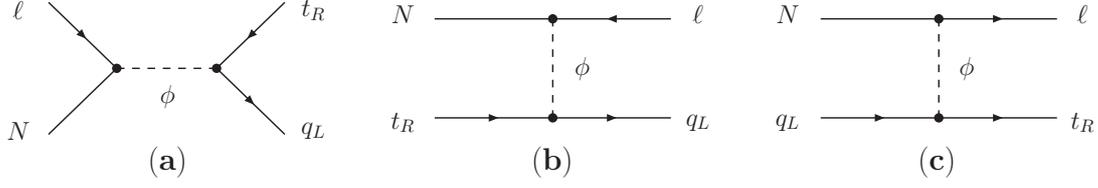}
\end{center} 
 \caption{The $\Delta L =\pm 1$ processes that can influence $n_{N_1}$ and $n_{B-L}$: {\bf (a)} $s$-channel scattering $N \ell \leftrightarrow q_L \bar{t}_R$, {\bf (b)} $t$-channel scattering $N t_R \leftrightarrow q_L \bar{\ell}$, {\bf (c)} $t$-channel scattering $N q_L \leftrightarrow t_R \ell$. Here $q_L$ denotes the 3rd generation of the quark doublet.}
 \label{fig1:lepto_std_s_t_scattering}
\end{figure}

\begin{figure}[t]
\begin{center}
 \includegraphics[width=0.75\textwidth]{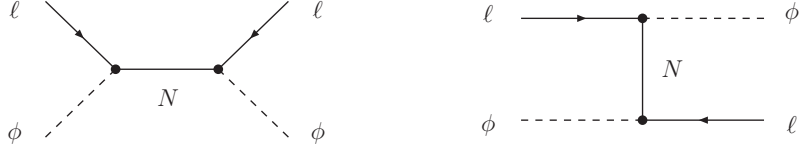}
\end{center} 
 \caption{The $\Delta L = \pm 2$ $s$- and $t$-channel  scattering processes mediated by $N$.}
 \label{fig1:lepto_std_N_scattering}
\end{figure}

Returning to the leptogenesis scenario, we identify that the quantities of interest are the number densities of $N_1$ and $B-L$ (with baryon and lepton number defined in the usual way). So, considering all processes up to second order, we note that besides the tree-level interaction ($N\leftrightarrow \ell \bar{\phi}$) of Fig.~\ref{fig1:lepto_std_decay_graphs}a, there are $s$-channel 
$N \ell \leftrightarrow q_L \bar{t}_R$ (Fig.~\ref{fig1:lepto_std_s_t_scattering}a) and $t$-channel
$N t_R \leftrightarrow q_L \bar{\ell}, N q_L \leftrightarrow t_R \ell$ (Fig.~\ref{fig1:lepto_std_s_t_scattering}b, c)  scattering processes that can alter the abundance of $N_1$. For the evolution of $B-L$, in addition to these, there are also $\Delta L = \pm 2$ scattering processes mediated by $N_1$ (Fig.~\ref{fig1:lepto_std_N_scattering}) which can be important.

Since $CP$ violation in the decay of $N_1$ is responsible for creating a nonzero $B-L$, it is convenient to parametrize the thermally averaged decay densities such that they encapsulate this information:
\begin{align}\label{eq1:lepto_std_BE_new_rate_D1}
 &\gamma(N_1\rightarrow \ell\bar{\phi}) \equiv \gamma(\bar{\ell}\phi\rightarrow N_1) = (1+\varepsilon_1) \gamma_D\;,
 \\
 &\gamma(N_1\rightarrow \bar{\ell}\phi) \equiv \gamma(\ell\bar{\phi}\rightarrow N_1)=
 (1-\varepsilon_1) \gamma_D\;,
 \label{eq1:lepto_std_BE_new_rate_D2}
\end{align}
where $\gamma_D$ is the tree-level decay density which is related to the rate given in (\ref{eq1:lepto_std_N1_tree}). It can be easily checked that 
\begin{equation}\label{eq1:lepto_std_BE_new_rate_D_check}
 \varepsilon_1 =
 \frac{\gamma(N_1\rightarrow \ell\bar{\phi})-\gamma(N_1\rightarrow \bar{\ell}\phi)}
 {\gamma(N_1\rightarrow \ell\bar{\phi})+\gamma(N_1\rightarrow \bar{\ell}\phi)}
 \;,
\end{equation}
which is consistent with the definition of $\varepsilon_1$ given in (\ref{eq1:lepto_std_CP_general_kj}).

Assuming that all particles except $N_1, \ell$ and $\bar{\ell}$ are in thermal equilibrium (i.e. $n_b \equiv n_b^\text{eq}$), and that for all scattering processes $\gamma(aX\rightarrow Y) \equiv \gamma(Y\rightarrow aX)$ to good approximation, then one can put all these together into (\ref{eq1:lepto_std_BE_final_form}) and obtain the following evolution equation for $n_{N_1}$: 
\begin{equation}\label{eq1:lepto_std_BE_N1_raw}
 \frac{d n_{N_1}}{d t} +3H n_{N_1} =
 -2 \left[
 \frac{n_{N_1}}{n_{N_1}^\text{eq}} -1
   \right] (\gamma_D + \gamma_{\phi,s} + 2 \gamma_{\phi,t})
   + \order{\varepsilon_1, \frac{\mu_\ell}{T}}
   \;,
\end{equation}
where $\gamma_{\phi,s}$ and $\gamma_{\phi,t}$ are thermally averaged reaction densities for the $s$-channel and $t$-channel processes respectively, and $\mu_\ell$ is the chemical potential for $\ell$. The overall factor of 2 in front reflects the fact that there are two channels linking $N_1$ to $\ell$ and $\phi$ for each graph, while the extra factor of 2 for $\gamma_{\phi,t}$ is due to the presence of two inequivalent $t$-diagrams \footnote{See Appendix~\ref{app:a_BE_N1} for more details on the derivation of Eq.~(\ref{eq1:lepto_std_BE_N1_raw}).}.

In the case of $n_{B-L}$, the situation is slightly more subtle as there is a non-negligible $CP$ violating contribution coming from the $\Delta L =\pm 2$ $s$-channel scattering mediated by an \emph{on-shell} $N_1$, which must be properly taken into account \cite{kolb_wolfram, BDP_pedestrain,Giudice_0310123}. To this end, one needs to subtract the term corresponding to the real intermediate state for $N_1$ from the \emph{full} cross section  of the $\ell \bar{\phi} \leftrightarrow \bar{\ell} \phi$ processes \footnote{Recent discussions on how to handle this can be found in \cite{BDP_pedestrain, Giudice_0310123}.}. Therefore, using the notations of (\ref{eq1:lepto_std_BE_new_rate_D1}) and (\ref{eq1:lepto_std_BE_new_rate_D2}), we can write down the reaction densities for these $\Delta L =\pm 2$ scatterings with the on-shell part subtracted as
\begin{align}
 \widetilde{\gamma}(\ell \bar{\phi} \rightarrow \bar{\ell} \phi)_s 
 &= \gamma(\ell \bar{\phi} \rightarrow \bar{\ell} \phi)_s + \varepsilon_1 \gamma_D
 \equiv \gamma_{N,s} + \varepsilon_1 \gamma_D
 \;,\label{eq1:lepto_std_BE_sub_Nrate1} \\
  \widetilde{\gamma}(\bar{\ell} \phi \rightarrow \ell \bar{\phi})_s 
 &= \gamma(\bar{\ell} \phi \rightarrow \ell \bar{\phi})_s - \varepsilon_1 \gamma_D
 \equiv \gamma_{N,s} - \varepsilon_1 \gamma_D
 \;,\label{eq1:lepto_std_BE_sub_Nrate2}
\end{align}
where subscript $s$ denotes $s$-channel. To obtain the expression for $n_{B-L}$, one must first write down the Boltzmann equation for $n_{\bar\ell}$ and then subtract from it the corresponding one for $n_\ell$. Applying the same assumptions as before and keeping all terms up to $\order{\varepsilon_1}$, the evolution equation for $n_{B-L}$ becomes \footnote{See Appendix~\ref{app:a_BE_B-L} for the complete derivations.}
\begin{equation}\label{eq1:lepto_std_BE_B-L_raw}
 \frac{d n_{B-L}}{d t} +3H n_{B-L} =
   -2\,\varepsilon_1 \left[\frac{n_{N_1}}{n_{N_1}^\text{eq}} -1\right]\gamma_D
   -\frac{n_{B-L}}{n_{\ell}^\text{eq}}\,
   \gamma_W
   + \order{\varepsilon_1^2 , \frac{\mu_\ell}{T}}
   \;,
\end{equation}
where the reaction density for ``washout'' is given by
\begin{equation}\label{eq1:lepto_std_BE_washout_raw}
 \gamma_W \equiv
  \gamma_D + \frac{n_{N_1}}{n_{N_1}^\text{eq}} \gamma_{\phi,s}
   + 2\gamma_{\phi,t} + 2\gamma_{N,s} + 2\gamma_{N,t}
   \;.
\end{equation}
The precise expressions for the amplitudes and cross sections which give rise to these reaction densities can be found in \cite{BDP_pedestrain, Giudice_0310123,luty, Plum_Zphy_NPB,BDP_0205349_CMB} \footnote{It should be mentioned that \cite{luty, Plum_Zphy_NPB,BDP_0205349_CMB} have miscalculated the $CP$ violation contribution from the real intermediate state term of the $\ell\bar{\phi} \rightarrow \bar{\ell}\phi$ processes, hence leading to a small overestimation of the washout effects.}. In calculations however, it is often better to consider the time evolution of the particle number $\mathcal{N}_b$ in some portion of the comoving volume $R^3(t)$ which contains only one photon before the onset of leptogenesis, rather than number density $n_b$. This change of variables will automatically take care of the effect due to expansion. Alternatively, one can instead consider the normalised quantity: $Y_b = n_b/s$, where $s$ is the entropy density, as many authors prefer to use (see for example \cite{kolb_turner}). 
So re-expressing (\ref{eq1:lepto_std_BE_N1_raw}) and (\ref{eq1:lepto_std_BE_B-L_raw}) in terms of $\mathcal{N}_{N_1}$, $\mathcal{N}_{B-L}$ while replacing the thermally averaged reaction densities with their corresponding reaction rates, we get \cite{BDP_pedestrain, BDP_0205349_CMB}
\begin{align}
 \frac{d\mathcal{N}_{N_1}}{dt} = -2(\Gamma_D + \Gamma_{\phi,s}+ 2\Gamma_{\phi,t})
  (\mathcal{N}_{N_1} - \mathcal{N}_{N_1}^\text{eq})
 \;, \label{eq1:lepto_std_BE_N1_dt}
  \\ 
  \frac{d\mathcal{N}_{B-L}}{dt} = -2\varepsilon_1 \Gamma_D
          (\mathcal{N}_{N_1} - \mathcal{N}_{N_1}^\text{eq})
   - \Gamma_W \mathcal{N}_{B-L}
 \;,
  \label{eq1:lepto_std_BE_B-L_dt}
\end{align}
where
\begin{equation}\label{eq1:lepto_std_BE_washout_dt}
\Gamma_W \equiv \Gamma_D + \frac{\mathcal{N}_{N_1}}{\mathcal{N}_{N_1}^\text{eq}} \Gamma_{\phi,s}  + 2\Gamma_{\phi,t} +2\Gamma_{N,s} +2\Gamma_{N,t}\;,
\end{equation}
and we have dropped the $\vev{..}$ around the rates for brevity. From Eq.~(\ref{eq1:lepto_std_BE_B-L_dt}), one can see that in thermal equilibrium, i.e. $\mathcal{N}_{N_1} =\mathcal{N}_{N_1}^\text{eq}$, any asymmetry in $B-L$ will be washed out, confirming the earlier claim. 

This coupled set of Boltzmann equations may be solved numerically or (semi-)analytically by asymptotic methods. Either way, the conclusion is that for a wide range of seesaw neutrino parameters, a nonzero excess of $B-L$ can be generated \cite{BDP_pedestrain, BDP_0205349_CMB}. Explicitly, if one expresses the maximum baryon-to-photon ratio generated as 
\begin{equation}\label{eq1:lepto_std_CMB_compare_basic}
  \eta_B^\text{max} \simeq 
   0.96\times 10^{-2} \,|\varepsilon_1|\, \kappa_\text{f}^\text{max}
   \;,
\end{equation}
with $\kappa_\text{f}^\text{max}$ denoting the maximum final efficiency factor obtained from solving the Boltzmann equations, and the pre-factor of $0.96\times 10^{-2}$ coming from the dilution due to imperfect sphaleron conversion and photon production before recombination~\footnote{See Chapter~\ref{ch_work_LV} for more details.}, then one may directly restrict the possible neutrino parameter space for successful baryogenesis via $|\varepsilon_1|$ (and to some degree $\kappa^\text{f}$ because the reaction rates depend on the mass of $N_1$ and the Yukawas). In the best case scenarios where a maximum efficiency factor of about $\kappa_\text{f}^\text{max} \approx 0.18$ is achieved \cite{BDP_pedestrain, BDP_0205349_CMB}, and assuming strongly hierarchical LH neutrinos, then one obtains a lower bound for the heavy RH neutrino mass $M_1$ as 
\begin{equation}\label{eq1:lepto_std_lower_bound_M1}
 M_1 \gtrsim 3.5 \times 10^9 \text{ GeV}
 \;,
\end{equation}
where we have used relation (\ref{eq1:lepto_std_eps_est_3}) and taken the value of $\eta_B$ given by (\ref{eq1:BAU_B-asmy_CMB}). 

More generally, in many situations with $M_1 \lesssim 10^{14}$~GeV, one has, to good approximation, $\kappa^\text{f} \simeq 2\times 10^{-2}$ \cite{BDP_pedestrain}. This then implies that a raw $CP$ asymmetry of about $|\varepsilon_1| \simeq 3\times 10^{-6}$ is required for baryogenesis to succeed \footnote{We shall make use of these ballpark figures in many of our analyses in this work.}.

In summary, we have highlighted some of the essential features in quantitatively understanding  the classic leptogenesis scenario of \cite{fukugita_yanagida_86} which has the type I seesaw setup as its backbone. Specifically, we have discussed the ``standard'' situation where the heavy RH Majorana neutrinos are strongly hierarchical. As a result, only the lightest of the three RH neutrinos, $N_1$, is expected to contribute significantly to the final asymmetry. This is because the $B-L$ violating interactions mediated by $N_1$ would still be in thermal equilibrium when $N_{2,3}$ decayed away, and therefore any excess $B-L$ produced by $N_{2,3}$ would be erased. When the $N_1$'s eventually decay out-of-equilibrium, an excess of $B-L$ is created through $CP$ violating loop effects. Subsequently, this excess is converted into a $B$ asymmetry by SM sphalerons. 

The exact amount of $B$ generated in this way depends crucially on the interplay between the decay and washout processes, as well as the raw $CP$ asymmetry the neutrino model under consideration contains. By studying the Boltzmann evolution of the particle species and the explicitly calculating the loop diagrams, both of these crucial ingredients may be conveniently encapsulated into the efficient factor ($\kappa^\text{f}$) and $CP$ asymmetry ($\varepsilon_1$) respectively. Consequently, variations to the standard scenario can be quantified by changes in these values.

Over the years, there has been a dramatic increase in the sophistication of the quantitative analysis of leptogenesis. Many previously neglected effects such as thermal corrections \cite{Giudice_0310123}, spectator processes \cite{spectator1, spectator2} and, above all, flavor effects \cite{blanchet_zeno, barbieri_flavor,
endoh_flavor,
nardi_flavor,
abada_flavor_1,
abada_flavor_2,
simone_riotto_flv,
blanchet_flavor} have been considered in recent analyses. Other variations to the general scheme, including asymmetry production dominated by the decays of the second lightest RH neutrino $N_2$ \cite{PD_seesaw_geom}, resonant leptogenesis \cite{main_resonant_1,
main_resonant_1b,
pilaf_resonant_2,
flanz_resonant,
pilaf_underwood_TeV,
pilaf_int_j_mod} and models with more than three heavy RH neutrinos \cite{4RH_models}, have also received attention.
In the next few subsections, we will briefly mention some of these ideas which go beyond the standard scenario, and hint on how they may broaden the class of neutrino models that will lead to successful leptogenesis.\\

\subsection{Flavor effects in leptogenesis}\label{subsec:lepto_flavor}

So far, when discussing the leptogenesis implications of the seesaw Lagrangian (\ref{eq1:lepto_std_Lag}) on page~\pageref{eq1:lepto_std_Lag}, we have been exclusively concentrating on the effects from the last two terms and largely ignored the presence of the first---~the charged lepton Yukawa coupling: $y_{\alpha\beta}\, \overline{\ell}_\alpha\, \widetilde{\phi}\, e_{\beta}$. This approach of neglecting the charged lepton Yukawas \footnote{Strictly speaking, we have not completely ignored their effects as the conversion rate of SM sphalerons depends highly on whether this term is present or not.} in the analysis of leptogenesis is sensible as long as the ambient temperature in which the significant part of the $N_1$ evolution takes place is above $10^{12}$~GeV \footnote{The more precise condition is given in \cite{blanchet_zeno}.}. Otherwise, the Higgs mediated charged lepton processes in equilibrium will destroy the coherent evolution of the lepton doublets produced in the decays of $N_1$, leading to \emph{flavor effects} \cite{blanchet_zeno, barbieri_flavor,
endoh_flavor,
nardi_flavor,
abada_flavor_1,
abada_flavor_2,
simone_riotto_flv,
blanchet_flavor}.

To understand the issue, let us consider a generic lepton state $\ket{\ell_{(1)}}$ produced by the out-of-equilibrium decay of $N_1$. In the previous section, what we have assumed is that the lepton state  and its antiparticle counterpart which are some superpositions of flavor eigenstates $\ket{\ell_\alpha}$, i.e. $\ket{\ell_{(1)}} = \sum_\alpha c_\alpha \ket{\ell_\alpha}$, will undergo \emph{coherent} evolution during the leptogenesis era, and as a result we have summed over the final state lepton  flavors in arriving at $\varepsilon_{1}$. But in the presence of rapid Yukawa interactions \footnote{Without loss of generality, we will work in the charged lepton diagonal basis.} for each lepton flavor $\alpha$, the state $\ket{\ell_{(1)}}$ would quickly decohere and be projected onto $\ket{\ell_\alpha}$. Consequently, it is the evolution of $\ket{\ell_\alpha}$ (and not $\ket{\ell_{(1)}}$) that matters, which in turn implies that the efficiency factor ($\kappa_\text{f}^\alpha$) is now flavor dependent as we must write down a set of Boltzmann equations \footnote{It should be noted that using the classical Boltzmann equations to analyse flavor effects due to decoherence is only appropriate when we are in the ``fully-flavored'' regime, i.e. the charged lepton Yukawa interactions are either fully in (or out) of thermal equilibrium. Otherwise, the quantum decoherence effects must be studied using the density matrix formalism (See for example \cite{barbieri_flavor, abada_flavor_1,
abada_flavor_2, simone_riotto_flv}). For our purposes though the classical approximation will be sufficient.} for each lepton flavor $\alpha$:
\begin{align}
 \frac{d\mathcal{N}_{N_1}}{dt} &= -\Gamma_D (\mathcal{N}_{N_1} - \mathcal{N}_{N_1}^\text{eq})   \;,
\label{eq1:lepto_flv_BE1}\\
 \frac{d\mathcal{N}_{Q_\alpha}}{dt} &= -\varepsilon_{1\alpha} \Gamma_D 
 (\mathcal{N}_{N_1} - \mathcal{N}_{N_1}^\text{eq})   
 -P^0_{1\alpha} \Gamma_W \mathcal{N}_{Q_\alpha}
 \;,
\label{eq1:lepto_flv_BE2}
\end{align}
where $Q_\alpha \equiv B/3 -L_\alpha$ and 
$P^0_{1\alpha} \simeq |\bra{\ell_{(1)}}\ket{\ell_\alpha}|^2$ is the tree-level contribution to the flavor projector \footnote{We delay the more detailed treatment of this to Chapter~\ref{ch_work_LV} where we study the implications of flavor effects on a specific neutrino model.} which signifies the $\alpha$-dependence in the washout term $P^0_{1\alpha} \Gamma_W$ in (\ref{eq1:lepto_flv_BE2}). In fact, we see that the raw $CP$ asymmetry $\varepsilon_{1\alpha}$, which is defined through (\ref{eq1:lepto_std_CP_general_kj}), and related to $\varepsilon_1$ via
\begin{equation}\label{eq1:lepto_flv_eps1_2_eps1a}
  \varepsilon_1 = \sum_{\alpha} \varepsilon_{1\alpha}\;.
\end{equation}
need not be the same for different $\alpha$. 

All these extra features can now change the original leptogenesis picture in a non-trivial way. Firstly, one may imagine the situation where each individual $\varepsilon_{1\alpha}$ is nonzero while their sum vanishes: $\varepsilon_{1}=0$. As a result, in the one-flavor approximation (i.e. the standard scenario) no asymmetry can be generated, while in the case with flavor effects included, each lepton asymmetry in flavor $\alpha$ may evolve differently and conspire to give a nonzero overall result.
This feature of $\varepsilon_{1}=0$ but $\varepsilon_{1\alpha}\neq 0$ appears, for example, in models with a $CP$ conserving RH neutrino sector. To see this, it is helpful to work in the basis where the charged lepton Yukawa and RH neutrino mass matrices are real and diagonal. Then, the neutrino Yukawa matrix $h$ (see Lagrangian (\ref{eq1:lepto_std_Lag})) contains all the necessary $CP$ violating phases that may manifest themselves in leptogenesis. Using the single value decomposition, one can express matrix $h$ as 
\begin{equation}\label{eq1:lepto_flv_h_decomp}
 h = V_L \, \text{diag}(x_1, x_2, x_3)\, V_R^\dagger \;, \qquad x_i \in \mathbb{R}\;,
\end{equation}
where $V_{L,R}$ are an arbitrary unitary matrices. This implies that the term of interest: $(h^\dagger h)_{1m}$ is given by $\left(V_R \,\text{diag}(x_1^2, x_2^2, x_3^2)\, V_R^\dagger\right)_{1m}$, where $V_L$ has noticeably disappeared. Since $V_R$ is directly related to the RH neutrino mixings, the assumption that there is no $CP$ violating phases in the RH sector implies that $V_R$ must be real. Therefore, $\text{Im}\left[(h^\dagger h)_{1m}^2\right]=0$ and $\varepsilon_1$ vanishes by virtue of result (\ref{eq1:lepto_std_CP_general_N1}). However, for $\varepsilon_{1\alpha}$, one does not sum over flavors and instead of $(h^\dagger h)_{1m}$ the term of interest becomes (in index form)
\begin{equation}\label{eq1:lepto_flv_h_h_alpha}
 h^*_{\alpha 1} h_{\alpha m} = 
 \sum_{j,k} \left(V_L^*\right)_{\alpha j} \, x_j \, \left(V_R\right)_{j1}\,
 \left(V_L\right)_{\alpha k} \, x_k \, \left(V_R\right)_{km}
  \;,
\end{equation}
which is in general complex due to the presence of $CP$ violating phases in $V_L$, and thus $\varepsilon_{1\alpha}$ need not be zero. Moreover, since the leptonic mixing matrix $\Upmns$ is tightly connected to $V_L$, this result indicates that for such model, successful leptogenesis must stem from low energy $CP$ violations in the lepton sector, in contrast to the standard scenario where this direct link is absent for all viable neutrino models. 

More generally, even in models where both the LH and RH sectors contain $CP$ phases that can lead to a lepton asymmetry, the dependence on the $\Upmns$ remains as long as flavor effects are important. This opens up the possibility of ``directly'' probing some of the parameters of leptogenesis in low energy neutrino experiments. By the same token, if leptogenesis is the correct theory for explaining the baryon asymmetry of the universe, then this flavored scenario can provide important constraints on the low energy Majorana phases beyond those imposed by neutrinoless double beta decay. In the wake of this connection, many authors have recently studied the implications of the phases in $\Upmns$ on this leptogenesis scenario with or without any assumptions on the $CP$ violating RH sector \cite{blanchet_flavor,
recent_flavor_low_energy,
recent_flavor_low_energy_2,
dirac-phase_lepto}.

There are of course other modifications to the standard setup one may expect from the inclusion of flavor effects. For instance, the criteria for which initial conditions become important in the leptogenesis predictions may alter \cite{blanchet_flavor} and, in certain regimes, the lower bound on $M_1$ shown in (\ref{eq1:lepto_std_lower_bound_M1}) gets relaxed by a factor of about 3 \cite{abada_flavor_1}. But one general conclusion is that by turning on flavor effects, the parameter space for which leptogenesis is successful becomes somewhat larger, although in some cases the predictions are highly dependent on the neutrino models employed.\\


\subsection{$N_2$-dominated scenario}\label{subsec:lepto_N2}

One major assumption in the previous discussions on leptogenesis (with or without flavor effects) is that the final asymmetry is predominantly produced by the $CP$ violating decay of the lightest heavy RH neutrinos, $N_1$. The justification for this relies on having a strongly hierarchical RH neutrino mass spectrum ($M_1 \ll M_2 \ll M_3$) so that one typically has a situation where either the lepton asymmetry generated by $N_{2,3}$ is effectively washed out by processes mediated by $N_1$ (i.e. $\kappa^\text{f}_{2,3} \ll \kappa^\text{f}_{1}$), or the raw $CP$ asymmetry of the $N_{2,3}$ decay is naturally suppressed compare to that of $N_1$ (i.e. $|\varepsilon_{2,3}| \ll |\varepsilon_{1}|$), or both. However, this conclusion of $N_1$ dominance implied by the strong hierarchy in the RH neutrino masses is not actually universal for all neutrino Yukawa matrix structures, and there exists notable special cases where the asymmetry is primarily due to $N_2$ \cite{PD_seesaw_geom} \footnote{It turns out that for hierarchical RH neutrinos, the $N_3$-dominate scenario is not actually possible (within the regime where initial conditions are unimportant) since any asymmetry produced by $N_3$ will be washed out by either $N_1$ or $N_2$ mediated processes. If one tries to circumvent this by considering the case of very weak washout from all processes, then the model loses a lot of its predictive powers as the outcome will be strongly sensitive to initial conditions and the precise thermal history of the evolution.}.

In order to identify the Yukawa structures which can simultaneously circumvent the suppression of $|\varepsilon_{2}| \ll |\varepsilon_{1}|$, as well as the washout problem due to $N_1$, it is helpful to recast the Yukawa matrix $h$ in terms of a complex orthogonal matrix, $\Omega$, using the Casas-Ibarra parametrisation \cite{Casas-Ibarra}:
\begin{equation}\label{eq1:omega_defn}
  h \equiv \frac{1}{\vev{\phi}}\,\Upmns\,D_m^{1/2}\,\Omega\,D_M^{1/2}\;,
     \qquad \textrm{with}\quad \Omega^T\Omega=I \;,
\end{equation} 
where $D_m^{1/2} = \text{diag}(\sqrt{m_1} , \sqrt{m_2}, \sqrt{m_3})$ and $D_M^{1/2} = \text{diag}(\sqrt{M_1},\sqrt{M_2},\sqrt{M_3})$ with $m_{1,2,3}$ and $M_{1,2,3}$ denoting the LH and RH neutrino masses respectively, while $\vev{\phi}$ is the Higgs VEV. In deriving (\ref{eq1:omega_defn}), we have used the type I seesaw relation in the basis where the charged lepton and heavy neutrino mass matrices are real and diagonal, i.e. $m_\nu \simeq -m_D\,D_M^{-1}\,m_D^T$ with $m_\nu = - \Upmns^*\,D_m\,\Upmns^\dagger$ and $m_D = h \vev{\phi}$.
From definition (\ref{eq1:omega_defn}), we then have
\begin{equation}\label{eq1:h_h_in_omega}
  (h^\dagger h)_{jk} = 
  \frac{\sqrt{M_j M_k}}{\vev{\phi}^2}\, \sum_{l=1}^3 \,m_l\,\Omega^*_{lj}\,\Omega_{lk} 
   \;.
\end{equation} 
Another quantity of interest is the so-called effective neutrino mass associated with the $N_k$ processes \cite{BDP_pedestrain,Plum_Zphy_NPB,BDP_0205349_CMB,PD_seesaw_geom}: 
\begin{equation}\label{eq1:eff_nu_mass}
  \widetilde{m}_k \equiv \frac{(m_D^\dagger m_D)_{kk}}{M_k}
  = \sum_{l=1}^3 m_l\, |\Omega_{lk}^2|
  \;, \quad
  k= 1,2 \text{ or } 3
   \;,
\end{equation} 
which governs the rate at which washout operates. It turns out that for $M_1 < 10^{14}$~GeV, all $N_1$ related washout processes are to good approximation proportional to $\widetilde{m}_1$ \cite{BDP_pedestrain,BDP_0205349_CMB,Giudice_0310123}, and hence it is a good measure as to whether washout due to $N_1$ interactions are strong ($\widetilde{m}_1 \gg \order{10^{-3}}$~eV) or weak ($\widetilde{m}_1 \ll \order{10^{-3}}$~eV) \footnote{We shall say more about this in Chapter~\ref{ch_work_LV}.}. Therefore, by
using definitions (\ref{eq1:h_h_in_omega}) and (\ref{eq1:eff_nu_mass}), it is easy to study the general behavior of $\varepsilon_{1,2}$ and $N_{1,2}$ washout effects as $\Omega$ (or  the neutrino Yukawa matrix structure) changes. 

Note that after applying the approximation of (\ref{eq1:lepto_std_eps_fVS_limit}), the $CP$ 
asymmetry due to $N_1$ decays becomes
\begin{equation}\label{eq1:lepto_N2_eps1}
|\varepsilon_{1}|  
  \simeq 
  \frac{3}{16\pi}\left|\sum_{m\neq 1} 
              \frac{\textrm{Im}\left[(h^\dagger h)^2_{1m}\right]}
               {(h^\dagger h)_{11}}
             \,\frac{M_1}{M_m}\right|\;.
\end{equation}             
Analogously, the case for $N_2$ is
\begin{align}
|\varepsilon_{2}|  
  &= 
  \frac{1}{8\pi}
    \left|
     \sum_{m\neq 2} 
              \frac{\textrm{Im}\left[(h^\dagger h)^2_{2m}\right]}
               {(h^\dagger h)_{22}}
             \left\{f_V\left(\frac{M_m^2}{M_2^2}\right)+f_S\left(\frac{M_m^2}{M_2^2}\right)
             \right\}
    \right|         
       \;, \nonumber\\
  &\simeq     
    \frac{1}{8\pi (h^\dagger h)_{22}}
    \left|
       \frac{2M_1}{M_2}
              \left(1 + \ln \left[\frac{M_1}{M_2}\right] \right)
               \textrm{Im}\left[(h^\dagger h)^2_{21}\right]
      -\frac{3M_2}{2M_3}
             \textrm{Im}\left[(h^\dagger h)^2_{23}\right]       
    \right|           \;,
       \label{eq1:lepto_N2_eps2}
\end{align}             
where the first term in (\ref{eq1:lepto_N2_eps2}) which involves $x\equiv M_1^2/M_2^2 \ll 1$ comes from the approximation: 
\begin{align}
 f_V(x)+f_S(x) 
 &= \sqrt{x}\left[1-(1+x)\ln\left(\frac{1+x}{x}\right)\right]+\frac{\sqrt{x}}{1-x}
 \;,
 \nonumber\\
 &\simeq 2 \sqrt{x}\, (1+ \ln \sqrt{x}) \;, \qquad \text{for } x \ll 1
 \;\label{eq1:lepto_N2_fVS_approx_x_small},
\end{align}
and we have employed (\ref{eq1:lepto_std_eps_fVS_limit}) like before for the second term which has $x\equiv M_1^3/M_2^2 \gg 1$.

With these expressions in hand, one can simply look for an $\Omega$ pattern which can minimise $|\varepsilon_1|$ while keeping $|\varepsilon_2|$ non-negligible. It has been shown that for a hierarchical LH neutrino spectrum (i.e. $m_1 \approx 0$), the following structure can achieve this aim \cite{PD_seesaw_geom}:
\begin{equation}\label{eq1:lepto_N2_omega_R23}
 \Omega = \ththMat
   {1}{0}{0}
   {0}{\Omega_{22}}{\sqrt{1-\Omega_{22}^2}}
   {0}{-\sqrt{1-\Omega_{22}^2}}{\Omega_{22}}\;,
\end{equation}
where $\Omega_{22}$ is an arbitrary complex number. Substituting this into (\ref{eq1:h_h_in_omega}), one immediately finds that $(h^\dagger h)_{1m} = 0$, and hence $|\varepsilon_1| = 0$. But it can be checked that this form of $\Omega$ does not lead to a vanishing $|\varepsilon_2|$ in general, and in fact a sufficiently large $CP$ asymmetry is possible \cite{PD_seesaw_geom}. More importantly though, this asymmetry from $N_2$ would not be washed out by processes involving the lighter $N_1$. This can be seen from directly evaluating  (\ref{eq1:eff_nu_mass}) for $N_1$:
\begin{align}
 \widetilde{m}_1
  &= m_1 |\Omega_{11}^2| +m_2 |\Omega_{21}^2|+m_3 |\Omega_{31}^2|\;,\nonumber\\
  &= m_1 \;,\nonumber\\
  &\approx 0 \;, \qquad \text{since }\;\; m_1 \ll m_2 \ll m_3\;.
\end{align}
In other words, $\widetilde{m}_1 \ll \order{10^{-3}}$~eV, which means washout is very weak. So, with the neutrino Yukawa structure implied by (\ref{eq1:lepto_N2_omega_R23}) and hierarchical LH neutrino masses, $N_2$ decays can be the main source of the final lepton asymmetry produced, leading to the so-called $N_2$-dominated scenario.

One important consequence of this is that the lower bound (\ref{eq1:lepto_std_lower_bound_M1}) on $M_1$ is no longer present and it is replaced by an analogous bound on $M_2$ which can significantly change the constraints on neutrino model building. In addition, when flavor effects are included, the prospect of a successful $N_2$-dominated asymmetry generation will increase. This is because typically washout processes from $N_1$ are less effective as certain flavor projections of the asymmetry may be ``protected'' because of the flavor dependent property of the efficiency factors (see for example \cite{vives_flavor_eg}). Secondly, in the flavored version of the raw $CP$ asymmetry, $|\varepsilon_{2\alpha}|$'s are not necessarily suppressed compare to  $|\varepsilon_{1\alpha}|$'s even in general \cite{blanchet_flavor}. As a result of this enlargement of the applicable parameter space  when flavor effects are considered, $N_2$-leptogenesis could be more important than previously thought and perhaps as relevant as the $N_1$ scenario.


\subsection{Resonant leptogenesis}\label{subsec:lepto_resonant}

Another way to go beyond the standard scenario of leptogenesis is to relax the condition that the heavy RH neutrinos are hierarchical. This possibility is logical as quasi-degenerate RH neutrinos are not excluded by any existing experimental data nor are they forbidden by the generic type~I seesaw setup. Furthermore, as we shall explain below, 
an effect know as \emph{resonant leptogenesis} \cite{main_resonant_1,main_resonant_1b, pilaf_resonant_2, pilaf_underwood_TeV, flanz_resonant,pilaf_int_j_mod} can occur when the mass splitting between two RH neutrinos becomes small enough, leading to enhancement of the $CP$ asymmetry $\varepsilon_j$, and consequently, opening up new domains of applicability for thermal leptogenesis in general.

Similar to other analyses of leptogenesis, the two main issues of concern here are the size of the $CP$ asymmetry and the final efficiency factor. When considering the situation of $M_j \simeq M_k$ for $j\neq k$ more closely, we first realise that, qualitatively, the  washout rate must increase at $T \simeq M_{j,k}$ because $L$ violating scattering processes mediated by $M_j$ and $M_k$ would both be active, providing more ways to erase the generated asymmetry. Secondly, we note that when we analyse the expression for $\varepsilon_{j}$ previously, we have either employed the approximation of $M_k/M_j \gg 1$ as in (\ref{eq1:lepto_std_eps_fVS_limit}) or $M_k/M_j \ll 1$ in (\ref{eq1:lepto_N2_fVS_approx_x_small}). However, a quasi-degenerate RH neutrino spectrum demands the condition of $M_k/M_j = \order{1}$, and hence the limits on $\varepsilon_{j}$ must be re-studied. 

By inspecting the form of the loop functions in (\ref{eq1:lepto_std_CP_general_N1_VS_fn}) which define $\varepsilon_j$, we see that the most interesting behavior must come from the self-energy correction term, $f_S(x)$ as $M_j \rightarrow M_k$ since naively
\begin{align}
 \lim_{x \rightarrow 1} f_S(x) = \lim_{x \rightarrow 1} \frac{\sqrt{x}}{1-x}
 &= \lim_{M_j \rightarrow M_k} \frac{M_j\,M_k}{M_j^2-M_k^2}\;,
 \qquad \text{ with } x \equiv M_k^2/M_j^2\;,
 \nonumber\\
  &\stackrel{?}{=} \infty \;.\label{eq1:res_lepto_fS_first_glance}
\end{align}
The apparent erroneous conclusion suggested by (\ref{eq1:res_lepto_fS_first_glance}) is a result of the fact that conventional finite-order perturbation theory, which this formula was originally derived from, does not take into account the unstable nature of the two RH Majorana neutrinos. To resolve this, one may follow the resummation approach of \cite{main_resonant_1,pilaf_int_j_mod,main_resonant_1b} where an additional regulating absorptive term due to the finite decay width of  $M_{j,k}$ naturally emerges to eliminate such an unphysical outcome. The self-energy contribution to the $CP$ violation parameter is then modified to \cite{main_resonant_1,pilaf_int_j_mod,main_resonant_1b}
\begin{equation}\label{eq1:res_lepto_resum_formula}
 \varepsilon_{j} \simeq 
 \frac{\text{Im}\left[(h^\dagger h)_{jk}^2 \right]}{(h^\dagger h)_{jj}(h^\dagger h)_{kk}}
 \,\frac{2(M_j^2-M_k^2)\,M_j\, \Gamma_j}{(M_j^2-M_k^2)^2 + 4 M_j^2\, \Gamma_j^2}\;,
\end{equation}
where $j,k=1,2 \text{ or } 2,3 \;(j\neq k)$ and $\Gamma_j = (h^\dagger h)_{jj}\, M_j/16\pi$ is the generalisation of the tree-level decay rate as defined in (\ref{eq1:lepto_std_N1_tree}) \footnote{It should be noted that Eq.~(\ref{eq1:res_lepto_resum_formula}) is only valid for a mixing system of two RH neutrinos. The generalisation to the three neutrinos case can be found in \cite{main_resonant_1b}.}. From the expression of (\ref{eq1:res_lepto_resum_formula}), one can see that $\varepsilon_j \rightarrow 0$ when $M_j \rightarrow M_k$ in accordance with the observation that the RH neutrino running in the loop must be different from the decaying one in order to generate an asymmetry (see discussion in Sec.~\ref{subsec:CP-violation}) \footnote{The situation for the vertex contribution as $M_j\rightarrow M_k$ is more subtle as each individual $\varepsilon_j^\text{vertex} \not\rightarrow 0$. However, it can be shown that $\varepsilon_j^\text{vertex}$ and $\varepsilon_k^\text{vertex}$ become exactly equal and opposite when $M_j\rightarrow M_k$, and therefore the overall $CP$ asymmetry will be zero as expected (see e.g. \cite{pilaf_int_j_mod}). In addition, this implies that the vertex contribution cannot exhibit resonant behaviors for any region of the mass parameter space.}. 

More importantly, Eq.~(\ref{eq1:res_lepto_resum_formula}) indicates that the $CP$ asymmetry will be enhanced provided that the mass splitting between the two RH neutrinos coincides with the region of mass parameters about which the $\varepsilon_j$ function peaks. Specifically, one requires
\begin{equation}\label{eq1:res_lepto_peak_condition}
 |M_j - M_k| \sim \Gamma_{j,k} \;,
\end{equation}
to maximise the resonant effect \footnote{Currently, there is a controversy surrounding the exact region of validity for this perturbative resummation approach. It has been argued in \cite{resonant_ABP} that the exact resonant effect cannot be reached without violating the perturbative assumption. As a result, a different formula to (\ref{eq1:res_lepto_resum_formula}) has also been suggested, which changes the prediction of the $CP$ asymmetry by typically a factor of $\order{10}$.}. With this, one can see that if the Yukawa couplings are such that 
\begin{equation}\label{eq1:res_lepto_peak_cond_Yuk}
 \frac{\text{Im}\left[(h^\dagger h)_{jk}^2 \right]}{(h^\dagger h)_{jj}(h^\dagger h)_{kk}} =
 \order{1}
  \;,
\end{equation}
then $\varepsilon_j$ can be as large as $\order{1}$, hence provide a lot more leverage for successful leptogenesis. Indeed, the increase in washout due to the tiny mass gap between $N_j$'s will eventually saturate when the degenerate limit reaches a certain point \cite{blanchet_DL}, and the enhancement from resonant effects will be able to dominate the outcome. Consequently, given the substantial enhancement by resonant leptogenesis, some of the stringent constraints on the neutrino properties imposed by the standard hierarchical scenario may be evaded. Most notably, the lower bound (\ref{eq1:lepto_std_lower_bound_M1}) on $M_1$ is completely removed, leading to the possibility of TeV scale RH neutrinos \footnote{Of course, a slightly different realisation of the seesaw mechanism may then be required \cite{TeV_seesaw_strings}.} and TeV leptogenesis \cite{pilaf_underwood_TeV}. In SUSY leptogenesis theories, this is particularly advantageous as the upper bound on the reheating temperature ($T_\text{reh}$) due to BBN constraints on gravitino over-production, is often in conflict with the condition, $T_\text{reh} \gtrsim M_j$, normally required for the sufficient thermal generation of $N_j$'s which participate in $L$ creation.
Furthermore, $N_{2}$- and even $N_{3}$-leptogenesis are now easily achievable under this scenario, and hence the set of applicable seesaw models is significantly expanded. 

If flavor effects are included, yet more possibilities can be accommodated. For example, one may envisage a situation where one of the flavors (say $\alpha$) for which the lepton asymmetry is predominantly generated in, is very weakly washed out while the Yukawa couplings between the other lepton families and some of the RH neutrinos can remain very large \cite{pilaf_resonant_2}. Then, this model predicts that lepton flavor violating interactions are potentially observable in the next generation of experiments. The reason that resonant leptogenesis is required in this setup is because normally (even with flavor effects) the $CP$ asymmetry $\varepsilon_{1\alpha}$ is necessarily suppressed if its associated projector $P_{1\alpha}$ is very small (i.e. very weak washout in $\alpha$)  \cite{abada_flavor_2}. 

Certainly, this particular model and many that employ resonant leptogenesis can have the RH Majorana neutrinos to be as small as 1~TeV \footnote{The only lower bound on $M_j$ here comes from the freeze-out temperature ($\simeq \order{10^2}$~GeV) of sphaleron processes, below which an $L$ asymmetry cannot be converted into a $B$ asymmetry.} and depending on their couplings to SM particles, collider signatures of them may also be accessible in the near future \cite{pilaf_resonant_2,pilaf_underwood_TeV, RH_collider_ph,Atwood:2007zza}.\\

\section{Outline of our work}\label{sec:thesis_outline}

As we have illustrated throughout this chapter, the observational evidence for nonzero neutrino masses and cosmological matter-antimatter asymmetry provides a strong indication for physics beyond the SM. Although many proposals have been suggested, a particularly attractive way (in our opinion) of explaining both phenomena \emph{simultaneously} is the inclusion of heavy RH Majorana neutrinos which are electroweak singlets, to the SM. As a result, tiny neutrino masses can be generated by type I seesaw \cite{type1_seesaw} while the problem of the cosmic baryon asymmetry is solved by thermal leptogenesis \cite{fukugita_yanagida_86}.

Given this setup, there is then an intricate connection between neutrinos properties and leptogenesis whereby the requirement for successful asymmetry generation naturally leads to limits on the masses and mixing in the lepton (or sometimes quark) sector. Therefore, it is of great interest to explore possible neutrino models based on type I seesaw that will allow leptogenesis to succeed. By the same token, it may be fruitful to investigate the implications of a given model, which has been specifically designed to address a different issue, in the leptogenesis context. 

With this in mind, our work involves studying several classes of neutrino models in the type I seesaw framework, which  contain either new symmetries or interactions. The main focus will of course be on their leptogenesis credentials,  however, other important phenomenological issues such as implications on neutrino masses and mixings, as well as collider signatures will also be discussed where appropriate.

The first part of our work is a comprehensive analysis on special types of seesaw models with abelian family symmetries as proposed in \cite{Low:2005yc}. The original motivation for these models is that they contain fewer overall free parameters than the default type I seesaw setup, hence making the theory more economical. However, no previous studies have been done on the possible cosmological implications of these models. Therefore, we shall investigate in Chapter~\ref{ch_work_LV} the viability of these models in leptogenesis \footnote{This part of the work is related to Publication~1 listed on page~\pageref{listofpub}.}.

Continuing the theme of flavor symmetries, in Chapter~\ref{ch_work_HLV} we explore how they can help constraining the seesaw sector in neutrino mass models such that it is completely determined in terms of low-energy mass, mixing angle and $CP$ violating phase observables \footnote{This part of the work is related to Publication~3.}. Given that leptogenesis is highly dependent on the properties of the RH neutrinos, having a model that can predict the otherwise arbitrary parameters in this sector is obviously beneficial. In addition, their connections to low-energy oscillation parameters provide a direct way to probe the elements of leptogenesis through current and future experiments. After explicitly building some representative models, we will discuss their phenomenological consequences, including the conditions under which leptogenesis can be successful.

In Chapter~\ref{ch_work_BKL}, we shift our attention to the electromagnetic interactions between the LH and RH neutrinos \footnote{This part of the work is related to Publication~2.}. It is known that neutrino masses and mixings imply the existence of neutrino electromagnetic dipole moments. So, the inclusion of heavy RH neutrinos to the SM as in type I seesaw then naturally gives rise to new transition electromagnetic moments involving both LH and RH neutrinos. Since these new interactions are potentially $CP$ violating, a lepton asymmetry may be generated through the out-of-equilibrium decays of the heavy RH neutrinos via this channel, in analogy to the standard mechanism mediated by  Yukawas. Consequently, our aim is to show that leptogenesis via such electromagnetic processes is possible by explicitly calculating the $CP$ asymmetry coming from the transition moments. Also, a comparison between this ``electromagnetic'' version of leptogenesis with the standard scenario will be included.

Finally, we conclude our entire work in Chapter~\ref{ch_conc}.

\subsubsection{Important comment on notations}
 
It should be noted that the same symbol appearing in different chapters/sections may \textbf{not} always have a common meaning. Owning to the fact that there are only a finite number of conventional symbols, letters or accents available, and that only certain symbols are meaningful for a particular purpose (e.g. $i$ to denote $\sqrt{-1}$ or $p$ to denote 4-momentum etc.), it is inevitable that some of them will have to be re-used in a different context. This is unfortunate but often necessary in order to ensure clarity of presentation within each chapter/section. For example, we often use $m_i$ to denote the mass of the $i$th LH neutrino while in some generic context it may simply mean the mass of particle species $i$. In order to avoid confusion, we will often reiterate the definition a symbol even though it may still have the exact meaning as previously defined or it may seem obvious from the context. 

Moreover, in some cases, the same quantity may be represented differently in different chapters because of the proliferation of similar symbols. We have made such conscious decisions to ensure the presentation is less confusing within each context. 


%
%
%
%
%



\chapter{Abelian family symmetry and leptogenesis}\label{ch_work_LV}

\ECap{W}{h~i~l~e}{7}{0}
the SM (with or without neutrino masses) has been very successful in explaining the dynamics of subatomic particles, it is not without its shortcomings. A typical example is its inability to predict quark and lepton masses. Such parameters are put in by hand using data from experiments. Altogether, there are almost 40 free parameters in the SM extended to include three RH neutrinos, with some 21 of them in the lepton mass sector alone \cite{SM_param_count}. Therefore, from the model building point of view, it is natural to look for ways to reduce the number of variables.

Given that the proliferation of masses and mixing parameters is the result of quarks and leptons having more than one family, a symmetry that governs the inter-family relation provides an excellent starting point in understanding their values. Moreover, since masses are generated by the Higgs mechanism, scalar fields that couple to the SM fermions, and their subsequent spontaneous symmetry breakings can play a role in determining the free parameters of interest. Hence, it is not unexpected that both the enlargement of the SM symmetry and of the Higgs sector are the cornerstones of many models which try to explain the masses and mixings \cite{TB_a4_list, TB_other_list, Low:2005yc, Low01+thesis, Low02, family_sym_eg}, even though not all of them end up being more economical than the minimally extended SM.

Our focus here is on a very special class of type I seesaw models which has an \emph{abelian} family symmetry and an extra real Higgs singlet added to the SM \cite{Low:2005yc}. Because of the interplay between the new ingredients, these models predict a fully hierarchical LH neutrino spectrum, and $\theta_{13}=0$ in the PMNS matrix as implied by the ``bi-large'' mixing pattern (see Sec.~\ref{subsec_osc_TB}). More importantly, they contain \textit{fewer} free parameters than the default seesaw setup \cite{Low:2005yc} \footnote{It should be noted that depending on which sub-class of the models in \cite{Low:2005yc} we are considering, the reduction in parameters will be different (varying between one to seven). Also, in one instance, there are only two RH neutrinos being added to the SM. For this case, the overall number of variables is one less than the standard seesaw with two RH neutrinos.}. Thus, given their relatively economical nature, it is interesting to see if these models carry additional benefits. Specifically, our aim is to check whether these models can maintain the ability of the standard seesaw to solve the baryogenesis problem via leptogenesis.

With this motivation in mind, our investigation begins with an outline of these neutrino models with abelian flavor symmetry. It serves to highlight all the key features, as well as to recast them in a form suitable for later discussions. For definiteness and ease of comparison with standard leptogenesis, we assume that the heavy RH Majorana neutrino masses, which are unspecified by these models, are hierarchical (i.e. $M_1\ll M_2 \ll M_3$). As a result, the final $L$ asymmetry will be assumed to have come predominantly  from the decays of $N_1$, the lightest RH neutrino. Our analysis will make use of various existing results and attempts to draw comparisons where appropriate with the aim of deducing whether these specific neutrino models may favor certain regimes or predict significant deviation from the standard scenarios of leptogenesis with and without flavor effects.\\

\section{Models with abelian family symmetry}\label{sec:LV_low_model}

As hinted earlier, the first step towards reducing the number of free variables in the seesaw setup may be achieved by increasing the symmetry of the Lagrangian. To this end, the SM group is extended to:
\begin{equation}\label{eq2:model_group}
 G = SU(3)_c \otimes SU(2)_L \otimes U(1)_Y \times G_{\textrm{family}},
\end{equation}
where $G_{\textrm{family}}$ is a leptonic family/flavor symmetry. However, it has been shown in \cite{Low01+thesis} that models with an unbroken family symmetry and just the SM Higgs doublet do not have more predictive powers than the default seesaw. Therefore, one sensible move is to expand the Higgs sector at the same time although it should be noted that other ways to circumvent this problem exist. In the work of \cite{Low:2005yc, Low02}, effects of abelian family symmetries with additional Higgs singlets, doublets or triplets on certain features of the $\Upmns$ matrix~\footnote{For brevity, we often write $U$ instead of $\Upmns$ in this chapter.} were studied. It was found that the simplest models that can predict $\theta_{13}=0$ contain one extra real Higgs singlet which transforms non-trivially in family space. The setup of these models in the leptonic sector is summarised as follows. Suppose the lepton Yukawa and mass terms are given by
\begin{equation}\label{eq2:model_lag1}
 \mathcal{L}_{\textrm{mass}} = -\overline{L}\,Y_\ell\,\Phi\, \ell_R
                               -\overline{L}\,Y_\nu\,\widetilde{\Phi}\,\nu_R
                               -\frac{1}{2}\,\overline{(\nu_R)^c}\,Y_{\chi}\,\chi\;\nu_R
                               -\frac{1}{2}\,\overline{(\nu_R)^c}\,M_\textrm{bare}\,\nu_R
                               + \textrm{h.c.}\;,
\end{equation}
where $Y_\ell, Y_\nu, Y_\chi$ and $\Mbare$ are complex coupling matrices in the flavor basis. $L=(\nu_L, \ell_L)^T$ is the LH lepton doublet while $\ell_R$ and $\nu_R$ are the RH charged and neutral lepton singlets respectively. The SM Higgs doublet is denoted by $\Phi=(\phi^+,\phi^0)^T$, with its charged conjugate as $\widetilde{\Phi}\equiv i\sigma_2 \Phi^*$, while $\chi$ is the newly included real singlet scalar field. Then, the action of the family symmetry, $G_{\textrm{family}}$, on the SM fields and the new singlet demands that the full Lagrangian is invariant under the unitary transformations:
\begin{equation}\label{eq2:model_transf}
 L\rightarrow S_L L\;,
 \quad \ell_R\rightarrow S_{\ell_R}\ell_R\;,
 \quad 
 \nu_R\rightarrow S_{\nu_R}\nu_R\;,
 \quad \Phi\rightarrow S_\Phi\Phi
 \quad\textrm{and}\quad
 \chi\rightarrow S_\chi\chi \;,
\end{equation}
in family space. These transformations have the ability to restrict the coupling matrices to certain forms such that the $e3$-component of the $\Upmns$ matrix is zero, or in other words, $\theta_{13} =0$ \cite{Low:2005yc} \footnote{We have assumed the parametrisation of  Eq.~(\ref{eq1:UPMNS_param}) on page~\pageref{eq1:UPMNS_param} for the $\Upmns$ matrix.}. It should be emphasized that in (\ref{eq2:model_lag1}), the inclusion of the $\Mbare$-term, as well as the $\chi$-term for generating the RH Majorana masses is essential (assuming the symmetry transformation properties displayed in (\ref{eq2:model_transf})) for it turns out that the absence of either of them will lead to mixing parameters that are ruled out by current experiments. For instance, removing the $\Mbare$-term from (\ref{eq2:model_lag1}) will force the solar mixing angle $\theta_{12}$ to maximal, which is incompatible with the best-fit data (see (\ref{eq1:best-fit_angle}) on page~\pageref{eq1:best-fit_angle}). 

One possible representation of the set of transformations in (\ref{eq2:model_transf})  and their associated abelian flavor symmetry is shown in Table~\ref{table2:tranf_matrix}. The corresponding coupling matrices (written in the same basis) induced by these transformations
are presented in Table~\ref{table2:coupling_matrix}. It can be seen from Table~\ref{table2:tranf_matrix} and \ref{table2:coupling_matrix} that the desire property of $U_{e3} =0$ can be achieved by models with only two RH neutrinos. This is evident from the vanishing third column of $Y_\nu$ in Case 1 of Table~\ref{table2:coupling_matrix}. Therefore, in this case, the third RH neutrino ($N_3$) is actually decoupled from the LH sector, and it may be removed from the theory without changing the overall prediction. In fact, this (Case 1) is then reduced to Case 3. As a result of this equivalence between Case 1 and 3, we can safely assumed that all of these models contain three heavy RH neutrinos when conducting our analysis in leptogenesis as we will be interested in the $N_1$-dominated scenario only.

%
\begin{table}[t]
\begin{center}
\begin{tabular}{|c|c|c|c|c|c|}
\hline
 &$S_L=S_{\ell_R}$ &$S_{\nu_R}$ &$S_\Phi, S_\chi$ &$G_\textrm{family}$&
 reduction of parameters\\
\hline
1&$\ththMat{1}{0}{0}{0}{-1}{0}{0}{0}{-1}$
 &$\ththMat{1}{0}{0}{0}{-1}{0}{0}{0}{i}$
 &$\MatCase{\Phi\rightarrow\Phi\,,}{\chi\rightarrow -\chi}$
 &$\Zfour$
 &\textbf{7} (cf. default with  3 $\nu_R$'s)\\
\hline
2&$\ththMat{1}{0}{0}{0}{-1}{0}{0}{0}{-1}$
 &$\ththMat{1}{0}{0}{0}{-1}{0}{0}{0}{1}$
 &$\MatCase{\Phi\rightarrow\Phi\,,}{\chi\rightarrow -\chi}$
 &$\Ztwo$
 &\textbf{3} (cf. default with  3 $\nu_R$'s)\\ 
\hline
3&$\ththMat{1}{0}{0}{0}{-1}{0}{0}{0}{-1}$
 &$\twtwMat{1}{0}{0}{-1}$
 &$\MatCase{\Phi\rightarrow\Phi\,,}{\chi\rightarrow -\chi}$
 &$\Ztwo$
 &\textbf{1} (cf. default with  2 $\nu_R$'s)\\
\hline
\end{tabular}\caption{Diagonal representation of the transformations in flavor space that gives $U_{e3}=0$ for the model with three or two (in Case 3) RH neutrinos. Here $\Ztwo$ and $\Zfour$ denote the discrete cyclic group of order 2 and 4 respectively. In the last column, the reduction of free parameters as compared to the standard seesaw setup for each model is stated.}
\label{table2:tranf_matrix}
\end{center}
\end{table}
%
%
\begin{table}[t]
\begin{center}
\begin{tabular}{|c|c|c|c|c|}
\hline
 &$Y_\ell$ &$Y_\nu$ &$Y_\chi$ &$M_\textrm{bare}$\\
\hline
1&$\ththMat{\times}{0}{0}{0}{\times}{\times}{0}{\times}{\times}$
 &$\ththMat{\times}{0}{0}{0}{\times}{0}{0}{\times}{0}$
 &$\ththMat{0}{\times}{0}{\times}{0}{0}{0}{0}{\times}$
 &$\ththMat{\times}{0}{0}{0}{\times}{0}{0}{0}{0}$\\
\hline
2&$\ththMat{\times}{0}{0}{0}{\times}{\times}{0}{\times}{\times}$
 &$\ththMat{\times}{0}{\times}{0}{\times}{0}{0}{\times}{0}$
 &$\ththMat{0}{\times}{0}{\times}{0}{\times}{0}{\times}{0}$
 &$\ththMat{\times}{0}{\times}{0}{\times}{0}{\times}{0}{\times}$\\ 
\hline
3&$\ththMat{\times}{0}{0}{0}{\times}{\times}{0}{\times}{\times}$
 &$\thtwMat{\times}{0}{0}{\times}{0}{\times}$
 &$\twtwMat{0}{\times}{\times}{0}$
 &$\twtwMat{\times}{0}{0}{\times}$\\
\hline
\end{tabular}\caption{Coupling matrices generated by the transformations in Table~\ref{table2:tranf_matrix} where ``$\times$'' denotes an arbitrary complex entry. Note that $Y_\chi$ and $M_\textrm{bare}$ are symmetric matrices.}
\label{table2:coupling_matrix}
\end{center}
\end{table}

Another note is that while these models keep the atmospheric mixing angle, $\theta_{23}$ and the solar mixing angle, $\theta_{12}$ as arbitrary inputs, overall they contain less free parameters than the standard seesaw picture despite the more complicated Higgs sector. One reason for this is that the texture zeros in the coupling matrices together with the seesaw formula~\footnote{This was first discussed in Sec.~\ref{subsec:seesaw1}.}
\begin{equation}\label{eq2:seesaw_formula}
 m_\nu \simeq -\widehat{m}_D M_R^{-1} (\widehat{m}_D)^{T}
 \;,
\end{equation}
give rise to a light neutrino mass matrix $m_\nu$ (in any basis choice for $m_\ell$ and $M_R$) which has the following form:
\begin{equation}\label{eq2:M_nu}
 m_\nu = \ththMat{a_1}{a_2 b_1}{a_2 b_2}
                 {a_2 b_1}{a_3 b_1^2}{a_3 b_1 b_2}
                 {a_2 b_2}{a_3 b_1 b_2}{a_3 b_2^2}\;, 
 \quad\textrm{where }\; a_1, a_2, a_3, b_1, b_2\in \mathbb{C}\;,
\end{equation}
which then predicts that one of the Majorana mass eigenstates of the light neutrinos to be massless. In (\ref{eq2:seesaw_formula}), we have $\widehat{m}_D=Y_\nu\Dvev$ and $M_R = Y_\chi\Svev + \Mbare$, where $\Dvev$ and $\Svev$ are the VEV of fields $\Phi$ and $\chi$ respectively.

For better illustration and subsequent discussions, it is convenient to rewrite the Lagrangian of (\ref{eq2:model_lag1}) in the mass eigenbasis of the charged leptons and heavy RH Majorana neutrinos:
\begin{equation}\label{eq2:model_lag2}
 \mathcal{L}_{\textrm{mass}} = -\frac{1}{2}\overline{N}\,D_M\,N
                               -\overline{\ell}\,h_\ell\,\Phi\, e
                               -\overline{\ell}\,h_\nu\,\widetilde{\Phi}\,N 
                               +\textrm{h.c.},
\end{equation}
where $\ell =({\nu}_L',{\ell}_L')^T$ and $e$ are the charged lepton doublet and singlet respectively. We have defined the heavy Majorana neutrino field:~\footnote{In general, $\nu_L'\neq\nu$ (the mass eigenstate for light neutrinos). $\nu_L'$ and $\nu_R'$ are new fields from the change of basis.} $N=(\nu_R'+{\nu_R'}^c)/\sqrt{2}\,$. Subsequently, the charged lepton mass matrix is given by $D_\ell\equiv \textrm{diag}(m_e,m_\mu,m_\tau)=h_\ell\Dvev$, while for the heavy neutrinos, the diagonalised mass matrix is $D_M=\textrm{diag}(M_1,M_2,M_3)$. Although $m_\nu$ produced via the seesaw formula (\ref{eq2:seesaw_formula}) is unaffected by the basis change, in general, $h_\nu$ in (\ref{eq2:model_lag2}) would have texture zeros different from $Y_\nu$ (in fact, the texture zeros would disappear in most cases; see Table~\ref{table2:new_texture}). Using $U \equiv \Upmns$ to diagonalise $m_\nu$ and choosing the sign convention that $D_m=-U^\dagger m_\nu U^*$, one obtains $D_m=\textrm{diag}(m_1, m_2, m_3)$, where $m_i$ denotes the $i$th light neutrino mass.

%
\begin{table}[t]
\begin{center}
\begin{tabular}{|c|c|c|c|}
\hline
 &1 &2 &3 \\
\hline
$Y_\nu$
 &$\ththMat{\times}{0}{0}{0}{\times}{0}{0}{\times}{0}$
 &$\ththMat{\times}{0}{\times}{0}{\times}{0}{0}{\times}{0}$
 &$\thtwMat{\times}{0}{0}{\times}{0}{\times}$\\
\hline
$h_\nu$
 &$\ththMat{\times}{\times}{0}{\times}{\times}{0}{\times}{\times}{0}$
 &$\ththMat{\times}{\times}{\times}{\times}{\times}{\times}{\times}{\times}{\times}$
 &$\thtwMat{\times}{\times}{\times}{\times}{\times}{\times}$\\
\hline
\end{tabular}\caption{The structure of the coupling matrices: $Y_\nu$ in the family basis and $h_\nu$ in the mass eigenbasis of the charged leptons and RH Majorana neutrinos for Case 1 to 3. In general, ``$\times$'' denotes an arbitrary complex number.}
\label{table2:new_texture}
\end{center}
\end{table}

Since $m_1 =0$ (or $m_3 =0$ for inverted hierarchy) in these models, we have a strongly hierarchical LH neutrino mass spectrum, and the values for $m_{i}$'s can be evaluated using Eqs.~(\ref{eq1:NH_m2_m3}) and (\ref{eq1:IH_m1_m2}) as
\begin{align}
  &m_2 = \msol\;, \quad\quad\quad\quad\;\;
  \qquad m_3 = \matm \;, \quad (m_1 =0, \text{normal hierarchy})
  \label{eq2:m2_m3_normal}\\
  &m_1 = \sqrt{\matm^2 - \msol^2}\;,\qquad m_2 = \matm \;, \quad 
  (m_3 =0, \text{inverted hierarchy})
  \label{eq2:m1_m2_invert}
\end{align}
where $\msol = \sqrt{\left|\Delta\msol^2\right|} \approx 9\times 10^{-3}$~eV and $\matm = \sqrt{\left|\Delta\matm^2\right|} \approx 0.05$~eV from the best-fit values shown in (\ref{eq1:best-fit_mass}). It must be added that all the results presented so far remain valid under one-loop renormalisation group running \cite{Low:2005yc}.

\subsection{Parameters fine-tuning}\label{subsec:LV_finetune}

In order to make these models workable, the Higgs sector was expanded to accommodate the real singlet $\chi$. Its addition has inevitably introduced a new energy scale, $\Svev$, to the theory. Since $M_R = Y_\chi\Svev + \Mbare$ depends on this, it is essential to understand the implications of the scale of $\Svev$ in relation to other parameters in the model. The most general and renormalisable Higgs potential incorporating $\chi$ is given by
\begin{equation}\label{eq2:model_Higgs_pot}
 V(\Phi, \chi)=\frac{1}{2}\mu_\Phi (\Phi^\dagger\Phi)
              +\frac{1}{4}\lambda_\Phi (\Phi^\dagger\Phi)^2
              +\frac{1}{2}\mu_\chi \chi^2
              +\frac{1}{4}\lambda_\chi \chi^4
              +\mu_{\Phi\chi}(\Phi^\dagger\Phi)\chi^2 \;,
\end{equation}
where $\mu_\Phi$ and $\mu_\chi$ are in general functions of temperature $(T)$. Also, if the potential is to be bounded from below, then $\mu_{\Phi\chi} > -\sqrt{\lambda_\Phi\lambda_\chi}/2$. 

From the seesaw mechanism, we expect that $\Svev\gg\Dvev$, and hence, $T_{c,\chi}\gg T_{c,\Phi}\simeq\order{10^2}$~GeV, where $T_{c}$ denotes the critical temperature for symmetry restoration. In fact, one would need $T_{c,\chi} > T_\textrm{reh}$, the reheating temperature, so that any topological defects (domain walls) created by the spontaneous breaking of the $\Ztwo$ discrete symmetry of $\chi$ can be eliminated via inflation. The required hierarchy, $\Svev\gg\Dvev$, is ensured (to tree-level) if $\mu_{\Phi\chi}\rightarrow 0$. This can be seen from the tree-level minimum condition for (\ref{eq2:model_Higgs_pot}):
\begin{align}
 \mu_\Phi &= -\lambda_\Phi\Dvev^2 - 2\mu_{\Phi\chi}\Svev^2\;,\label{eq2:model_Higgs_min1}\\
 \mu_\chi &= -\lambda_\chi\Svev^2 - 2\mu_{\Phi\chi}\Dvev^2\;.\label{eq2:model_Higgs_min2}
\end{align} 
If $\order{\lambda_\Phi}\simeq \order{\lambda_\chi}=\order{1}$, then $\mu_{\Phi\chi}\rightarrow 0$ guarantees that $\mu_\Phi$ remains at $\order{\Dvev^2}$. In this limit, it also means that $\Phi$ and $\chi$ fields are decoupled from each other.

Because in the typical thermal leptogenesis analysis it is usually assumed that the reheating temperature $(T_\textrm{reh})$ after inflation is larger than the mass of the decaying heavy neutrino $M_j$, so we demand: \footnote{This restriction is only important if one wants to minimise theoretical uncertainties. In certain regimes, this may be relaxed without any appreciable change to the predictions. The decaying neutrino $M_j$ refers to the one that dominates the $L$ asymmetry generation which is $N_1$ for our investigations here.}
\begin{equation}\label{eq2:model_Temp_rel}
 T_{c,\chi} > T_\textrm{reh}> M_{j}  \gg T_{c,\phi}\;.
\end{equation}
This relation implies that $Y_\chi$ must be suitably fine-tuned so that $Y_\chi\Svev\simeq \order{M_j}$. In other words, elements of $Y_\chi$ cannot be arbitrarily small, and as a result, radiative corrections to the potential $V(\Phi,\chi)$ due to $\chi$'s coupling to other fields in the model may destroy the hierarchy between $\Dvev$ and $\Svev$. This presents a notable drawback for this class of models and thus the benefit of parameter reduction is \emph{not} unconditional.
Nonetheless, for the purpose of our investigation, we shall work in the assumption that such stability problem and any additional fine-tuning of the framework will be inconsequential to our main discussion.\\


\section{Implications in standard leptogenesis}\label{sec:LV_std_lepto}

In order to understand the potential implications of our specific models on leptogenesis predictions, it is helpful to first recall all the elements of the standard scenario and recast them, where appropriate, into the notations introduced in this chapter \footnote{It also serves to provide more details on the quantitative issues regarding the leptogenesis analysis, some of which we have skimmed through earlier.}. The key relation that captures the dependence of the predicted baryon-to-photon number ratio $(\eta_B)$ at recombination time on the elements of thermal leptogenesis can be written as
\begin{align}
 \eta_B &= \tilde{d}\, 
            \sum_{j=1}^{3} \varepsilon_j\, \kappa_j^\textrm{f}\;, \nonumber\\
        &\simeq \tilde{d}\, \varepsilon_1\, \kappa_1^\textrm{f} \;,
        \qquad (N_1\textrm{-dominated case})
        \label{eq2:std_etaB}
\end{align}
where $\tilde{d}$ is the dilution factor that accounts for the partial conversion of the generated excess $B-L$ into $\eta_B$ through sphaleron processes, as well as the increase of photon number per comoving volume from the onset of leptogenesis to recombination; $\varepsilon_j$ measures the $CP$ asymmetry in the decays of $N_j$ and is defined by:
\begin{equation}\label{eq2:std_cp1}
 \varepsilon_j = 
   \frac{\Gamma(N_j\rightarrow \ell\,\Phi)-\Gamma(N_j\rightarrow \bar{\ell}\,\bar{\Phi})}
        {\Gamma(N_j\rightarrow \ell\,\Phi)+\Gamma(N_j\rightarrow \bar{\ell}\,\bar{\Phi})}
        \equiv
        \frac{\Gamma_j - \overline{\Gamma}_j}{\Gamma_j + \overline{\Gamma}_j}\;,
\end{equation}
while $\kappa_j^\textrm{f}$ represents the (final) efficiency factor for $B-L$ production from $N_j$ decays which takes into account the initial conditions and the dynamics of particle interactions in the leptogenesis era, in particular, the interplay between decays, inverse decays and $\Delta L\neq 0$ scatterings. Hence, there are potentially three places where our models may modify the overall $\eta_B$ prediction. We shall discuss each of them as follows.

\subsection{Dilution factor}\label{subsec:LV_dilution}

As hinted towards the later part of Sec.~\ref{subsec:lepto_N1}, to track the time evolution of the number density of a quantity, $X$ (eg. $N_j$, $B-L$ or $B$), in an expanding universe, it is convenient to consider the number of particles, $\curlN_X(t)$, in a portion of comoving volume $(R^3(t))$ that contains one photon at some time $t' \ll t_\textrm{lepto}$, where $t_\textrm{lepto}$ denotes the time at the onset of leptogenesis \footnote{This is usually corresponds to the time when temperature $T \simeq M_1$ for the $N_1$-dominated scenarios.}. The conventional number density is then related to this via $\curlN_X(t) = n_X(t) R^3(t)$. We choose the normalization for $R^3(t)$ such that in relativistic thermal equilibrium, it contains on average $\curlN_{N_j}^{\textrm{eq}}(t\ll t_\textrm{lepto})=1$ heavy RH neutrino. So, for the baryon-to-photon ratio at recombination time (i.e. the ratio observed today), we have the relation:
\begin{equation}\label{eq2:LV_etaB_culNB_Nrec}
 \eta_B = \frac{\curlN_B^\textrm{f}}{\curlN_\gamma(t_\textrm{rec})}
 \;, 
\end{equation}
where $t_\textrm{rec}$ is the recombination time while $\curlN_B^\textrm{f}$ denotes the final $B$ excess after the leptogenesis era and with sphaleron effects already accounted for.
In the simplest case of constant entropy and assuming standard photons production from $t'$ to $t_\textrm{rec}$, one has 
\begin{equation}\label{eq2:LV_culN_Nrec}
 \curlN_\gamma(t_\textrm{rec})=
 \frac{\curlN_\gamma(t_\textrm{rec})}{\curlN_\gamma(t')}
  = \frac{4}{3}\times \frac{g^*_s(t')}{g^*_s(t_\textrm{rec})}
  = \frac{4}{3}\times \frac{434/4}{43/11}\approx 37 
  \;. 
\end{equation}
Note that $\curlN_\gamma(t')=1$ by definition. The pre-factor of 4/3 originates from our choice of normalisation and $g^*_s(t)$ is the relativistic degrees of freedom at time $t$ . In (\ref{eq2:LV_culN_Nrec}), we have already assumed the $N_1$-dominated scenario for leptogenesis in taking $g^*_s(t') = 427/4 + 7/4 = 434/4$ which is the degrees of freedom from all SM particles and $N_1$ only. $g^*_s(t_\textrm{rec})$ is equivalent to $g^*_s(t_\text{today})$ which is given by $2+21/11 = 43/11$ and accounts for the contribution coming from the relativistic photons and the three generations of light neutrinos (see for example \cite{kolb_turner}).

Up to this point, the calculation is as per usual because the only new particle in our models is the physical Higgs $H_\chi$ which is expected to gain mass at a very high energy $(T_{c,\chi})$, and by the time $t'$, it would have become non-relativistic. 
Although one can argue that since $M_{H_\chi}$ depends on $\lambda_\chi$ (which is unspecified), the mass of $H_\chi$ cannot be determined. However, as in Sec.~\ref{subsec:LV_finetune}, we may assume that $\order{\lambda_\chi}\simeq\order{\lambda_\phi}=\order{1}$, which is not unreasonable if one expects to find the Higgs $(H_\phi)$ at the TeV scale. As a result, this implies that $M_{H_\chi}\simeq\order{\Svev}$ would be very heavy. Besides, even if $H_\chi$ is included in the calculation of $g^*_s(t')$, the numerical value only differs from (\ref{eq2:LV_culN_Nrec}) by less than 2\%.

Another important element that contributes to the dilution factor comes from the imperfect conversion of $\curlN_{B-L}^\textrm{f}$ into $\curlN_{B}^\textrm{f}$. In the simplified picture which assumes that sphaleron processes are mostly active after the leptogenesis era, one can use the standard sphaleron conversion factor of $a_\textrm{sph}=\curlN_{B}^\textrm{f}/\curlN_{B-L}^\textrm{f} = 28/79$ in models with only one Higgs doublet \cite{khle_shaposh_et_al,harvey_turner_PRD42} \footnote{This factor is somewhat different if electroweak sphalerons remain in equilibrium until slightly after $T_{c,\Phi}$ \cite{Laine:1999wv}. Note that this case is highly probable because the electroweak phase transition seems to be not strongly first order. However, this is an issue that will affect not just our neutrino models but \emph{all} models, and so it will be inconsequential to our comparison here.}.
Then, the dilution factor is given by
\begin{equation}\label{eq2:LV_dilution_overall}
 \tilde{d} = \frac{28}{79}\times \frac{1}{37} \approx 0.96 \times 10^{-2}\;,
\end{equation}
where we have combined with the result from (\ref{eq2:LV_culN_Nrec}).
For a more thorough analysis when the combined effect of all spectator processes \cite{spectator1, spectator2} in the plasma  (eg. Yukawa interactions, QCD and electroweak sphalerons) is taken into account, the resultant value receives a 20\% to 40\% enchancement or suppression \cite{spectator2} depending on the specific leptogenesis temperature $T_\textrm{lepto}$ assumed \footnote{The dependence on temperature is originated from the fact that an increasing number of Yukawa or sphaleron processes comes into equilibrium as $T$ decreases.}.
The astute reader may have realised that this temperature $T_\textrm{lepto}$ is the same that will dictate the importance of flavor effects in leptogenesis . But in order to avoid confusion, we shall not elaborate on it here. It suffices to say that the change in the conversion rate comes about because the chemical potential for particles in thermal equilibrium are modified during leptogenesis, and not all of these potentials are independent as there are SM and sphaleron interactions relating them \cite{harvey_turner_PRD42,khle_shaposh_et_al}. 
As a result, the final value for $a_\textrm{sph}$ can be slightly different from the naive calculation.

\begin{figure}[tb]
\begin{center}
 \includegraphics[width=0.8\textwidth]{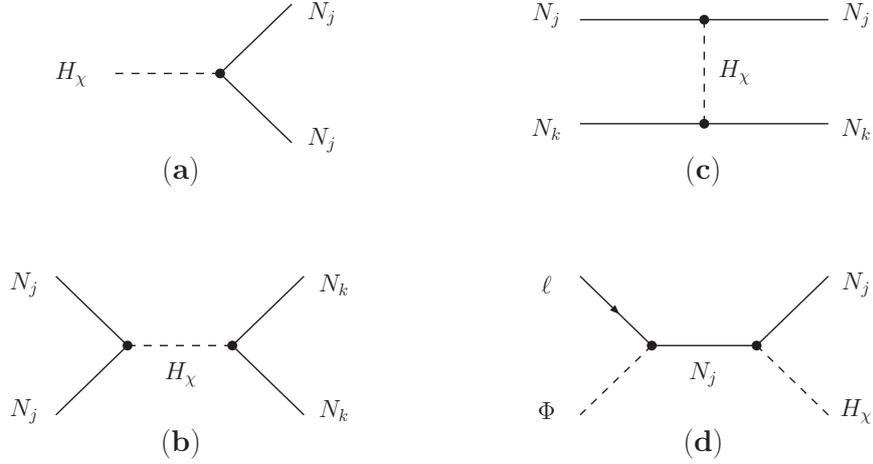}
\end{center} 
 \caption{Examples of new interactions involving $H_\chi$, $N_j$ and other SM particles.}
 \label{fig2:Nj_chi_processes}
\end{figure}

Applying these ideas to our models, it is not hard to see that $a_\textrm{sph}$ receives no significant modifications from the existence of $H_\chi$. This is because interactions such as $H_\chi\leftrightarrow N_j N_j$ and $N_j N_j\leftrightarrow N_k N_k$ (see Fig~\ref{fig2:Nj_chi_processes}a--c) do not change $\curlN_L$ or $\curlN_\Phi$ in the plasma, while processes like: $\ell \Phi \leftrightarrow H_\chi N_j$ (Fig.~\ref{fig2:Nj_chi_processes}d) cannot be in equilibrium for most temperatures due to the heaviness of $N_j$ and $H_\chi$. 
Furthermore, even if $H_\chi$ is light, the potentially relevant four-particle interaction $H_\chi H_\chi\leftrightarrow \Phi\,\bar{\Phi}$ which can change $\curlN_\Phi$ (and hence $\curlN_L$) is impotent since $\mu_{\Phi\chi}\rightarrow 0$. 
Therefore, we can conclude that our specific models have essentially the same dilution factor $\tilde{d}$ as the standard seesaw model whether or not spectator processes are considered.

\subsection{$CP$ asymmetry}\label{subsec:LV_CP_asym}

While dilution $\tilde{d}$ and efficiency factor $\kappa_j^\textrm{f}$ govern the portion of the generated asymmetry that would survive after the entire process, it is the $CP$ and $L$ violating decays of the heavy neutrino ($N_j\rightarrow l\Phi$ or $\overline{l}\Phi^\dagger$) that give rise to such asymmetry in the first place. The size of this quantity is controlled by the so-called \emph{seesaw geometry} \cite{PD_seesaw_geom} and in general highly dependent on the neutrino Yukawa structure. Using the notation of (\ref{eq2:model_lag2}), we first note that the tree-level particle and antiparticle decay rates are identical (as expected):
\begin{equation}\label{eq2:LV_tree_decay}
 \Gamma_j = \overline{\Gamma}_j = \frac{(h_\nu^\dagger h_\nu)_{jj}}{16\pi}M_j \;,
\end{equation}
and so no asymmetry is generated at this level. Recall from Chapter~\ref{ch_intro} that the leading contribution to $\varepsilon_j$ must come from the interference of the one-loop vertex and self-energy corrections with the tree-level coupling (see Fig.~\ref{fig1:lepto_std_decay_graphs} on page~\pageref{fig1:lepto_std_decay_graphs}). With the appropriate change in notations from Eq.~(\ref{eq1:lepto_std_CP_general_N1}), we have \cite{Covi:1996wh,
CP_calculations_Flanz,
CP_calculations_BP}
\begin{equation}\label{eq2:LV_cp2}
 \varepsilon_j = \frac{1}{8\pi}\sum_{k\neq j} 
              \frac{\textrm{Im}\left[(h_\nu^\dagger h_\nu)^2_{jk}\right]}
               {(h_\nu^\dagger h_\nu)_{jj}}
             \left\{f_V\left(\frac{M_k^2}{M_j^2}\right)+f_S\left(\frac{M_k^2}{M_j^2}\right)
             \right\}
             \;,
\end{equation} 
where $f_V(x)$ and $f_S(x)$ are given by
\begin{equation}\label{eq2:LV_fVS_defn}
 f_V(x) = \sqrt{x}\left[1-(1+x)\ln\left(\frac{1+x}{x}\right)\right]\quad\textrm{and}\quad
 f_S(x) = \frac{\sqrt{x}}{1-x}
\end{equation}
which denote the vertex and self-energy contributions respectively. For $j=1$ and in the limit of hierarchical RH neutrino with $M_1\ll M_{2,3}$ (i.e. $x\gg 1$), we have
\begin{equation}\label{eq2:LV_fVS_approx}
 f_V(x)+f_S(x)\simeq -\frac{3}{2\sqrt{x}}\;.
\end{equation}
Therefore, the $CP$ asymmetry for $N_1$ decays in this limit is given by
\begin{align}
 \varepsilon_1 &\simeq -\frac{3 M_1}{16\pi (h_\nu^\dagger h_\nu)_{11}}\sum_{k\neq 1}\, 
             \textrm{Im}\left[(h_\nu^\dagger h_\nu)^2_{1k}\right]
             \frac{1}{M_k}
             \;,
             \label{eq2:LV_cp3a}\\
            &= -\frac{3 M_1}{16\pi (h_\nu^\dagger h_\nu)_{11}}\,\textrm{Im}
            \left[(h_\nu^\dagger h_\nu D_M^{-1} h_\nu^T h_\nu^*)_{11}\right]
            \;.
            \label{eq2:LV_cp3b}
\end{align} 
Invoking the Casas-Ibarra parametrisation \cite{Casas-Ibarra}, $h_\nu$ in (\ref{eq2:LV_cp3b}) can be expressed in terms of the complex orthogonal matrix $\Omega$:
\begin{equation}\label{eq2:LV_omega1}
 \Omega = \Dvev\, D_m^{-1/2}\, U^\dagger\, h_\nu\, D_M^{-1/2} 
        \equiv D_m^{-1/2}\, U^\dagger\, m_D\, D_M^{-1/2}\quad \textrm{with}\quad \Omega^T\Omega=I \;,
\end{equation} 
where $m_D = h_\nu \Dvev$, $D_m^{1/2} = \text{diag}(\sqrt{m_1} , \sqrt{m_2}, \sqrt{m_3})$ and $D_M^{1/2} = \text{diag}(\sqrt{M_1},\sqrt{M_2},\sqrt{M_3})$, with all other symbols as defined in Sec.~\ref{sec:LV_low_model}. As in Chapter~\ref{ch_intro}, we shall adopt the sign convention such that $D_m =-U^\dagger m_\nu U^*$. Using (\ref{eq2:LV_omega1}) in (\ref{eq2:LV_cp3b}) and after some manipulations, one gets \cite{Casas-Ibarra}
\begin{align}
 \varepsilon_1 &\simeq \frac{3 M_1}{16\pi\Dvev^2}\frac{\sum_k m_k^2\,  
      \textrm{Im}(\Omega_{k1}^2)}
             {\sum_k m_k\, |\Omega_{k1}^2|}
             \;,\label{eq2:LV_omega_form}\\
           &=\frac{3 M_1 \matm}{16\pi\Dvev^2}
           \,\beta(m_1,\widetilde{m}_1,\Omega_{k1}^2)
           \;,\label{eq2:LV_cp4}
\end{align}
where we have introduced the dimensionless quantity \cite{PD_seesaw_geom}
\begin{align}
 \beta(m_1,\widetilde{m}_1,\Omega_{k1}^2) 
        &\equiv \frac{\sum_k m_k^2\, \textrm{Im}(\Omega_{k1}^2)}
             {\matm \sum_k m_k\, |\Omega_{k1}^2|}
             \;,\label{eq2:LV_beta1}\\
        &=\frac{\sum_k m_k^2\, \textrm{Im}(\Omega_{k1}^2)}
             {\matm\, \widetilde{m}_1}
             \;.\label{eq2:LV_beta2}
\end{align}
Here $\widetilde{m}_1$ is the effective neutrino mass associated with $N_1$ decays and its definition is shown in (\ref{eq1:eff_nu_mass}) on page~\pageref{eq1:eff_nu_mass}. For the purpose of the current analysis, we may assume that $\widetilde{m}_1$ and $M_1$ have been fixed due to other considerations such as the washout regime selected or the potency of flavor effects. Then, it can be easily seen that the size of the $CP$ asymmetry is controlled only by $m_1$ and the configuration of the three $\Omega_{k1}^2$'s (which are  neutrino model dependent). Note that even though the $\Omega_{k1}^2$'s are generally complex, the orthogonality condition $(\sum_k \Omega_{k1}^2=1)$ means that only 3 real independent parameters are needed to specify them. Moreover, since the numerator of (\ref{eq2:LV_beta2}) depends on the imaginary part of $\Omega_{k1}^2$, only 2 of these 3 parameters will manifest itself in $\beta(m_1,\widetilde{m}_1,\Omega_{k1}^2)$ when $\widetilde{m}_1 = \sum_k m_k |\Omega_{k1}^2|$ is a constant.

In the light of this, it is convenient to define $\Omega_{k1}^2=X_{k}+iY_{k}$, where $X_k, Y_k \in \mathbb{R}$. Moreover, in order to analyse both the normal and inverted hierarchy cases at the same time, we introduce a new subscript labeling system for the light neutrino masses: $m_{(k)}$ such that $m_{(1)} < m_{(2)} < m_{(3)}$ is always obeyed. In other words, we identify 
\begin{align}
&m_{(1)} = m_1\;\;,\;\; m_{(2)} = m_2\;\;,\;\; m_{(3)} = m_3\;, 
\qquad\text{for normal hierarchy}\;, \label{eq2:LV_k_label_NH}
 \\
&m_{(1)} = m_3\;\;,\;\; m_{(2)} = m_1\;\;,\;\; m_{(3)} = m_2\;, 
\qquad\text{for inverted hierarchy} \;. \label{eq2:LV_k_label_IH}
\end{align}
The corresponding squared orthogonal matrices are rewritten as $\Omega_{(k)1}^2 \equiv X_{(k)}+iY_{(k)}$ so that we have $\widetilde{m}_1 = \sum_{(k)} m_{(k)} |\Omega_{(k)1}^2|$ and $\sum_{(k)} \Omega_{(k)1}^2=1$. 

In this new notations and using the orthogonality condition, (\ref{eq2:LV_beta2}) becomes
\begin{equation}\label{eq2:LV_beta3}
 \beta(m_{(1)},\widetilde{m}_1,\Omega_{(k)1}^2) 
 = \frac{Y_2(m_{(2)}^2-m_{(1)}^2)+Y_3(m_{(3)}^2-m_{(1)}^2)}{\matm\,\widetilde{m}_1}
 \;.
\end{equation}
For a typical analysis, one is interested in which form of $\Omega_{(k)1}^2$'s will yield a maximum $CP$ asymmetry given fixed values for $m_{(1)}$ and $\widetilde{m}_1$. Thus, it is useful to ascertain the upper bound \footnote{Without loss of generality, we may adopt the convention that the $CP$ asymmetry is positive and hence the lower bound is given by $0 \leq\beta(m_{(1)},\widetilde{m}_1,\Omega_{(k)1}^2)$.} 
 on $\beta(m_{(1)},\widetilde{m}_1,\Omega_{(k)1}^2)$. To this end, we break up the dependence on $m_{(1)}, \widetilde{m}_1$ and $\Omega_{(k)1}^2$ and introduce an effective leptogenesis phase $\delta_L$ as follows \cite{PD_seesaw_geom, Asaka_Hamaguchi, Buchmuller:2003gz, davidson_ibarra_eps_bound}
\begin{align}
 \beta(m_{(1)},\widetilde{m}_1,\Omega_{(k)1}^2) 
 &=\beta_\textrm{max}(m_{(1)},\widetilde{m}_1)\sin \delta_L
 \;,
  \label{eq2:LV_beta4}\\
 &=\beta_1(m_{(1)})\,\beta_2(m_{(1)},\widetilde{m}_1)
  \sin \delta_L
 \;,
 \label{eq2:LV_beta5}
\end{align}
where $\delta_L$ is in general a complicated function of $m_{(1)},\widetilde{m}_1$ and $\Omega_{(k)1}^2$. $\beta_\textrm{max} \leq 1$ represents the maximal value for $\beta$ given a particular $m_{(1)}$ and $\widetilde{m}_1$, while $\beta_1$ is the upper bound for $\beta$ when $m_{(1)}$ is fixed. If $\widetilde{m}_1$ is also fixed, then $\beta_2$ provides the necessary correction to $\beta_1$. By carefully analysing (\ref{eq2:LV_beta3}), one finds that \cite{PD_seesaw_geom, davidson_ibarra_eps_bound}
\begin{equation}\label{eq2:LV_b_1}
 \beta_1(m_{(1)}) = \frac{m_{(3)}-m_{(1)}}{\matm} \;.
\end{equation}
To get $\beta_2$, one observes that for a generic $\Omega$ matrix, a configuration that maximises $\beta$ while keeping $\widetilde{m}_1$ the same is achieved when $\Omega_{(2)1}^2=0$ \cite{PD_seesaw_geom}. This allows one to rewrite (\ref{eq2:LV_beta3}) as 
\begin{align}
 \beta &= \frac{Y_{(3)}(m_{(3)}^2-m_{(1)}^2)}{\matm\,\widetilde{m}_1}
   \;,\label{eq2:LV_beta6}\\
       &= \frac{Y_{(3)\textrm{m}}\sin\delta_L (m_{(3)}^2-m_{(1)}^2)}{\matm\,\widetilde{m}_1}
       \;,\label{eq2:LV_beta7}
\end{align}
where $Y_{(3)\textrm{m}}$ is the maximum value of $Y_{(3)}$ when $\Omega_{(2)1}^2=0$. By putting (\ref{eq2:LV_beta5}), (\ref{eq2:LV_b_1}) and (\ref{eq2:LV_beta7}) together, one gets
\begin{equation}\label{eq2:LV_b_2}
 \beta_2 
 = \frac{\beta}{\beta_1\,\sin\delta_L} = \frac{m_{(1)}+m_{(3)}}{\widetilde{m}_1}\,Y_{(3)\textrm{m}}
 \;.
\end{equation}
In general, $Y_{(3)\textrm{m}}$ will depend on the light neutrino masses. For a fully hierarchical neutrino spectrum, i.e. $m_{(1)}=0$ (like our specific models), we have $\beta_1=m_{(3)}/\matm$ and $\beta_2=1$ (with $X_{(2)}=Y_{(2)}=X_{(3)}=0$ and $Y_{(3)\textrm{m}}=\widetilde{m}_1/m_{(3)}$). Furthermore, if the hierarchy is normal, $\beta_1=1$ as $m_{(3)}=\matm$ (see (\ref{eq2:m2_m3_normal})).

Applying this decomposition of the $CP$ asymmetry to our models, it is then straight forward to deduce the specific predictions coming from the structures of $h_\nu$ presented in Table~\ref{table2:new_texture} \footnote{We shall ignore Case 3 for it is effectively identical to Case 1.}. 
To begin with, we note that in Case 2, the corresponding $\Omega$ matrix is arbitrary since all entries of $h_\nu$ in this case are unconstrained by the symmetry. As a result, from the above general analysis, it is certain that Case 2 has enough parameter freedom to produce a sufficient $CP$ asymmetry for the leptogenesis purpose, and in fact it predicts no additional restrictions other than those implied by the model independent analysis on fully hierarchical LH neutrinos.

The situation for Case 1 is slightly different as there are texture zeros in the matrix $h_\nu$. To study the behaviors of $\Omega$ given this $h_\nu$ structure, it is better to replace all zero entries with an infinitesimal parameter $\varrho$, and so we have
\begin{equation}\label{eq2:LV_h_rho}
 h_\nu = \ththMat{w_{11}}{w_{12}}{\varrho}{w_{21}}{w_{22}}{\varrho}{w_{31}}{w_{32}}{\varrho}
 \;, \quad \text{ where } \; w_{ij} \in \mathbb{C}\;.
\end{equation}
Using (\ref{eq2:LV_omega1}), the complex orthogonal matrix then has the form \footnote{Note that we have momentarily switched back to the original notations to avoid the need to displaying the two hierarchical cases separately.}
\begin{equation}\label{eq2:LV_omega_rho}
 \Omega=\ththMat
 {w_{11}'/\sqrt{m_{1} M_1}}{w_{12}'/\sqrt{m_{1} M_2}}{w_{13}'\,\varrho/\sqrt{m_{1} M_3}}
 {w_{21}'/\sqrt{m_2 M_1}}{w_{22}'/\sqrt{m_2 M_2}}{w_{23}'\,\varrho/\sqrt{m_2 M_3}}
 {w_{31}'/\sqrt{m_3 M_1}}{w_{32}'/\sqrt{m_3 M_2}}{w_{33}'\,\varrho/\sqrt{m_3 M_3}}
 \;,
\end{equation}
where $w_{ij}'$ is an arbitrary complex number. Note that we have kept all $m_j$'s and $M_j$'s as variables and only assumed $U_{e3}=0$ in deriving this. For definiteness, let's assume the normal hierarchy~\footnote{The corresponding conclusion for the inverted case will be the same with the only change coming from the relocation of the 0's and 1 in $\Omega$.}, and 
so in the limit $\varrho\rightarrow 0$, one can immediately conclude that entries $\Omega_{23}, \Omega_{33} \rightarrow 0$ because $m_{2,3}$ and $M_3$ are finite and nonzero. On the other hand, since this model predicts $m_{1}=0$, hence at first glance, $\Omega_{13}$ is indeterminate while $\Omega_{11}$ and $\Omega_{12}$ are infinite as $m_{1}, \varrho\rightarrow 0$. But by appealing to the orthogonality condition, both issues can be resolved and $\Omega$ simplifies to
\begin{equation}\label{eq2:LV_omega_3}
 \Omega=\ththMat{0}{0}{1}
                {\sqrt{1-\Omega_{31}^2}}{-\Omega_{31}}{0}
                {\Omega_{31}}{\sqrt{1-\Omega_{31}^2}}{0}\;,
                \quad \text{ where } \; \Omega_{31} \in \mathbb{C}\;.
\end{equation}
It is interesting to note that this orthogonal seesaw matrix is identical to the special form derived from models with $M_3 \rightarrow \infty$ \cite{2RHnu_1, 2RHnu_2, Ibarra:2003xp}, or analogously models with only two RH neutrinos. Therefore, this further illustrates that Case 1 and 3 of our models are phenomenologically equivalent.

The only distinction between these generic models and ours is that (\ref{eq2:LV_omega_3}) is originated from the Yukawa couplings and family symmetry (which gives $m_{1}=0$ as a by-product) rather than the imposition of $M_3 \rightarrow \infty$. Hence, the constraints from our Case 1 on the $CP$ asymmetry and their subsequent implications in thermal leptogenesis would be very similar to those models with only two RH neutrinos \cite{Ibarra:2003up}.

Firstly, for this case, $\widetilde{m}_1$ is bounded from below as
\begin{equation}\label{eq2:LV_m1eff_case1}
 \widetilde{m}_1 = m_{(2)} |1-\Omega_{(3)1}^2|+m_{(3)}|\Omega_{(3)1}^2|\;,
\end{equation}
where we have used the generalised version of (\ref{eq2:LV_omega_3}), which takes care of both normal and  inverted cases, in the definition for $\widetilde{m}_1$. So when $|\Omega_{(3)1}^2|\ll 1$, then $\widetilde{m}_1\simeq m_{(2)}$ whereas when $|\Omega_{(3)1}^2|\gg 1$, then $\widetilde{m}_1\gg m_{(2)}$. Thus, we have $\widetilde{m}_1 \gtrsim m_{(2)}$. It is worth mentioning that in the previous case, we can safely set $X_{(2)}=Y_{(2)}=X_{(3)}=0$ in order to maximise $\beta_2$ which leads to the relation: $\widetilde{m}_1 = m_{(3)} Y_{(3)\textrm{m}}$, and hence no bounds on $\widetilde{m}_1$, whereas in the current case, $X_{(2)}$ and $Y_{(2)}$ are no longer arbitrary. This is due to the fact that $\Omega_{(1)1}=0$ and thus the freedom to change $X_{(2)}$ and $Y_{(2)}$ while keeping $\widetilde{m}_1$ constant has been lost.

Secondly, the reduction in free variables in $\Omega$ allows us to consider the function $\beta$ for this model directly which has become
\begin{equation}
 \beta =\frac{(m_{(3)}^2-m_{(2)}^2)\,Y_{(3)}}{\matm\,\widetilde{m}_1}
 \;,\label{eq2:LV_beta_case1}
\end{equation}
with
\begin{equation}
 \widetilde{m}_1 =m_{(2)}\sqrt{(1-X_{(3)})^2+Y_{(3)}^2}+m_{(3)}\sqrt{X_{(3)}^2+Y_{(3)}^2}
 \;,
\end{equation}
where we have imposed $X_{(2)}=1-X_{(3)}$ and $Y_{(2)}=-Y_{(3)}$ (from orthogonality). As a result of the bound: $\widetilde{m}_1\gtrsim m_{(2)}$, the size of $\beta$ (and hence the $CP$ asymmetry) is very sensitive to the size of $Y_{(3)}$. This is because when $\widetilde{m}_1$ is unbounded, $Y_{(3)}\gg 1$  and $Y_{(3)}\ll 1$ lead to $\widetilde{m}_1\gg1$ and $\widetilde{m}_1\ll 1$ respectively and so $\beta$ (which is proportional to $Y_{(3)}/\widetilde{m}_1$) approaches the same limit ($\simeq 1$) for both situations, whereas 
in (\ref{eq2:LV_beta_case1}) we have instead
\begin{equation}\label{eq2:LV_beta_limit}
 \beta = \frac{(m_{(3)}^2-m_{(2)}^2)}{\matm}\times
 \begin{cases}
  \displaystyle{\quad\frac{Y_{(3)}}{m_{(2)}}}\\
  \displaystyle{\frac{Y_{(3)}}{(m_{(3)}+m_{(2)})\,Y_{(3)}}}
 \end{cases}
 \longrightarrow 
 \begin{cases}
  \quad 0 & \textrm{if}\; Y_{(3)}\ll X_{(3)}\ll 1 \;,\\
  \displaystyle{\frac{m_{(3)}-m_{(2)}}{\matm}}& \textrm{if}\; Y_{(3)}\gg X_{(3)}\gg 1 \;.
 \end{cases}
\end{equation}
Therefore in order to obtain maximum $CP$ asymmetry, $\widetilde{m}_1$ must be very large. However, this is potentially detrimental to the success of thermal leptogenesis because $\widetilde{m}_1 \gg 1$ also implies large washout rates (see next subsection). Therefore, $\widetilde{m}_1$ would actually be upper bounded. Fortunately, $M_1$ also dictates the final efficiency factor and the net result is that in order to circumvent the problem, one requires a larger lower bound for $M_1$ at the same time \cite{BDP_0205349_CMB}. Within the non-supersymmetric context, the mass of $M_1$, which is related to $T_\textrm{reh}$, is not too tightly constrained and thus Case 1 of our model will be workable in leptogenesis.

Another observation from (\ref{eq2:LV_beta_limit}) is that the maximum attainable $\beta$ value in the limit of $Y_{(3)}\gg X_{(3)}\gg 1$ is drastically different depending on the hierarchy scheme assumed for the light neutrinos. For normal hierarchy, $\beta_\textrm{max} \simeq (\matm-\msol)/\matm \approx 0.82$ whereas in the inverted case $\beta_\textrm{max} \simeq \msol/\matm \approx 0.18$. Although this result alone is not sufficient to rule out the inverted case, it is clear that, for Case 1 at least, the inverted hierarchy is strongly disfavored.

Overall, we have demonstrated in this subsection that our specific neutrino models with family symmetry can naturally generate the required raw $CP$ asymmetry necessary for successful leptogenesis. The parameter space in these models is not overly restrictive, and the predictions are identical to a couple of special cases in the default seesaw, namely, the fully hierarchical light neutrinos limit (Case 2) and the two RH neutrinos scenario (Case 1 or 3).

\subsection{Efficiency factor}\label{subsec:LV_eff}

In the $N_1$-dominated scenario, the dynamical generation of a $B-L$ asymmetry in the leptogenesis era depends on the out-of-equilibrium decays of the heavy $N_1$'s, as well as other interactions in the thermal plasma. These non-equilibrium processes which control the evolution of $\curlN_{N_1}$ and $\curlN_{B-L}$ are quantified by a system of two Boltzmann kinetic equations as first discussed in Sec.~\ref{sec:lepto_main}. An important issue in setting up these is the identification of all relevant interactions which can modify $\curlN_{N_1}$ and $\curlN_{B-L}$ (see Table~\ref{table2:washout_int} and Fig.~\ref{fig2:Nj_chi_processes}). Following Eqs.~(\ref{eq1:lepto_std_BE_N1_dt}) and (\ref{eq1:lepto_std_BE_B-L_dt}), and making a change of variable from $t$ to the dimensionless quantity $z=M_1/T$ (for convenience) \cite{luty,
Plum_Zphy_NPB,BDP_0205349_CMB,kolb_turner, BDP_pedestrain}, we have 
\begin{align}
 \odifone{\curlN_{N_1}}{z} &= -(D+S)(\curlN_{N_1}-\curlN_{N_1}^\textrm{eq})\;,
 \label{eq2:LV_BE1}\\
 \odifone{\curlN_{B-L}}{z} &= -\varepsilon_1\,D\,(\curlN_{N_1}-\curlN_{N_1}^\textrm{eq})
    -W\,\curlN_{B-L}
    \;,
    \label{eq2:LV_BE2}
\end{align}
where $\curlN_{N_1}^\textrm{eq}$ denotes the equilibrium value for $\curlN_{N_1}$ which is now a function of $z$. The term $D,S,W=\Gamma_{D,S,W}/(z H)$ encapsulate the reaction rate of the various processes with $\Gamma_{D,S,W}$ denoting the decay, scattering and washout (thermally averaged) rates respectively, and $H$ being the Hubble parameter which is given by
\begin{equation}
 H \simeq 1.66 \sqrt{g_s^*}\, \frac{M_1^2}{M_\textrm{Pl}}\frac{1}{z^2}\;,
 \label{eq2:LV_hubble}
\end{equation}
where $g_s^*(z\simeq 1) =106.75$ is the number of relativistic degrees of freedom~\footnote{This includes all SM particles only. The $N_1$ degrees of freedom is not included because in the preferred \textit{strong} washout regime (see e.g. \cite{BDP_pedestrain}), they are non-relativistic at $t_\textrm{lepto}$ (corresponding to $z_\textrm{lepto}=M_1/T_\textrm{lepto}\simeq 1$).},
$M_\textrm{Pl}\approx 1.22\times 10^{19}$~GeV is the Planck mass. The thermally averaged (total) decay rate $\Gamma_D$ which accounts for decays and inverse decays ($N_1\leftrightarrow \ell\Phi$), is related to the zero-temperature decay rate $\Gamma_D^{(T=0)}\equiv\Gamma_1+\overline{\Gamma}_1$ via~\footnote{This redefinition of $\Gamma_D$ absorbs the pre-factor of 2 appeared in (\ref{eq1:lepto_std_BE_N1_dt}) and (\ref{eq1:lepto_std_BE_B-L_dt}). Subsequently, $\Gamma_{S,W}$ have also been redefined accordingly.}
\begin{equation}\label{eq2:LV_T=0_decay_rate}
 \Gamma_D = \Gamma_D^{(T=0)} \frac{\mathcal{K}_1(z)}{\mathcal{K}_2(z)}\;,
\end{equation}
where $\mathcal{K}_n(z)$ is the $n$th order modified Bessel function of the second kind.
$\Gamma_S$ represents the $\Delta L =\pm 1$ scatterings (eg. $N_1 \ell\leftrightarrow \overline{t} q\,$) but ignores scatterings which involve gauge bosons $V_\mu$ to first approximation~\footnote{It turns out that they are only critical if one considers the \textit{weak} washout regime. See for example \cite{Giudice_0310123, Covi:1997dr}.}.
The washout rate $\Gamma_W$, which incorporates everything that tends to erase the $B-L$ asymmetry, is dependent on the rates for inverse decay ($\ell\Phi\rightarrow N_1$), all $\Delta L=\pm 1$ scatterings (except those with $V_\mu$), as well as, the $\Delta L=\pm 2$ processes mediated by $N_1$ (eg. $\ell \Phi\leftrightarrow \bar{\ell} \bar{\Phi}$)~\footnote{$\Delta L=\pm 2$ processes mediated by $N_{2,3}$ are suppressed at $z\simeq 1$ because $M_{2,3}\gg M_1$.}.

%
\begin{table}[t]
\begin{center}
\begin{tabular}{|l||l|}
\hline
$\underline{\Delta L=0}$ & $\underline{\Delta L=\pm 1}$\\ 
$\quad H_\chi\leftrightarrow N_j N_j$ (tree + loop)$\star$
      & $\quad N_j\leftrightarrow \ell \Phi$ (tree + loop)\\
$\quad N_j N_j\leftrightarrow N_k N_k$ ($H_\chi, s$)$\star$
      & $\quad N_j\;\ell\leftrightarrow \overline{t}\; q$ ($\Phi, s$)\\
      & $\quad N_j\;t\leftrightarrow \overline{\ell}\; q$ ($\Phi, t$)\\
$\quad N_j N_j\leftrightarrow N_j N_j$ ($H_\chi, s$)$\star$
      & $\quad N_j\;q\leftrightarrow \ell\; t$ ($\Phi, t$)\\
$\quad N_j N_j\leftrightarrow N_j N_j$ ($H_\chi, t$)$\star$
      &\\
$\quad N_j N_k\leftrightarrow N_j N_k$ ($H_\chi, t$)$\star$
      & $\quad N_j\;\overline{\ell}\leftrightarrow \Phi\; V_\mu$ ($\Phi, s$)\\
      & $\quad N_j\;\overline{\ell}\leftrightarrow \Phi\; V_\mu$ ($\ell, t$)\\
$\quad N_j N_j\leftrightarrow H_\chi\bar{\Phi}\Phi$ ($H_\chi, s$)$\star$
      & $\quad N_j\, V_\mu\leftrightarrow \Phi\; \ell$ ($\Phi, t$)\\
$\quad N_j H_\chi\leftrightarrow N_j\bar{\Phi}\Phi$ ($H_\chi, t$)$\star$
      & $\quad N_j\, V_\mu\leftrightarrow \Phi\; \ell$ ($\ell, t$)\\
      & $\quad N_j\bar{\Phi}\leftrightarrow V_\mu\, \ell$ ($\Phi, t$)\\      
$\underline{\Delta L=\pm 2}$
      & $\quad N_j\bar{\Phi}\leftrightarrow V_\mu\, \ell$ ($\ell, s$)\\
$\qquad \ell\Phi\leftrightarrow \overline{\ell}\bar{\Phi}$ ($N_j, s$)
      &\\
$\qquad \ell\Phi\leftrightarrow \overline{\ell}\bar{\Phi}$ ($N_j, t$)
      & $\quad \ell\,\Phi\leftrightarrow N_j H_\chi$ ($N_j, s$)$\star$\\
$\qquad \ell\,\ell\leftrightarrow \bar{\Phi}\bar{\Phi}$ ($N_j, t$)
      & $\quad N_j\,\ell\leftrightarrow H_\chi \bar{\Phi}$ ($N_j, t$)$\star$\\
      & $\quad H_\chi\, \ell\leftrightarrow N_j \bar{\Phi}$ ($N_j, t$)$\star$\\
\hline
\end{tabular}\caption{A collection of potentially important processes in leptogenesis. The type of process (eg. tree-level, vertex/self-energy loop, $s$- or $t$-channel) and, where applicable, the mediating particle are in brackets. Interactions that are not in the default see-saw are marked by a ``$\star$''. $q$, $t$ and $V_\mu$ denote the (3rd generation) quark doublet, top quark singlet and gauge boson respectively.}
\label{table2:washout_int}
\end{center}
\end{table}

To describe the behavior of the solutions to (\ref{eq2:LV_BE1}) and (\ref{eq2:LV_BE2}), it is customary to introduce the decay parameter \cite{kolb_turner, BDP_0205349_CMB, luty, Plum_Zphy_NPB,BDP_pedestrain}
\begin{equation}\label{eq2:LV_decayK}
 K_1 = \frac{\Gamma_D^{(T=0)}}{H(z=1)} \equiv \frac{\widetilde{m}_1}{m_*}
 \;,
\end{equation}
where
\begin{equation}\label{eq2:LV_mstar}
 m_* =\frac{16}{3}\sqrt{\frac{\pi^5 g^*_s}{5}}\,\frac{\Dvev^2}{M_\textrm{Pl}} = \order{10^{-3}}\;
      \textrm{eV}
      \;,
\end{equation}
is the equilibrium neutrino mass. $\widetilde{m}_1$ is the effective neutrino mass as defined in the previous section which measures how strongly coupled $N_1$ is to the thermal plasma. It should be noted that the size of $\widetilde{m}_1$ relative to $m_*$ marks the boundary between the so-called weak ($\widetilde{m}_1 < m_*$) and strong ($\widetilde{m}_1 > m_*$) washout regimes where the corresponding analyses are qualitatively different \cite{BDP_pedestrain}.
This dependence of the solutions comes about because the interaction terms $D, S$ and $W_1$ (defined as $W_1 = W - \delta W$, where $\delta W = (\Gamma_{N,s}+\Gamma_{N,t})/(zH)$ represents the contribution from \emph{non-resonant} $\Delta L=\pm 2$ processes, i.e. on-shell component of the propagator of $N_1$ has been properly subtracted) are proportional to $\widetilde{m}_1$ \cite{BDP_0205349_CMB}:
\begin{equation}\label{eq2:LV_DSW_m1eff}
 D, S, W_1 \propto \frac{\Mpl\,\widetilde{m}_1}{\Dvev^2}
 \;,\quad\textrm{while}\quad
 \delta W\propto \frac{\Mpl\, M_1\, \overline{m}^2}{\Dvev^4}
 \;,
\end{equation}
where $\overline{m}^2 = m_1^2+m_2^2+m_3^2$. For hierarchical light neutrinos and $M_1\ll 10^{14}$ GeV (note that both of these are satisfied in our models~\footnote{The only restriction on $M_1$ is coming from (\ref{eq2:model_Temp_rel}) which does not limit the possibility of $M_1\ll 10^{14}$ GeV.}), it turns out that contribution from $\delta W$ can be safely neglected \cite{BDP_0205349_CMB, BDP_pedestrain}, and therefore the generated $B-L$ asymmetry is to a good approximation independent of $M_1$. In the interesting case of \textit{strong} washout, the $W_1$ term is dominated by inverse decays \cite{Giudice_0310123, BDP_pedestrain}. Moreover, in this regime, the heavy $N_1$'s in the plasma can reach thermal abundance before $t_\textrm{lepto}$ even if the scattering term $S$ is turned off. As a result, the details of $N_1$'s production prior to their decays become irrelevant and all subsequent analyses are greatly simplified. In the light of this, we shall concentrate on the strong washout regime in much of our discussions here.

With the above simplifications, (\ref{eq2:LV_BE1}) and (\ref{eq2:LV_BE2}) reduce to
\begin{align}
 \odifone{\curlN_{N_1}}{z} &= -D\,(\curlN_{N_1}-\curlN_{N_1}^\textrm{eq})
 \;,\label{eq2:LV_BE3}\\
 \odifone{\curlN_{B-L}}{z} &= -\varepsilon_1\,D\,(\curlN_{N_1}-\curlN_{N_1}^\textrm{eq})
    -W_1^\textrm{ID}\curlN_{B-L}
    \;,
    \label{eq2:LV_BE4}
\end{align}
where $W_1^\textrm{ID}(	\simeq W_1)$ is the dominant piece in the washout term which originates from inverse decays. The solution for $\curlN_{B-L}$ can be expressed in an integral form \cite{BDP_pedestrain, Kolb:1983ni}
\begin{equation}\label{eq2:LV_integ_std}
 \curlN_{B-L}(z) = \curlN_{B-L}^{\textrm{i}}\;e^{-\int_{z_\textrm{i}}^z dz'\, W_1^\textrm{ID}(z')} 
  -\varepsilon_1\,\kappa_1(z)
  \;,
\end{equation}
where $\curlN_{B-L}^{\textrm{i}}$ and $z_\textrm{i}$ denote the initial values for $\curlN_{B-L}$ and $z$ respectively. The corresponding efficiency factor is given by
\begin{equation}\label{eq2:LV_kappa_1}
 \kappa_1(z)= -\int_{z_\textrm{i}}^{z}\,dz'\,\odifone{\curlN_{N_1}}{z'}\; 
   e^{-\int_{z'}^{z}\,dz''\,W_1^\textrm{ID}(z'')}
   \;.
\end{equation}
Since in the strong washout regime one can invoke the approximation $d\curlN_{N_1}/dz'\simeq d\curlN_{N_1}^\textrm{eq}/dz'$, the final value as $z\rightarrow\infty$ may be readily worked out and one obtains \cite{BDP_pedestrain}
\begin{equation}\label{eq2:LV_kappa_2}
 \kappa^\textrm{f}_1(K_1)\simeq 
   \frac{2}{K_1\,z_B(K_1)}\left[1-e^{-\frac{K_1\,z_B(K_1)}{2}}\right]\;,
   \quad\textrm{for } K_1\gtrsim 1
   \;,
\end{equation}
where we have expressed $\kappa^\textrm{f}_1$ as a function of the decay parameter $K_1$, and $z_B$ ($\gg z_\textrm{lepto}\simeq 1$) is the value around which the asymptotic expansion of the $z'$ integral in (\ref{eq2:LV_kappa_1}) receives a dominant contribution. The temperature $T_B$ that corresponds to $z_B$ ($\equiv M_1/T_B$) is referred to as the \emph{baryogenesis temperature}. For practical purposes, (\ref{eq2:LV_kappa_2}) is well approximated by the simple power law \cite{BDP_pedestrain}:
\begin{equation}\label{eq2:LV_kappa_3}
 \kappa^\textrm{f}_1 \simeq (2\pm 1)\times 10^{-2}\left[
    \frac{0.01\,\textrm{eV}}{\widetilde{m}_1}\right]^{1.1\pm 0.1} \;,
   \quad\textrm{for}\; \widetilde{m}_1 > m_* \;.
\end{equation}
Assuming we are working in the strong washout region, we will now argue that the standard results described above are directly applicable to our models with family symmetry. Of the $\Delta L =0$ or $\Delta L =\pm 1$ new interactions originating from coupling to $H_\chi$, all involving an external $H_\chi$ in the initial state (e.g. $H_\chi\rightarrow N_j N_j$) can be disregarded because the reaction density for $H_\chi$ is almost zero at $z\simeq 1$ while the reverse processes are kinematically forbidden due to the heaviness of $H_\chi$~\footnote{It should be noted that the $2\leftrightarrow 3$ reactions such as $N_j N_j\leftrightarrow H_\chi\Phi^\dagger\Phi$ are in any case impotent because of $\mu_{\phi\chi}\rightarrow 0$.}. Furthermore, the $N_j$-$N_k$ scatterings mediated by a virtual $H_\chi$ do not play a role because in the \textit{strong} washout regime, the analysis is insensitive to the initial $N_j$ abundance in the thermal plasma. Hence, there is no new important contribution from the list of starred ($\star$) reactions in Table~\ref{table2:washout_int} to the standard scattering and washout terms, and we can conclude that our models do not predict a modification to the efficiency factor $\kappa_1^\textrm{f}$ given by the default seesaw case.

\subsubsection{Weak washout regime}

We end this section by including a short discussion of the phenomenologies that might be essential to the analysis of our extended seesaw models in the context of weak washout~\footnote{For a complete review of the analysis in the weak washout regime, see for example \cite{BDP_pedestrain,
Giudice_0310123}.}. While the efficiency factor is typically enhanced in this regime, one obvious drawback is that the prediction is no longer independent of initial conditions as in the scenario for strong washout. As a result, the $N_j$-$N_k$ scatterings via a $H_\chi$ may be significant since these can actively modify $\curlN_{N_1}^{\textrm{i}}$ in the plasma while inverse decays are no longer strong enough to ensure thermal abundance before the onset of leptogenesis.

However, upon closer inspection and assuming a hierarchical RH neutrino mass spectrum, one would not expect a sizable change to $\curlN_{N_1}^{\textrm{i}}$. This is because in Table~\ref{table2:washout_int} there is no ($\Delta L =0$) creation or annihilation process which involves $N_1$'s and some \textit{lighter} particles. Note that one does not need to be concerned with $\Delta L \neq 0$ processes in this analysis because they are in general too weak to bring $N_j$'s into equilibrium at high temperature ($T_\textrm{lepto} \ll T < T_\textrm{reh}$). Thus only processes such as $N_1 N_1 \leftrightarrow N_k N_k\; (k=2,3)$ can bring $N_1$'s into equilibrium. But as has been shown in a similar model in \cite{Plum_Zphy_NPB} where the additional $N_j$-$N_k$ interactions come from couplings to massive neutral gauge bosons (related to GUT breaking), these scatterings have very little effect on the final asymmetry prediction if one has a pronounced RH neutrino mass hierarchy~\footnote{The model presented in \cite{Plum_Zphy_NPB} provides a good guide to the phenomenology expected in ours with the exception that our models do not possess interactions that link $N_j$ to SM particles. Consequently, our models would predict even less modifications than in the case of \cite{Plum_Zphy_NPB}.}. In addition, the $N_1$-dominated approximation remains valid.

It should be pointed out that all interactions involving an external $H_\chi$ remain unimportant. When $z\ll 1$, processes that can produce $N_j$ (e.g. $H_\chi\rightarrow N_j N_j$) are irrelevant because they are only potent before the inflationary stage (see (\ref{eq2:model_Temp_rel})) and any excess $N_j$ created will be diluted away. On the other hand, for $z\simeq 1$, reaction density for $H_\chi$ is almost zero as in the strong washout regime.

Other issues that are potentially important in the weak washout regime include the effects of scattering processes that couple to gauge bosons $V_\mu$  \cite{Giudice_0310123} (see Table~\ref{table2:washout_int}), as well as thermal corrections to $\Gamma_D$ and $\Gamma_S$ \cite{Giudice_0310123, Covi:1997dr}. These elements provides an additional source of theoretical uncertainty to the overall analysis and hence making the weak washout scenario less attractive than the strong washout case. For the purpose of our work though, these effects unnecessarily complicate the analysis and therefore will be ignored.

Although we have not shown by explicit calculations that predictions of our models will not significantly deviate from (or bias towards certain parameter space within) the standard case with weak washout, it is apparent from the above discussion that a lot of it will be highly dependent on the assumptions made on the mass of $N_j$, $H_\chi$ and their Yukawa couplings. Given that all of these are free parameters in our models, any deviations with respect to the standard see-saw case will therefore stay within the amount of these uncertainties. Hence, we can conclude that their predictions are effectively the same as the default case.\\

\section{Implications in flavored leptogenesis}\label{sec:LV_flv_lepto}

As outlined in Sec.~\ref{subsec:lepto_flavor}, flavor effects become important when the assumption on the temperature range in which leptogenesis happens is altered \cite{blanchet_zeno,
barbieri_flavor,
endoh_flavor,
nardi_flavor,
abada_flavor_1,
abada_flavor_2,
simone_riotto_flv,
blanchet_flavor}. This arises because Yukawa interactions become more prominent at temperatures $T\lesssim 10^{12}$ GeV and as a result they can destroy the coherent evolution of the lepton doublets, $\ell$'s in the early Universe.

To elaborate this a bit more, consider the lepton states, $|\,\ell_{(j)}\rangle$'s~\footnote{The subscript $(j)$ highlights the fact that the flavor decomposition of $\ell_{(j)}$ can be different for each $N_j$.} generated by the decays of $N_j$'s. Normally, in the one-flavor approximation,  $|\,\ell_{(j)}\rangle$ will evolve \emph{coherently} as the Universe cools. But the presence of charged lepton Yukawa interactions in equilibrium essentially introduces a source of decoherence whereby the $|\,\ell_{(j)}\rangle$'s are projected onto one of the three flavor eigenstates, $|\,\ell_{\alpha}\rangle$'s ($\alpha = e, \mu, \tau$) with probability $|\langle \ell_{(j)}|\,\ell_{\alpha}\rangle|^2$. 

Because of the difference in size between the tauon and muon Yukawa couplings, there exist two temperature thresholds: $T^\textrm{eq}_\tau$ and $T^\textrm{eq}_\mu$, which govern the range where tauon and muon interactions come into equilibrium respectively. When $T > T^\textrm{eq}_\tau\simeq 10^{12}$~GeV, both the $\tau$- and $\mu$-Yukawa interactions are out-of-equilibrium and flavor effects can be ignored. For the temperature range $T^\textrm{eq}_\tau \gtrsim T\gtrsim T^\textrm{eq}_\mu\simeq 10^{9}$~GeV \cite{barbieri_flavor, abada_flavor_1,
abada_flavor_2}, only $\tau$-Yukawas are in equilibrium and one effectively has a two-flavor ($\tau$ and a linear combination of $\mu$ and $e$) problem. Finally, in the case of $T^\textrm{eq}_\mu \gtrsim T$, both reactions are strong enough to instigate a full three-flavor system. In our discussion here, we do not explicitly distinguish between the two situations and simply assume the three-flavor regime. Furthermore, spectator effects in the form discussed in \cite{spectator1, spectator2} will be ignored for brevity~\footnote{These effects typically change the result by a factor of $\order{1}$ in most scenarios, besides they affect both our models and the standard framework in the same way, hence, they can be safely ignored.}.

To illustrate the origin of the flavor affected $CP$ asymmetry, we begin by rewriting  Lagrangian (\ref{eq2:model_lag2}) with sub-indices~\footnote{A proper discussion of this should be done within the density matrix framework \cite{barbieri_flavor, abada_flavor_1}. But for our purpose, it is enough to follow the more intuitive approach as in \cite{nardi_flavor}.}
\begin{equation}\label{eq2:model_lag3}
 \mathcal{L}_{\textrm{mass}} = -\frac{1}{2}\overline{N}_j\,D_{M_j}\,N_j
                               -\overline{\ell}_\alpha\,(h_\ell)_\alpha\,\Phi\, e_\alpha
     -\overline{\ell}_\alpha\,(h_\nu)_{\alpha j}\,\widetilde{\Phi}\,N_j 
                               +\textrm{h.c.}\;,
\end{equation}
where $j=1,2,3$ and $\alpha=e,\mu,\tau$. It is then clear that the state $|\,\ell_{(j)}\rangle$ and $|\,\overline{\ell}_{(j)}\rangle$ in (\ref{eq2:std_cp1}) are in general \textit{not} a $CP$ conjugate of each other owning to the complex entries in matrix $h_\nu$. If we are considering the temperature ranges where the relevant lepton Yukawa interactions are either fully in equilibrium or out-of-equilibrium but not in between such that state $|\,\ell_{(j)}\rangle$ will  quickly decohere into one of the $|\,\ell_{\alpha}\rangle$ states available, then we can effectively think of $\Gamma_j$ as an incoherent sum (over $\alpha$) of the partial decay rates
$\Gamma_{j\alpha}\equiv \Gamma(N_j\rightarrow \ell_\alpha \Phi)$. Likewise for $|\,\overline{\ell}_{(j)}\rangle, |\,\overline{\ell}_\alpha\rangle, \overline{\Gamma}_j$ and $\overline{\Gamma}_{j\alpha}\equiv\Gamma(N_j\rightarrow \overline{\ell}_\alpha \bar{\Phi})$. Note here that $|\,\ell_{\alpha}\rangle$ and $|\,\overline{\ell}_\alpha\rangle$ are $CP$ conjugate states of each other. 

From this, it is straightforward to see that $CP$ violation in $N_j$ decays can manifest itself in two places:
\begin{enumerate}
 \item[(A).] The amount of $|\,\ell_\alpha\rangle$ and $|\,\overline{\ell}_\alpha\rangle$ produced are not the same because $|\,\ell_{(j)}\rangle$ and $|\,\overline{\ell}_{(j)}\rangle$ are produced at different rates which corresponds to $\Gamma_j\neq \overline{\Gamma}_j$. 
 \item[(B).] The amount of $|\,\ell_\alpha\rangle$ and $|\,\overline{\ell}_\alpha\rangle$ produced are not the same because $\Gamma_{j\alpha}\neq \overline{\Gamma}_{j\alpha}$ (regardless of the relation between $\Gamma_j,  \overline{\Gamma}_j$).
\end{enumerate}
Obviously, the second effect is only relevant when one is considering the evolution of the individual lepton flavor asymmetry $L_\alpha$. For the one-flavor approximation, one tracks the evolution of $L=\sum_\alpha L_\alpha$ instead, and hence only the first effect comes into play. However, an important corollary is that when flavor effects are included, the associated \textit{flavored} $CP$ asymmetry, $\varepsilon_{j\alpha}$, can be nonzero even if $\Gamma_j =\overline{\Gamma}_j$ (i.e. $\varepsilon_j =0$)~\footnote{See Sec.~\ref{subsec:lepto_flavor} for related discussion on this.}. To properly quantify all these, it is convenient to introduce the flavor projectors \cite{nardi_flavor}
\begin{align}
 P_{j\alpha} &\equiv \frac{\Gamma_{j\alpha}}{\Gamma_{j}} 
  = |\langle \ell_{(j)}|\,\ell_{\alpha}\rangle|^2
 \;,
 \label{eq2:LV_P1}\\
 \qquad\qquad\qquad\qquad
 \overline{P}_{j\alpha} &\equiv \frac{\overline{\Gamma}_{j\alpha}}{\overline{\Gamma}_{j}} 
 = |\langle \overline{\ell}_{(j)}|\,\overline{\ell}_{\alpha}\rangle|^2
 \;,\qquad
  (j=1,2,3;\, \alpha=e,\mu,\tau)
 \label{eq2:LV_P2}
\end{align}
where $\Gamma_j =\sum_\alpha\, \Gamma_{j\alpha}$ and $\overline{\Gamma}_j =\sum_\alpha\, \overline{\Gamma}_{j\alpha}$. Therefore, by definition, $\sum_\alpha\, P_{j\alpha} =\sum_\alpha\, \overline{P}_{j\alpha}=1$. The associated $\alpha$ flavor $CP$ asymmetry is given by
\cite{barbieri_flavor, nardi_flavor}
\begin{align}
 \varepsilon_{j\alpha} \equiv \frac{\Gamma_{j\alpha}-\overline{\Gamma}_{j\alpha}}
                       {\Gamma_{j}+\overline{\Gamma}_j}
           &=\frac{\Gamma_{j}P_{j\alpha}-\overline{\Gamma}_j\overline{P}_{j\alpha}}
                       {\Gamma_{j}+\overline{\Gamma}_j}
                       \;,\label{eq2:LV_fcp1}\\
           &=\frac{P_{j\alpha}+\overline{P}_{j\alpha}}{2}\,\varepsilon_j
                        +\frac{P_{j\alpha}-\overline{P}_{j\alpha}}{2}\;,
                       \label{eq2:LV_fcp2}\\
           &\simeq P^0_{j\alpha}\,\varepsilon_j+\frac{\delta P_{j\alpha}}{2}\;,
                       \label{eq2:LV_fcp3}
\end{align}
where $P^0_{j\alpha}$ is the tree-level contribution to the projector $P_{j\alpha}$ with $P^0_{j\alpha}=\overline{P}^0_{j\alpha}$, while $\delta P_{j\alpha}\equiv P_{j\alpha} -\overline{P}_{j\alpha}$ is the quantity that characterises the $CP$ violating effect of type (B). From (\ref{eq2:LV_fcp3}), it is clear that even if $\varepsilon_j=0$ and hence $P^0_{j\alpha}\,\varepsilon_j$  vanishes~\footnote{This term corresponds to type (A) $CP$ violating effect.}, $\varepsilon_{j\alpha}$ does not necessarily go to zero. In addition, 
\begin{equation}\label{eq2:LV_proj_sum2zero}
 \sum_\alpha\,\delta P_{j\alpha} = \sum_\alpha\,(P_{j\alpha} -\overline{P}_{j\alpha})
  = 1- 1=0
  \;,
\end{equation}
demonstrating that $\varepsilon_j\equiv \sum_\alpha\,\varepsilon_{j\alpha}$ will not depend on $\delta P_{j\alpha}$, and hence it is consistent with our claim that only type (A) effect can contribute to the overall asymmetry when flavor effects is neglected.

Putting these quantities in terms of parameters in (\ref{eq2:model_lag3}) and setting $j=1$ for the $N_1$-dominated scenario, we get \cite{nardi_flavor}
\begin{align}
  P_{1\alpha}^0 =\overline{P}^0_{1\alpha} &= \frac{(h_\nu^*)_{\alpha 1}(h_\nu)_{\alpha 1}}
  {(h_\nu^\dagger h_\nu)_{11}}
  \;,\label{eq2:LV_flavorP0a}\\
  &=\frac{|\sum_k\,\sqrt{m_k}\,U_{\alpha 1} \Omega_{k1}|^2}{\sum_k\,m_k |\Omega^2_{k1}|}
  \;,
  \label{eq2:LV_flavorP0b}
\end{align}
and \cite{Covi:1996wh}
\begin{equation}\label{eq2:LV_flavorCP}
 \varepsilon_{1\alpha}=
  -\frac{1}{8\pi(h_\nu^\dagger h_\nu)_{11}}
   \sum_{k\neq 1}\textrm{Im}\left\{ (h_\nu^*)_{\alpha 1}(h_\nu)_{\alpha k}\frac{M_1}{M_k}
   \left[\,\frac{3}{2}(h_\nu^\dagger h_\nu)_{1k}
      +\frac{M_1}{M_k}(h_\nu^\dagger h_\nu)_{k1}\right]
   \right\}\;,
\end{equation}
with $\varepsilon_1$ given by (\ref{eq2:LV_cp3a}). Using these definitions, one can then calculate $\delta P_{1\alpha}$ via (\ref{eq2:LV_fcp3}). 
It is worth highlighting that in (\ref{eq2:LV_flavorCP}), the term proportional to $(h_\nu^\dagger h_\nu)_{k1}$ drops out upon summing over $\alpha$. This is because when combined with the pre-factor outside the brackets, it becomes $(h_\nu^\dagger h_\nu)_{1k}(h_\nu^\dagger h_\nu)_{k1}$ which is real. As a result, it indicates that this term actually corresponds to $\delta P_{1\alpha}$ type $CP$ effect.

To derive the network of Boltzmann equations, first we note that the final asymmetry $\curlN^\textrm{f}_{B-L}$ would now depend on the evolution of all individual flavor asymmetries $\curlN_{Q_\alpha}$ where $Q_\alpha \equiv B/3-L_\alpha$ and $\curlN^\textrm{f}_{B-L}=\sum_\alpha\,\curlN^\textrm{f}_{Q_\alpha}$. 
Secondly, the washout terms would be modified in the multi-flavor case because $\Delta L \neq 0$ interactions involving $\Phi$ couple to state $|\,\ell_{(1)}\rangle$ and not $|\,\ell_\alpha\rangle$. Hence, washout effects will be dependent on $P_{1\alpha}$ (or approximately $P_{1\alpha}^0$).  

If the Yukawa interactions are either strongly in equilibrium or out-of-equilibrium but not in the transition region~\footnote{See \cite{blanchet_zeno} for a thorough discussion on the location of this region for various setups.}, then the kinetic equations are greatly simplified as flavor dynamics due to the coherences (i.e. off-diagonal terms of the relevant density operator) may be neglected. 
Otherwise, the analysis must be done in the density matrix formalism \cite{barbieri_flavor,abada_flavor_1} so that effects from partial losses of coherence and correlations in flavor space can be fully accounted for.

Appealing to the approximation  (as in Sec.~\ref{sec:LV_std_lepto} for the unflavored case) that decays and inverse decays dominate over the washout and scattering term, then overall we get \cite{blanchet_flavor,nardi_flavor} 
\begin{align}
 \odifone{\curlN_{N_1}}{z} &= -D\,(\curlN_{N_1}-\curlN_{N_1}^\textrm{eq})
 \;,\label{eq2:LV_fBE1}\\
 \odifone{\curlN_{Q_\alpha}}{z} &= -\varepsilon_{1\alpha}\,D\,(\curlN_{N_1}-\curlN_{N_1}^\textrm{eq})
    -P^0_{1\alpha}\,W_1^\textrm{ID}\,\curlN_{Q_\alpha}
    \;,\label{eq2:LV_fBE2}
\end{align}
which are very similar to (\ref{eq2:LV_BE3}) and (\ref{eq2:LV_BE4}). So, just like (\ref{eq2:LV_integ_std}), the solution to $\curlN_{Q_\alpha}$ can be expressed in an integral form:
\begin{equation}    \label{eq2:LV_fsoln1}
 \curlN_{Q_\alpha}(z) = 
 \curlN_{Q_\alpha}^{\textrm{i}}\,
   e^{-P^0_{1\alpha}\int_{z_\textrm{i}}^z dz'\, W_1^\textrm{ID}(z')} 
  -\varepsilon_{1\alpha}\,\kappa_{1\alpha}(z)
  \;,
\end{equation}
where the flavor dependent efficiency factor is given by
\begin{equation}\label{eq2:LV_fkappa1}
 \kappa_{1\alpha}(z)= -\int_{z_\textrm{i}}^{z}\,dz'\,\odifone{\curlN_{N_1}}{z'}\; 
   e^{-P^0_{1\alpha}\int_{z'}^{z}\,dz''\,W_1^\textrm{ID}(z'')}
   \;.
\end{equation}
In the strong washout regime, one then gets (cf. (\ref{eq2:LV_kappa_2})) \cite{blanchet_flavor}
\begin{equation}\label{eq2:LV_fkappa2}
 \kappa^\textrm{f}_{1\alpha}(K_{1\alpha})\simeq 
   \frac{2}{K_{1\alpha}\,z_B(K_{1\alpha})}
   \left[1-e^{-\frac{K_{1\alpha}\,z_B(K_{1\alpha})}{2}}\right]\;,
\end{equation}
where $K_{1\alpha}\equiv P^0_{1\alpha} K_{1}$. Since $P^0_{1\alpha}\leq 1$, we have $K_{1\alpha} \leq K_{1}$, and so (\ref{eq2:LV_fkappa2}) implies that washout is in general reduced because of flavor effects. As a consequence of this, one finds that the region of parameter space corresponding to the strong washout regime is also reduced \cite{blanchet_flavor}.

From the above discussion, it is clear that much of the analysis done  in Sec.~\ref{sec:LV_std_lepto} for our models will be applicable to the case when flavor effects are included. Comparing (\ref{eq2:LV_fBE1}) and (\ref{eq2:LV_fBE2}) with (\ref{eq2:LV_BE3}) and (\ref{eq2:LV_BE4}), we observe that $P^0_{1\alpha}$ and $\varepsilon_{1\alpha}$ are the only two new ingredients which contain all the additional model dependent information originating from flavor effects. Hence, it suffices to investigate the implications of our specific models on these two quantities.

To begin with, we note that both $P^0_{1\alpha}$ and $\varepsilon_{1\alpha}$ depends explicitly on the mixing matrix $U$, unlike in the one-flavor approximation where everything can be written in terms of the seesaw orthogonal matrix $\Omega$. This can be seen from (\ref{eq2:LV_flavorP0b}) and by expanding the two terms containing $h_\nu$'s in (\ref{eq2:LV_flavorCP}) using (\ref{eq2:LV_omega1}) to obtain
\begin{equation}\label{eq2:LV_fCPterm1}
 (h_\nu^*)_{\alpha 1}(h_\nu)_{\alpha k}(h_\nu^\dagger h_\nu)_{1k}
 = \frac{M_1 M_k}{\Dvev^4}
    \sum_n \, m_n \Omega^*_{n1}\Omega_{nk}\,
    \sum_{\ell,\,m}\sqrt{m_\ell\,m_m}\: 
    \Omega^*_{\ell 1}\Omega_{mk}U^*_{\alpha\ell}U_{\alpha m}
    \;,
\end{equation}
and
\begin{equation}\label{eq2:LV_fCPterm2}
 (h_\nu^*)_{\alpha 1}(h_\nu)_{\alpha k}(h_\nu^\dagger h_\nu)_{k1}
 = \frac{M_1 M_k}{\Dvev^4}
    \sum_n \, m_n \Omega_{n1}\Omega^*_{nk}\,
    \sum_{\ell,\,m}\sqrt{m_\ell\,m_m}\: 
    \Omega^*_{\ell 1}\Omega_{mk}U^*_{\alpha\ell}U_{\alpha m}
    \;.
\end{equation}
These general expressions  illustrate   how the low energy $CP$ violating phases enter explicitly into the predictions of leptogenesis when flavor effects are important. But it is also clear from them that such effects are usually masked by the complex phases in $\Omega$ (or $h_\nu$) and can only have a direct impact when $\Omega$ is real as mentioned in Sec.~\ref{subsec:lepto_flavor}. 
In the context of our models with family symmetry, since the prediction is $U_{e3}=0$, the Dirac phase $\delta$ from $\Upmns$ does not enter into the theory at all, and hence $CP$ violating effects arising from low energy parameters would entirely come from the \textit{Majorana} phases. 

Regarding whether our models would give rise to a significant departure in the predictions of leptogenesis (with flavors), there are some general observations we can make~\footnote{Note that there is nothing in all cases of our models which specifically indicates whether flavor effects should be included or not as $M_1$ is a free parameter in the theory. Thus, leptogenesis with flavor effects is phenomenologically not excluded by our models.}. Firstly, by examining (\ref{eq2:LV_flavorP0a}) and (\ref{eq2:LV_flavorCP}), we can safely conclude that for Case 2 of our models (see Table~\ref{table2:new_texture}) there is essentially no modifications to the standard seesaw scenario. This is due to the fact that all entries in $h_\nu$ for this case are unconstrained by the symmetry and therefore it can, in principle, accommodate all specific scenarios the default seesaw allows (except $m_{(1)}=0$ must be obeyed). 

Secondly, using expression (\ref{eq2:LV_flavorP0b}) for $P^0_{1\alpha}$, we see that both Case 1 and 2 do not provide any predictions or restrictions on the projectors $P^0_{1\alpha}$'s~\footnote{Again Case 3 is automatically accounted for when considering Case 1.}. This is because $P^0_{1\alpha}$ depends either on the 21- and 31-entries (normal hierarchy) or the 11- and 21-entries (inverted hierarchy)~\footnote{No dependence on the 11-entry in the normal mass spectrum because $m_1=0$ and contribution to the sum in (\ref{eq2:LV_flavorP0b}) automatically disappears. Similarly for the inverted spectrum when $m_3=0$.} in both $U$ and $\Omega$ and we know from before that these are unconstrained by the symmetry (see for example matrix (\ref{eq2:LV_omega_3}) for the normal hierarchy).

To study the implications of Case~1 (normal hierarchy) on $\varepsilon_{1\alpha}$, we can apply $m_1=0$, $U_{e3}=0$ and $\Omega$ given in (\ref{eq2:LV_omega_3}) to expressions (\ref{eq2:LV_fCPterm1}) and (\ref{eq2:LV_fCPterm2}). A key feature which results is that both (\ref{eq2:LV_fCPterm1}) and (\ref{eq2:LV_fCPterm2}) vanishes when $k=3$, which is in accordance with the previous conclusion that Case~1 corresponds to the seesaw case with only two RH neutrinos. In fact, this is the only definite prediction from the model. Although one gets significant simplification to (\ref{eq2:LV_fCPterm1}) and (\ref{eq2:LV_fCPterm2}) for particular values of $k$ and $\alpha$, the resulting expressions are still a function of unconstrained parameters, hence one has to impose additional conditions in order to restrict the result. For instance, when $k=2$ and $\alpha=e$, (\ref{eq2:LV_fCPterm2}) simplifies to
\begin{equation}\label{eq2:LV_fterm1}
 (h_\nu^*)_{e 1}(h_\nu)_{e 2}(h_\nu^\dagger h_\nu)_{21}
 = \frac{m_2|U_{e2}|^2M_1 M_2}{\Dvev^4}\, 
 \left[m_2 |\Omega_{31}^2| |1-\Omega_{31}^2|-m_3  \Omega_{31}^2(1-\Omega^2_{31})^*
 \right]
 \;,
\end{equation}
where we have used $m_1=0, U_{e3}=0$ and (\ref{eq2:LV_omega_3}). Plugging this into (\ref{eq2:LV_flavorCP}) and defining $\Omega_{31}^2=X_3+iY_3$, we simply get (ignoring $\order{(M_1/M_2)^2}$ terms):
\begin{equation}
 \varepsilon_{1e} \simeq -\frac{3 M_1}{16\pi \Dvev^2\, \widetilde{m}_1}\,
  m_2^2\,|U_{e2}|^2\, Y_3. 
\end{equation}
This example illustrates that although the texture zeros can lead to partial simplification, in the end, one has to specify $\Omega_{31}^2$ and $U_{e2}$ (in this case) in order to make any predictions about $\varepsilon_{1e}$. Besides, the final asymmetry would in general depends on all of the $\varepsilon_{1\alpha}$'s and not just $\varepsilon_{1e}$. Therefore, all we may conclude is that Case~1 of these models will possess the phenomenologies given by the standard seesaw with two RH neutrinos. An extensive study on this particular situation in the context of leptogenesis with flavor effects has been performed recently in \cite{abada_flavor_2}.

So far, we have assumed the strong washout scenario exclusively. The reason for that is in the weak washout regime, the calculation of the final asymmetry relies on a good knowledge of the thermal history of all RH neutrinos in the plasma. Since our models do not provide specific information on that front and flavor effects cannot actually alter the phenomena we discussed in Sec.~\ref{subsec:LV_eff}, therefore, our conclusion on this regard would be similar. In the light of all these, it is clear that implications of our models in leptogenesis with flavor effects are basically the same as those from the normal seesaw with one massless LH neutrino.

\section{Summary of results}\label{sec:LV_summary}

Given the intricate connection between neutrino properties and leptogenesis, it is natural to ask whether a specific class of constrained seesaw models which has less free parameters than the default setup, can solve the baryogenesis problem. In this chapter, we have performed a thorough check in this context on such neutrino models as those proposed in \cite{Low:2005yc}. These models obey certain abelian family symmetries and contain one additional singlet in the Higgs sector such that the total number of arbitrary parameters in the seesaw theory is reduced. In particular, they predict that $\theta_{13}=0$ in the $\Upmns$ matrix and a fully hierarchical light LH neutrino spectrum.

By dissecting the dependence of the final asymmetry on sphalerons, $CP$ violating decays and washouts, we have identified the key elements that can modify the leptogenesis predictions. Consequently, the implications of different seesaw neutrino models can be easily compared. It was found that in all cases of our models, leptogenesis predictions are almost identical to those allowed by the default seesaw model, with the exception that our models naturally select the hierarchical light neutrino solution (for $m_{1\,\text{or}\,3}=0$ is one of the main features). This conclusion is true for both the one-flavor approximation, as well as when flavor effects are included, since all of the essential elements that can change the final asymmetry turn out to be unconstrained in our models. In one case, the phenomenology is very much the same as the standard situation with three RH neutrinos whereas in the two other cases, they correspond to the scenario with only two RH neutrinos. Furthermore, for leptogenesis with flavor effects, we have found that Majorana phases in the light neutrino mixing matrix can play an important role since in our models the Dirac phase disappears due to the fact that $\theta_{13}=0$.

It should be noted that the entire investigation has been done assuming the limit of hierarchical RH neutrinos with emphasis on the $N_1$-dominated scenario in all situations. However, it is not hard to see that both the $N_2$-dominated scenario (see Sec.~\ref{subsec:lepto_N2}) and resonant leptogenesis (see Sec.~\ref{subsec:lepto_resonant}) are viable alternatives  as the weakly constrained nature of our models on the standard leptogenesis parameter space suggests.

Although in much of this work the analyses are done in the strong washout regime, there are clear indications that our general conclusion can be extended to the weak washout case. A more precise calculation on this front, however, cannot be carried out since our models do not provide any specific predictions on the RH neutrino sector. On one hand, this means that nothing new is predicted by our models, but on the other, this non-existence of heavy constraints ensures that these models can lead to successful leptogenesis in most scenarios. Consequently, the compatibility with thermal leptogenesis highlights another benefit of these simple seesaw neutrino models with abelian family symmetry.



%
%
%
%
%



\chapter{Seesaw sector and low-energy observables}\label{ch_work_HLV}

\ECap{A}{s}{7}{0}
we have seen in Chapter~\ref{ch_work_LV}, enlarging the SM symmetry (as well as the Higgs sector) can give rise to a theory that contains less arbitrary inputs and, in some cases, also provides definite predictions. Therefore, within the framework of type I seesaw, it is natural to ask whether the largely unrestricted RH neutrino sector can be determined through symmetries. This possibility is especially attractive because of
the connection between seesaw parameters and leptogenesis. As a result, cosmological constraints may provide important hints on the selection of the underlying symmetries employed in model building.

To understand the role symmetries can play in this regard, we first notice that through the seesaw mechanism \cite{type1_seesaw}, a relation between the heavy RH neutrino mass matrix and the low-energy neutrino data can be established although this link is incomplete as there are many parameters remaining to be fixed. Secondly, the neutrino Dirac  mass matrix structure may have some resemblances to those of quarks and charged leptons such as when the SM originates from a GUT or partial unification scheme. Hence, there are key regions in the parameter space where the powers of symmetry can be used to bridge the gap between low-energy observables and the unknown seesaw sector.

In the following, we begin by elucidating all the model building strategies while establishing the notations. Then, we construct a few representative models which can realise this connection between low- and high-energy parameters, before presenting a phenomenological study of them.\\

\section{Seesaw structure and relation to the low-energy sector}\label{sec:HLV_low-high_link}

The basic link between the heavy RH seesaw and the low-energy sectors comes from the seesaw relation (first discussed in Sec.~\ref{subsec:seesaw1}) involving the effective light neutrino mass matrix $m_\nu$ which is defined through 
\begin{equation}
\frac{1}{2}\; \overline{\nu} \; \mnu \, \nu^c + \text{h.c.} \;,
\end{equation}
and the heavy RH Majorana mass matrix $M_R$ coming from
\begin{equation}
\frac{1}{2} \,\overline{(\nu_R)^c}\, M_R\, \nu_R + \text{h.c.} \;,
\end{equation}
where $\nu$ and $\nu_R$ are the effective light and RH Majorana neutrino field respectively. Explicitly, the formula is given by~\footnote{We have ignored terms of $\order{(\mnud)^3/M_R^2}$ or higher.}
\begin{equation}
\mnu \simeq -\mnud\, M_R^{-1} (\mnud)^T \;,
   \label{eq3:HLV_seesaw_main}
\end{equation}
where $\mnud$ is the neutrino Dirac mass matrix, defined through
\begin{equation}
 \overline{\nu}_L \,\mnud\, \nu_R + \text{h.c.} \;,
\end{equation}
with $\nu_L$ being the LH neutrino field in the weak eigenbasis~\footnote{The relationship between $\nu$ and $\nu_L$ was discussed in Sec.~\ref{subsec:seesaw1}}. Let $\nu^m = V_{\nu} \;\nu$ be the mass eigenstates for LH Majorana neutrinos, where $V_{\nu}$ is the unitary diagonalization matrix for $\mnu$. Then the diagonal light neutrino mass matrix, $\mnuhat$ is given by~\footnote{Diagonalized matrices will always be denoted by a carat in this chapter.}
\begin{equation}
\mnuhat \equiv {\rm diag}(m_1, m_2, m_3) = -V_{\nu} \,\mnu\, V^{T}_{\nu}\;,
\label{eq3:HLV_ML_diag}
\end{equation}
where $m_i$ denotes the $i$th light neutrino mass. With this, Eq.~(\ref{eq3:HLV_seesaw_main}) implies that
\begin{equation}
\mnuhat \simeq  V_{\nu}\, \mnud \,M_R^{-1} (\mnud)^T V^{T}_{\nu}\;.
\label{eq3:HLV_seesaw_form2}
\end{equation}
The matrix $\mnuhat$ has been experimentally determined up to an absolute light neutrino mass scale, and as a result we shall conveniently parametrize it through the lightest $\mnu$ eigenvalue $m_1$ (normal hierarchy) or $m_3$ (inverted hierarchy). Other neutrino mass eigenvalues are then expressed in terms of this and the experimentally measured square-mass differences: $\Delta\msol^2$ and $\Delta\matm^2$ \cite{nu_best-fit}, as in (\ref{eq1:NH_m2_m3}) and (\ref{eq1:IH_m1_m2}) on page~\pageref{eq1:NH_m2_m3}. 

If we suppose further that $e_L^m = V_{eL}\,e_L$ are the mass eigenstates for LH charged leptons, then the leptonic weak interaction Lagrangian becomes
\begin{equation}\label{eq3:HLV_weak_lag}
 \mathcal{L}_\text{weak} =
  i g\, \overline{e}^m_L\, \slashed{W}\, V_{eL} V_\nu^\dagger\, \nu_m
  + \text{h.c.}
  \;,
\end{equation}
where $V_{eL} V_\nu^\dagger \equiv \Upmns$ is identified to be the leptonic mixing matrix, and as discussed in Chapter~\ref{ch_intro}, its entries are constrained by low-energy neutrino experiments. Similarly, one can define the quark mixing matrix as $\Uckm \equiv V_{uL} V_{dL}^\dagger$, where $V_{uL}$ and $V_{dL}$ are the left-diagonalization matrices for the up- and down-type quark matrices respectively~\footnote{CKM stands for Cabibbo-Kobayashi-Maskawa.}.

In order to connect the high- and low-energy sectors, one must have $M_R$ completely determined by known quantities. One way to achieve this is to have $M_R$ constructed from some combination of $\mnuhat$, the charged fermion mass matrices, $\widehat{m}_f$ with $f=e,d,u$, and the mixing matrices ($\Upmns$ and $\Uckm$) respectively. As a consequence, Eq.~(\ref{eq3:HLV_seesaw_form2}) demands that 
\begin{equation}
\fbox{
The neutrino Dirac mass matrix, \mnud, must be predicted by the theory.}  
\label{eq3:HLV_cond1}
\end{equation}
Suppose that the use of symmetries allows one to impose
\begin{equation}
\mnud = m_f\ \ {\rm for\ one\ of}\ f=e,\ f=d\ {\rm or}\ f=u\;,
\label{eq3:HLV_mass_rel}
\end{equation}
then (\ref{eq3:HLV_seesaw_form2}) can be written as
\begin{equation}
\mnuhat \simeq  V_{\nu}\, m_{f}\, M_R^{-1}\, m_{f}^T\, V^{T}_{\nu}\;.
\label{eq3:HLV_seesaw_form3}
\end{equation}
Note that there are custodial $SU(2)$, unification and quark-lepton symmetries that can enforce each of the conditions in (\ref{eq3:HLV_mass_rel}) at tree-level, as we shall explain later. But if we simply take (\ref{eq3:HLV_mass_rel}) as an ansatz at this point, then by diagonalizing the fermion mass matrix:
\begin{equation}
\widehat{m}_f = V_{fL}\, m_f\, V_{fR}^{\dagger}\;,
  \label{eq3:HLV_hatmf}
\end{equation}
where $V_{fL}$ and $V_{fR}$ are the left- and right-diagonalization matrices for $m_f$ respectively, we can put (\ref{eq3:HLV_seesaw_form3}) into a more convenient form:
\begin{align}
\mnuhat &\simeq 
 V_{\nu}\, V_{fL}^{\dagger}\, \widehat{m}_{f}\, V_{fR}\, M_R^{-1}\, V_{fR}^{T}\, \widehat{m}_{f}\, V_{fL}^*\, V^{T}_{\nu}\;,\nonumber\\
 &=
 (V_{fL}\,V_{\nu}^{\dagger})^{\dagger}\, \widehat{m}_{f}\, V_{fR}\, M_R^{-1}\, V_{fR}^{T}\, \widehat{m}_{f}\, (V_{fL}\,V_{\nu}^{\dagger})^{*}\;.
\label{eq3:HLV_seesaw_form4}
\end{align}
From this, we see that another necessary condition to have $M_R$ completely determined is:
\begin{equation}
\fbox{The matrix product, $V_{fL} V^{\dagger}_{\nu}$ 
 and the right-diagonalization matrix, $V_{fR}$ must be known.}
 \label{eq3:HLV_cond2}
\end{equation}

However, since the $V_{fR}$ cannot be measured as the weak interaction is known to be left-handed~\footnote{The discovery of
right-handed weak interactions would of course change this situation. But we shall not consider that possibility here.}, it must be predicted from the theory. As will be elaborated later, this issue can be resolved by including a flavor (or family) symmetry in the model so that fully determined diagonalisation matrices with their entries only made up of known constants can result. These constants are usually related to the Clebsch-Gordon coefficients associated with the flavor symmetry group under consideration.

Regarding the matrix product $V_{fL} V^{\dagger}_{\nu}$ in (\ref{eq3:HLV_seesaw_form4}), it may be a directly measurable quantity or partly provided by the theory or both. A distinctive feature of this product is that its form is similar to the definitions of both $\Upmns$ and $\Uckm$ (re-stated here for convenience):
\begin{equation}\label{eq3:HLV_PMNS_CKM}
\Upmns = V_{eL} V_{\nu}^{\dagger}
\qquad,\qquad \Uckm = V_{uL} V_{dL}^{\dagger}\;.
\end{equation}
The simplest possibility is perhaps when $f=e$ which leads to $V_{fL} V^{\dagger}_{\nu} = \Upmns$. If at the same time, the model predicts $V_{fR} = V_{eR}=I$, then we have a very special case where $M_R$ is completely determined through
\begin{equation}
 M_R \simeq \widehat{m}_e\, \Upmns^* \,\mnuhat^{-1}\, \Upmns^{\dagger}\, \widehat{m}_e 
\;.
 \label{eq3:HLV_MR_e_PMNS}
\end{equation}
Two other possibilities, arising from the enforcement of ($\nu \leftrightarrow d,u$) and the appropriate flavor symmetries, are that
\begin{eqnarray}
& f = d\;, \qquad\text{  with  }\;\; V_{dR} = I\quad,\quad V_{dL} = V_{eL}\;, & 
\label{eq3:HLV_d_PMNS}\\
\text{and  }
& f = u\;, \qquad\text{  with  }\;\; V_{uR} = I\quad,\quad V_{uL} = V_{eL}\;,& \label{eq3:HLV_u_PMNS}
\end{eqnarray}
leading to
\begin{equation}
M_R \simeq  \widehat{m}_{d,u}\, \Upmns^*\, \mnuhat^{-1}\, \Upmns^{\dagger}\, \widehat{m}_{d,u}\;.
\label{eq3:HLV_MR_ud_PMNS}
\end{equation}
Because of the automatic presence of $V_{\nu}$ in the formula for $M_R$, it is relatively straightforward to find
symmetries leading to Eqs.~(\ref{eq3:HLV_MR_e_PMNS}) and (\ref{eq3:HLV_MR_ud_PMNS}) where the leptonic PMNS mixing matrix is a key feature. However, it may also be of interest to consider symmetry structures that can lead to the PMNS matrix being replaced by the CKM matrix (or a product of the two). One possibility is to have the symmetries to dictate that $V_{uL}=V_{dR}=V_\nu =I$ always holds and so when we identify $f=d$ in (\ref{eq3:HLV_seesaw_form4}), this will give rise to the relation
\begin{equation}
M_R \simeq  \widehat{m}_{d}\, \Uckm^T\, \mnuhat^{-1}\, \Uckm\, \widehat{m}_{d}\;,
\label{eq3:HLV_MR_d_CKM}
\end{equation}
where $\Uckm = V_{dL}^\dagger$ in this case. 
Alternatively, one can consider the case where $d$ and $u$ have their roles interchanged. 
Other options which could equally well be contemplated include enforcing $V_{dL}^\dagger  V_{eL}=V_{uR}=I$ while keeping $V_{uL}$ and $V_\nu^\dagger$ arbitrary, leading to the situation with $M_R$ given by
\begin{equation}\label{eq3:HLV_MR_PMNSxCKM}
 M_R \simeq  \widehat{m}_{u}\, \Uckm^*\Upmns^*
   \, \mnuhat^{-1}\, \Upmns^\dagger \Uckm^\dagger\, \widehat{m}_{u}
 \;,
\end{equation}
with $\Uckm = V_{uL}V_{dL}^\dagger$ and $\Upmns = V_{eL}V_{\nu}^\dagger$.

Finally, there is the relatively mundane case where all of the diagonalization matrices in the formula for $M_R$ are equal to the identity, so that one simply gets
\begin{equation}
M_R \equiv \widehat{M}_R  \simeq \text{diag}\left(\frac{m_{f1}^2}{m_1},\frac{m_{f2}^2}{m_2},\frac{m_{f3}^2}{m_3}
  \right)\;.
\label{eq3:HLV_mundane}
\end{equation}
Interestingly, this is not possible for the $f = e$ choice, because the PMNS matrix is known to be very dissimilar to the identity.  However, flavor symmetries allowing, Eq.(\ref{eq3:HLV_mundane}) can in principle be achieved for $f= d$ or $u$. In these situations, one would then get $\Upmns = V_{eL}$ and $\Uckm = V_{uL}\;(\text{if } f=d)$ or $V_{dL}^\dagger\;(\text{if } f=u)$.

In summary, the general properties of enforcing a ($\nu \leftrightarrow e,d,u$) symmetry in parallel with some flavor symmetries motivate relations of the form~\footnote{Note that although the analysis was framed in terms of the leading seesaw expression 
$\mnu \simeq - \mnud M_R^{-1} (\mnud)^T$, it
generalizes to cases where additional terms on the right-hand side are kept, because the higher-order terms contain {\it a priori} the
same unknowns as does the leading term.}
\begin{equation}
M_R = M_R(\widehat{m}_e\,, \widehat{m}_d\,, \widehat{m}_u\,, \Upmns\,, \Uckm)
\label{eq3:HLV_general_MR}
\end{equation}
of which Eqs.(\ref{eq3:HLV_MR_e_PMNS}), (\ref{eq3:HLV_MR_ud_PMNS}), (\ref{eq3:HLV_MR_d_CKM}), (\ref{eq3:HLV_MR_PMNSxCKM}) and (\ref{eq3:HLV_mundane}) are important examples.\\

\section{The use of symmetries}
\label{sec:HLV_sym_basics}

One of the key assumption leading to result (\ref{eq3:HLV_seesaw_form4}) is that the neutrino Dirac mass matrix must be equal to one of the quark or charged lepton matrices as indicated in (\ref{eq3:HLV_mass_rel}). In addition, completely deriving $M_R$ requires a good knowledge of the diagonalisation matrices because of condition (\ref{eq3:HLV_cond2}). 
So, the aim of this section is to briefly illustrate how mass relations of the type:
\begin{equation}\label{eq3:HLV_massrel_std_form}
\mnud = K \,\widehat{m}_{e,\,d \text{ or } u}\;,
 \quad \text{(where $K$ is a known matrix)}
\end{equation}
may be enforced, as well as the role of flavor symmetry in determining the diagonalisation matrices of interest. We will present some concrete examples that utilise these ideas to good effect in the next section.

It is well known that in a minimal $SO(10)$ framework one obtains the mass relations $\mnud = m_e = m_d = m_u$, because all fermions are in the same multiplet and the electroweak Higgs lies in a real fundamental of $SO(10)$.  However, these relations are too strong from a phenomenological perspective. While we desire the equality between the neutrino Dirac mass matrix and that of one other fermion, the rest of the mass predictions $m_e = m_d = m_u$ are clearly ruled out by experiments. Therefore, the aim is to search for gauge groups that contain the SM as a subgroup and have enough power to establish the mass relation we seek without violating any observations. Indeed, given the partial success of the minimal $SO(10)$ setup, subgroups that are contained in $SO(10)$ will make excellent starting points in the search for a workable model. Furthermore, the use of discrete rather than continuous symmetries to relate different multiplets constitutes another sensible strategy outside of the $SO(10)$-like ideas.

To implement the strategy of using $SO(10)$-inspired models, let us highlight some of the key features of several popular unification groups (which are subgroups of $SO(10)$) that may be of relevance. Firstly, we have the standard $SU(5)$  setup \cite{Georgi:1974sy} where the LH charged leptons ($\nu_L, e_L$) and down antiquarks ($d_R^c$) are packaged into the $\overline{5}$ representation (or $\overline{5}$-rep for short), while the other SM fermions ($u_R^c, u_L, d_L$ and $e_R^c$) are placed in the 10-rep. In the SM with heavy singlet neutrinos, the $\nu_R^c$ can be included as the 1-rep of $SU(5)$. Then, given a minimal Higgs sector (with $\Phi$ in the 5-rep) which accompanies these, it can seen that the Yukawa term that links the $\overline{5}$-rep and 10-rep together will lead to the relation $m_e = m_d$, whereas $m_u$ and $\mnud$ remain unrelated due to the fact that the respective fermion fields are in separate representations, leading to independent Yukawa couplings in general.

Secondly, there is the flipped-$SU(5)$ ($\equiv SU(5)\otimes U(1)_X$) setup \cite{Barr:1981qv} where the down antiquarks ($d_R^c$) and charged antileptons ($e_R^c$) flip roles with the up antiquarks ($u_R^c$) and singlet antineutrinos ($\nu_R^c$) respectively. As a result of these new particle assignments, the model in its minimal version can then induce $\mnud = m_u$ without enforcing the unwanted mass relation of $m_e = m_d$. This observation makes the flipped-$SU(5)$ group a prime candidate for our purpose, unlike both the minimal $SO(10)$ and $SU(5)$ unification schemes~\footnote{Alternatively, a Pati-Salam-like \cite{Pati:1974yy} subgroup $SU(4) \otimes SU(2)_L \otimes SU(2)_R$ can also be used to enforce $\mnud = m_u$.}.

Thirdly, we have the left-right symmetry group: $SU(3)_c \otimes SU(2)_L \otimes SU(2)_R \otimes U(1)_{B-L}$ \cite{LR_Mohapatra.Pati} which has the power to enforce mass degeneracy between weak isospin partners: $m_d = m_u$ and $m_e = \mnud$ \cite{Volkas:1995yn}. Such a degeneracy follows from requiring a bidoublet Higgs to be \emph{real}~\footnote{This basically causes $SU(2)_R$ to become custodial $SU(2)$.}, which at the $SO(10)$ level is equivalent to having a real Higgs 10-plet. Hence, we see that both the flipped-$SU(5)$ (Sec.~\ref{subsec:HLV_fSU5}) and left-right (Sec.~\ref{subsec:HLV_LR}) setup provide the basic starting points in achieving relations of the type of (\ref{eq3:HLV_massrel_std_form}).

To obtain the remaining possibility of $\mnud = m_d$, one can employ the other strategy of introducing a discrete symmetry. In particular, as we shall show in Sec.~\ref{subsec:HLV_QL}, a discrete quark-lepton symmetry \cite{QL_symmodel} would be appropriate. This is because the extended gauge structure contains a new $SU(3)_\ell$ leptonic color group which now permits a discrete interchange symmetry between the SM quarks and (generalised or colored) leptons. After spontaneous symmetry breaking, leptons of one of the colors are identified as the SM leptons, and as a result $\mnud = m_d$ can follow by virtue of the discrete transformation giving these fermions common Yukawa couplings.

It is worth mentioning that one can also contemplate the situations where the model directly provides the generic relation (\ref{eq3:HLV_massrel_std_form}) as the starting point. In such cases, the matrix $K$ is normally a matrix of Clebsch-Gordan coefficients related to the gauge group being employed. For instance, at the $SO(10)$ level, this can come about when the electroweak Higgs doublet is embedded in a higher-dimensional representation rather than the usual 10-rep or there are multiple Higgses that contribute to the Yukawa couplings. This situation is analogous to the well-known Georgi-Jarlskog \cite{Georgi:1979df}  modification of the $m_e$ to $m_d$ relation in $SU(5)$ unification.
Although this may be an interesting alternative, it is not the approach we pursue in this work.

Once the appropriate fermion-mass-constraining group is selected, the remaining challenges are twofold. The first, as well-illustrated
by minimal $SO(10)$, is the removal of byproducts such as unwanted mass relations or interactions. The second is the need to have predictable diagonalization matrices. Quite frequently, it is possible to meet both of these challenges by introducing a \emph{flavor symmetry} and a non-minimal Higgs sector. For the former, the presence of a larger symmetry can naturally restrict the type of coupling one can have, hence removing some or all of the undesirable terms.  In cases where this is not sufficient, unbroken global non-flavor symmetries may also be imposed. To resolve the latter issue, the key concept is that of a ``form-diagonalizable matrix'' \cite{Low01+thesis}.  This is a matrix containing relations amongst its elements and perhaps also texture zeros so as to make the diagonalization matrices fully determined while leaving the eigenvalues arbitrary.

It is interesting to note that for the SM with RH neutrinos but no mass terms, the largest flavor symmetry possible is $U(3)_{q_L}\otimes U(3)_{u_R}\otimes U(3)_{d_R}\otimes U(3)_{\ell_L}\otimes U(3)_{e_R}\otimes U(3)_{\nu_R}$, which corresponds to one independent $U(3)_f$ for each fermion multiplet. If the SM originates from some GUT, then this underlying flavor group is necessarily smaller. For instance in $SO(10)$ GUT, the largest of such symmetry is $U(3)$ as all SM fermions (including $\nu_R$) are placed in the fundamental 16-rep. Eventually though, when all fermions (including LH and RH neutrinos) gain mass, such an underlying group is in general completely broken.
But the point is that in model building, there are often many possible flavor (sub)groups and their subsequent breaking patterns one may select given a gauge structure. 

Although the restrictions coming from the gauge group may differ, one sensible choice is to pick groups with 3-dimensional irreducible representations such as $SU(3)$ \cite{SU3-eg}, $SO(3)$ \cite{SO3-eg}, $\mathbb{Z}_7\rtimes \mathbb{Z}_3$ \cite{Z7Z3-eg}, $\Delta(27)$ \cite{D27-eg}, $S_4$ \cite{S4-eg} or $A_4$ \cite{TB_a4_list} to be the starting point, because there are three generations of fermions~\footnote{It is worth adding that all of these groups have in recent years been widely used to try to understand the ``tribimaximal'' form  for leptonic mixing \cite{TBmixing, TB_a4_list, TB_other_list}. See Sec.~\ref{subsec_osc_TB}.}. Then, the challenge is to ensure that the subsequent breaking of the underlying flavor symmetry chosen can enforce form-diagonalizability for a particular setup. In principle, all of the groups mentioned above can be viable candidates for this purpose, and which one to take is often merely a model building decision.

For the representative models presented in the next section, we select the discrete group  $A_4$ to be the underlying flavor symmetry. This choice is motivated by the relatively simple nature of group, as well as the fact that it has been well-studied \cite{TB_a4_list, He:2006dk, more_a4_eg1, more_a4_eg2}. Using this and the mass-relating symmetry discussed earlier, we will demonstrate that relation (\ref{eq3:HLV_massrel_std_form}) can be readily implemented.

\section{Some representative models}\label{sec:HLV_models}

In this section, we illustrate the ideas discussed previously by constructing three realistic models that can enforce 
$\mnud = K\,\widehat{m}_{e,d \text{ or } u}$, and subsequently lead to relations (\ref{eq3:HLV_MR_e_PMNS}) and (\ref{eq3:HLV_MR_ud_PMNS}) respectively.

\subsection{Relating $\mnud$ to $\widehat{m}_u$ via a flipped-$SU(5)$ model}\label{subsec:HLV_fSU5}

As hinted before, we can relate the neutrino Dirac mass and the up-quark mass matrices using a flipped-$SU(5)$ framework. To this end, we consider the following group structure \cite{Barr:1981qv} augmented by a $A_4$ flavor symmetry \cite{TB_a4_list, He:2006dk}:
\begin{align}
 G_1 &= SU(5) \otimes U(1)_X \times A_4\;,
 \label{eq3:HLV_su5_1}\\
 &\supset SU(3)_c \otimes SU(2)_L \otimes \underbrace{U(1)_T \otimes U(1)_X}_{U(1)_Y} 
  \times \,A_4 \;.
 \label{eq3:HLV_su5_2}
\end{align}
We shall choose the normalization such that hypercharge $Y$ is given by:
\begin{equation}\label{eq3:HLV_su5_Y}
 Y= -\frac{1}{5}\,T + \frac{2}{5}\, X
 \;,
\end{equation}
where $T = \text{diag}(-2/3,-2/3,-2/3,1,1)$ is the generator for $U(1)_T$. The advantage of this setup is that we can also avoid the bad mass-relation of $m_e=m_d$ as we will demonstrate below.
For this model, the particle contents and their transformation properties under $G_1$ are given by:
\begin{eqnarray}
 &\psi_{L\alpha} = \fivecMat{u_{R}^{1c}}{u_{R}^{2c}}{u_{R}^{3c}}{e_L}{-\nu_L}
                  \;\sim\; (\overline{5}, -3)(\rept)
\;\; ; \qquad\qquad\qquad
  e_R^c \sim\; (1,5)(\rep \oplus\repp\oplus\reppp) \;\; ;
 \nonumber\\
 &\chi_L^{\alpha\beta} = \frac{1}{\sqrt{2}}
  \begin{pmatrix}
    0       & d_{R}^{3c}  & -d_{R}^{2c}  & -u_{L}^1 & -d_{L}^1 \\
  -d_{R}^{3c} &    0      & d_{R}^{1c}   & -u_{L}^2 & -d_{L}^2 \\
   d_{R}^{2c} & -d_{R}^{1c} &    0       & -u_{L}^3 & -d_{L}^3 \\
   u_{L}^1   & u_{L}^2    & u_{L}^3     &    0    & -\nu_R^c\\
   d_{L}^1   & d_{L}^2    & d_{L}^3     & \nu_R^c &    0    \\
  \end{pmatrix}
  \;\sim\; (10,1)(\rep \oplus\repp\oplus\reppp) \;;
    \nonumber
\end{eqnarray}  
\begin{eqnarray}  
&\Phi_{(3)}^\sigma = \fivecMat{h_{d}^1}{h_{d}^2}{h_{d}^3}{\phi_{(3)}^{0*}}{-\phi_{(3)}^{+}}
  \;\sim\;(5,-2)(\rept)\;\; ;
  \qquad\qquad
  \Phi_{(1\oplus 1' \oplus 1'')}^\sigma \;\sim\;
   (5,-2)(\rep \oplus\repp\oplus\reppp)\;\;;
   \nonumber\\
  &\Delta^{\alpha\beta\gamma\delta} \;\sim\; 
  (\overline{50},2)(\rep \oplus\repp\oplus\reppp)\;,
  \label{eq3:HLV_su5_content}
\end{eqnarray}
where the superscripts 1,2 and 3 and Greek letters are the color and $SU(5)$ indices respectively~\footnote{For the branching rules of $SU(5)$ and other groups see for example \cite{slansky}.}. $\rep,\repp,\reppp$ and $\rept$ denote the irreducible representations of $A_4$ (see Appendix~\ref{app:b_A4_def}). In matrix form, the $G_1$ invariant interaction Lagrangian then contains the following terms:
\begin{align}
 -\mathcal{L}_1 &=
    Y_{\lambda 1}\, \overline{\psi}_L\, \Phi_{(3)}^*\, e_R
  + \sqrt{2}\;Y_{\lambda 2}\; \overline{\psi}_L\, \chi_L^c \,\Phi_{(3)}
  + \frac{Y_{\lambda 3}}{4}\, (\overline{\chi}_L)_{\alpha\beta} (\chi_L^c)_{\gamma\delta}
   \left(\Phi^*_{(1\oplus 1'\oplus 1'')}\right)_\sigma
    \,\epsilon^{\alpha\beta\gamma\delta\sigma} \nonumber\\
  &\qquad
  + Y_{\lambda 4}  (\overline{\chi}_L)_{\alpha\beta} (\chi_L^c)_{\gamma\delta}
    \,\Delta^{\alpha\beta\gamma\delta} + \text{h.c.}\;,
    \label{eq3:HLV_SU5_Lag1}
\intertext{where $\epsilon^{\alpha\beta\gamma\delta\sigma}$ is the Levi-Civita tensor. When the neutral components of $\Phi$ and $\Delta$ obtain nonzero vacuum expectation values (VEVs), one gets mass terms of the form}
 &= Y_{\lambda 1}\, \overline{e}_L\, \vev{\phi^0_{(3)}}\, e_R
   - Y_{\lambda 2}\, (
   \overline{u}_L\,\vev{\phi^{0*}_{(3)}}\,u_R
    +\overline{\nu}_L\,\vev{\phi^{0*}_{(3)}}\,\nu_R)
\nonumber\\
 & \qquad
   + \frac{Y_{\lambda 3}}{2}\, \left(\overline{d^c_R}\,d^c_L + \overline{d}_L\,d_R \right)\vev{\phi^0_{(1\oplus 1'\oplus 1'')}} 
   + Y_{\lambda 4}\,\overline{\nu_R^c}\,\nu_R \vev{\Delta^0_{(1\oplus 1'\oplus 1'')}} + \text{h.c.}\;,
   \label{eq3:HLV_SU5_Lag2}
\end{align}
where we have used $\overline{u}_R^c  \vev{\phi^{0*}_{(3)}} u_L^c \equiv
 \overline{u}_L\,\vev{\phi^{0*}_{(3)}}\,u_R $.
Note that $\vev{\Delta^0}$, which provides the heavy Majorana mass, breaks $G_1$ down to the SM, and is expected to be at a much higher energy scale than $\vev{\Phi}$ which breaks electroweak symmetry. 
 
Writing out the $A_4$ structure of the $Y_{\lambda 1}$- and $Y_{\lambda 2}$-terms in Eq.~(\ref{eq3:HLV_SU5_Lag2}) with the vacuum $\vev{\phi_{(3)}^0} \equiv \vev{\phi_{(3)}^{0*}}=  (v_{(3)},v_{(3)},v_{(3)})$ where $v_{(3)} \in \mathbb{R}$, one gets
\begin{align}
 m_e:\,\quad
 & \lambda_1\, (\overline{e}_L\,\vev{\phi_{(3)}^0})_{\underline{1}}\,e_R
 + \lambda_1'\,(\overline{e}_L\,\vev{\phi_{(3)}^0})_{\underline{1}'}\,e_R''
 + \lambda_1''\,(\overline{e}_L\,\vev{\phi_{(3)}^0})_{\underline{1}''}\,e_R' 
 + \text{h.c.}\;,
 \label{eq3:HLV_SU5_me}
 \\
 m_u:\,\quad
 &
 - \lambda_2\, \overline{u}_L (\vev{\phi_{(3)}^{0*}}\,u_R)_{\underline{1}}
 - \lambda_2'\, \overline{u}_L'' (\vev{\phi_{(3)}^{0*}}\,u_R)_{\underline{1}'}
 - \lambda_2''\,\overline{u}_L' (\vev{\phi_{(3)}^{0*}}\,u_R)_{\underline{1}''}
   + \text{h.c.}\;,\label{eq3:HLV_SU5_mu}
\\
 \mnud\,:\quad
 &
 -\lambda_2\, (\overline{\nu}_L\,\vev{\phi_{(3)}^{0*}})_{\underline{1}}\,\nu_R
 - \lambda_2'\,(\overline{\nu}_L\,\vev{\phi_{(3)}^{0*}})_{\underline{1}'}\,\nu_R''
 - \lambda_2''\,(\overline{\nu}_L\,\vev{\phi_{(3)}^{0*}})_{\underline{1}''}\,\nu_R' 
 + \text{h.c.}\;.\label{eq3:HLV_SU5_mnud}
\end{align}
By expanding out the $A_4$ invariants using the tensor product rules in Appendix~\ref{app:b_A4_def} and collect them into matrix form, one can easily examine the structure of these mass matrices:

\noindent 
$\boxed{m_e}\;\;$ \textbf{charged lepton mass matrix} 
\begin{align}
  &\lambda_1\, (\overline{e}_L\,\vev{\phi_{(3)}^0})_{\underline{1}}\,e_R
 + \lambda_1'\,(\overline{e}_L\,\vev{\phi_{(3)}^0})_{\underline{1}'}\,e_R''
 + \lambda_1''\,(\overline{e}_L\,\vev{\phi_{(3)}^0})_{\underline{1}''}\,e_R' 
 + \text{h.c.}\;,\nonumber\\
 =& \threerMat{\overline{e}_{L1}}{\overline{e}_{L2}}{\overline{e}_{L3}}
  \ththMat
  {\lambda_1\, v_{(3)} }{\lambda_1'\, v_{(3)} }{\lambda_1''\, v_{(3)} }
  {\lambda_1\, v_{(3)} }{\omega\lambda_1'\, v_{(3)} }{\omega^2\lambda_1''\, v_{(3)} }
  {\lambda_1\, v_{(3)} }{\omega^2\lambda_1'\, v_{(3)} }{\omega\lambda_1''\, v_{(3)} }
  \threecMat{e_R}{e_R''}{e_R'}+ \text{h.c.}\;,
  \\
 =&
  \threerMat{\overline{e}_{L1}}{\overline{e}_{L2}}{\overline{e}_{L3}}
 \underbrace{\MatU}_{\Uw} 
 \underbrace{\sqrt{3}\,v_{(3)}\Matdiag {\lambda_1}
  {\lambda_1'}{\lambda_1''}
 }_{\widehat{m}_e}
%
  \threecMat{e_R}{e_R''}{e_R'} + \text{h.c.}\;,  \label{eq3:HLV_su5_me_mat}
\end{align}
where subscripts 1,2,3 of $e_L$ are the flavor indices and $\omega = e^{2\pi i/3}$. From this, we can readily deduce that
\begin{equation}\label{eq3:HLV_su5_VeLR}
 V_{eL}^\dagger = \Uw \quad\text{and}\quad V_{eR} = I\;.
\end{equation}\\
~
$\boxed{m_u}\;\;$ \textbf{up-type quark mass matrix} 
\begin{align}
 & - \lambda_2\, \overline{u}_L (\vev{\phi_{(3)}^{0*}}\,u_R)_{\underline{1}}
 - \lambda_2'\, \overline{u}_L'' (\vev{\phi_{(3)}^{0*}}\,u_R)_{\underline{1}'}
 - \lambda_2''\,\overline{u}_L' (\vev{\phi_{(3)}^{0*}}\,u_R)_{\underline{1}''}
    + \text{h.c.}\;,\nonumber\\
=&
 - \threerMat{\overline{u}_L}{\overline{u}_L''}{\overline{u}_L'}
  \underbrace{\sqrt{3}\,v_{(3)}
  \Matdiag  {\lambda_2}{\lambda_2'}{\lambda_2''}
  }_{\widehat{m}_u}
  \MatU
  \threecMat{u_{R1}}{u_{R2}}{u_{R3}}
  + \text{h.c.}\;. \label{eq3:HLV_su5_mu_mat}
\end{align}\\
~
$\boxed{\mnud}\;\;$ \textbf{neutrino Dirac mass matrix} 
\begin{align}
  & -\lambda_2\, (\overline{\nu}_L\,\vev{\phi_{(3)}^{0*}})_{\underline{1}}\,\nu_R
 - \lambda_2'\,(\overline{\nu}_L\,\vev{\phi_{(3)}^{0*}})_{\underline{1}'}\,\nu_R''
 - \lambda_2''\,(\overline{\nu}_L\,\vev{\phi_{(3)}^{0*}})_{\underline{1}''}\,\nu_R'
 + \text{h.c.}\;,\nonumber\\
 =&
  -\threerMat{\overline{\nu}_{L1}}{\overline{\nu}_{L2}}{\overline{\nu}_{L3}}
  \MatU
  \underbrace{\sqrt{3}v_{(3)}
  \Matdiag  {\lambda_2}{\lambda_2'}{\lambda_2''}
  }_{\widehat{m}_u}
  \threecMat{\nu_R}{\nu_R''}{\nu_R'}
  + \text{h.c.}\;. \label{eq3:HLV_su5_mnud_mat}
\end{align}
Inspecting (\ref{eq3:HLV_su5_mu_mat}) and (\ref{eq3:HLV_su5_mnud_mat}), it immediately reveals the following relations:
\begin{equation}\label{eq3:HLV_su5_VuLR-mnud}
 V_{uL}^\dagger = I\;, \quad V_{uR} = -\Uw \equiv -V_{eL}^\dagger
  \quad\text{and}\quad
  \mnud = -\Uw\,\widehat{m}_u \;,
\end{equation}
which implies
\begin{equation}
 \mnud = -V_{eL}^\dagger\,\widehat{m}_u\;.
\end{equation} 
Therefore, we have successfully achieved the form we seek. Putting it into (\ref{eq3:HLV_seesaw_form2}) and using definition (\ref{eq3:HLV_PMNS_CKM}), we get
\begin{align}
\mnuhat 
 &\simeq V_{\nu}\, V_{eL}^\dagger\,\widehat{m}_u \,M_R^{-1} (V_{eL}^\dagger\,\widehat{m}_u)^T V^{T}_{\nu}\;,
\nonumber\\
 &= \Upmns^\dagger \,\widehat{m}_u \,M_R^{-1}\,\widehat{m}_u \,\Upmns^{*}\;,
 \nonumber\\
\Rightarrow\qquad 
M_R 
 &\simeq  \widehat{m}_{u}\, \Upmns^*\, \mnuhat^{-1}\, \Upmns^{\dagger}\, \widehat{m}_{u}\;.
\label{eq3:HLV_su5_main_result}
\end{align}
To check the predictions of the rest of the theory, we turn our attention to the $Y_{\lambda 3}$- and $Y_{\lambda 4}$-terms in (\ref{eq3:HLV_SU5_Lag2}):

\noindent 
$\boxed{m_d}\;\;$ \textbf{down-type quark mass matrix} 
\begin{align}
  \;\;&
  \frac{1}{2} \lambda_3^{(11)} (\overline{d^c_R} d_L^c + \overline{d}_L d_R) \vev{\phi^0_{(1)}}
 +\frac{1}{2} \lambda_3^{(22)} (\overline{d^c_R}'' (d_L^c)'' + \overline{d}_L'' d_R'') \vev{\phi^0_{(1'')}}  \nonumber\\
&\;\;
 +\frac{1}{2} \lambda_3^{(33)} (\overline{d^c_R}' (d_L^c)' + \overline{d}_L' d_R') \vev{\phi^0_{(1')}}
  +\frac{1}{2} \lambda_3^{(12)} (\overline{d^c_R}'' d_L^c + \overline{d}_L'' d_R) \vev{\phi^0_{(1')}} \nonumber\\
 &\;\;
 +\frac{1}{2} \lambda_3^{(13)} (\overline{d^c_R}' d_L^c + \overline{d}_L' d_R) \vev{\phi^0_{(1'')}}
 +\frac{1}{2} \lambda_3^{(23)} (\overline{d^c_R}' (d_L^c)'' + \overline{d}_L' (d_R)'') \vev{\phi^0_{(1)}} + \text{h.c.}\;, \nonumber\\
&\nonumber\\
 =\;\;&
   \lambda_3^{(11)} \overline{d}_L d_R \vev{\phi^0_{(1)}}
  + \lambda_3^{(22)} \overline{d}_L'' d_R'' \vev{\phi^0_{(1'')}}
  + \lambda_3^{(33)} \overline{d}_L' d_R' \vev{\phi^0_{(1')}}
\nonumber\\
 &\;\;\;
  + \frac{1}{2} \lambda_3^{(12)}
     (\overline{d}_L d_R'' + \overline{d}_L'' d_R) \vev{\phi^0_{(1')}}
  +  \frac{1}{2} \lambda_3^{(13)}
     (\overline{d}_L d_R' + \overline{d}_L' d_R) \vev{\phi^0_{(1'')}}
\nonumber\\
 &\;\;\;
  + \frac{1}{2} \lambda_3^{(12)}
     (\overline{d}_L'' d_R' + \overline{d}_L' d_R'') \vev{\phi^0_{(1)}}
  +\text{h.c.}\;,
\nonumber\\
=\;\;& \threerMat{\overline{d}_{L}}{\overline{d}_{L}''}{\overline{d}_{L}'}
 \underbrace{
 \ththMat
 {\lambda_3^{(11)}v_{(1)}}
 {\frac{1}{2}\lambda_3^{(12)}v_{(1)}'}
 {\frac{1}{2}\lambda_3^{(13)}v_{(1)}''}
 {\frac{1}{2}\lambda_3^{(12)}v_{(1)}'}
 {\lambda_3^{(22)}v_{(1)}''}
 {\frac{1}{2}\lambda_3^{(23)}v_{(1)}}
 {\frac{1}{2}\lambda_3^{(13)}v_{(1)}''}
 {\frac{1}{2}\lambda_3^{(23)}v_{(1)}}
 {\lambda_3^{(33)}v_{(1)}'}
 }_{m_d}
  \threecMat{d_R}{d_R''}{d_R'}+ \text{h.c.}\;,  
  \label{eq3:HLV_su5_md_mat}
\end{align}
where we have substituted $\vev{\phi^0_{(1)}} = v_{(1)}, \vev{\phi^0_{(1')}} = v_{(1)}', \vev{\phi^0_{(1'')}} = v_{(1)}''$. From (\ref{eq3:HLV_su5_md_mat}), it is clear that $m_d$ is a general complex symmetric matrix, and hence we have $V_{dL}^\dagger = V_{dR}^T$ is arbitrary. Similarly, for $M_R$:

\noindent 
$\boxed{M_R}\;\;$ \textbf{neutrino Majorana mass matrix}
\begin{align}
 &
   \lambda_4^{(11)} \overline{\nu_R^c} \nu_R \vev{\Delta^0_{(1)}}
 +\lambda_4^{(22)} \overline{\nu_R^c}'' \nu_R'' \vev{\Delta^0_{(1'')}}
 +\lambda_4^{(33)} \overline{\nu_R^c}' \nu_R' \vev{\Delta^0_{(1')}}
 +\lambda_4^{(12)}
    \left[\overline{\nu_R^c} \nu_R''
         +\overline{\nu_R^c}'' \nu_R\right] \vev{\Delta^0_{(1')}}\nonumber\\
 &\quad
 +\lambda_4^{(13)}
    \left[\overline{\nu_R^c} \nu_R'
         +\overline{\nu_R^c}' \nu_R\right] \vev{\Delta^0_{(1'')}}
 +\lambda_4^{(23)}
    \left[\overline{\nu_R^c}'' \nu_R'
         +\overline{\nu_R^c}' \nu_R''\right] \vev{\Delta^0_{(1)}}\;\;+\text{h.c.}\;,
\end{align}
\begin{align}
 =& \threerMat{\overline{\nu^c_R}}{\overline{\nu^c_R}''}{\overline{\nu^c_R}'}
  \underbrace{
  \ththMat
   {\lambda_4^{(11)}v_\delta}
   {\lambda_4^{(12)}v_\delta'}
   {\lambda_4^{(13)}v_\delta''}
   {\lambda_4^{(12)}v_\delta'}
   {\lambda_4^{(22)}v_\delta''}
   {\lambda_4^{(23)}v_\delta}
   {\lambda_4^{(13)}v_\delta''}
   {\lambda_4^{(23)}v_\delta}
   {\lambda_4^{(33)}v_\delta'}
  }_{M_R}
  \threecMat{\nu_R}{\nu_R''}{\nu_R'}\;\;+ \text{h.c.}\;, \label{eq3:HLV_su5_MR_mat}
\end{align}
where $\vev{\Delta^0_{(1)}} = v_\delta, \vev{\Delta^0_{(1')}} = v_\delta', \vev{\Delta^0_{(1'')}} = v_\delta''$. So as we can see, $M_R$ is also a general complex symmetric matrix. Consequently, the diagonalization matrix $V_\nu$ is this model will be arbitrary since $m_\nu$ is a function of $M_R$ via (\ref{eq3:HLV_su5_main_result}). The two results from $m_d$ and $M_R$ imply that this model places no restrictions on the neutrino and the quark mixing matrices, which are given by $\Upmns = V_{eL} V_\nu^\dagger =\Uw V_\nu^\dagger$ and $\Uckm = V_{uL} V_{dL}^\dagger =V_{dL}^\dagger$ respectively, and therefore one can set them to match the experimental values by tuning the $\lambda_{3,4}$'s.

\subsection{Relating $\mnud$ to $\widehat{m}_d$ via a quark-lepton symmetric model}\label{subsec:HLV_QL}

Next, we construct a slightly more complicated model within the framework of a discrete quark-lepton symmetry  \cite{QL_symmodel}. As well as the usual $A_4$ flavor symmetry, we also introduce an additional unbroken $Z_2$ global symmetry to forbid certain interaction terms in the Lagrangian. The symmetry group for this model is:
\begin{align}
 G_2
    &= G_{q\ell} \times A_4 \times Z_2\;,\nonumber\\
    &= \underbrace{SU(3)_\ell \otimes SU(3)_q}_{Z_\text{QL}} \otimes SU(2)_L \otimes U(1)_X 
    \times A_4 \times Z_2
    \;,\\
    &\supset (SU(2)_\ell\otimes U(1)_T) \otimes SU(3)_q \otimes SU(2)_L \otimes U(1)_X 
    \times A_4 \times Z_2\;, 
\end{align}
where $Z_\text{QL}$ is the discrete quark-lepton symmetry that relates the leptonic and quark color groups ($SU(3)_{\ell} \overset{Z_\text{QL}}{\longleftrightarrow} SU(3)_q$) while hypercharge $Y$ is given by
\begin{equation}
 Y= X +\frac{1}{3}\,T\;,
\end{equation}
where $T= \text{diag}(-2,1,1)$ is a generator of $SU(3)_\ell$. The field contents are
\begin{eqnarray}
F_L = \twocMat{N_L}{E_L} \sim (3,1,2,-1/3)(\rept)(1) \quad  &\overset{Z_\text{QL}}{\longleftrightarrow}& \quad
 Q_L = \twocMat{u_L}{d_L} \sim (1,3,2,1/3)(\rept)(1)\;,\nonumber\\
 E_R \sim (3,1,1,4/3)(\rep\oplus \repp\oplus \reppp)(1) \quad  &\longleftrightarrow& \quad
 u_R \sim (1,3,1,-4/3)(\rep\oplus \repp\oplus \reppp)(1)\;,\nonumber\\
 N_R \sim (3,1,1,2/3)(\rep\oplus \repp\oplus \reppp)(-1) \quad  &\longleftrightarrow& \quad
 d_R \sim (1,3,1,-2/3)(\rep\oplus \repp\oplus \reppp)(-1)\;,\nonumber
\end{eqnarray}

\begin{eqnarray}
 \chi_1^{(0)} \sim (3,1,1,2/3)(\rep\oplus \repp\oplus \reppp)(1) \quad 
 &\overset{Z_\text{QL}}{\longleftrightarrow}&\quad
 \chi_2^{(0)} \sim (1,3,1,-2/3)(\rep\oplus \repp\oplus \reppp)(1)\;,\nonumber\\
 \chi_1^{(1)} \sim (3,1,1,2/3)(\rep\oplus \repp\oplus \reppp)(-1) \quad 
 &\longleftrightarrow&\quad
 \chi_2^{(1)} \sim (1,3,1,-2/3)(\rep\oplus \repp\oplus \reppp)(-1)\;,\nonumber\\
 \phi_1 = \twocMat{\phi_1^0}{\phi_1^-} \sim (1,1,2,-1)(\rept)(1) \quad 
 &\longleftrightarrow&\quad
 \phi_2 = \twocMat{\phi_2^+}{\phi_2^0} \sim (1,1,2,1)(\rept)(1)\;,\nonumber\\
 \phi_2^c = \twocMat{\phi_2^{0*}}{-\phi_2^-} \sim (1,1,2,-1)(\rept)(1)\quad  
 &\longleftrightarrow&\quad
 \phi_1^c = \twocMat{\phi_1^+}{-\phi_1^{0*}} \sim (1,1,2,1)(\rept)(1)\;,\nonumber\\
 \phi_d^c = \twocMat{\phi_d^{0*}}{-\phi_d^-} \sim (1,1,2,-1)(\rept)(-1)\quad 
   &\longleftrightarrow&\quad
  \phi_d = \twocMat{\phi_d^+}{\phi_d^0} \sim (1,1,2,1)(\rept)(-1)\;,\nonumber\\
 \Delta_1 \sim (\overline{6}_s,1,1,-4/3)(\rep\oplus \repp\oplus \reppp)(1) \quad
  &\longleftrightarrow&\quad
 \Delta_2 \sim (1,\overline{6}_s,1,4/3)(\rep\oplus \repp\oplus \reppp)(1)\;,
 \nonumber\\
\end{eqnarray}
where 
\begin{equation}
E_{L,R} = \threecMat{E_{1L,R}}{E_{2L,R}}{e_{L,R}}\;\;,\;\;
N_{L,R} = \threecMat{N_{1L,R}}{N_{2L,R}}{\nu_{L,R}}\;\;\text{are triplets in $SU(3)_\ell$ space}\;.
\end{equation}
$E_{1L,R}, E_{2L,R}, N_{1L,R}, N_{2L,R}$ are exotic leptonic-color partners of the usual leptons. 
The discrete $Z_\text{QL}$ symmetry is broken and these exotic leptons gain mass when $\chi_1^{(0,1)}$ picks up a nonzero VEV:
\begin{equation}
 \vev{\chi_1^{(0,1)}} = \threecMat{0}{0}{v_\chi^{(0,1)}}
 \qquad\text{while }\quad \vev{\chi_2^{(0,1)}} = 0\;.
\end{equation}
We arrange $\vev{\Delta_1} \neq 0$ to give a large Majorana mass while keeping $\vev{\Delta_2}=0$. The $\phi$'s will break electroweak symmetry as usual. In order to avoid domain walls~\footnote{Cosmological domain walls will form when the discrete quark-lepton symmetry is spontaneously broken. Arranging for this breaking scale to be large allows these observationally unacceptable topological defects to be inflated away \cite{Lew:1992rr}.} and allow the implementation of the seesaw mechanism, we demand the following hierarchy for the energy scales:
\begin{equation}
 \vev{\chi_1^{(0,1)}} > T_\text{inflation} > \vev{\Delta_1} \gg \vev{\phi_1}
 \simeq \vev{\phi_2} \simeq \vev{\phi_d} = \order{10^2} \text{ GeV}\;.
 \label{eq3:HLV_energy_scale}
\end{equation}
Overall, the symmetry group $G_2$ will give rise to an interaction Lagrangian with the following terms ($\epsilon^{\alpha\beta\gamma}$ is the Levi-Civita tensor):
\begin{align}
 -\mathcal{L}_2 &=
  \left[\lambda_{f1} \left(\overline{F^c_L}_{\alpha}\, F_{L\beta}\, \chi_{1\gamma}^{(0)}
  +\overline{Q^c_L}_{\alpha}\, Q_{L\beta}\, \chi_{2\gamma}^{(0)}
  \right)
  +\lambda_{f2} \left(\overline{E^c_R}_{\alpha}\, N_{R\beta}\, \chi_{1\gamma}^{(1)}
  +\overline{u^c_R}_{\alpha}\, d_{R\beta}\, \chi_{2\gamma}^{(1)}
  \right)\right]\epsilon^{\alpha\beta\gamma}
  \nonumber\\
  &\qquad
  + \lambda_{g1}\left(\overline{Q}_L u_R \phi_1 +\overline{F}_L E_R \phi_2\right)
  + \lambda_{g2}\left(\overline{Q}_L u_R \phi_2^c +\overline{F}_L E_R \phi_1^c\right)
   \nonumber\\
  &\qquad
  + \lambda_{g3}\left(\overline{Q}_L d_R \phi_d +\overline{F}_L N_R \phi_d^c\right)
  + \lambda_{h1}\left(\overline{N^c_R}_{\alpha}\, N_{R\beta}\, \Delta_1^{\alpha\beta}
  +\overline{d^c_R}_{\alpha}\, d_{R\beta}\, \Delta_2^{\alpha\beta}\right)
  +\text{h.c.}\;,  \label{eq3:HLV_QL-Lag}
\end{align}
where $\alpha, \beta, \gamma$ are $SU(3)_{\ell \text{ or } q}$ indices and the terms proportional to $\lambda_{f1,2}$ are the mass terms for the exotic fermions. From (\ref{eq3:HLV_QL-Lag}) and taking $\vev{\phi_1^0} = v_1, \vev{\phi_2^0} = v_2$ and $\vev{\phi_d^0}\equiv \vev{\phi_d^{0*}} = v_d$, we expect the following mass relations from this model:
\begin{align}
 m_u &= \lambda_{g1} v_1 + \lambda_{g2} v_2^*\;, \; & m_d &= \lambda_{g3} v_d\;,\\
 m_e &= \lambda_{g1} v_2 - \lambda_{g2} v_1^*\;, \;  & \mnud &= \lambda_{g3} v_d \;.
\end{align}
So, in general, $m_e\neq m_u$ but $\mnud=m_d$. Writing out the $A_4$ structure of the relevant terms in (\ref{eq3:HLV_QL-Lag}) we have:
\begin{align}
m_e\;:\quad
 &g_1\, (\overline{e}_L\,\vev{\phi_{2}^0})_{\underline{1}}\,e_R
 + g_1'\,(\overline{e}_L\,\vev{\phi_{2}^0})_{\underline{1}'}\,e_R''
 + g_1''\,(\overline{e}_L\,\vev{\phi_{2}^0})_{\underline{1}''}\,e_R' \nonumber\\
&\quad
 -g_2\, (\overline{e}_L\,\vev{\phi_{1}^{0*}})_{\underline{1}}\,e_R
 - g_2'\,(\overline{e}_L\,\vev{\phi_{1}^{0*}})_{\underline{1}'}\,e_R''
 - g_2''\,(\overline{e}_L\,\vev{\phi_{1}^{0*}})_{\underline{1}''}\,e_R'
 + \text{h.c.}\;,\label{eq:QL_me}\\
m_u\;:\quad 
&g_1\, (\overline{u}_L\,\vev{\phi_{1}^0})_{\underline{1}}\,u_R
 + g_1'\,(\overline{u}_L\,\vev{\phi_{1}^0})_{\underline{1}'}\,u_R''
 + g_1''\,(\overline{u}_L\,\vev{\phi_{1}^0})_{\underline{1}''}\,u_R' \nonumber\\
&\quad
 +g_2\, (\overline{u}_L\,\vev{\phi_{2}^{0*}})_{\underline{1}}\,u_R
 + g_2'\,(\overline{u}_L\,\vev{\phi_{2}^{0*}})_{\underline{1}'}\,u_R''
 + g_2''\,(\overline{u}_L\,\vev{\phi_{2}^{0*}})_{\underline{1}''}\,u_R'
+ \text{h.c.}\;,\label{eq:QL_mu}\\
m_d\;:\quad 
  &g_3\, (\overline{d}_L\,\vev{\phi_{d}^0})_{\underline{1}}\,d_R
 + g_3'\,(\overline{d}_L\,\vev{\phi_{d}^0})_{\underline{1}'}\,d_R''
 + g_3''\,(\overline{d}_L\,\vev{\phi_{d}^0})_{\underline{1}''}\,d_R'+ \text{h.c.}\;,
 \label{eq:QL_md}\\
\mnud\;:\quad 
 &g_3\, (\overline{\nu}_L\,\vev{\phi_{d}^{0*}})_{\underline{1}}\,\nu_R
 + g_3'\,(\overline{\nu}_L\,\vev{\phi_{d}^{0*}})_{\underline{1}'}\,\nu_R''
 + g_3''\,(\overline{\nu}_L\,\vev{\phi_{d}^{0*}})_{\underline{1}''}\,\nu_R'+ \text{h.c.}\;.
\end{align}
Following a similar procedure as in Sec.~\ref{subsec:HLV_fSU5} and choosing the vacuum patterns: $\vev{\phi_{1,2}^{0(*)}} = (v_{1,2}^{(*)},v_{1,2}^{(*)},v_{1,2}^{(*)})\,, \vev{\phi_d^0}\equiv \vev{\phi_d^{0*}} =(v_d,v_d,v_d)$, we find that~\footnote{See Appendix~\ref{app:b_QL_details} for the detailed steps leading to these.} 
\begin{align}
&m_e   = \Uw \widehat{m}_e\;\;,\;\;
 m_u   = \Uw \widehat{m}_u \;\;,\;\;
 m_d   = \mnud = \Uw \widehat{m}_d  \;\;,
 \label{eq3:HLV_QL_all_m}\\
\text{i.e. }&\quad
 V_{eL}^\dagger = V_{uL}^\dagger = V_{dL}^\dagger = \Uw\;,\; 
 V_{eR}=V_{uR}=V_{dR}=I
 \;,\label{eq3:HLV_QL_Vs}
\end{align}
where 
\begin{equation}
\widehat{m}_{e} =
\sqrt{3}\;\text{diag}\left(
 g_1 v_2 -g_2 v_1^*\;,\;
 g_1' v_2 -g_2' v_1^*\;,\;
 g_1'' v_2 -g_2'' v_1^*
\right)
\;,
\end{equation}
\begin{align}
\widehat{m}_{u} &=
\sqrt{3}\;\text{diag}\left(
 g_1 v_1 +g_2 v_2^*\;,\;
 g_1' v_1 +g_2' v_2^*\;,\;
 g_1'' v_1 +g_2'' v_2^*
\right)
\;,\\
\widehat{m}_{d} &=
\sqrt{3}\;\text{diag}\left(
 g_3\, v_d\;,\;
 g_3'\, v_d\;,\;
 g_3''\, v_d
\right)
\;.
\end{align}
In addition, it can be shown that when the $A_4$ singlets \vev{{\Delta_1^{0}}}, \vev{{\Delta_1^{0}}'} and \vev{{\Delta_1^{0}}''} acquire nonzero VEVs, the resulting neutrino Majorana mass matrix, $M_R$ is an arbitrary complex symmetric matrix (see Appendix~\ref{app:b_QL_details}). 

Putting all these together we see that this model predicts the relation:
\begin{equation}
M_R 
 \simeq  \widehat{m}_{d}\, \Upmns^*\, \mnuhat^{-1}\, \Upmns^{\dagger}\, \widehat{m}_{d}\;,
\end{equation}
where $\Upmns = V_{eL} V_\nu^\dagger = \Uw^\dagger V_\nu^\dagger$ is unconstrained whereas $U_\text{CKM} = V_{uL}V_{dL}^\dagger = \Uw^\dagger\Uw =I$. Hence, at tree-level, there is no quark mixing. However, since the symmetry enforcing this result is now broken, radiative corrections can generate nonzero quark mixing~\footnote{We have not attempted to prove that realistic mixing angles can be obtained since the quark sector is not the main focus of our work here.}. Furthermore, it is interesting that the form of the mixing matrices predicted by this model is consistent with small quark mixing ($\Uckm \simeq I$), whereas neutrino mixing ($\Upmns = \Uw^\dagger V_\nu^\dagger$) is large \cite{He:2006dk}. This is because $\Uw^\dagger$ is a trimaximal mixing matrix, and so, unless $V_\nu^\dagger \approx \Uw$, one expects the product of the two would be very dissimilar to the identity.

\subsection{Relating $\mnud$ to $\widehat{m}_e$ via a left-right model}\label{subsec:HLV_LR}

Finally, we explore the possibility of relating $\mnud$ to the charged lepton mass matrix via a left-right model \cite{LR_Mohapatra.Pati} with $A_4$ flavor symmetry. The symmetry group under consideration is
\begin{equation}
 G_3
    = SU(3)_c \otimes SU(2)_L \otimes SU(2)_R \otimes U(1)_{B-L}
    \times A_4
    \;.
    \label{eq3:HLV_LR_gp}
\end{equation}
Here, the imposition of the discrete $L\leftrightarrow R$ parity symmetry is not necessary, and hence will be omitted for simplicity. The complete list of relevant particle contents for this setup is:
\begin{eqnarray}
 \ell_L =\twocMat{\nu_L}{e_L} \sim (1,2,1,-1)(\rept)\;;\quad\quad\quad\quad
 \ell_R =\twocMat{\nu_R}{e_R} \sim (1,1,2,-1)(\rep\oplus \repp\oplus \reppp)\;; 
 \nonumber
\end{eqnarray}
\begin{align}
 &q_L =\twocMat{u_L}{d_L} \sim (3,2,1,1/3)(\rept)\;;\quad\quad\quad\quad
 q_R =\twocMat{u_R}{d_R} \sim (3,1,2,1/3)(\rept)\;;\nonumber\\
&\Phi_\ell 
    = \begin{pmatrix} \phi^{0}& \phi^{+}\\ \phi^{-}& -\phi^{0*}\end{pmatrix}
  \sim (1,2,\overline{2},0)(\rept)\;;\quad
\widetilde{\Phi}_\ell 
    = \tau_2\Phi_\ell^* \tau_2 =
  \begin{pmatrix} -\phi^{0}& -\phi^{+}\\ -\phi^{-}& \phi^{0*}\end{pmatrix}
  \sim (1,2,\overline{2},0)(\rept)\;; \nonumber\\
 &\Phi_q = \begin{pmatrix} \phi^{0}_{A}& \phi^{+}_{B}\\
                              \phi^{-}_{A}& \phi^{0}_{B}\end{pmatrix}
  \sim (1,2,\overline{2},0)(\rep\oplus \repp\oplus \reppp)\;;
  \nonumber\\
 &
 \widetilde{\Phi}_q 
    = \tau_2\Phi_q^* \tau_2 
  =\begin{pmatrix} \phi^{0*}_{B}& -\phi^{+}_{A}\\
                  -\phi^{-}_{B}& \phi^{0*}_{A}\end{pmatrix}
  \sim (1,2,\overline{2},0)(\rep\oplus \repp\oplus \reppp)\;; \nonumber\\
 &
 \Delta_R = \begin{pmatrix} \delta^{+}/\sqrt{2}& \delta^{++}\\
                              \delta^{0}& -\delta^{+}/\sqrt{2}\end{pmatrix}
  \sim (1,1,3,2)(\rep\oplus \repp\oplus \reppp)\;, 
\end{align}
where we have deliberately embedded the same Higgs doublet into $\Phi_\ell$ to form a \emph{real} bidoublet. In matrix form, the $G_3$ invariant Lagrangian has the following terms:
\begin{align}
 -\mathcal{L}_3 &=
      \lambda_{y1}\,\overline{\ell}_L\,\Phi_\ell\,\ell_R
   + \widetilde{\lambda}_{y1}\,\overline{\ell}_L\,\widetilde{\Phi}_\ell\,\ell_R
   + \lambda_{y2}\,\overline{q}_L\,\Phi_\ell\,q_R
   + \widetilde{\lambda}_{y2}\,\overline{q}_L\,\widetilde{\Phi}_\ell\,q_R
   + \lambda_{y3}\,\overline{q}_L\,\Phi_q\,q_R
     \nonumber\\
   &\quad
   + \widetilde{\lambda}_{y3}\,\overline{q}_L\,\widetilde{\Phi}_q\,q_R
   + \lambda_{y4}\,\overline{\ell_R^c}\, i \tau_2\, \Delta_R\, \ell_R
  +\text{h.c.}\;,
   \label{eq3:HLV_LR-Lag}
\end{align}
where $(i\tau_2)_{12} = -(i\tau_2)_{21} =1$ and $(i\tau_2)_{11} = (i\tau_2)_{22} =0$. When the symmetry is broken spontaneously by the nonzero VEVs,
\begin{equation}
 \vev{\Phi_\ell} = \begin{pmatrix} v_\ell & 0\\ 0 & -v_\ell
                   \end{pmatrix}
                   \;;
  \quad
 \vev{\Phi_q} = \begin{pmatrix} v_A & 0\\ 0 & v_B
                   \end{pmatrix}\;;
  \quad
 \vev{\widetilde{\Phi}_q} = \begin{pmatrix} v_B^* & 0\\ 0 & v_A^*
                   \end{pmatrix}\;;
  \quad
 \vev{\Delta_R} = \begin{pmatrix} 0 & 0\\ v_\delta & 0
                   \end{pmatrix}\;,
\end{equation}
where $\vev{\Phi_\ell} \equiv - \vev{\widetilde{\Phi}_\ell}$, $v_\ell \in \mathbb{R}$ and $\order{v_\delta} \gg \order{v_{\ell,A,B}}$, we obtain mass relations of the form:
\begin{eqnarray}
 &m_u = (\lambda_{y2} -\widetilde{\lambda}_{y2}) \,v_\ell +
       \lambda_{y3} \,v_A +\widetilde{\lambda}_{y3} \,v_B^* \;,\qquad
 &\mnud = (\lambda_{y1} -\widetilde{\lambda}_{y1}) \,v_\ell\;,\\
 &m_d = -(\lambda_{y2} -\widetilde{\lambda}_{y2}) \,v_\ell +
       \lambda_{y3} \,v_B +\widetilde{\lambda}_{y3} \,v_A^* \;,\qquad
 &m_e = -(\lambda_{y1} -\widetilde{\lambda}_{y1}) \,v_\ell\;.
\end{eqnarray}
In flavor space, the charged-lepton and neutrino Dirac-mass terms become
\begin{align}
 m_e\;:\quad
 &-\left[
    y_1 \,(\overline{e}_L\,\vev{\phi^{0*}})_{\underline{1}}\,e_R
  + y_1' \,(\overline{e}_L\,\vev{\phi^{0*}})_{\underline{1}'}\,e_R''
  + y_1'' \,(\overline{e}_L\,\vev{\phi^{0*}})_{\underline{1}''}\,e_R'
 \right]\nonumber\\
 &\qquad
  + \widetilde{y}_1 \,(\overline{e}_L\,\vev{\phi^{0*}})_{\underline{1}}\,e_R
  + \widetilde{y}_1' \,(\overline{e}_L\,\vev{\phi^{0*}})_{\underline{1}'}\,e_R''
  + \widetilde{y}_1'' \,(\overline{e}_L\,\vev{\phi^{0*}})_{\underline{1}''}\,e_R'
  +\text{h.c.}\;,\label{eq3:HLV_LR_me}\\
 \mnud\;:\quad
 &
    y_1 \,(\overline{\nu}_L\,\vev{\phi^{0}})_{\underline{1}}\,\nu_R
  + y_1' \,(\overline{\nu}_L\,\vev{\phi^{0}})_{\underline{1}'}\,\nu_R''
  + y_1'' \,(\overline{\nu}_L\,\vev{\phi^{0}})_{\underline{1}''}\,\nu_R'
  \nonumber\\
&\qquad
  -\left[ \widetilde{y}_1 \,(\overline{\nu}_L\,\vev{\phi^{0}})_{\underline{1}}\,\nu_R
  + \widetilde{y}_1' \,(\overline{\nu}_L\,\vev{\phi^{0}})_{\underline{1}'}\,\nu_R''
  + \widetilde{y}_1'' \,(\overline{\nu}_L\,\vev{\phi^{0}})_{\underline{1}''}\,\nu_R'
  \right]
  +\text{h.c.}\;.\label{eq3:HLV_LR_mnud}
\end{align}  
%
Taking $\vev{\phi^{0*}} \equiv \vev{\phi^0}
  = (v_\ell, v_\ell, v_\ell)$ and then comparing Eqs.~(\ref{eq3:HLV_LR_me}) and (\ref{eq3:HLV_LR_mnud}), one gets~\footnote{See Appendix~\ref{app:b_LR_details} for details.}
\begin{equation}
 m_e = \Uw \widehat{m}_e\;,\quad \mnud = -\Uw \widehat{m}_e = -V_{eL}^\dagger \widehat{m}_e\;,\;
 \label{eq3:HLV_LR_relation}
\end{equation}
where $\widehat{m}_e = \text{diag}(\sqrt{3}(-y_1 +\widetilde{y}_1)v_\ell\;, \sqrt{3}(-y_1' +\widetilde{y}_1')v_\ell\;, \sqrt{3}(-y_1''+\widetilde{y}_1'')v_\ell)$. Whereas the neutrino Majorana mass matrix is a general complex symmetric just like in our other examples, the quark mass matrices have a special form. For $m_u$ the expanded Lagrangian,
\begin{align}
  &
   y_{2s}\,(\overline{u}_L\,u_R)_{\rept s} \,\vev{\phi^{0}}
  +y_{2a}\,(\overline{u}_L\,u_R)_{\rept a} \,\vev{\phi^{0}}
  -\widetilde{y}_{2s}\,(\overline{u}_L\,u_R)_{\rept s} \,\vev{\phi^{0}}
  -\widetilde{y}_{2a}\,(\overline{u}_L\,u_R)_{\rept a} \,\vev{\phi^{0}} 
\nonumber\\
  &\quad  
  +y_3\,(\overline{u}_L\,u_R)_{\rep} \,\vev{\phi^{0}_A}
  +y_3'\,(\overline{u}_L\,u_R)_{\repp} \,\vev{{\phi^{0}_A}''}
  +y_3''\,(\overline{u}_L\,u_R)_{\reppp} \,\vev{{\phi^{0}_A}'} 
  +\widetilde{y}_3\,(\overline{u}_L\,u_R)_{\rep} \,\vev{\phi^{0*}_B}
\nonumber\\
  &\quad\quad
  +\widetilde{y}_3'\,(\overline{u}_L\,u_R)_{\repp} \,\vev{{\phi^{0*}_B}''}
  +\widetilde{y}_3''\,(\overline{u}_L\,u_R)_{\reppp} \,\vev{{\phi^{0*}_B}'}+\text{h.c.}
\end{align}
gives rise to a mass matrix of the form (See Appendix~\ref{app:b_LR_details})
\begin{equation}
 m_u=\ththMat{Y_A^{(1)}}{Y^{+}_2}{Y^{-}_2}
             {Y^{-}_2}{Y_A^{(1')}}{Y^{+}_2}
             {Y^{+}_2}{Y^{-}_2}{Y_A^{(1'')}} \;, \label{eq3:HLV_LR_mu}
\end{equation}
while it can be shown that mass matrix $m_d$ also has a similar structure: 
\begin{equation}
 m_d=\ththMat{Y_B^{(1)}}{-Y^{+}_2}{-Y^{-}_2}
             {-Y^{-}_2}{Y_B^{(1')}}{-Y^{+}_2}
             {-Y^{+}_2}{-Y^{-}_2}{Y_B^{(1'')}} \;,\label{eq3:HLV_LR_md} 
\end{equation}
where $Y^{\pm}_2$, $Y_A^{(1),(1'),(1'')}$ and $Y_B^{(1),(1'),(1'')}$ are complicated functions of the VEVs and Yukawa couplings.
Equations (\ref{eq3:HLV_LR_mu}) and (\ref{eq3:HLV_LR_md}) imply that the diagonalization matrices $V_{uL}$ and $V_{dL}$ are not completely arbitrary. However, it is easy to see that there are enough degrees of freedom in the resulting $\Uckm = V_{uL}V_{dL}^\dagger$ such that experimental data can be fitted. 

Returning to (\ref{eq3:HLV_LR_relation}), it is clear that the main prediction of this model is 
\begin{equation}
M_R 
 \simeq  \widehat{m}_{e}\, \Upmns^*\, \mnuhat^{-1}\, \Upmns^{\dagger}\, \widehat{m}_{e}\;,
\end{equation}
where  $\Upmns = V_{eL} V_\nu^\dagger = \Uw^\dagger V_\nu^\dagger$ is \emph{a priori} arbitrary and to be fitted to the experimental observations.

\section{Phenomenology}\label{sec:HLV_phen}

The general conclusion from the previous section is that it is possible to use symmetries to construct the relation
\begin{equation}
 M_R \simeq	
  \widehat{m}_f\,\Upmns^*\, \mnuhat^{-1}\, \Upmns^{\dagger}\, \widehat{m}_{f}\;,\quad 
  f= e, d \text{ or } u\;,
\label{eq3:HLV_ph_MR}
\end{equation}
that links the high-energy seesaw sector to low-energy observables. Using the current experimental data on quarks and leptons, the properties of the heavy RH Majorana neutrinos in these models can therefore be inferred directly, and interesting consequences may arise.
So in this section, we discuss some of the implications of these models.

To begin with, we recall that $\Upmns$ can be conveniently parametrized by the generic form given in  (\ref{eq1:UPMNS_param}) on page~\pageref{eq1:UPMNS_param}, where the mixing matrix is completely defined by three mixing angles: $\theta_{23}, \theta_{12}, \theta_{13}$, one $CP$ violating Dirac phase: $\delta$, and two Majorana phases: $\alpha_1, \alpha_2$. However, in calculations, it is often convenient to absorb the Majorana phases into $\mnuhat$ in (\ref{eq3:HLV_ph_MR}) and allow the $m_i$'s to be complex masses instead. This way the expressions are much simpler without loss of generality, and one can always reintroduce the Majorana phases when needed.

In our numerical analyses, we use the best fit values for the mixing angles displayed in (\ref{eq1:best-fit_angle}) on page~\pageref{eq1:best-fit_angle}, while for the analytical work, we assume that $\Upmns$ has an exact tribimaximal form \cite{TBmixing} (see (\ref{eq1:osc_TriBi})). In other words, we take
\begin{equation}\label{eq3:HLV_TB_values}
 \sin^2\theta_{12} = \frac{1}{3}\;,\quad
 \sin^2\theta_{23} = \frac{1}{2}\;,\quad
 \sin^2\theta_{13} = 0\;.
\end{equation}
The inputs to the light neutrino mass matrix $\mnuhat$ will be governed by the square-mass differences of (\ref{eq1:best-fit_mass}) and the implied upper limit from cosmology \cite{wmap_numass,
sdss_numass} of
\begin{equation}\label{eq3:HLV_nu_mass_limit}
 |m_i| \lesssim 0.2 \text{ eV} \;, 
\end{equation}
for each $i$th LH neutrino~\footnote{See discussion on page~\pageref{eq1:numass_cosmo}.}.
Furthermore, we study (\ref{eq3:HLV_ph_MR}) by taking a generic form for $\widehat{m}_f \equiv \text{diag}(\xi_1,\xi_2,\xi_3)$ where $\xi_1 \ll \xi_2 \ll \xi_3$ is assumed. 
It is obvious that once $\widehat{m}_f$ has been chosen (i.e. $\xi_i$'s are known), only $\delta,\alpha_1,\alpha_2$ and $|m_1|$ (or $|m_3|$ for the inverted hierarchy case) can 
potentially change the form of $M_R$ and its eigenvalue spectrum. Moreover, if $\theta_{13} \simeq 0$, it is expected that the Dirac phase, $\delta$, would not play a significant role which will reduce the parameter space even further. But it should be pointed out that when the 13-mixing is nonzero (e.g. at the best fit value of $5.7^\circ$), the choice of Dirac phase can influence the $M_R$ mass eigenvalues significantly (almost 2 orders of magnitude) for tiny sets of Majorana phases and $|m_{1,3}|$ values as our numerical scan of the parameter space have indicated. Therefore, care must be taken when analysing $M_R$, especially in the study of their leptogenesis implications.

Using the prescription described above, we approximate $\Upmns$ with the tribimaximal form and absorb $\alpha_{1,2}$ into $m_{1,2}$ respectively. So after expanding out the RHS of (\ref{eq3:HLV_ph_MR}), we obtain
\begin{equation}\label{eq3:HLV_MR_exact}
 M_R\equiv M_R^T = 
 \ththMat
 {\frac{2\xi_1^2}{3m_1}+\frac{\xi_1^2}{3m_2}}
 {-\frac{\xi_1 \xi_2}{3m_1}+\frac{\xi_1\xi_2}{3m_2}}
 {-\frac{\xi_1 \xi_3}{3m_1}+\frac{\xi_1\xi_3}{3m_2}}
 {\cdots}
 {\frac{\xi_2^2}{6m_1}+\frac{\xi_2^2}{3m_2}+\frac{\xi_2^2}{2m_3}}
 {\frac{\xi_2\xi_3}{6m_1}+\frac{\xi_2\xi_3}{3m_2}-\frac{\xi_2\xi_3}{2m_3}}
 {\cdots}
 {\cdots}
 {\frac{\xi_3^2}{6m_1}+\frac{\xi_3^2}{3m_2}+\frac{\xi_3^2}{2m_3}}\;.
\end{equation}
The leading behaviors of the mass spectrum for $M_R$ can be studied by investigating the limiting cases of Eq.~(\ref{eq3:HLV_MR_exact}). Consequently, with the support of numerical analyses, important insights into the connection between the RH and LH seesaw sectors, as well as predictions in leptogenesis and collider phenomenologies can be gained. We present our studies on these in the next few subsections.

\subsection{Fully hierarchical light neutrinos}\label{subsec:HLV_H_nu}

For the normal hierarchy scheme, we have $|m_1| \rightarrow 0$ with $|m_{2,3}|$ related to $|m_1|$ via (\ref{eq1:NH_m2_m3}). Therefore, in this limit, we can write Eq.~(\ref{eq3:HLV_MR_exact}) as
\begin{equation}\label{eq3:HLV_MR_MR0_dM}
 M_R = M_{R0} + \Delta M_R\;, \qquad \text{where }\;\;
 M_{R0}\equiv \ththMat
 {\frac{2\xi_1^2}{3m_1}}
 {-\frac{\xi_1 \xi_2}{3m_1}}
 {-\frac{\xi_1 \xi_3}{3m_1}}
 {\cdots}
 {\frac{\xi_2^2}{6m_1}}
 {\frac{\xi_2\xi_3}{6m_1}}
 {\cdots}
 {\cdots}
 {\frac{\xi_3^2}{6m_1}}\;
\end{equation}
is the dominant part of the matrix as $|m_1| \rightarrow 0$, while $\Delta M_R$ is considered to be a small perturbation. Suppose that the eigenvalue equation for $M_{R0}$ reads
\begin{equation}\label{eq3:HLV_eval_eqn_MR0}
 M_{R0}\,u_{i0} = E_{i0}\,u_{i0}\;,
\end{equation}
where $E_{i0}$ and $u_{i0}$ denote the $i$th eigenvalues and eigenvectors respectively. Then, perturbation theory implies that the true solutions for $M_R$ may be expressed as
\begin{align}
 E_{i} &= E_{i0} + \Delta E_i \;,
\end{align}
\begin{align}
 u_{i} &= u_{i0} + \Delta u_i\;,
\end{align}
with the variation in the eigenvalues given by~\footnote{Here, $u_{i0}$'s have been chosen to be orthonormal to each other.}
\begin{equation}\label{eq3:HLV_MR_dE_eqn}
 \Delta E_i = u_{i0}^T \cdot (\Delta M_R) \cdot u_{i0}\;, \quad i=1,2 \text{ and } 3\;,
\end{equation}
to first order. Solving Eq.~(\ref{eq3:HLV_eval_eqn_MR0}) for $E_{i0}$, one immediately gets
\begin{equation}\label{eq3:HLV_MR_E0}
 E_{10} \;, E_{20} = 0\;,\quad E_{30} = \frac{4\xi_1^2+\xi_2^2+\xi_3^2}{6m_1}
 \simeq \frac{\xi_3^2}{6|m_1|}\;,
\end{equation}
and 
\begin{equation}\label{eq3:HLV_MR_u0}
 u_{10} = \frac{1}{k_1 k_2}
 \threecMat{\xi_3 k_1}
           {2\xi_1 \xi_2}
           {-2\xi_1 k_2}\;,\;\;
 u_{20} = \frac{1}{k_1}
 \threecMat{0}
           {k_2}
           {\xi_2}\;,\;\;
 u_{30} = \frac{1}{k_1 k_2}
 \threecMat{2\xi_1 k_1}
           {-\xi_2 \xi_3}
           {\xi_3 k_2}\;,                    
\end{equation}
where $k_1 = \sqrt{4\xi_1^2+\xi_2^2+ \xi_3^2}$ and $k_2 = \sqrt{4\xi_1^2+\xi_3^2}$.
Using these in (\ref{eq3:HLV_MR_dE_eqn}) and in the limit of $\xi_3 \gg \xi_{1,2}$ and $|m_3| \gg |m_2|$, one obtains
\begin{equation}\label{eq3:HLV_MR_dE_NH}
 \Delta E_1 \simeq \frac{3\xi_1^2}{m_2}\;,\qquad
 \Delta E_2 \simeq \frac{2\xi_2^2}{m_3}\;,\qquad
 \Delta E_3 \simeq \frac{2\xi_3^2}{m_2}\;.
\end{equation}
Hence to leading order, the heavy RH neutrino masses are given by
\begin{equation}\label{eq3:HLV_NH_MR_mass}
 |M_1| \simeq \frac{3\xi_1^2}{|m_2|}\;,\qquad
 |M_2| \simeq \frac{2\xi_2^2}{|m_3|}\;,\qquad
 |M_3| \simeq \frac{\xi_3^2}{6|m_1|}\;.
\end{equation}
It is interesting to note that due to the large neutrino mixing, the expected correspondence between $m_i$ and the Dirac masses, $m_i \propto \xi_i^2$, no longer holds and that only the largest RH neutrino mass is a function of $|m_1|$~\footnote{These results are consistent with those in references \cite{Akhmedov03, Branco02}.}. Substituting in the running fermion masses $m(\mu)$ at $\mu \simeq 10^9 \text{ GeV}$ \cite{running_mass} as typical values for $\xi_i$'s, the predictions of the RH neutrino masses for all cases: $f = u,d,e$ (and assuming normal hierarchy for light neutrinos) are
\begin{align}
 u\;: \quad
  &|M_1| \simeq 5.6\times 10^5 \text{ GeV}\;,\;\;
   |M_2| \simeq 5.5\times 10^{9} \text{ GeV}\;,\;\;
    |M_3| \gtrsim 2.0\times 10^{14} \text{ GeV}\;, \label{eq3:HLV_MR_u_case}\\
  d\;: \quad
  &|M_1| \simeq 2.3\times 10^6 \text{ GeV}\;,\;\;
   |M_2| \simeq 1.1\times 10^8 \text{ GeV}\;,\;\;
    |M_3| \gtrsim 3.8\times 10^{10} \text{ GeV}\;, \label{eq3:HLV_MR_d_case}\\
  e\;: \quad
  &|M_1| \simeq 9.0\times 10^4 \text{ GeV}\;,\;\;
   |M_2| \simeq 4.8\times 10^8 \text{ GeV}\;,\;\;
    |M_3| \gtrsim 5.7\times 10^{10} \text{ GeV}\;. \label{eq3:HLV_MR_e_case}        
\end{align}
The plots of $M_{1,2,3}$ as a function of $|m_1|$ for the case $\widehat{m}_f=\widehat{m}_u$ and for many different values of $\delta, \alpha_{1,2}$ are shown in Fig.~\ref{fig3:mu_M123} while similar plots for the $d$- and $e$-case can be found in Fig.~\ref{fig3:mdme_M123}. These numerical results validate the trend predicted by the theoretical analysis. The tallest spikes in the diagrams of Fig.~\ref{fig3:mu_M123} are locations where level crossing occurs ($M_{1,2}$ or $M_{2,3}$ are quasi-degenerate) for certain special values of Dirac and Majorana phases, an effect that has been previously studied in \cite{Akhmedov03, Nezri00}.

For the inverted hierarchy scheme ($|m_3| \ll |m_1| \simeq |m_2|$), we return to Eq.~(\ref{eq3:HLV_MR_exact}) and note that the dominant part of the matrix looks like
\begin{equation}\label{eq3:HLV_IH_MR0}
 M_{R0}\equiv 
 \ththMat
 {0}
 {0}
 {0}
 {\cdots}
 {\frac{\xi_2^2}{2m_3}}
 {\frac{-\xi_2\xi_3}{2m_3}}
 {\cdots}
 {\cdots}
 {\frac{\xi_3^2}{2m_3}}\;.
\end{equation}
Following the procedure outlined previously, we find that (\ref{eq3:HLV_IH_MR0}) gives rise to
\begin{equation}\label{eq3:HLV_MR_E0_IH}
 E_{10} \;, E_{20} = 0\;,\quad E_{30} = \frac{\xi_2^2+\xi_3^2}{2 m_3}
 \simeq \frac{\xi_3^2}{2|m_3|}\;,
\end{equation}
and
\begin{equation}\label{eq3:HLV_MR_u0_IH}
 u_{10} = \frac{1}{k_3}
 \threecMat{0}
           {\xi_3}
           {\xi_2}\;,\qquad
 u_{20} = 
 \threecMat{1}
           {0}
           {0}\;,\qquad
 u_{30} = \frac{1}{k_3}
 \threecMat{0}
           {-\xi_2}
           {\xi_3}\;,                    
\end{equation}
where $k_3 = \sqrt{\xi_2^2+ \xi_3^2}$. These results then
leads to the following expressions for the $M_R$ masses:
\begin{equation}\label{eq3:HLV_MR_mass_IH}
 |M_1| \simeq \frac{\xi_1^2}{|m_2|}\;,\quad
 |M_2| \simeq \frac{2\xi_2^2}{|m_2|}\;,\quad
 |M_3| \simeq \frac{\xi_3^2}{2|m_3|}+\frac{\xi_3^2}{2|m_2|}
       \simeq \frac{\xi_3^2}{2|m_3|}
 \;.
\end{equation}
Invoking relation (\ref{eq1:IH_m1_m2}) on page~\pageref{eq1:IH_m1_m2} and taking $|m_3|\rightarrow 0$, it can be shown that the resulting numerical values for (\ref{eq3:HLV_MR_mass_IH}) are very similar to those shown in Eqs.~(\ref{eq3:HLV_MR_u_case}) to (\ref{eq3:HLV_MR_e_case}).

\subsection{Quasi-degenerate light neutrinos}\label{subsec:HLV_QD_nu}

\begin{figure}[p]
\begin{center} 
\includegraphics[width=0.70\textwidth]{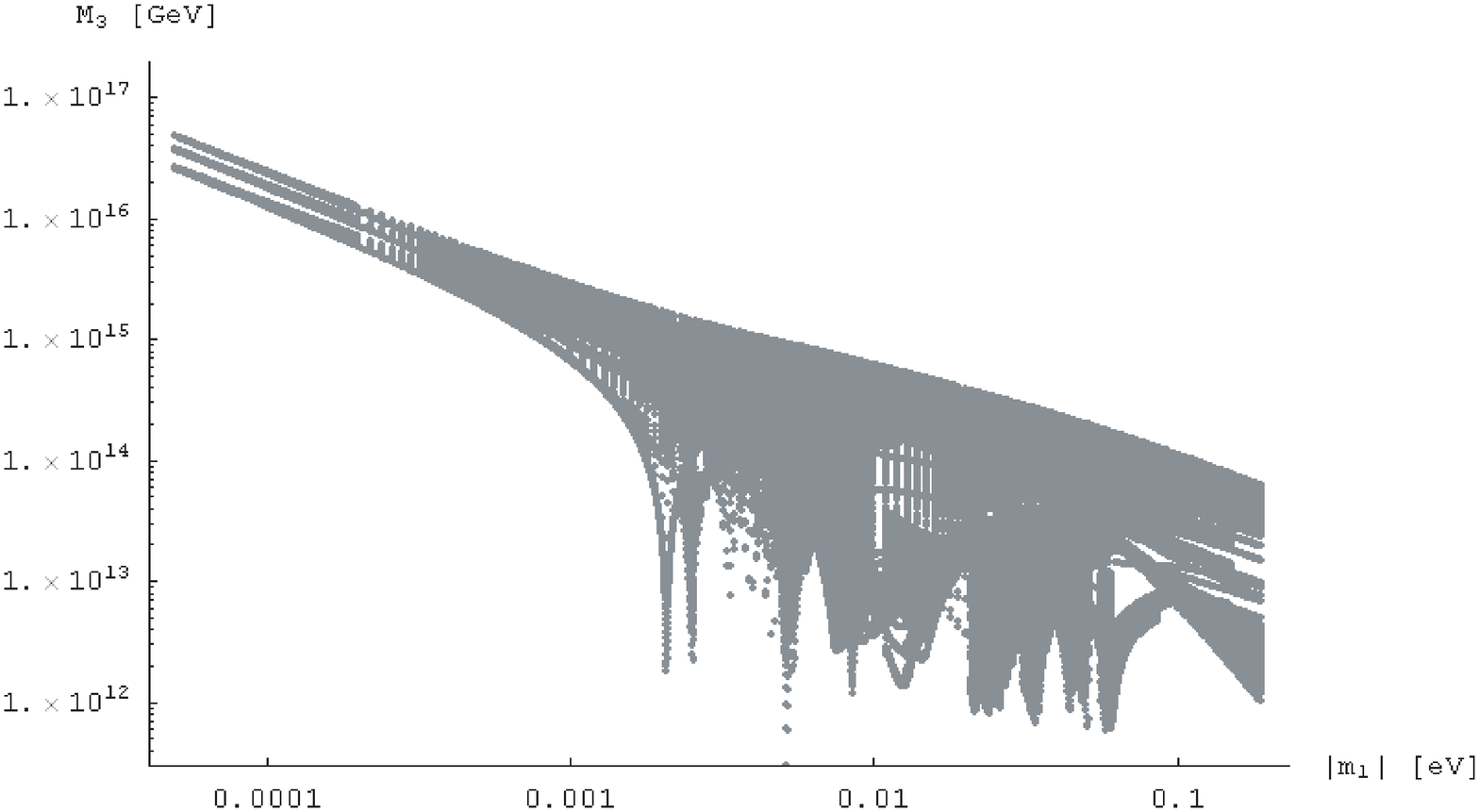}\\
%
\includegraphics[width=0.70\textwidth]{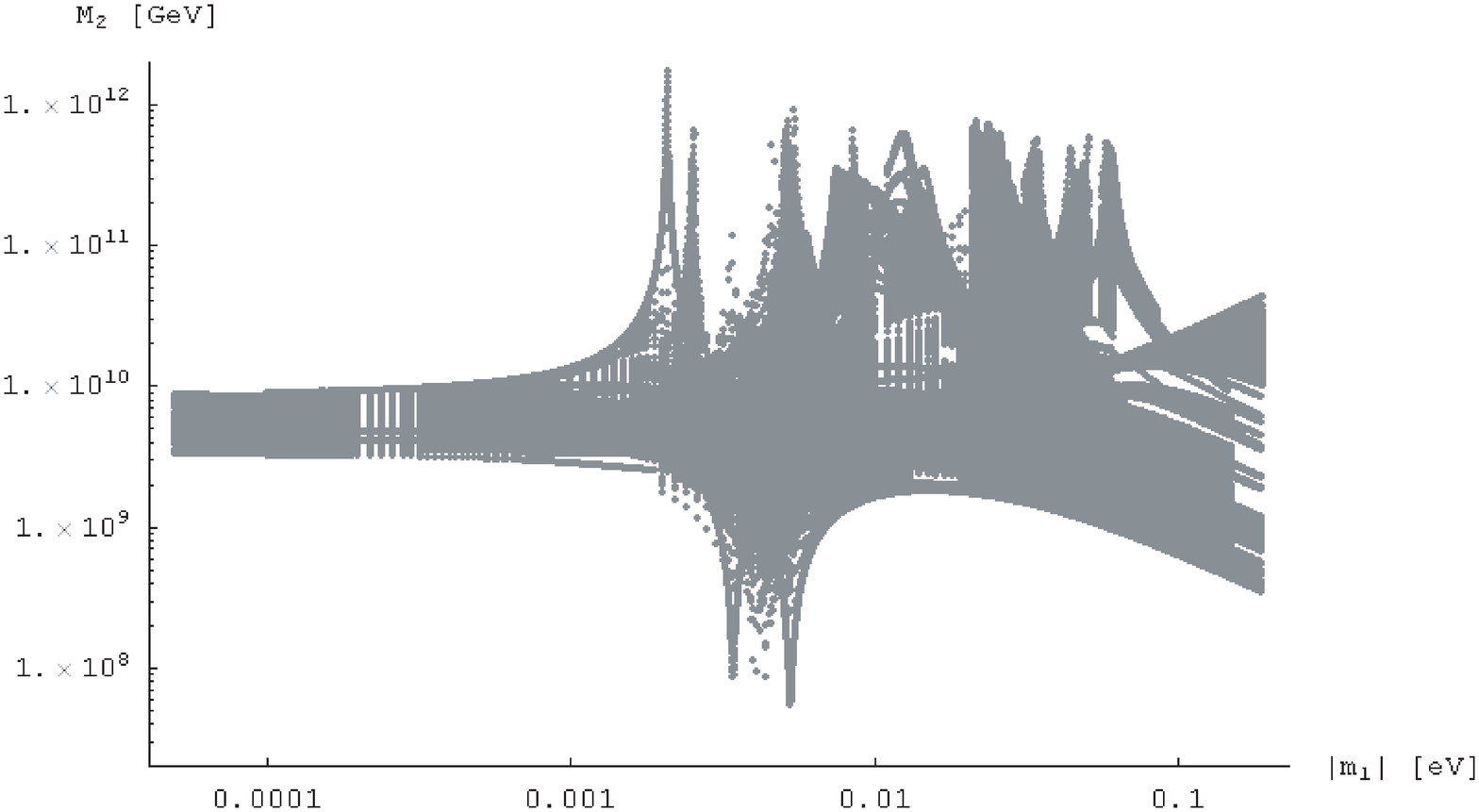}\\
%
\includegraphics[width=0.70\textwidth]{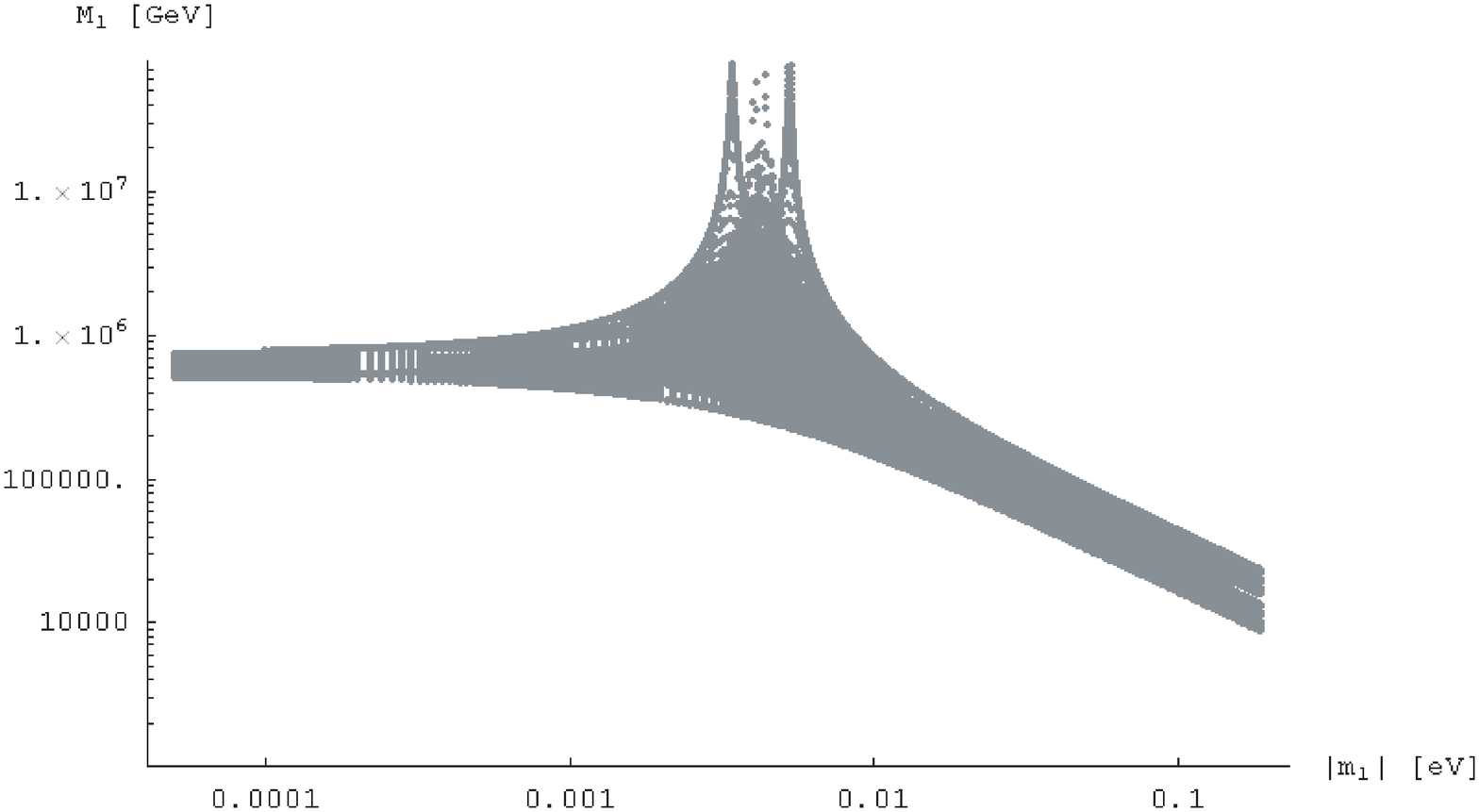}
\end{center}
\caption{Plots of $M_{1,2,3}$ vs. $|m_1|$ in the $\widehat{m}_f=\widehat{m}_u$ case with normal hierarchy for light neutrino masses assumed. 
Input running masses used: $m_u(\mu)=1.3$~MeV, $m_c(\mu)=0.37$~GeV, $m_t(\mu)=1.1\times 10^2$~GeV, where $\mu\simeq 10^9$~GeV.
Each plot contains approximately $3.18\times 10^5$ data points produced by systematically sweeping the $|m_1|$ and $\delta, \alpha_{1,2} \in (0, 2\pi)$ parameter space.}\label{fig3:mu_M123}
\end{figure}

\begin{figure}[t]
\begin{center}
\includegraphics[width=0.49\textwidth]{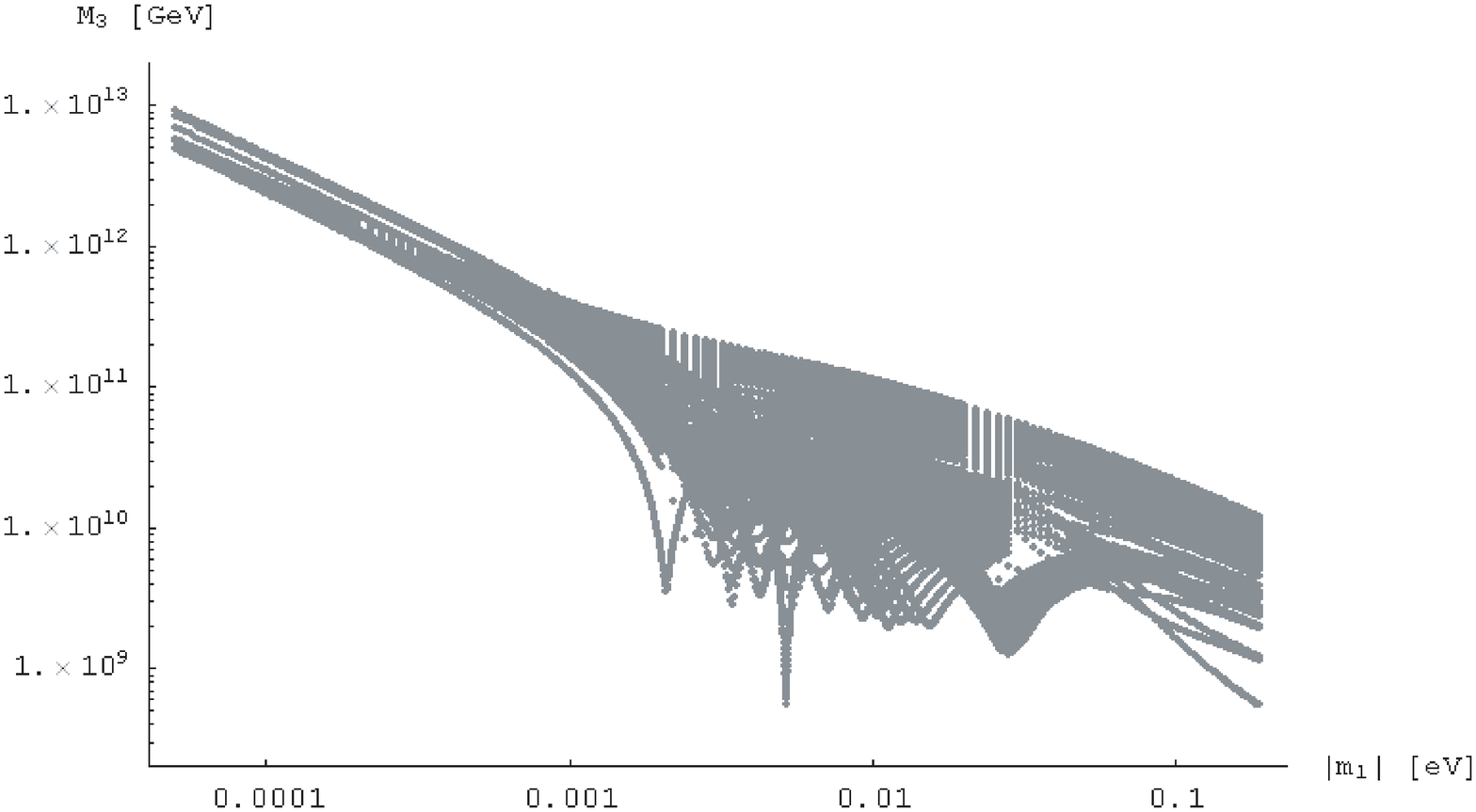}
\includegraphics[width=0.49\textwidth]{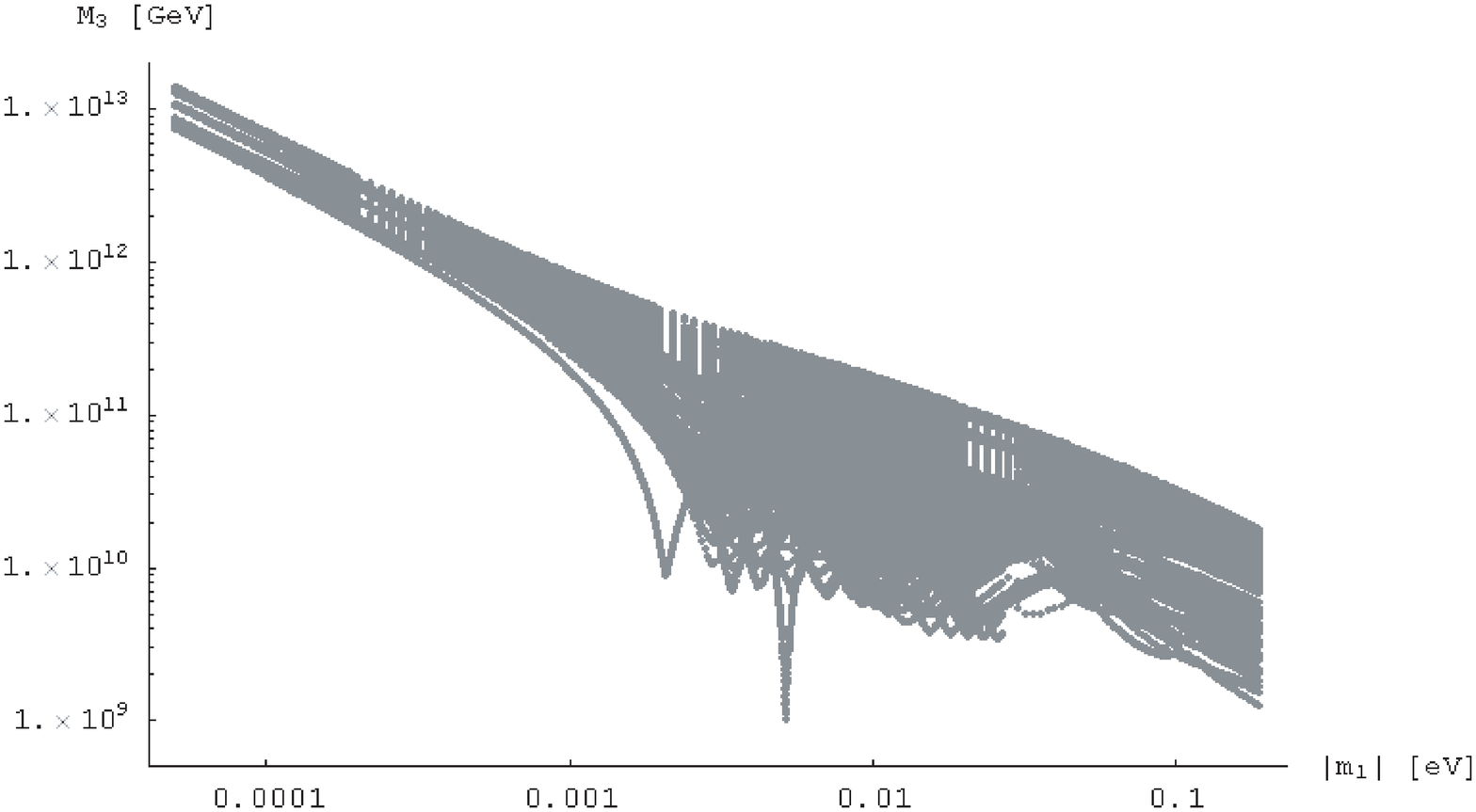}
\includegraphics[width=0.49\textwidth]{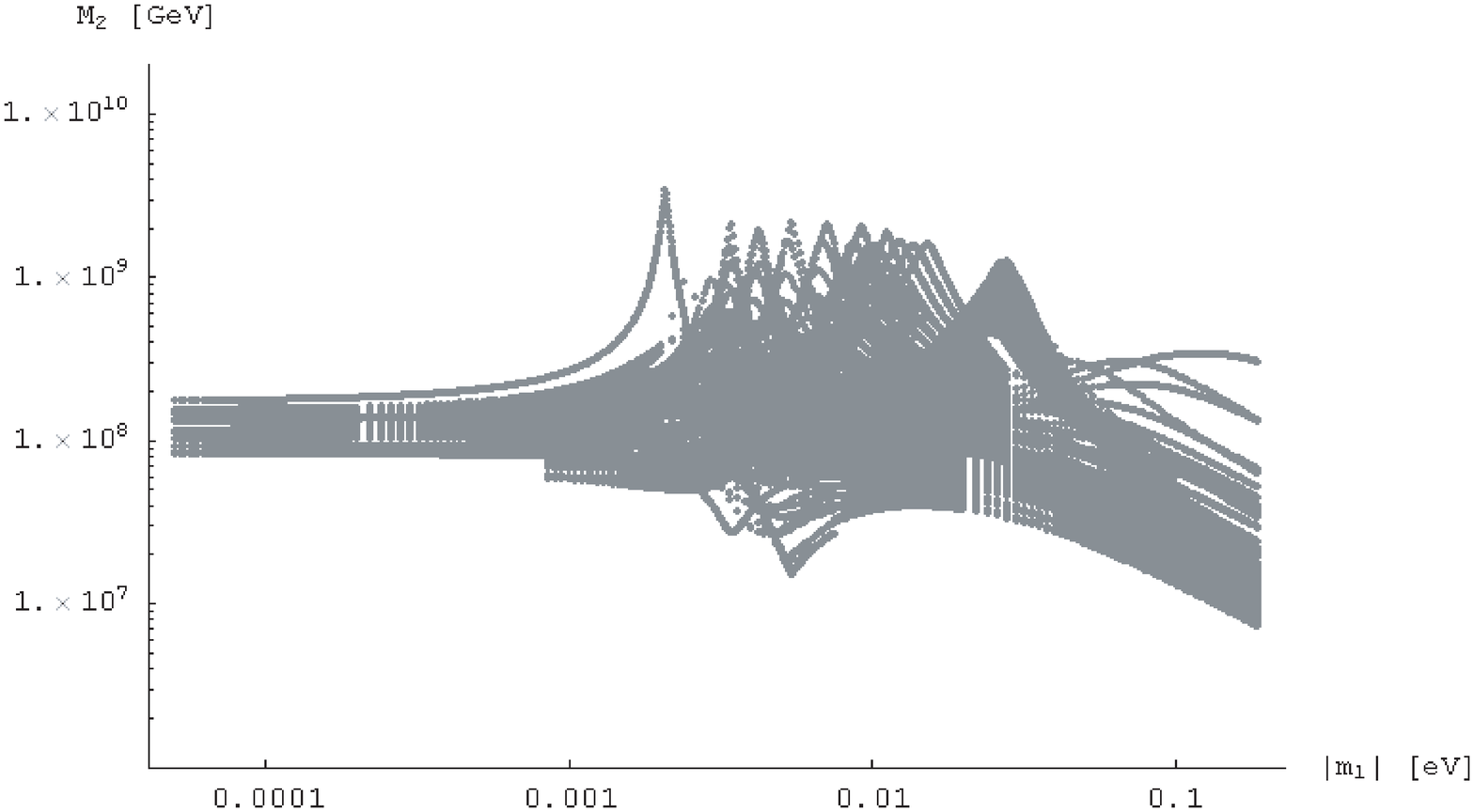}
\includegraphics[width=0.49\textwidth]{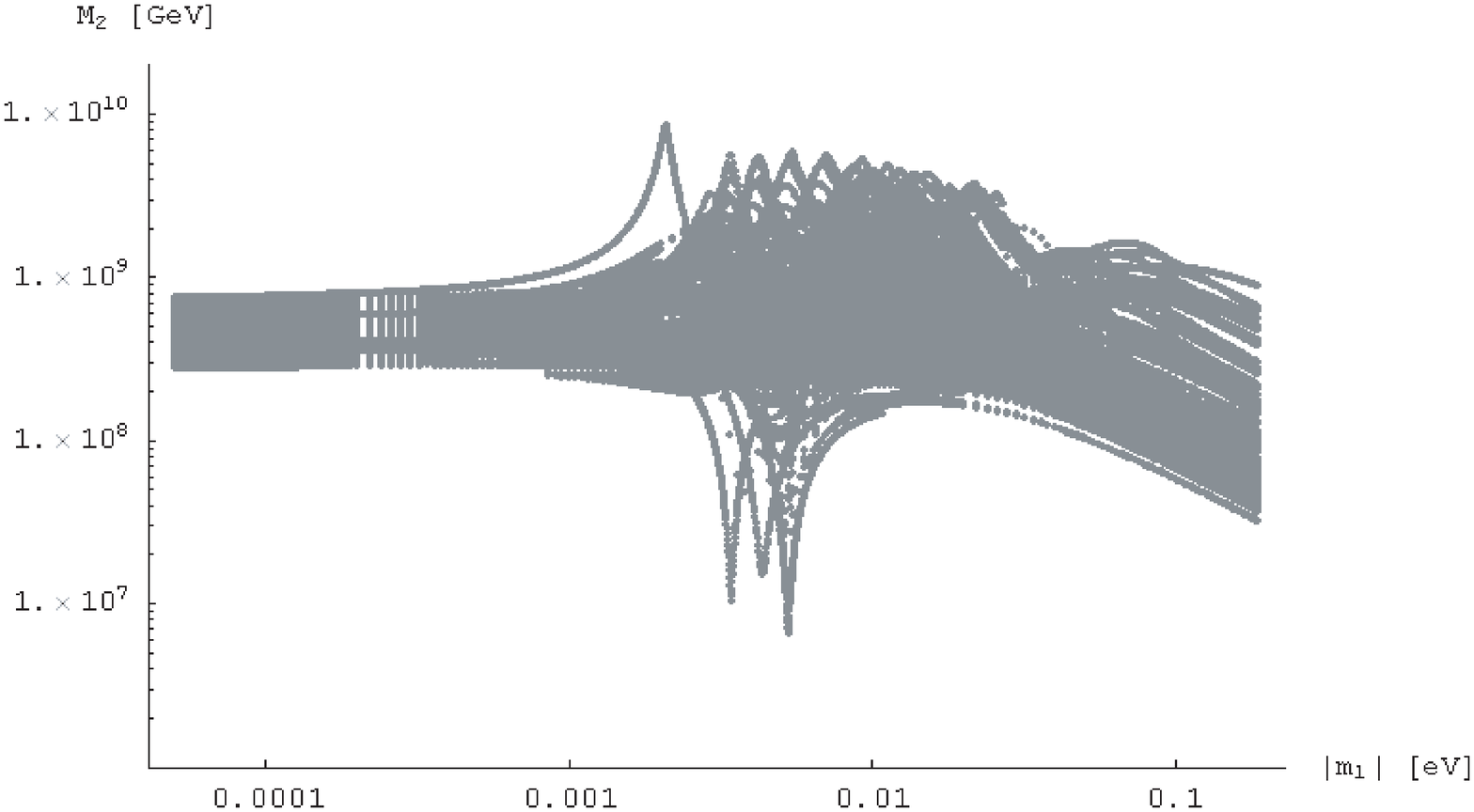}
\includegraphics[width=0.49\textwidth]{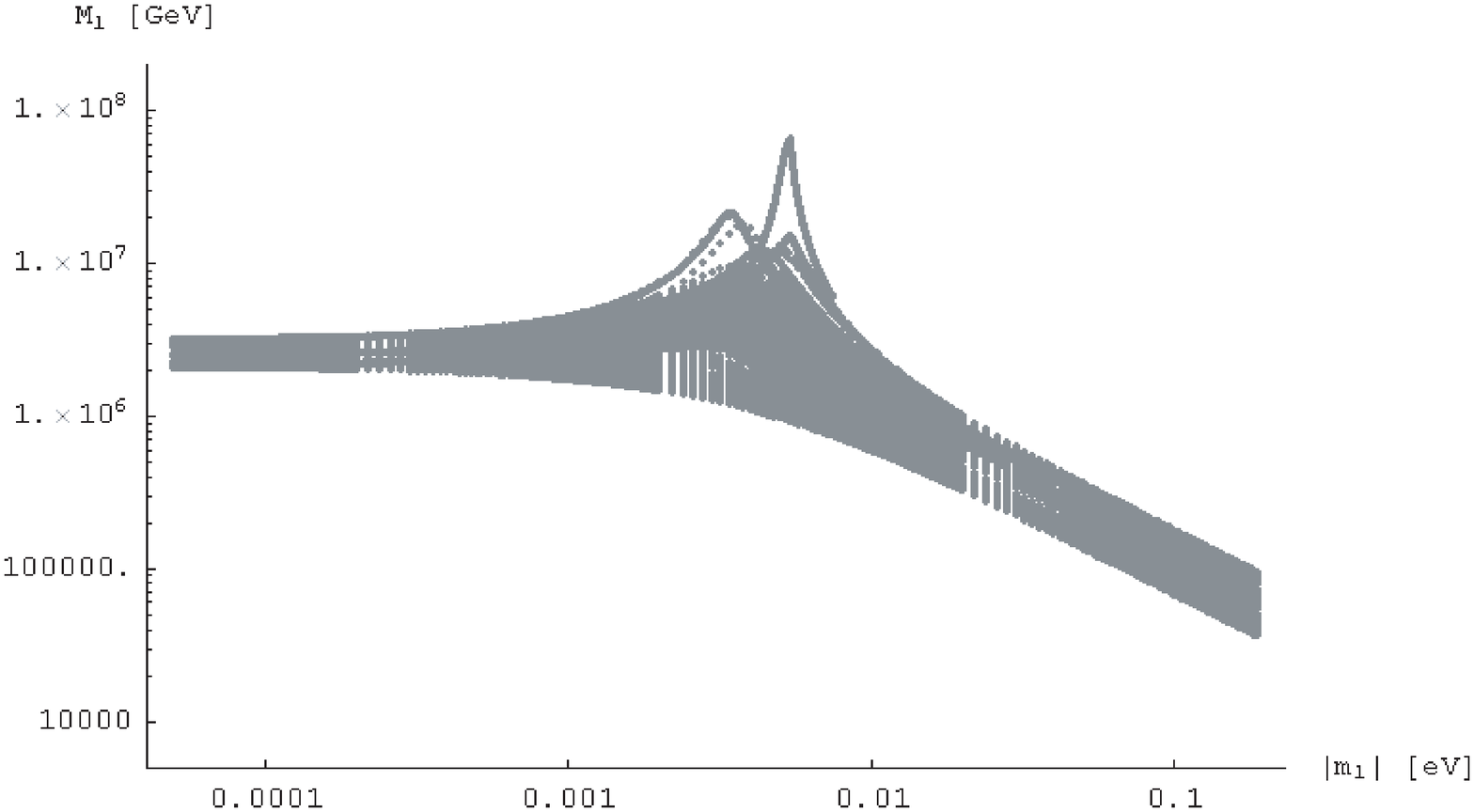}
\includegraphics[width=0.49\textwidth]{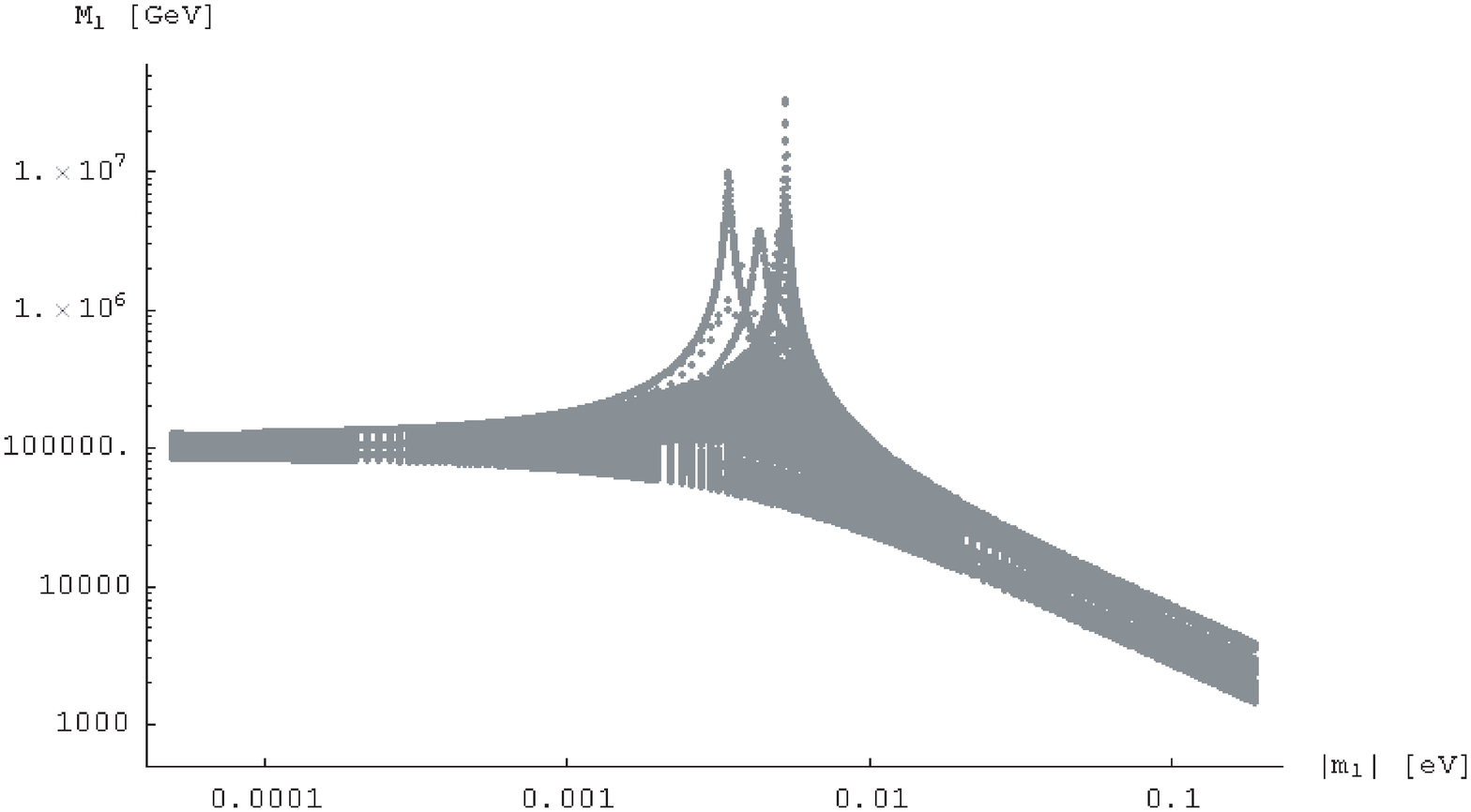}
\end{center}
\caption{Plots of $M_{1,2,3}$ vs. $|m_1|$ in the $\widehat{m}_f=\widehat{m}_d$ case (LEFT column) and $\widehat{m}_f=\widehat{m}_e$ case (RIGHT column) with normal hierarchy for light neutrino masses assumed. Input running masses used: (LEFT) $m_d(\mu)=2.6$~MeV, $m_s(\mu)=52$~MeV, $m_b(\mu)=1.5$~GeV, and (RIGHT) $m_e(\mu)=0.52$~MeV, $m_\mu(\mu)=1.1\times 10^2$~MeV, $m_\tau(\mu)=1.8$~GeV, where $\mu\simeq 10^9$~GeV. Each plot contains approximately $1.0\times 10^5$ data points produced by systematically sweeping $|m_1|$ and the $\delta, \alpha_{1,2} \in (0, 2\pi)$ parameter space.}
\label{fig3:mdme_M123}
\end{figure}

When the lightest neutrino mass approaches the upper bound of (\ref{eq3:HLV_nu_mass_limit}), we have $|m_1|\simeq |m_2| \simeq|m_3|$. Furthermore, if we assume that the Majorana phases $\alpha_{1,2}$ are negligible, then Eq.~(\ref{eq3:HLV_MR_exact}) has a simple form:
\begin{equation}\label{eq3:HLV_QD_Mexact}
 M_R \simeq 
 \ththMat
 {\frac{\xi_1^2}{|m_1|}}{0}{0}
 {\cdots}{\frac{\xi_2^2}{|m_1|}}{0}
 {\cdots}{\cdots}{\frac{\xi_3^2}{|m_1|}}\;.
\end{equation}
So, it follows immediately that the approximate scale for the $M_i$'s for quasi-degenerate light neutrinos is given by
\begin{align}
 u\;: \;\;
  &|M_1| \simeq 8.5\times 10^3 \text{ GeV}\;,\quad
   |M_2| \simeq 6.8\times 10^{8} \text{ GeV}\;,\quad
    |M_3| \simeq 5.9\times 10^{13} \text{ GeV}\;, \label{eq3:HLV_QD_MR_u}\\
  d\;: \;\;
  &|M_1| \simeq 3.4\times 10^4 \text{ GeV}\;,\quad
   |M_2| \simeq 1.3\times 10^7 \text{ GeV}\;,\quad
    |M_3| \simeq 1.1\times 10^{10} \text{ GeV}\;, \label{eq3:HLV_QD_MR_d}\\
  e\;: \;\;
  &|M_1| \simeq 1.4\times 10^3 \text{ GeV}\;,\quad
   |M_2| \simeq 5.9\times 10^7 \text{ GeV}\;,\quad
    |M_3| \simeq 1.7\times 10^{10} \text{ GeV}\;. \label{eq3:HLV_QD_MR_e}        
\end{align}
Inspecting the numerical results for $M_R$ near $|m_1| \simeq 0.1$ in Fig.~\ref{fig3:mu_M123} and \ref{fig3:mdme_M123}, we see that the above estimates agree well (within about 1 order of magnitude) with the more encompassing treatment where Majorana phases $\alpha_{1,2}$ are not ignored.

%

%

\subsection{Thermal leptogenesis}\label{subsec:HLV_lepto}

Using the $M_R$ mass spectrum information presented above, several general comments on the possibility of baryon asymmetry generation via thermal leptogenesis for the models discussed in Section~\ref{sec:HLV_models} can be made.
First of all, we note that for both the hierarchical and quasi-degenerate cases, $M_1$ is typically in the range of $10^3 - 10^6$ GeV. In addition, it is clear that the RH Majorana neutrinos are strongly hierarchical ($M_1\ll M_2\ll M_3$) in all these situations. Therefore, based on our discussion in Sec.~\ref{subsec:lepto_N1}, and in particular, the resultant lower bound on $M_1$ (see (\ref{eq1:lepto_std_lower_bound_M1})), conventional leptogenesis where the asymmetry is generated predominantly by the decay of $N_1$'s would \emph{not} be successful in these models~\footnote{A similar conclusion was reached in \cite{Akhmedov03} where they studied the implications of having a neutrino Dirac mass matrix with strongly hierarchical eigenvalues.}. However, as we have illustrated in Sec.~\ref{subsec:lepto_flavor} to Sec.~\ref{subsec:lepto_resonant}, there exists other special solutions to the leptogenesis scenario which can circumvent those restrictions imposed by the standard setup.

As was pointed out earlier, the tall spikes in the plots of Figs.~\ref{fig3:mu_M123} and \ref{fig3:mdme_M123} indicate that there are regions in the parameter space for these models where $M_{1}$ and $M_2$ become almost degenerate. Consequently, it has been shown in a similar model in \cite{Akhmedov03} that a sufficient baryon asymmetry can be generated from resonant enhancement \cite{main_resonant_1,
main_resonant_1b,
pilaf_resonant_2,
flanz_resonant,
pilaf_underwood_TeV,
pilaf_int_j_mod} to the raw $CP$ asymmetry in the decays of $N_1$'s (see Sec.~\ref{subsec:lepto_resonant}). Furthermore, a similar enhancement to the decay of the next-to-the-lightest RH neutrino $N_2$, when $M_2$ and $M_3$ become degenerate, can also produce the desired asymmetry in principle, as long as washout effects mediated by the lighter $N_1$'s are insufficient \cite{PD_seesaw_geom}. In other words, both resonant $N_1$- and $N_2$-leptogenesis are realistic possibilities for the representative models discussed in Sec.~\ref{sec:HLV_models}.

Another interesting observation is that, recently, Ref.~\cite{Bari_Riotto} investigated the possibility of successful leptogenesis (without the need for resonant enhancement) in models with $SO(10)$-inspired mass relations which have properties similar to those presented here (see also \cite{Abada:2008gs}). In the analysis of \cite{Bari_Riotto}, they explored the situation where the asymmetry is predominantly generated by $N_2$ decays \emph{and} when flavor effects \cite{blanchet_zeno,
barbieri_flavor,
endoh_flavor,
nardi_flavor,
abada_flavor_1,
abada_flavor_2,
simone_riotto_flv,
blanchet_flavor} are important. Specifically, $N_2$-leptogenesis (see Sec.~\ref{subsec:lepto_N2}) in the relevant range of $10^9 \lesssim M_2 \lesssim 10^{12}$~GeV was studied. 
This 
results in a two-flavor regime where the lepton asymmetry is stored in the $\tau$-component, as well as a coherent superposition of $(e,\mu)$-components.
Subsequently, flavor dependent washout effects (see Sec.~\ref{subsec:lepto_flavor}) coming from interactions with $N_1$'s would  not completely erase all components of the asymmetry generated by the $N_2$'s under certain situations. One central conclusion in \cite{Bari_Riotto} is that, for this mechanism to generate enough asymmetry, the mass of the next-to-the-lightest RH neutrino must be about $M_2 \simeq 10^{11}$~GeV.

Inspecting the $M_2$-plot of Fig.~\ref{fig3:mu_M123} (corresponding to the $\widehat{m}_f = \widehat{m}_u$ case), we can see that the condition of $M_2 \simeq 10^{11}$~GeV can be marginally met by a small region of the parameter space (near the various spikes for $|m_1|$ values between $2\times 10^{-3}$ and $8\times 10^{-2}$~eV), whereas the $\widehat{m}_f = \widehat{m}_{d,e}$ cases are definitely ruled out for this scenario due to the smallness of $M_2$. Therefore, it appears that for some special values of $|m_1|$ with certain sets of phases $(\delta,\alpha_{1,2})$, leptogenesis via $N_2$ decays taking into account the effects of flavor is also possible (for the $\widehat{m}_f =\widehat{m}_u$ model) in addition to resonant leptogenesis.

Moreover, if this picture of flavored $N_2$-leptogenesis is indeed the mechanism responsible for generating the baryon asymmetry of the Universe, then the corresponding sets of low energy phases  in our model $(\delta,\alpha_{1,2})$ which make this possible will generally lead to modifications of the effective Majorana neutrino mass, $m_{\beta\beta}$ (see Sec.~\ref{subsec_abs_numass} and (\ref{eq1:numass_0vbb}) on page~\pageref{eq1:numass_0vbb}) responsible for controlling the neutrinoless double beta decay rate.
For example, taking $|m_1|= 0.070$~eV and assuming normal hierarchy, the phases implied by $N_2$-leptogenesis will lead to $m_{\beta\beta} \approx 0.047$~eV, which is a noticeable reduction from 0.070~eV in cases where both Majorana phases are turned off~\footnote{The reason we have picked $|m_1|= 0.070$~eV in this discussion is because so far we have not found any set of phases for $|m_1|\gtrsim 0.08$ in which $N_2$-leptogenesis is actually viable.}.
However, present experimental upper limits on $m_{\beta\beta}$ lie somewhere between 0.16 and 0.68~eV \cite{CUORICINO}, and so it is difficult to distinguish such differences. The detection of this may only be possible in future experiments such as CUORE~\cite{CUORE}, GERDA~\cite{GERDA} and \textsl{Majorana}~\cite{MAJORANA} which have a projected sensitivity down to about 0.05~eV~\footnote{See Sec.~\ref{subsec_abs_numass} when this was first discussed.}. 

In summary, while the models presented in Sec.~\ref{sec:HLV_models} do not generically lead to successful baryon asymmetry generation via thermal leptogenesis, some fine-tuned special cases do exist. Furthermore, it is possible that the enlargement of the workable parameter space for leptogenesis can result from modifications to the Higgs sector of these models, however, such analyses are beyond the scope of this work.

\subsection[Collider signatures]{Collider signatures~\protect\footnote{X.~G.~He has made significant contribution to this subsection.}}\label{subsec:HLV_collider} 

It is interesting to note that in the case with $\widehat{m}_f = \widehat{m}_e$, the lightest heavy Majorana neutrino mass, $M_1$ can be as low as about 1 TeV, opening up the possibility of seeing signals of such a particle at the Large Hadron Collider (LHC) and or a future International Linear Collider (ILC). But as we shall demonstrate below, it turns out that such signal are too weak to be observable.

Firstly recall that through the type I seesaw mechanism, the heavy Majorana neutrinos,~$N$ (where we have dropped the subscript for brevity) may interact with the SM gauge bosons because of their mixing between light neutrinos (see Eqs.~(\ref{eq1:L_nu_matrix}) to (\ref{eq1:L_nu_Maj_field})). We parametrize this mixing through the quantity $V_{\ell N}$ ($\ell=e,\mu,\tau$), and the interaction Lagrangians look like
\begin{align}
\mathcal{L}_W &= -\frac{g}{\sqrt{2}}\, V_{\ell N}\, \overline{\ell}\, \slashed{W}\, 
             P_L\, N  + \text{h.c.}\;,\nonumber\\
\mathcal{L}_Z &= -\frac{g}{2 \cos\theta_W}\, V_{\ell N}\, \overline{\nu}\,\slashed{Z}\,
             P_L\, N+ \text{h.c.}\;,
\end{align}
where $P_{L,R} = (1 \mp \gamma_5)/2$. With these interactions, it is possible to produce signals for heavy neutral leptons through
$q \overline{q}' \rightarrow W^{*} \rightarrow \ell N$ followed by $N \rightarrow \ell\, W$ or $\nu Z$. Note that the production of $N$ by
$q \overline{q} \rightarrow Z^* \rightarrow \nu N$ is much harder to study due to large backgrounds.
However, since the dominant component of the heavy neutral leptons, $N$ is the RH electroweak singlets $\nu_R$, the amplitude of the mixing quantity, $V_{\ell N}$ is in general very tiny and is of order $m_\ell\,M_N^{-1}$, where $m_\ell$ and $M_N$ denote the masses of charged lepton $\ell$ and heavy neutrino $N$ respectively. So typically, one has about $V_{\ell N} \simeq \order{10^{-7}}$.

Moreover in \cite{delAguila:2008cj}, a model-independent study assuming only $V_{\ell N} \simeq \order{10^{-2}}$ was done for such mechanism, and it was found that in order
to lead to a detectable heavy neutral lepton signal, the mass of $N$ must be of 
order 100 GeV or less, for the initial stage of LHC running with
luminosity of order 10 $\text{fb}^{-1}$. Hence, even if our models can satisfy the condition for $V_{\ell N}$, the $N_1$ masses are too large for this purpose.
Besides, our numerical results have indicated that the amplitudes of $V_{\ell N}$ are actually too tiny even in the best case scenarios. For instance, assuming $|m_1| \simeq 0.2$~eV which will saturate the bound of (\ref{eq3:HLV_nu_mass_limit}), and special choice of phases, one obtains $|V_{eN}| \simeq 2.3\times 10^{-7}$ (with $M_1 \simeq 1.2\times 10^{3}$~GeV in the inverted hierarchy scheme). This is much less than the minimum \order{10^{-2}} required to produce an observable signal in any of the channels \cite{Aguila0502189}. The suppression is even greater for the $\mu$ or $\tau$ flavor because of the hierarchical structure of $M_R$. As a result,
it is very difficult to detect the heavy neutral leptons through this mechanism even with an integrated luminosity up to 300 $\text{fb}^{-1}$.

It is worth mentioning that if there is only one Higgs doublet, there is also a light neutrino and heavy neutral lepton interaction with the Higgs particle given by
\begin{equation}
\mathcal{L}_H = -\frac{g M_N}{2 M_w}\,\left(V_{\ell N}\, \overline{\nu}\, H\, P_R\, N 
 + \text{h.c.}\right)\;.
\end{equation}
This interaction, although not of much help in the production of heavy neutral leptons through $q\bar q \rightarrow H^* \rightarrow \nu N$, 
does provide another channel for $N$ decay. If the Higgs mass is not too much larger than the $W$ boson mass, the decay rate is similar to that for
$N \rightarrow \ell \,W$  or $\nu Z$.

In the models we are considering, there are several Higgs doublets. The neutral Higgs couplings to light neutrinos and heavy neutral leptons 
are then not necessarily proportional to $M_NV_{\ell N}$ and can increase the decay rate. Besides, in our models there are
charged Higgs bosons interacting with light neutrinos and heavy neutral leptons which  provide additional channels for detection of the $N$'s. 
Unfortunately, given the smallness of the mixing quantity $V_{\ell N}$ mentioned above,
it is still very difficult to detect a heavy neutral lepton with mass of order 1 TeV at the LHC even with 300 $\text{fb}^{-1}$ of luminosity.

Finally, charged Higgs couplings to charged leptons and heavy neutral leptons may have interesting signals at the ILC through 
$e^+e^-\rightarrow H^+ H^-$ with $t$-channel heavy Higgs exchange, and $e^\pm e^\pm \rightarrow H^\pm H^\pm$ with $u$-channel $N$ exchange \cite{Atwood:2007zza}. In particular, the processes: $e^\pm e^\pm \rightarrow H^\pm H^\pm$ are very sensitive to the heavy neutral lepton mass. It has been shown in \cite{Atwood:2007zza} that 
if $|V_{\ell N}|$ is in the range of $10^{-2}$ to $10^{-4}$, 
the ILC with an energy of 500~GeV can probe heavy neutral lepton masses up to $10^4$ TeV. In our case, the charged Higgs coupling to charged leptons and heavy neutral leptons can be larger than $V_{\ell N}\sim m_\ell M_N^{-1}$, but still too small to be probed using the processes mentioned above.\\

\section{Summary and outlook}
\label{sec:HLV_conc}

If the type I seesaw mechanism is the correct theory in explaining how ordinary neutrinos gain their tiny but nonzero mass, then the extended SM necessarily contains a high-energy RH neutrino sector. But owing to the largely unconstrained nature of the heavy seesaw parameter space, such a framework per se lacks the genuine predictive powers one would need for testing it in low-energy experiments. Indeed, the seesaw relation implies that the scale of $M_R$ must be much higher than the electroweak scale, and hence direct measurements of this sector are beyond the reach of current and next-generation colliders. 

In the light of this missing link, we explored the possibility of constraining the heavy seesaw sector through symmetries such that the RH neutrino mass matrix is given in terms of only low-energy fermion masses, mixing angles and $CP$ phases. The general strategy employed in this work involves, firstly, expanding the symmetry of the gauge group so that the Yukawa couplings of the neutrinos are forced to be the same as some of the charged leptons or quarks, and secondly, introducing an underlying flavor symmetry relating the different generations in such a way that the relevant diagonalisation matrices are completely known. We have shown by explicit examples that such symmetries exist and the simplest models of this type yield a relation: $M_R \simeq \widehat{m}_{f}\, \Upmns^* \,\mnuhat^{-1}\, \Upmns^{\dagger}\, \widehat{m}_{f}$ where $f=e,d,u$. 

Since the connection between the high- and low-energy sectors is complete in these models, it is essential to examine their improved predictability and testability. Most notably, the strong dependence of leptogenesis on the RH neutrino sector provides an important avenue to rule in or out models of this type based on studies from cosmology. In particular, for our representative models, we have found that successful leptogenesis is only possible in the $f=u$ case and within certain fine-tuned regions of the parameter space. As a result, future precision measurements on the low-energy $CP$ phases can be a direct way to distinguishing these models with others in the context of the leptogenesis.

Furthermore, we have discovered that the $f=e$ case can supply a heavy neutral lepton with a mass being as low as a few TeV, opening the prospect for directly probing them in colliders. Although the detailed investigation has concluded that the signals coming from the heavy neutral leptons are far too weak to be detectable at either the LHC or a future ILC, it is nonetheless an illustrative example of the new possibilities such models can provide.

Potential extension to the work presented here may include exploring the role of the CKM matrix in place of $\Upmns$ in the current models, or the effect of having both mixing matrices in the formula for $M_R$. Secondly, given that many of the diagonalization matrices in our explicit models are forced to be the identity by the selected flavor symmetry, it may be fruitful to examine other symmetry groups that can give rise to a richer structure. Finally, another obvious line of investigation could be to generalize the relationship between the neutrino Dirac mass matrix and $\widehat{m}_f$ away from being a strict equality by the use of Clebsch-Gordan coefficients and a more complicated Higgs sector.


%
%
%
%
%



\chapter{Electromagnetic leptogenesis}\label{ch_work_BKL}

\ECap{R}{e~c~a~l~l}{7}{0}
from our discussion in Sec.~\ref{subsec:numass_emdm} that the existence of neutrino electromagnetic dipole moments (EMDM) is a major consequence of neutrinos having a nonzero mass. Although in cases where neutrinos are Majorana particles (e.g. in the type~I seesaw mechanism) only their \emph{transition} dipole moments are nonzero due to the Hermiticity of the Lagrangian \cite{SV_nu_dm, Wolf_nu_dm}, the role played by these 
 interactions are important in their own right. For instance, the active-active neutrino transitions through such moments can have interesting implications in astrophysics when  they are resonantly enhanced in matter \cite{nu_emdm_matter_Akhmedov, GG_book}. Also, it is known that neutrino EMDM can provide crucial hints on the nature of neutrinos (Majorana or otherwise) \cite{Bell_mdm_PRL,
Bell_mdm_PLB}. Therefore, further studies into their effects are well-motivated.

Our interest here is the active-sterile interactions between the LH light and RH heavy neutrinos via the effective transition dipole operator and their potential implications in cosmology.  Specifically, we would like to investigate whether the lepton number violating radiative decay of the heavy sterile neutrinos ($N \rightarrow \nu\,\gamma$) can give rise to a baryon asymmetry, in analogy to the  standard leptogenesis scenario where $N$-decays are mediated by the Yukawa couplings ($N \rightarrow \ell\,\bar\phi$). This issue is worth exploring not only because the presence of these transition moments are very likely, but also from the neutrino model building point of view, such new processes may lead to significant change in the region of applicability for successful leptogenesis.

In the following, we outline the relevant properties of the EMDM couplings and present explicit calculations of the $CP$ asymmetry induced by the decays of $N$ through such effective operators, with the aim to demonstrate the viability of lepton number creation in this newly added sector. To this end, we also revisit the topic on the necessary requirements for a decay process to manifestly violate $CP$ and discuss the simplified Majorana Feynman rules employed in the calculations. We shall conclude by comparing this scenario of \emph{electromagnetic leptogenesis} with the standard Yukawa-mediated case, as well as comment on the connection between EMDM operators and neutrino mass.\\

\section{EMDM coupling between light and heavy neutrinos} \label{sec:BKL_EDMD_properties}

In order to deduce the potential implications of the EMDM operators in leptogenesis, it is imperative to understand the properties of the transition form factors $\mu_{jk}$ and $d_{jk}$ in the generic dipole moment coupling between light ($\nu$) and heavy ($N$) neutrinos:~\footnote{See also Eq.~(\ref{eq1:emdm_op_generic}) on page~\pageref{eq1:emdm_op_generic}.}
\begin{equation}\label{eq4:BKL_emdm_op_massbasis}
 \mathcal{L}_\text{EM} =
  \overline{\nu}_j\, (\mu_{jk} +i\gamma^5 d_{jk})\, \sigma_{\alpha\beta}\, N_k \,F^{\alpha\beta}
  +\text{h.c.}
  \;,
\end{equation}
where $\nu_j = e^{i\vartheta_j}\, \nu_j^c$ and $N_k = e^{i\varphi_k}\, N_k^c$ ($j,k$ are the mass labels) are Majorana neutrino fields~\footnote{Since we shall work within the type I seesaw framework (with three RH neutrinos) throughout, neutrinos will always assumed to be Majorana with $e^{i\vartheta_j}$ and $e^{i\varphi_k}$ denoting the charge conjugation phase factors.} while $F^{\alpha\beta}$ denotes the photon field tensor as usual. Also, we use the definition: $\sigma_{\alpha\beta} = i\,[\gamma^\alpha, \gamma^\beta]/2$.
Rewriting $\nu_j$ and $N_k$ using the Majorana condition, we obtain
\begin{align}
 \mathcal{L}_\text{EM} &\equiv 
  \overline{(e^{i\vartheta_j}\nu_j^c)}\, (\mu_{jk} +i\gamma^5 d_{jk})\, \sigma_{\alpha\beta}\, e^{i\varphi_k}\, N_k^c
   \,F^{\alpha\beta}
  +\text{h.c.}\;,\nonumber\\
 &= 
  -e^{-i(\vartheta_j-\varphi_k)}
  \nu_j^T C^{-1}\, (\mu_{jk} +i\gamma^5 d_{jk})\,
   \sigma_{\alpha\beta}\, C \overline{N}_k^T
   \,F^{\alpha\beta}
  +\text{h.c.}\;,
  \label{eq4:BKL_emdm_op_mass_rewrite}
\end{align}
where $C$ is the charge conjugation operator with the following conventions:
\begin{gather}
 \psi^c = C\overline{\psi}^T \;,\;\;
 C^\dagger = C^{-1} \;,\;\;
 C^T = -C \;,\;\;
 C^\dagger C^T = C^* C = -I \;,\;\;
 C^{-1} \gamma^5 C = (\gamma^5)^T \;,\nonumber\\
 C^{-1} \gamma^\mu C = (-\gamma^\mu)^T \;,\;\;
 C^{-1} \sigma^{\mu\nu} C = (-\sigma^{\mu\nu})^T \;,\;\;
 C^{-1} P_{R,L} C = (P_{R,L})^T \;,
 \label{eq4:BKL_C_conventions}
\end{gather}
where $P_{R,L} \equiv (1 \pm\gamma^5)/2$. Transposing the first term in (\ref{eq4:BKL_emdm_op_mass_rewrite}) and using (\ref{eq4:BKL_C_conventions}) to simplify the expression, one eventually gets after some algebra
\begin{equation}\label{eq4:BKL_emdm_op_mass_1st-term}
\mathcal{L}_\text{EM} =
-e^{-i(\vartheta_j-\varphi_k)}
  \overline{N}_k \, (\mu_{jk} +i\gamma^5 d_{jk})\,
   \sigma_{\alpha\beta}\, \nu_j
   \,F^{\alpha\beta}
   +\text{h.c.}
   \;.
\end{equation}
If we write out the h.c.~term (which is $\overline{N}_k \, (\mu_{jk}^* +i\gamma^5 d_{jk}^*)\,\sigma_{\alpha\beta}\, \nu_j\,F^{\alpha\beta}$) and compare it with the first term in (\ref{eq4:BKL_emdm_op_mass_1st-term}), we can conclude that
\begin{equation}\label{eq4:BKL_emdm_op_mu_d-rel}
 \mu_{jk}= -e^{i(\vartheta_j-\varphi_k)} \mu_{jk}^* 
 \quad
 \text{ and }
 \quad
 d_{jk}= -e^{i(\vartheta_j-\varphi_k)} d_{jk}^* \;.
\end{equation}
From this, we get
\begin{align}
 \mu_{jk}^2 &= |\mu_{jk}|^2 \,e^{i(\vartheta_j-\varphi_k+\pi)} \;,\\
\Rightarrow\quad
 \mu_{jk} &= |\mu_{jk}|\, i\, e^{i(\vartheta_j-\varphi_k)/2}\;.\label{eq4:BKL_mu_cond}
\end{align}
Similarly, we have the analogous expression for $d_{jk}$. An important note on this is that although the relations between $\mu_{jk}$ and $\mu_{jk}^*$, as well as $d_{jk}$ and $d_{jk}^*$ depends on the choice of the charge conjugation phase factor, once $\vartheta_j$ and $\varphi_k$ are chosen, they are fixed. In particular, when $\vartheta_j = \varphi_k$, we have the situation where $\mu_{jk}$ and $d_{jk}$ must be purely imaginary. 
Furthermore, it is worth mentioning that if Lagrangian (\ref{eq4:BKL_emdm_op_massbasis}) is $CP$ invariant, then only one of $\mu_{jk}$ and $d_{jk}$ survives \cite{SV_nu_dm, Wolf_nu_dm}. But in our work here we do not impose such condition and the only assumptions we shall make are Hermiticity and $CPT$ invariance.

In calculations, it is often much simpler to consider the EMDM coupling between the associated chiral components of the $\nu$ and $N$ (instead of using the form written in (\ref{eq4:BKL_emdm_op_massbasis})) because the resultant Lagrangian contains only one type of electromagnetic dipole moment coupling rather than distinct magnetic ($\mu_{jk}$) and electric ($\gamma^5 d_{jk}$) moment terms as $\gamma^5 P_{R,L} =\pm P_{R,L}$. 
Letting $\nu_j = \nu_{Lj} + e^{i\vartheta_j}\nu_{Lj}^c$ and $N_k = N_{Rk} + e^{i\varphi_k}N_{Rk}^c$ where $\nu_{L}$ and $N_R$ are the usual LH and RH neutrino states, then (\ref{eq4:BKL_emdm_op_massbasis}) can be rewritten into
\begin{align}
 \mathcal{L}_\text{EM} &=
  \overline{\nu}_{Lj}\, (\mu_{jk} +i d_{jk})\, \sigma_{\alpha\beta}\, 
     N_{Rk}\,F^{\alpha\beta}
  + e^{-i(\vartheta_j-\varphi_k)}
    \overline{(\nu_{Lj})^c}\, (\mu_{jk} -i d_{jk})\, \sigma_{\alpha\beta}\, 
     N_{Rk}^c \,F^{\alpha\beta} \nonumber\\
  &\;\;\;\;
  + e^{i(\vartheta_j-\varphi_k)}
   \overline{(N_{Rk})^c}\, (\mu_{jk}^* +i d_{jk}^*)\, \sigma_{\alpha\beta}\, 
     \nu_{Lj}^c \,F^{\alpha\beta}
  +\overline{N}_{Rk}\, (\mu_{jk}^* -i d_{jk}^*)\, \sigma_{\alpha\beta}\, 
     \nu_{Lj} \,F^{\alpha\beta}   
   \;,\nonumber\\
 &=
  \overline{\nu}_{Lj}\, (\mu_{jk} +i d_{jk})\, \sigma_{\alpha\beta}\, 
     N_{Rk}\,F^{\alpha\beta}
  - e^{-i(\vartheta_j-\varphi_k)}
    \overline{N}_{Rk}\, (\mu_{jk} -i d_{jk})\, \sigma_{\alpha\beta}\, 
     \nu_{Lj} \,F^{\alpha\beta} \nonumber\\
  &\;\;\;\;
  - e^{i(\vartheta_j-\varphi_k)}
   \overline{\nu}_{Lj}\, (\mu_{jk}^* +i d_{jk}^*)\, \sigma_{\alpha\beta}\, 
     N_{Rk} \,F^{\alpha\beta}
  +\overline{N}_{Rk}\, (\mu_{jk}^* -i d_{jk}^*)\, \sigma_{\alpha\beta}\, 
     \nu_{Lj} \,F^{\alpha\beta}   
   \;,
\end{align}
where in the last step we have followed the same procedure as that leading to (\ref{eq4:BKL_emdm_op_mass_1st-term}). Using (\ref{eq4:BKL_mu_cond}) and the analogous form for $d_{jk}$, the Lagrangian simplifies to the form (after absorbing the common factor of 2 into the definitions of $\mu$ and $d$):
\begin{align}
 \mathcal{L}'_\text{EM} &=
  \overline{\nu}_{Lj}\, (\mu_{jk} +i d_{jk})\, \sigma_{\alpha\beta}\, 
     N_{Rk} \,F^{\alpha\beta}
     +\text{h.c.}\;,\label{eq4:BKL_emdm_chiral_formA} \\
     &=
  \overline{\nu}_{Lj}\, \lambda_{jk}\, \sigma_{\alpha\beta}\,P_R\, 
    N_{k} \,F^{\alpha\beta}
         +\text{h.c.}\;,
     \label{eq4:BKL_emdm_chiral_form}
\end{align}
where we have defined the EMDM coupling, $\lambda_{jk} \equiv \mu_{jk} +i d_{jk}$. It is crucial to realise that although both $\mu_{jk}$ and $d_{jk}$ are individually restricted by relations like (\ref{eq4:BKL_mu_cond}), the quantity $\lambda_{jk}$ appearing in the effective operator is in general complex. As a result, we can assume that the EMDM coupling matrix $\lambda$ is completely arbitrary in all of our subsequent analyses.\\

\section{Feynman rules and other useful tools}\label{sec:BKL_Feynman_rules}

The purpose of this section is to highlight all the relevant results and graphical rules used in the subsequent sections. In particular, the simplified set of Feynman rules for Majorana fermions adopted in our calculations will be discussed. Although we shall not attempt to prove these rules from first principles, the Lagrangian from which they are derived and their proper usage will be explained. In addition, explicit examples are presented (in this section and in Appendix~\ref{app:c_chapter}) to demonstrate the ability of these rules to reproduce known results, as well as to elucidate all the intricate steps involved.

\subsection{Simplified rules for Majorana fermions}\label{subsec:BKL_simp_Maj-rules}

Some of the major difficulties in a typical calculation of Feynman graphs involving Majorana fermions originate from the fact that there are several different vertices and propagators one may need to consider. This multiplicity and other ambiguities such as spinor assignments for external lines and factor of $1/2$ in loops are direct consequences of the self-conjugacy of Majorana particles. Although these subtleties have been properly addressed in the literature \cite{
Jones:1983eh,
Haber:1984rc,
Gates:1987ay,
Denner_FeynRules,
Gluza:1991wj
}
(see also \cite{Fuk_Yan_book, massive_nu_book}), the resulting graphical rules that attempt to resolve the issues are often very complicated themselves. Besides, in some extreme cases, one is probably better off to forego these rules and revert to Wick's theorem for an unequivocal treatment. Therefore, it is sensible to develop a simplified approach in handling the Majorana fermions which captures many of the essential features without the over-complications that plague the conventional methods. This is especially appropriate for performing calculations of our type, where absolute precision is not a paramount requirement. Indeed, most parameters in leptogenesis carry a high degree of uncertainty, hence a multiplicative constant or an overall sign error in the expression for the $CP$ asymmetry is manifestly unimportant.

With this in mind, we shall follow the approach outlined in \cite{luty} and write down the corresponding rules for Majorana fermions based on a four-component version (rather than the usual two) of the Weyl spinor field, $\Psi \equiv \Psi_R + e^{i\phi} \Psi_R^c$ (i.e. the Majorana field). The advantage of this method is that it yields only \emph{one} type of propagator for the Majorana fermion, and similarly a single vertex factor is sufficient for each physical process. While it has been claimed in \cite{Gates:1987ay} that not all ambiguities may be resolved in this way due to the lack of a conserved quantum number carried by the Majorana fermions, and that extra rules are required to fix the fermion-flow problems, we have found that (by experimenting with many different graphs related to our work) such subtleties are inconsequential for our calculations, and the simplified approach of \cite{luty} is self-consistent. Furthermore, the (partially) simplified rules presented in \cite{Gluza:1991wj, Gates:1987ay, Denner_FeynRules}, while universal, do not specifically discuss the best way to handle loop graphs, thus we have made a conscious decision to adopt the approach of \cite{luty} which naturally eliminates the confusion coming from loops with both Dirac and Majorana fermions---~a situation which often appears in leptogenesis diagrams.

In the following, we first write down the list of key rules and observations resulting from this simplified formulation, before presenting two examples from standard leptogenesis to demonstrate their usage.
\subsubsection{Majorana fermion propagator}\label{subsubsec:4_propagator}
Since the Majorana fermion of interest in leptogenesis is the RH neutrino $\nu_R$, it makes sense to base our discussion of the Majorana propagator on it. To begin with, we write down the theory in terms of the two-component RH neutrino field, $\nu_R = (\nu_{R1}, \nu_{R2},\nu_{R3})^T$, where the subscripts are indices in flavor space:
\begin{equation}\label{eq4:prop_nu_R_Lag}
 \mathcal{L}_{\nu_R} 
  = 
    i \,\overline{\nu}_{R}\, \dslash\, \nu_{R}
    -\frac{1}{2}\, \overline{(\nu_{R})^c}\, M_{R} \,\nu_{R}  
    -\frac{1}{2}\, \overline{\nu}_{R} \,M_{R}^* \,\nu_{R}^c
   \;.
\end{equation}
To diagonalise $M_R$, we let $\nu_R =\eta^* V^\dagger N_R$, where $\eta= \text{diag}(e^{i\varphi_1/2},e^{i\varphi_2/2},e^{i\varphi_3/2})$ and $V$ is a unitary matrix. Note that one can always select $V$ in such a way that the eigenvalues for $M_R$ are all real and positive. We have pulled out the phase $\varphi_k$, and will identify it as the charge conjugation phase factor later. So, $ \mathcal{L}_{\nu_R}$ becomes
\begin{equation}\label{eq4:prop_N_R_Lag}
 \mathcal{L}_{N_R} 
  = 
    i\, \overline{N}_R\, \dslash\, N_R
    -\frac{1}{2}\, \overline{(N_R)^c}\, D_M\, (\eta^*)^2\, N_R  
    -\frac{1}{2}\, \overline{N}_R  \, D_M\, \eta^2 \,N_R^c
   \;,
\end{equation}
where $D_M = \text{diag}(M_1,M_2,M_3)$ is the diagonal mass matrix for the RH neutrinos. At this point, it is convenient to switch to index form and rewrite $\mathcal{L}_{N_R}$ as follows:
\begin{equation}
\mathcal{L}_{N_R} =
  \frac{1}{2}\left[
   i\, \overline{N}_{Rk}\, \dslash\, N_{Rk}  
  + i\, \overline{(N_{Rk})^c}\, \dslash\, N_{Rk}^c
  - M_k\,e^{-i\varphi_k} \overline{(N_{Rk})^c} N_{Rk}
  - M_k\,e^{i\varphi_k}\overline{N}_{Rk} N_{Rk}^c
  \right] \;,\nonumber
\end{equation}
\begin{align}
  &=
  \frac{1}{2}\left[
  i\, (\overline{N}_{Rk}+ e^{-i\varphi_k}\overline{(N_{Rk})^c})\, \dslash\, N_{Rk}
  +i\,(\overline{N}_{Rk} + e^{-i\varphi_k} \overline{(N_{Rk})^c})\, \dslash\, 
      e^{i\varphi_k} N_{Rk}^c \right. \nonumber\\
  &\left.\qquad
   - M_k\,(\overline{N}_{Rk}+ e^{-i\varphi_k} \overline{(N_{Rk})^c}) N_{Rk}   
   - M_k\,(\overline{N}_{Rk}+e^{-i\varphi_k}\overline{(N_{Rk})^c}) e^{i\varphi_k} N_{Rk}^c
  \right] \;,\nonumber\\
  &= \frac{1}{2}\,
  \left[
  i \,\overline{N}_k \, \dslash\,N_k
  -M_k\,\overline{N}_k\, N_k
  \right]\;,\label{eq4:prop_N_k_form}
\end{align}
where we have introduced the four-component Majorana field, 
$N_k = N_{Rk}+ e^{i\varphi_k}N_{Rk}^c$ which satisfies $N_k \equiv e^{i\varphi_k} N_{k}^c$. Using the charge conjugation conventions of (\ref{eq4:BKL_C_conventions}), we note that $\overline{N}_k = \overline{e^{i\varphi_k} N_k^c} = -e^{-i\varphi_k} N_k^T C^\dagger$. Therefore, one may rewrite (\ref{eq4:prop_N_k_form}) as
\begin{equation}\label{eq4:prop_N_k_readoff}
\mathcal{L}_{N_R} = -\frac{1}{2}\,
  e^{-i\varphi_k} N_k^T C^\dagger
   \left[
  i \, \dslash
  -M_k 
  \right]N_k\;.
\end{equation}
From this, the Majorana propagator for $N_k$ can be readily read off as~\footnote{While we have been carefully carrying around the phase $\varphi_k$ up to this point, it is not actually needed as it can always be eliminated via appropriate redefinition of fields and couplings constants. In addition, it will be apparent later (after we have discussed the vertex factors) that they do not enter into the final expression in any calculation because $e^{i\varphi_k}$ and its complex conjugate always appear together. Hence, for convenience, we shall drop this phase factor from our Feynman rules.}
\begin{equation}\label{eq4:prop_Maj_prop}
 \SetScale{0.6}
    \begin{picture}(75,0)(40,-73) 
    \SetWidth{0.8}
    \SetColor{Black}
    \Line(88,-118)(170,-118)
    \SetWidth{0.5}
    \Vertex(170,-118){2.83}
    \Vertex(88,-118){2.83}
    \SetWidth{0.8}
    \LongArrow(109,-104)(149,-104)
    \Text(80,-55)[]{\Black{$p$}}  
    \Text(48,-80)[]{\Black{$B$}}   
    \Text(105,-80)[]{\Black{$A$}}  
  \end{picture}
:\qquad\quad
 \left[S_{N_k}(p)\right]_{AB} = 
 \left[\frac{-i\, (\pslash + M_k)\, C}{p^2 - M_k^2 +i\epsilon}\right]_{AB}
  \;,\rule{0pt}{30pt}
\end{equation}
where $A,B$ are spinor indices and $p$ is the four-momentum. Note that this is the one and only  Majorana fermion propagator arising in this approach. 
Because the Majorana particle does not carry any conserved quantum number, we denote it by a solid line with no arrows.
Consequently, it is often the case that the direction of fermion-flow through the propagator in a diagram is ambiguous (an issue we have alluded to earlier). To resolve this, we set the flow direction to be the same as the \emph{arbitrarily} defined internal momentum $p$ for a given diagram. In other words, if $p$ flows from $B$ to $A$ as in (\ref{eq4:prop_Maj_prop}), we write down the matrix element in the order shown. Otherwise, for $p$ flowing from $A$ to $B$, we use $\left[\cdots\right]_{BA}$ instead. This scheme is consistent as long as when we draw the ``conjugate'' version of a given graph (i.e. the graph with all Dirac fermion arrows reversed), we reverse the direction of the internal momentum $p$ for the Majorana propagator while keeping all other momenta the same as before~\footnote{See the example of ``vertex contribution to the $CP$ asymmetry'' in Sec.~\ref{subsec:BKL_maj_feyn_examples} on page~\pageref{subsubsec:BKL_vertex_eg} for further discussion of the fermion-flow issue.}. This way, the relative sign between the two graphs will be correct with their amplitudes being complex conjugates of each other~\footnote{It is highly probable that an ambiguity in the overall sign may remain, but fortunately it is not a crucial issue for us.}.

\subsubsection{Vertex factors involving a Majorana fermion}\label{subsubsec:4_vertex_fac}

There are basically two types of interactions which are relevant to our discussion of leptogenesis. Firstly, we have the Yukawa coupling between $\ell_L$ and $N_R$, and secondly, we have the electromagnetic dipole interaction of (\ref{eq4:BKL_emdm_chiral_form}). To be consistent with the notation used in the last section, let us again begin by writing down the interaction Lagrangian in terms of the chiral field $\nu_R$
\begin{equation}\label{eq4:vertex_lag_nu_R}
 \mathcal{L}_\text{int}
  = 
  -\overline{\ell}_L\,Y\,\nu_R\,\phi
  -\overline{\nu}_L\,\tilde{\lambda}\,\sigma_{\alpha\beta}\,\nu_R\,F^{\alpha\beta}
  +\text{h.c.}\;,
\end{equation}
where $\ell_L= (\nu_L, e_L)^T$ and $\phi= (\phi^0, \phi^-)^T$ are doublets of $SU(2)_L$. Using $\nu_R =\eta^* V^\dagger N_R$ to write (\ref{eq4:vertex_lag_nu_R}) in the mass eigenbasis for the RH neutrinos, where all symbols are as defined in the previous section, the Lagrangian becomes
\begin{equation}\label{eq4:vertex_lag_N_R}
  \mathcal{L}_\text{int}
  =
  -\eta^*\,\overline{\ell}_L\,h\,N_R\,\phi
  -\eta^*\,
  \overline{\nu}_L\,\lambda\,\sigma_{\alpha\beta}\,N_R\,F^{\alpha\beta}
  +\text{h.c.}\;,
\end{equation}
where we have set $h = Y V^\dagger$ and $\lambda = \tilde{\lambda} V^\dagger$. Writing this in index form and introducing the four-component Majorana field, $N_k = N_{Rk}+ e^{i\varphi_k}N_{Rk}^c$, we then get
\begin{align}
 \mathcal{L}_\text{int}
  &=
  -e^{-i\varphi_k}\, h_{jk}\, \overline{\ell}_{Lj} \,P_R\, N_{k}\, \phi
  -e^{i\varphi_k}\, h_{jk}^*\, \overline{N}_{k}\, P_L\, \ell_{Lj}\, \phi^\dagger
  \nonumber\\
 &\qquad 
  -e^{-i\varphi_k}\, \lambda_{jk}\,\overline{\nu}_{Lj} \,\sigma_{\alpha\beta}
      \,P_R \,N_{k}\, F^{\alpha\beta}
   -e^{i\varphi_k}\, \lambda_{jk}^*\,\overline{N}_{k}\,\sigma_{\alpha\beta}
      \,P_L \,\nu_{Lj}\, F^{\alpha\beta}\;, \label{eq4:vertex_2ndfinal_form}    \\
  &=
  e^{-i\varphi_k}\left[
  - h_{jk}\, \overline{\ell}_{Lj} \,P_R\, N_{k}\, \phi
  + h_{jk}^*\, N_k^T\,C^\dagger\, P_L\, \ell_{Lj}\, \phi^\dagger
  \right.
  \nonumber\\
 &\qquad\qquad\;\; \left.
   -2\lambda_{jk}\,\overline{\nu}_{Lj} \,\sigma_{\alpha\beta}
      \,P_R \,N_{k}\, \partial^{\alpha}A^{\beta}
  +2\lambda_{jk}^*\,N_{k}^T\,C^\dagger\,\sigma_{\alpha\beta}
      \,P_L \,\nu_{Lj}\,\partial^{\alpha}A^{\beta}
      \right]\;,  \label{eq4:vertex_final_form}   
\end{align}
where in the last step we have used the fact that $F^{\alpha\beta} = \partial^{\alpha}A^{\beta}-\partial^{\beta}A^{\alpha}$ with $A$ being the photon field, and $\sigma_{\alpha\beta} = - \sigma_{\beta\alpha}$, to simplify the expression. It is important to note that the transition EMDM term displayed in (\ref{eq4:vertex_2ndfinal_form}) has the same form as Eq.~(\ref{eq4:BKL_emdm_chiral_form}), hence  everything that we have discussed in Sec.~\ref{sec:BKL_EDMD_properties} regarding the coupling $\lambda_{jk}$ remains valid.

Returning to (\ref{eq4:vertex_final_form}), the vertex factors for the four processes are given by :
\begin{align}
 N_k \rightarrow \ell_{Lj}\, \bar\phi\;:\;\; 
\qquad
 \SetScale{0.6}
   \begin{picture}(75,0)(30,-88)  
    \SetWidth{0.8}
    \SetColor{Black}
    \Line(90,-141)(125,-141)
    \ArrowLine(125,-141)(155,-115)
    \DashLine(125,-141)(155,-167){2.6}
    \SetWidth{0.5}
    \Vertex(125,-141){3}
    \Text(45,-85)[]{\small{\Black{$N$}}}
    \Text(100,-70)[]{\small{\Black{$\ell$}}}
    \Text(100,-98)[]{\small{\Black{$\bar\phi$}}}
  \end{picture}
\qquad &= -i\,h_{jk}\,P_R 
 \label{eq4:vertex_rule_std1}\\
 N_k \rightarrow \bar{\ell}_{Lj}\, \phi\;:\;\; 
\qquad
 \SetScale{0.6}
   \begin{picture}(75,0)(30,-88)  
    \SetWidth{0.8}
    \SetColor{Black}
    \Line(90,-141)(125,-141)
    \ArrowLine(155,-115)(125,-141)
    \DashLine(125,-141)(155,-167){2.6}
    \SetWidth{0.5}
    \Vertex(125,-141){3}
    \Text(45,-85)[]{\small{\Black{$N$}}}
    \Text(100,-70)[]{\small{\Black{$\ell$}}}
    \Text(100,-98)[]{\small{\Black{$\phi$}}}
  \end{picture}
\qquad 
 &= i\,h_{jk}^*\,C^\dagger P_L \rule{0pt}{35pt}
  \label{eq4:vertex_rule_std2}
    \\
 N_k \rightarrow \nu_{Lj}\, A^{\rho}\;:\;\; 
\qquad 
 \SetScale{0.6}
   \begin{picture}(75,0)(30,-80)  
    \SetWidth{0.8}
    \SetColor{Black}
    \Line(90,-128)(125,-128)
    \ArrowLine(125,-128)(155,-102)
    \SetWidth{0.5}
    \SetWidth{0.8}
    \Photon(125,-128)(155,-154){3}{3}
    \LongArrow(121,-139)(141,-157)
    \Vertex(125,-128){3}
    \Text(45,-77)[]{\small{\Black{$N$}}}
    \Text(100,-62)[]{\small{\Black{$\nu$}}}
    \Text(100,-90)[]{\small{\Black{$\gamma$}}}  
    \Text(71,-94)[]{\small{\Black{$q$}}}
    \end{picture}
\qquad  
  &= 2 \lambda_{jk}\,P_R\,\sigma^{\alpha\rho} q_\alpha \rule{0pt}{35pt}
  \label{eq4:vertex_rule_emdm1}
%
\end{align}
\begin{equation}\label{eq4:vertex_rule_emdm2}
 N_k \rightarrow \bar{\nu}_{Lj}\, A^{\rho}\;:\;\; 
\qquad 
 \SetScale{0.6}
   \begin{picture}(75,0)(30,-80)  
    \SetWidth{0.8}
    \SetColor{Black}
    \Line(90,-128)(125,-128)
    \ArrowLine(155,-102)(125,-128)
    \SetWidth{0.5}
    \SetWidth{0.8}
    \Photon(125,-128)(155,-154){3}{3}
    \LongArrow(121,-139)(141,-157)
    \Vertex(125,-128){3}
    \Text(45,-77)[]{\small{\Black{$N$}}}
    \Text(100,-62)[]{\small{\Black{$\nu$}}}
    \Text(100,-90)[]{\small{\Black{$\gamma$}}}  
    \Text(71,-94)[]{\small{\Black{$q$}}}
    \end{picture}
\qquad  
  = -2 \lambda_{jk}^*\,C^\dagger \sigma^{\alpha\rho} q_\alpha P_L \rule{0pt}{25pt}
\end{equation}
where\rule{0pt}{35pt} 
we have again dropped the phase factor for convenience~\footnote{The $i$ is missing from (\ref{eq4:vertex_rule_emdm1}) and (\ref{eq4:vertex_rule_emdm2}) because when differentiating $A^\rho$ one gets a factor of $i q_\rho$, and this extra $i$ will cancel with the $i$ originating from the Dyson series.}. As stated before, this method gives rise to only one vertex factor per process. In addition, because of the presence of the helicity projection operators, $P_{R,L}$, in the definition of these vertices, the number of spin states propagating in any closed Majorana fermion loop will automatically be correct. Therefore, unlike other approaches, this eliminates the need to include a factor of 1/2 for loops, and it is especially advantageous in situations where both Dirac and Majorana fermions appear together.

\subsubsection{External lines for Majorana fermion}\label{subsubsec:4_spinors}

Because of the self-conjugacy of Majorana fermions, there are several possible choices in assigning spinor wave functions to the external lines. To avoid confusion, we select one convention that is consistent and use it for all diagrams. Specifically, our assignment is as follows
\begin{align}
\text{incoming } N\;:\;\; 
\quad
 \SetScale{0.6}
  \begin{picture}(90,0)(30,-92) 
    \SetWidth{0.8}
    \SetColor{Black}
    \Line(90,-147)(140,-147)
    \LongArrow(99,-137)(130,-137)
    \GOval(166,-147)(14,25)(0){0.882}
    \Text(46,-88)[]{\small{\Black{$N$}}}
    \Text(70,-75)[]{\small{\Black{$p$}}}
   \end{picture}
\qquad
&=u^c(p)
\label{eq4:ext-lines_in}\\
\text{outgoing } N\;:\;\; 
\quad
 \SetScale{0.6}
  \begin{picture}(90,0)(30,-92) 
    \SetWidth{0.8}
    \SetColor{Black}
    \Line(128,-147)(178,-147)
    \LongArrow(137,-137)(168,-137)
    \GOval(102,-147)(14,25)(0){0.882}
    \Text(115,-88)[]{\small{\Black{$N$}}}
    \Text(94,-75)[]{\small{\Black{$p$}}}
  \end{picture}
\qquad
&=u(p) \rule{0pt}{35pt}
\label{eq4:ext-lines_out}
\end{align}
~\vspace{0.5em}

\subsection{Frequently used results and cutting rules}
\label{subsec:BKL_std_rules_results}

For completeness and to establish the notations, in this subsection, we concisely state all the standard Feynman rules and other tools (e.g. polarization sums, decay rates and cutting rules) which are relevant for our calculations~\footnote{For a complete treatment, see \cite{QFT_books}.}.

\subsubsection{Propagators and external fields}

\begin{equation}\label{eq4:Feyn_scalar_prop}
\text{scalar particle } \phi\;:\qquad
 \SetScale{0.6}
    \begin{picture}(75,0)(40,-73) 
    \SetWidth{0.8}
    \SetColor{Black}
    \DashLine(88,-118)(170,-118){4}
    \SetWidth{0.5}
    \Vertex(170,-118){2.83}
    \Vertex(88,-118){2.83}
    \SetWidth{0.8}
    \LongArrow(109,-104)(149,-104)
    \Text(80,-55)[]{\Black{$p$}}  
  \end{picture}
\qquad
 D(p)=
 \frac{i}{p^2 - m_\phi^2 +i\epsilon}
\end{equation}
\begin{align}
\text{massless spin-1 particle}\;:\qquad
 \SetScale{0.6}
    \begin{picture}(75,0)(40,-73) 
    \SetWidth{0.8}
    \SetColor{Black}
    \Photon(88,-118)(170,-118){3}{5.5}
    \SetWidth{0.5}
    \Vertex(170,-118){2.83}
    \Vertex(88,-118){2.83}
    \SetWidth{0.8}
    \LongArrow(109,-104)(149,-104)
    \Text(80,-55)[]{\Black{$p$}}  
    \Text(48,-80)[]{\Black{$\nu$}}   
    \Text(105,-80)[]{\Black{$\mu$}}  
  \end{picture}
\qquad
 &D_{\mu\nu}(p) =
 \frac{-i g_{\mu\nu}}{p^2 +i\epsilon}
 \label{eq4:Feyn_photon_prop}
 \\
\text{Dirac fermion } \ell\;:\qquad
 \SetScale{0.6}
    \begin{picture}(75,0)(40,-73) 
    \SetWidth{0.8}
    \SetColor{Black}
    \ArrowLine(88,-118)(170,-118)
    \SetWidth{0.5}
    \Vertex(170,-118){2.83}
    \Vertex(88,-118){2.83}
    \SetWidth{0.8}
    \LongArrow(109,-104)(149,-104)
    \Text(80,-55)[]{\Black{$p$}}  
    \Text(48,-80)[]{\Black{$B$}}   
    \Text(105,-80)[]{\Black{$A$}}  
  \end{picture}
\qquad
 &\left[S_\ell(p)\right]_{AB}
  =
  \left[\frac{i(\pslash +m_\ell)}{p^2 - m_\ell^2 +i\epsilon}\right]_{AB}
  \rule{0pt}{25pt}
 \label{eq4:Feyn_Dirac_prop}\\
\text{external scalar particle}\;:\;\; 
\qquad
\qquad
&1 \\
 \text{incoming/outgoing photon}\;:\;\;\qquad\qquad 
 &\varepsilon_\mu(p)\,/\,\varepsilon^*_\mu(p)\\
 \text{incoming/outgoing Dirac fermion}\;:\;\; \qquad\qquad 
 &u(p)\,/\, \overline{u}(p)\\
 \text{incoming/outgoing Dirac antifermion}\;:\;\; \qquad\qquad 
 &\overline{v}(p)\,/\, v(p)
\end{align}
In the above, $p$ denotes four-momentum as usual.

\subsubsection{Polarization sums and decay rates}

In calculations, the following results are often useful:
\begin{gather}
 \sum_s\, u\overline{u} = \pslash + m \;,\quad
 \sum_s\, v\overline{v} = \pslash - m \;,\quad
 \sum_\text{pol} \varepsilon^*_\mu \varepsilon_\nu = - g_{\mu\nu}\;, 
\nonumber
\\
C\,\left[\sum_s\, u\overline{u}\right]^T C^\dagger 
  = C\left(\pslash^T + m\right)C^\dagger = -\pslash + m \;,
\nonumber
\\
 (u^c)^T = \overline{u}\,C^T\;,\quad 
  (u^c)^\dagger = - u^T C^\dagger \gamma^0\;,\quad
  (u^c)^* = C^* \gammazero u\;,\;\;
\nonumber
\\
  \gamma^{\mu\dagger} = \gammazero\gammamu\gammazero\;,\quad     
   \sigma^{\mu\nu\dagger} = \gammazero\sigma^{\mu\nu}\gammazero\;,\rule{0pt}{20pt}
   \label{eq4:freq_used_spin_tools}
\end{gather} 
where $p$ and $m$ are momentum and mass respectively, and ``pol'' stands for polarizations. The charge conjugation conventions are as in (\ref{eq4:BKL_C_conventions}) while the signature for $g_{\mu\nu}$ is $(+,-,-,-)$. Other important formulas include the 2-body decay rate in the centre-of-mass frame:
\begin{equation}\label{eq4:freq_used_cm_decayrate}
 \Gamma_\text{cm}=\frac{|\vec q|}{8\pi E_\text{cm}^2}\,|\mathcal{\overline{M}}|^2 \;,
\end{equation}
where $\vec q$ is the momentum of one of the final state particle, $E_\text{cm}$ is the centre-of-mass energy and $|\mathcal{\overline{M}}|^2$ is the decay amplitude averaged over initial and summed over final degrees of freedom, and the 3-body ($\Psi_1 \rightarrow \Psi_2 + \Psi_3 + \Psi_4$) differential decay rate:
\begin{equation}\label{eq4:freq_used_3-body-rate}
 d\Gamma_{1\rightarrow 234}= \frac{1}{2E_1}\,
  \frac{d^3 p_2}{(2\pi)^3 2E_2}\,
  \frac{d^3 p_3}{(2\pi)^3 2E_3}\,
  \frac{d^3 p_4}{(2\pi)^3 2E_4}\,
  (2\pi)^4 \delta^{(4)}(p_1-p_2-p_3-p_4)|\mathcal{\overline{M}}|^2  \;.
\end{equation}

\subsubsection{Cutkosky's cutting rules}

It has been shown in the toy model of Sec.~\ref{subsec:CP-violation} that the $CP$ asymmetry coming from a particle decay is directly proportional to the \emph{imaginary part} of the interference term of the decay amplitude. As a result, a salient feature of any Feynman graph calculation in leptogenesis involves isolating and computing the imaginary component. However, it is often quite difficult to evaluate this quantity directly, so one usually resorts to other methods of extracting it.

A typical way of handling this is to invoke the optical theorem for Feynman diagrams and the associated cutting rules~\cite{cutkosky,smatrix}. It is well-known that a Feynman graph will give rise to an imaginary part for the amplitude $\mathcal{M}$ only when the internal particles running in a loop go on-shell. The reason for this is that at the momentum values for which the virtual particles are on-shell, the complex function corresponding to  the amplitude has a branch cut singularity. Such discontinuity across the cut is related to the imaginary part of the amplitude via
\begin{equation}\label{eq4:cutting_disc-Im_rel}
 \text{Disc}(\mathcal{M}) = 2i\, \text{Im}(\mathcal{M})
 \;,
\end{equation}
where Disc$(\mathcal{M})$ denotes the discontinuity of $\mathcal{M}$. Since it has been proved by Cutkosky~\cite{cutkosky} that there is a simple algorithm to evaluate Disc$(\mathcal{M})$ for any Feynman diagram, the computation of Im$(\mathcal{M})$ can be simplified. This method of extracting the discontinuity involves applying the following \emph{cutting rules} to a given Feynman graph
\begin{enumerate}
 \item Deduce all possible ways to cut the graph such that all cut propagators can simultaneously put on-shell.
 \item For each case, the inverse of the denominator of all cut propagators are replaced by the mass-shell $\delta$-function, i.e.
   \begin{equation}\label{eq4:cutting_rules_delta-function}
    \frac{1}{p^2-m^2 +i\epsilon} \;\rightarrow\; -2\pi i\,\delta(p^2-m^2)\,\Theta(\pm |E|)
    \;,
   \end{equation} 
   where $p^2 = E^2 - |\vec p|^2$ with $E$ denoting total energy, while $\delta(x)$ and $\Theta(x)$ are the Dirac-delta and unit step functions respectively~\footnote{The choice of $+|E|$ or $-|E|$ in (\ref{eq4:cutting_rules_delta-function}) depends on how the momentum-flow is defined in the diagram. We select the $+|E|$ when four-momentum is flowing forward in time, otherwise we pick $-|E|$.}.
 \item Perform the loop integrals in each case.
 \item Sum up all contributions.
\end{enumerate}
The imaginary component can then easily be obtained from (\ref{eq4:cutting_disc-Im_rel}).

\subsection{Some illustrative examples}\label{subsec:BKL_maj_feyn_examples}

The aim of this section is to provide some concrete examples~\footnote{More examples are presented in Appendix~\ref{app:c_std_eg}.} showing the application of the simplified Majorana Feynman rules introduced in Sec.~\ref{subsec:BKL_simp_Maj-rules}. In doing so, we demonstrate that these rules are capable of reproducing the standard results while highlighting all the intricacies involved. It will also serve as a guide for our calculations in electromagnetic leptogenesis later.

\subsubsection{Tree-level contribution to $N_k\rightarrow \ell\, \bar\phi$}

\begin{figure}[t]
\begin{center}
 \includegraphics[width=0.75\textwidth]{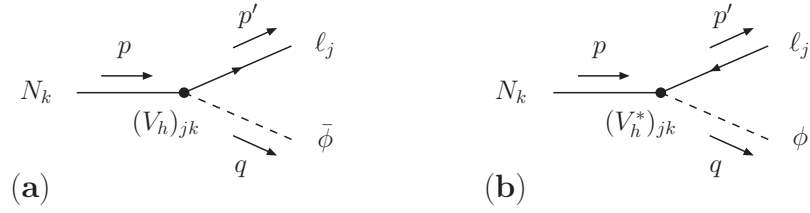}
\end{center} 
 \caption{{\bf (a)}~The Feynman graph for the process $N_k\rightarrow \ell_j \bar\phi$.
 {\bf (b)}~The graph for $N_k\rightarrow \bar{\ell}_j \phi$. Here $q=p-p'$, and $(V_h)_{jk} \equiv -i\,h_{jk}\,P_R$ and $(V_h^*)_{jk} \equiv i\,h_{jk}^*\,C^\dagger P_L$ are the vertex factors.}
 \label{fig4:BKL_std_tree_cal_graphs}
\end{figure}

Our first example is to calculate the tree-level contribution to the decay rate of $N_k\rightarrow \ell\,\bar\phi$, using the rules and conventions described in the previous two subsections. The corresponding Feynman diagram for this process is shown in Fig.~\ref{fig4:BKL_std_tree_cal_graphs}a.

Following the rules outlined in Sec.~\ref{subsec:BKL_simp_Maj-rules} and Sec.~\ref{subsec:BKL_std_rules_results}, we can immediately write down the amplitude for this decay as
\begin{align}
 -i\mmat &= \ubarj (-i \, h_{jk} P_R)u_k^c \;, \nonumber\\
         &= \ubarj (-i \, h_{jk} P_R) C \ubark^T \;. \\
 |\mmat|^2 &= \ubarj (-i \, h_{jk} P_R) C \ubark^T
            \left[\ubarj (-i \, h_{jk} P_R) u_k^c \right]^\dagger\;,\nonumber\\
           &= \ubarj (-i \, h_{jk} P_R) C \ubark^T
            (i \, h_{jk}^*) (-u_k^T C^\dagger P_L u_j)\;,\nonumber\\ 
           &= -(h_{jk}^* h_{jk})\ubarj P_R C \ubark^T u_k^T C^\dagger P_L u_j\;,
          \\            
 |\overline{\mmat}|^2 
   &= -(h_{jk}^* h_{jk}) \, P_R\, 
      C\left[
       \frac{1}{2}\sum_s u_k \ubark\right]^T C^\dagger
       \, P_L\, \sum_{s'}u_j\ubarj \quad 
      \text{(index form)} \;.\nonumber\\
\intertext{When the universe was hot enough, $\ell_j$ and $\phi$ are strongly relativistic, so $m_{\ell_j} ,m_\phi \approx 0$ and}
|\overline{\mmat}|^2
   &= -\frac{(h_{jk}^* h_{jk})}{2}\,
   \text{Tr}\left[P_R (-\pslash + M_{k}) P_L (\ppslash)\right]
    \;,\\
   &= \frac{1}{2}(h_{jk}^* h_{jk})\,\text{Tr}\left[P_R \pslash\,\ppslash\right] \;,\\
   &= (h_{jk}^* h_{jk}) (p\cdot p') \;.\label{eq4:examples_result_1}
\end{align}
The four-momenta in the centre-of-mass frame are given by
\begin{equation}
 p =(M_k\;,\;\vec 0)\;,\quad
 p' = (M_k/2\;,\;-\vec q)\;,\quad
 q = (M_k/2\;,\;\vec q)\;,
  \label{eq4:examples_std_cm_momenta}
\end{equation}
and one can quickly deduce that 
$|\vec q| = M_k/2$ and $p\cdot p' = p\cdot q= p' \cdot q=M_k^2/2$.

Therefore, we obtain
\begin{align}
 |\overline{\mmat}|^2 
   &= (h_{jk}^* h_{jk}) \frac{M_k^2}{2} \;,\nonumber\\
   &= (h^\dagger h)_{kk}\frac{M_k^2}{2}\;. \qquad \text{ (after summing over $j$)}
\end{align}
Using Eq.~(\ref{eq4:freq_used_cm_decayrate}), the decay rate for $N_k\rightarrow \ell\, \bar\phi$ is then
\begin{align}
 \Gamma(N_k\rightarrow \ell\, \bar\phi)   
  &= 2\times \frac{|\vec q|}{8\pi E_\text{cm}^2} |\overline{\mmat}|^2 \;,\nonumber\\
  &= 2(h^\dagger h)_{kk}
    \frac{M_k^2}{2}\frac{1}{8\pi}\frac{M_k}{2} \frac{1}{M_k^2} \;,
    \qquad(E_\text{cm} \equiv M_k)\;,
    \nonumber\\
  &= \frac{(h^\dagger h)_{kk}}{16\pi} \,M_k\;,\label{eq4:examples_std_tree_half_rate}
\end{align}
where the factor of 2 comes from the fact that there are two possible decay channels: $N_k\rightarrow \nu \phi^0$ and $N_k \rightarrow e^- \phi^+$. When $k=1$, result (\ref{eq4:examples_std_tree_half_rate}) is identical to the rate quoted earlier in (\ref{eq1:lepto_std_N1_tree}) on page~\pageref{eq1:lepto_std_N1_tree}.

\subsubsection{Tree-level contribution to $N_k\rightarrow \bar\ell\, \phi$}

We shall repeat the exercise the for the antiparticle decay, $N_k\rightarrow \bar\ell\, \phi$ and check for consistency. The Feynman graph for this is depicted in Fig.~\ref{fig4:BKL_std_tree_cal_graphs}b. From this we have for the amplitude
\begin{align}
 -i\mmat &= (u_k^c)^T (i\, h_{jk}^* \,C^\dagger \,P_L)\,v_j \;, \nonumber\\
         &= -i h_{jk}^* \ubark P_L v_j\;. \\
 |\mmat|^2 &=-i \, h_{jk}^*\, \ubark\, P_L\, v_j
              \left[-i\, h_{jk}^*\, \ubark \,P_L \,v_j\right]^\dagger\;,\nonumber\\
           &= (h_{jk}^* h_{jk})\, \ubark P_L\, v_j\, \vbarj \,P_R\, u_k\;.          
\end{align}
Averaging over initial and summing up final spin states, it then becomes
\begin{align}
 |\overline{\mmat}|^2 
   &= (h_{jk}^* h_{jk}) P_L \sum_{s'}v_j\vbarj \,P_R\,
       \frac{1}{2}\sum_s u_k \ubark
       \, P_L\,\quad 
      \text{(index form)} \;,\nonumber\\
   &= \frac{(h_{jk}^* h_{jk})}{2}\,
   \text{Tr}\left[P_L \ppslash P_R (\pslash+ M_k)\right]
    \;,\quad m_{\ell_j} ,m_\phi \rightarrow 0 \;,\nonumber\\
   &= \frac{1}{2}(h_{jk}^* h_{jk})\,\text{Tr}\left[P_L \ppslash\,\pslash\right] \;,
   \nonumber\\
   &= (h_{jk}^* h_{jk}) (p\cdot p')\;,
\end{align}
which is exactly (\ref{eq4:examples_result_1}). Hence, we must have 
$\Gamma(N_k\rightarrow \bar\ell\, \phi)=\Gamma(N_k\rightarrow \ell\, \bar\phi)$.
This should come as no surprise since there can be no $CP$ violation at tree-level as mentioned in Sec.~\ref{subsec:CP-violation}, therefore, the rates are necessarily equal.

\subsubsection{Vertex contribution to the $CP$ asymmetry}\label{subsubsec:BKL_vertex_eg}

Before concluding this subsection, we present the explicit computation of the vertex contribution to the $CP$ asymmetry in standard leptogenesis. This is extremely instructive for our purposes here (and later) as it can put all the tools and formulas we have mentioned previously to the test.

\begin{figure}[t]
\begin{center}
 \includegraphics[width=0.80\textwidth]{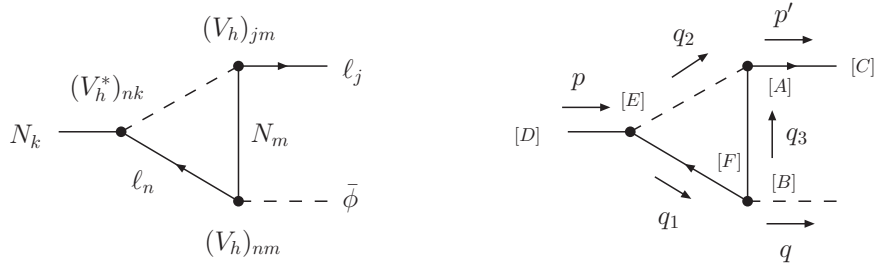}
\end{center} 
 \caption{One-loop vertex correction graph for the process $N_k\rightarrow \ell_j \bar\phi$. LEFT: $(V_h)_{ab} \equiv -i\,h_{ab}\,P_R$ and $(V_h^*)_{ab} \equiv i\,h_{ab}^*\,C^\dagger P_L$ are the vertex factors; RIGHT: we have included the momentum flows and spinor indices~$[X]$, where $q=p-p'$, $q_2=p-q_1$ and $q_3=q_1-q$.}
 \label{fig4:BKL_std_cpvloop_cal_graphs}
\end{figure}

To begin with, we note that the required contribution comes from the interference between the one-loop vertex graph in Fig.~\ref{fig4:BKL_std_cpvloop_cal_graphs} and its tree-level counterpart in Fig.~\ref{fig4:BKL_std_tree_cal_graphs}a. So from the result displayed in (\ref{eq1:CP_loop_expand_X}) on page~\pageref{eq1:CP_loop_expand_X}, it is clear that one needs to compute the imaginary part of the interference of the two amplitudes (schematically speaking):
\begin{equation}\label{eq4:examples_std_vloop_basic-form}
 \text{Im}\left[(\mmat_\text{tree})^\dagger \, \mmat_\text{loop}\right]
 \;\propto\; \varepsilon_\text{vertex}
 \;.
\end{equation}
Looking at Fig.~\ref{fig4:BKL_std_cpvloop_cal_graphs}, the first challenge appears to be writing down the matrix elements for the amplitude in the correct order. Indeed, this is one of the major issues with Majorana Feynman rules discussed in Sec.~\ref{subsec:BKL_simp_Maj-rules}. For our simplified scheme, we use the following conventions:
\begin{itemize}
 \item We always assign the external Majorana fermions to flow in the direction which ensures that fermion-flow at the vertex it connects to is not broken. For example, if the flow from another fermion is into the vertex, then the external Majorana must be chosen to flow out of it and vice-versa. Note that there are no vertices involving three fermions in any of the models that interest us, as a result there is no potential contradiction with this convention.
 \item If the fermion-flow through the entire diagram is unbroken (this usually happens when the flow through the Majorana propagators are \emph{unambiguous}), then use the normal convention, i.e. follow the reverse fermion-arrow direction in writing down all matrix elements.
 \item When some of the Majorana propagators have ambiguous fermion-flow, follow the convention used in (\ref{eq4:prop_Maj_prop}) for these, and use the normal convention for all other fermions and vertices (including those at the two ends of the Majorana propagators).  
\end{itemize}
%

Thus, for the diagram in Fig.~\ref{fig4:BKL_std_cpvloop_cal_graphs}, the orderings are:
\begin{equation}
[C]\rightarrow [A]\rightarrow [B]
\quad\text{ and }\quad
[D]\rightarrow [E] \rightarrow [F] \rightarrow [B]
\;.
\end{equation}
From this, the amplitude of the interference term in index form is given by
\begin{align}
 I_\text{vertex}'
 &= \int \frac{d^4 q_1}{(2\pi)^4}
 (-i  h_{jm})
 (-i  h_{nm})(i  h_{nk}^*)
 \left[\ubarj\right]_{1C} 
 \left[P_R\right]_{CA} 
 \left[S_{N_m}(q_3)\right]_{AB}  
  \left[P_R\right]_{FB} 
  \nonumber\\
 &\qquad\times
 \left[S_{\ell}(-q_1)\right]_{EF} 
 [C^\dagger P_L]_{DE} 
 \left[u_k^c\right]_{D1} 
 \left[D(q_2)\right]_{11}
 \underbrace{
 \left[-i h_{jk}^* u_k^T C^\dagger P_L u_j\right]_{11}}_{(\mmat_\text{tree})^\dagger}
 \;,
 \label{eq4:examples_std_vloop_index-form}
\end{align}
where all symbols are as defined previously. So letting
$A_h = h_{jk}^* h_{jm} h_{nm} h_{nk}^*$, this becomes (in matrix notation)
\begin{align}
 I_\text{vertex}' 
 &=
  A_h 
  \int\frac{d^4 q_1}{(2\pi)^4}
  \frac{\ubarj\, P_R (-i)(\qthreeslash+M_m)C\,P_R^T \, i(-\qoneslash)^T P_L^T C^* u_k^c (i)
  (-1)u_k^T C^\dagger P_L u_j}
  {(q_3^2-M_{m}^2 +i\epsilon)(q_1^2+i\epsilon)(q_2^2+i\epsilon)}\;, \nonumber\\
 &=
  -i A_h\int\frac{d^4 q_1}{(2\pi)^4}
  \frac{\ubarj\, P_R (\qthreeslash+M_m)C\,C^\dagger P_R C \, 
  C^\dagger\qoneslash C C^\dagger P_L C C^*\; C\ubark^T 
  u_k^T C^\dagger P_L u_j}
  {(q_3^2-M_{m}^2 +i\epsilon)(q_1^2+i\epsilon)(q_2^2+i\epsilon)}\;, \nonumber\\
\intertext{and using the tools introduced earlier to simplify, we obtain}
 I_\text{vertex}'
  &=
    \frac{i A_h}{2}\int\frac{d^4 q_1}{(2\pi)^4}
  \frac{P_R (\qthreeslash+M_m)\,P_R \qoneslash P_L \; 
  C\left(\sum_s u_k\ubark\right)^T C^\dagger P_L \sum_{s'}u_j\ubarj}
  {(q_3^2-M_m^2 +i\epsilon)(q_1^2+i\epsilon)(q_2^2+i\epsilon)}\;,\nonumber\\
   &=
  \frac{i A_h}{2}\int\frac{d^4 q_1}{(2\pi)^4}
  \frac{\text{Tr}\left[
  P_R (\qthreeslash+M_m)\,P_R \qoneslash P_L\,
  C\left(\pslash^T +M_k\right)C^\dagger P_L \ppslash
  \right]}
  {(q_3^2-M_m^2 +i\epsilon)(q_1^2+i\epsilon)(q_2^2+i\epsilon)}\;,\nonumber\\
  &\hspace{140pt} \vdots \nonumber\\
 &=
  i A_h\,M_k\,M_m\int\frac{d^4 q_1}{(2\pi)^4}
  \frac{q_1\cdot p'}
  {(q_3^2-M_m^2 +i\epsilon)(q_1^2+i\epsilon)(q_2^2+i\epsilon)}\;.  
\end{align}
Let us concentrate on the integral:
\begin{equation}\label{eq4:exmaples_vloop_integral_main}
  I' =
  i\,M_k\,M_m \int\frac{d^4 q_1}{(2\pi)^4}
  \frac{q_1\cdot p'}
  {(q_3^2-M_m^2 +i\epsilon)(q_1^2+i\epsilon)(q_2^2+i\epsilon)}\;.
\end{equation}
To pick out the discontinuity, we apply the cutting rules discussed on page~\pageref{eq4:cutting_rules_delta-function}. Firstly, we note that of the three possible ways to cut the diagram, only one  of them can simultaneously put both cut propagators on-shell, due to the heaviness of $N_m$:
\begin{equation}
 \SetScale{0.5}
   \begin{picture}(300,0)(40,-102)
    \SetWidth{0.8}
    \SetColor{Black}
    \Line(371,-205)(371,-133)
    \Line(276,-168)(309,-168)
    \ArrowLine(371,-205)(309,-168)
    \ArrowLine(371,-133)(418,-133)
    \DashLine(309,-168)(371,-133){6}
    \DashLine(418,-205)(371,-205){6}
    \SetWidth{0.5}
    \Vertex(309,-168){2.83}
    \Vertex(371,-133){2.83}
    \Vertex(371,-205){2.83}
    \SetWidth{0.8}
    \Line(571,-205)(571,-133)
    \Line(476,-168)(509,-168)
    \ArrowLine(571,-205)(509,-168)
    \ArrowLine(571,-133)(618,-133)
    \DashLine(509,-168)(571,-133){6}
    \DashLine(618,-205)(571,-205){6}
    \SetWidth{0.5}
    \Vertex(509,-168){2.83}
    \Vertex(571,-133){2.83}
    \Vertex(571,-205){2.83}
    \SetWidth{0.8}
    \Line(171,-205)(171,-133)
    \Line(76,-168)(109,-168)
    \ArrowLine(171,-205)(109,-168)
    \ArrowLine(171,-133)(218,-133)
    \DashLine(109,-168)(171,-133){6}
    \DashLine(218,-205)(171,-205){6}
    \SetWidth{0.5}
    \Vertex(109,-168){2.83}
    \Vertex(171,-133){2.83}
    \Vertex(171,-205){2.83}
    \SetWidth{0.8}
    \DashLine(146,-116)(146,-225){2} 
    \DashLine(398,-177)(306,-134){2} 
    \DashLine(512,-195)(601,-155){2} 
    \Text(50,-109)[]{\LARGE{\Black{\ding{52}}}}
    \Text(152,-109)[]{\LARGE{\Black{\ding{56}}}}
    \Text(255,-109)[]{\LARGE{\Black{\ding{56}}}}  
  \end{picture}
  \rule{0pt}{40pt}
\end{equation}
This only way (leftmost diagram) corresponds to cutting through the propagators associated with momenta $q_1$ and $q_2$ (see Fig.~\ref{fig4:BKL_std_cpvloop_cal_graphs}). Thus, we make the replacement
\begin{align}
 \frac{1}{q_1^2 + i\epsilon} \;&\rightarrow\; -2\pi i \delta(q_1^2)\Theta(E_{1})\;,
 \nonumber\\
 \frac{1}{q_2^2 + i\epsilon} \;&\rightarrow\; -2\pi i \delta(q_2^2)\Theta(E_{2})
  = -2\pi i \delta((p-q_1)^2)\Theta(M_k - E_{1})\;,\label{eq4:examples_vloop_cut-replace}
\end{align}
in (\ref{eq4:exmaples_vloop_integral_main}), where $q_1 = (E_{1},\vec q_1)$ and $q_2 = (E_{2},\vec q_2)$. Using the definitions in (\ref{eq4:examples_std_cm_momenta}) for $p,p'$ and $q$, we can evaluate
\begin{equation}
 q_1\cdot p' = E_{1} \frac{M_k}{2} - \vec q_1 \cdot(-\vec q)
  = E_{1} \frac{M_k}{2} + |\vec q_1| |\vec q| \cos \theta
  = \frac{M_k}{2} (E_{1} +|\vec q_1| \cos \theta)\;,
\end{equation}
where $\theta$ is the smaller angle between $\vec q_1$ and $\vec q$. Putting all these together and substituting $q_3=q_1 -q$, we obtain ($\epsilon \rightarrow 0$):
\begin{equation}
  \text{Disc}(I') 
  =
  \frac{i\,M_k^2M_m}{2}\int\frac{d^4 q_1}{(2\pi)^4}
  \frac{
   (-2\pi i)^2
   (E_{1} +|\vec q_1| \cos \theta) 
   \delta(q_1^2)
   \delta((p-q_1)^2)\Theta(E_{1})\Theta(M_k - E_{1})
   }
  {(q_1-q)^2-M_m^2}\;,\nonumber
\end{equation}
\begin{align}
   &=
  \frac{-i\,M_k^2M_m}{8\pi^2}\int dE_1 d^3 q_1
 \;
  \delta(E_{1}^2-|\vec q_1|^2)\,
  \delta\left[(M_k-E_{1})^2-|\vec q_1|^2\right]
  \Theta(E_{1})\Theta(M_k - E_{1})
 \nonumber\\
  &\qquad\qquad\qquad\quad \times  
  \frac{
   (E_{1} +|\vec q_1| \cos \theta)}
  {(E_1-\halfmNk)^2 -|\vec q_1-\vec q|^2-M_m^2}\;.
\end{align}
Applying the identity: $\delta(x^2-a^2)=\left[\delta(x-a)+\delta(x+a)\right]/2|a|$, we can rewrite $\delta(E_{1}^2-|\vec q_1|^2)$ as
\begin{equation}
 \delta(E_{1}^2-|\vec q_1|^2) = \frac{1}{2\modq1}\left[\delta(E_1-\modq1)
  +\delta(E_1+\modq1)\right]\;.
\end{equation}
Integrating over $E_1$, the terms corresponding to the unphysical energy option of $E_1 = -\modq1$ will drop out automatically because of the step function $\Theta(-\modq1)=0$ and one obtains after some algebra
\begin{align}
 \text{Disc}(I')
  &=
  \frac{-i\,M_k^2M_m}{16\pi^2}\int \modq1^2 d\modq1 d\Omega\,
  \frac{1}{|-2M_k|}
  \;
  \delta\left[|\vec q_1|-\halfmNk\right]
  \Theta(M_k - \modq1)
 \nonumber\\
  &\hspace{80pt} \times
  \frac{
   1+ \cos \theta} 
    {(\modq1-\halfmNk)^2 -\modq1^2 -\frac{M_k^2}{4}+ \modq1 M_k \costh - M_m^2}\;, 
    \label{eq4:examples_vloop_after_E1_intg} 
\end{align}
where we have used the identity $\delta(ax)=\delta(x)/|a|$ and the fact that
\begin{align}
 -|\vec q_1-\vec q|^2
 &= -\left[
   |\vec q_1|^2 +|\vec q|^2 - 2|\vec q_1||\vec q|\cos (\theta)
   \right]
   = -|\vec q_1|^2 -\frac{M_k^2}{4} + |\vec q_1|M_k \cos \theta\;.
\end{align}
Note that $\theta$ is again the smaller angle between $\vec q_1$ and $\vec q$. We perform the $d\modq1$ integral in (\ref{eq4:examples_vloop_after_E1_intg}) to obtain
\begin{align}
  \text{Disc}(I')  
  &=
  \frac{-i\,M_k \,M_m}{32\pi^2}\int d\Omega\,
  \frac{M_k^2}{4}
  \frac{
   1+ \cos \theta 
  }
  {-\frac{M_k^2}{4} -\frac{M_k^2}{4}+ \frac{M_k^2}{2}\costh - M_m^2}\;,\nonumber\\
  &=
  \frac{-i\,M_k\,M_m}{32\pi^2}\int d\Omega\,
  \frac{
   1+ \cos \theta 
  }
  {-1 -1+ 2\costh - 4z}\;,
\end{align}
where $z\equiv M_m^2/M_k^2$. We then evaluating $\int d\Omega$:
\begin{align}
  \text{Disc}(I')  
  &=
  \frac{-i\,M_kM_m}{32\pi^2}
  \int d\phi \int d(\costh)
  \frac{
   1+ \cos \theta 
  }
  {-2(1- \costh) - 4z}\;,\nonumber\\  
  &=
  \frac{i\,M_kM_m}{32\pi^2}\,
  (2\pi) \int_{-1}^{1} dx\,
  \frac{
   1+ x 
  }
  {2(1- x) + 4z}\;,\nonumber\\
  &=
  \frac{i\,M_k M_m}{16\pi} \left[
  -z\ln(-2z) -\ln(-2z)\right. \nonumber\\ 
  &\hspace{70pt}\left.
  +z\ln(-2(z+1)) + \ln(-2(z+1)) -1
  \right]\;,\nonumber\\
   &=
  \frac{-i\,M_k^2}{16\pi}\; \sqrt{z}\left(
  1-(z+1)\ln\left[\frac{z+1}{z}\right] 
  \right)\;.
\end{align}
Therefore, the imaginary part of $I'$ is given by
\begin{align}
 \text{Im}\left(I'\right)
 &= \frac{1}{2i}\;
 \text{Disc}\left(I'\right)\;, \nonumber\\
 &=
  -\frac{M_k^2}{32\pi}\; \sqrt{z}\left(
  1-(z+1)\ln\left[\frac{z+1}{z}\right] 
  \right)\;.
\end{align}
The two last pieces of information we require before evaluating the $CP$ asymmetry is the total decay rate, which we can get from (\ref{eq4:examples_std_tree_half_rate})
\begin{equation}
 \Gamma_\text{tot} = \Gamma + \overline{\Gamma} = 2 \times \frac{(h^\dagger h)_{kk}}{16\pi}  \,M_k
 = \frac{(h^\dagger h)_{kk}}{8\pi} \,M_k\;,
\end{equation}
and the 2-body phase space factor which may be readily read off using (\ref{eq4:freq_used_cm_decayrate}) as
\begin{equation}
 V_\varphi = \underbrace{2\;\;\;\;\times}_{\text{two channels}} 
 \frac{|\vec q|}{8\pi E_\text{cm}^2} 
   = 2\times \frac{1}{8\pi}\;\frac{M_k}{2\,M_k^2} 
   = \frac{1}{8\pi\,M_k} \;.
\end{equation}
Putting all these together into the general formula derived in (\ref{eq1:CP_aysm_final_form}) on page~\pageref{eq1:CP_aysm_final_form}, and summing over all heavy Majorana neutrino species $m\neq k$, as well as the internal lepton species $n$, the expression for the $CP$ asymmetry due to the vertex contribution is therefore
\begin{align}
 \varepsilon_\text{vertex}
 &= -\frac{4}{\Gamma_\text{tot}} \;
 \sum_{m\neq k}\sum_{n}
 \text{Im} (A_h)\;
    \text{Im} (I'\, V_\varphi)  \;,
   \\
\intertext{where we have used
    $\text{Im}(I'\, V_\varphi)\equiv \text{Im} (I')V_\varphi$ as
     $V_\varphi\in \mathbb{R}$, and so}
 \varepsilon_\text{vertex}
 &=
  4\times \frac{8\pi}{(h^\dagger h)_{kk}\;M_k}
  \sum_{m\neq k}\sum_{n}
  \text{Im} (h_{jk}^* h_{jm} h_{nm} h_{nk}^*)
\nonumber
\\
&\hspace{110pt}  
  \times
  \frac{M_k^2}{32\pi}\; \sqrt{z}\left(
  1-(z+1)\ln\left[\frac{z+1}{z}\right] 
  \right)
  \frac{1}{8\pi\,M_k}\;, \nonumber\\
 &=
  \frac{1}{8\pi}
  \sum_{m\neq k}
  \frac{
  \text{Im} \left[h_{jk}^* h_{jm} (h^\dagger h)_{km}\right]}
  {(h^\dagger h)_{kk}}
  \sqrt{z}\left(
  1-(z+1)\ln\left[\frac{z+1}{z}\right] 
  \right)
  \;,\label{eq4:examples_vloop_final_result}
\end{align}  
with $z \equiv M_m^2/M_k^2$. Eq.~(\ref{eq4:examples_vloop_final_result}) is identical to the standard result given in \cite{Covi:1996wh, CP_calculations_BP}~\footnote{See Appendix~\ref{app:c_std_eg} for the calculation of the self-energy contributions.}. Upon summing over flavor $j$, this then becomes the usual expression mentioned in Chapters~\ref{ch_intro} and \ref{ch_work_LV}.
Hence, we have successfully demonstrated that these simplified Feynman rules can give rise to the correct result while avoiding most of the subtleties in other approaches, and we shall use them extensively in our calculations here on electromagnetic leptogenesis.

\section{Lepton number creation from EMDM interactions}\label{sec:BKL_emdm_models}

The stage has now been set for us to tackle the main theme of this work--- to investigate the viability of electromagnetic leptogenesis. In doing so, we must first check that the out-of-equilibrium decay of the RH neutrinos can give rise to a nonzero $CP$ asymmetry under the most general situations. In addition, because of the constraints from other sectors of the theory, it is also necessary to examine whether the parameter space has enough degrees of freedom to produce an asymmetry of the correct magnitude.

Below, after revisiting the issue of $CP$ violation in particle decays where a general proof of the need to go beyond the lowest order is presented, we calculate the decay rates and $CP$ asymmetry induced in a toy model with EMDM interactions between the light and heavy neutrinos of the form shown in (\ref{eq4:BKL_emdm_chiral_form}). The aim is to obtain explicit expressions for the relevant quantities so that a comparison with the standard scenario can be made. Later, we shall modify this toy model to ensure that the EMDM interaction is compatible with the SM gauge symmetry. We then repeat some of the calculations and discuss the implications of this extension.

\subsection{$CP$ violation in decays: revisited}\label{subsec:BKL_CP_needs_loops}

As was mentioned in Sec.~\ref{subsec:CP-violation}, $CP$ violating rate differences between $CP$ conjugate decays of a heavy particle can only appear once higher order terms are included. While it was demonstrated in our examples from the last section that this is indeed the case for standard leptogenesis processes, it is important to note that the result can be generalised to models with any type of interactions as long as we have $CPT$ invariance. 

At first glance, this may not seem obvious in models where the same particles can interact via \emph{more than one} distinct couplings. For instance, if one considers the dipole Lagrangian of (\ref{eq4:BKL_emdm_op_massbasis}) where there could be independent magnetic ($\mu_{jk}$) and electric ($d_{jk}$) transition moments linking $\nu_j$ and $N_k$ (for a given $j$ and $k$), then one may suspect that the tree-level interference between the amplitudes from $\mu_{jk}$ and $d_{jk}$ would be sufficient to create a difference between the decay rates for $N_k \rightarrow \nu_j\,\gamma$ and its $CP$ conjugate. But as we prove below~\footnote{This proof is due to B.~J.~Kayser.}, the two rates can never be different until one goes beyond the first order in the underlying Hamiltonian. 

Suppose we have a particle $X$ with the decay process, $X \rightarrow a_1 + a_2 +\cdots$ (in the rest frame of $X$). If $CPT$ invariance
holds, the amplitude for this decay obeys the constraint
\begin{equation}
 |\langle a_1(\vec p_1, s_1)\, a_2(\vec p_2, s_2)\cdots |\mathcal{T}|\,X(\widehat m) \rangle|^2
 =
   |\langle \overline{a}_1(\vec p_1, -s_1)\, 
   \overline{a}_2(\vec p_2, -s_2)\cdots |\mathcal{T}^\dagger|\,\overline{X}(-\widehat m) \rangle|^2\;,
   \label{eq4:BKL_proof_CPT-cond}
\end{equation}
where $\vec p_i$ and $s_i$ are, respectively, the momentum and
helicity of the daughter particle $a_i$.  $\widehat m$ is the $z$-axis
projection of the spin of $X$, and $\mathcal{T}$ is the transition operator for the decay with $\mathcal{T}=i(S-I)$, where $S$ is the corresponding $S$-matrix operator. Since to first order, we have $\mathcal{H}=\mathcal{T}$ for the system, where $\mathcal{H}$ denotes the Hamiltonian, it follows that $\mathcal{T}^\dagger = \mathcal{T}$. Using this with (\ref{eq4:BKL_proof_CPT-cond}), and after summing over the final helicities and integrating over the outgoing momenta, we obtain
\begin{equation}
 \Gamma\left[\,\overline{X} \rightarrow \overline{a}_1 +\overline{a}_2 +\cdots\, \right]=
  \Gamma\left[\,X \rightarrow a_1 +a_2+\cdots\, \right]\;.
\end{equation}
This equality must hold to first order in $\mathcal{H}$ regardless of
whether $\mathcal{H}$ contains numerous terms and $CP$ violating
coupling constants.

In the special case of a two-body decay, $X \rightarrow a_1 + a_2$, we
have $\vec p_1 = -\vec p_2 \equiv \vec p$. For this case, let us rotate the
system of particles on the right-hand side of Eq.~(\ref{eq4:BKL_proof_CPT-cond}) by
$180^\circ$ about the axis perpendicular to the $z$-axis and to $\vec p$. Eq.~(\ref{eq4:BKL_proof_CPT-cond}) then states that, to first order in
$\mathcal{H}$ (so that $\mathcal{T} = \mathcal{T}^\dagger$),
\begin{equation}
 |\langle a_1(\vec p, s_1)\, a_2(-\vec p, s_2)|\mathcal{T}|\,X(\widehat m) \rangle|^2
= 
  |\langle \overline{a}_1(-\vec p, -s_1)\, 
   \overline{a}_2(\vec p, -s_2)|\mathcal{T}|\,\overline{X}(\widehat m) \rangle|^2\;.
\end{equation}
The processes whose amplitudes appear on the two sides of this
constraint are the $CP$ mirror images of each other. Thus, in two-body
decays, to first order in $\mathcal{H}$, the rates for
$CP$ mirror-image decay processes must be equal even before one sums
over final helicities and integrates over outgoing momenta. 

As a result, we can conclude that the $CP$ conjugate rates must be the same at tree-level and our proof is complete.

\subsection{An EMDM toy model}\label{subsec:BKL_toy_model_5D}

In this section we explore, by means of a toy model, the possibility of generating a lepton asymmetry through the EMDM interactions described earlier. Since we are interested in leptogenesis energy scales above the electroweak phase transition, we shall identify the light neutrino in (\ref{eq4:BKL_emdm_chiral_form}) 
%
%
to be a massless LH state (the same $\nu_L$ as appears in the SM lepton doublet), while $N$ is assumed to have a large Majorana mass as in type~I seesaw.

The simplistic toy model that we are considering contains the minimally extended SM Lagrangian with three heavy RH neutrinos~\footnote{For simplicity, we shall ignore the effects of the neutrino Yukawa terms and consider standard leptogenesis as being switched off.} augmented by dimension-5 EMDM operators of the form of (\ref{eq4:BKL_emdm_chiral_form}). We assume that these EMDM couplings are generated by some new physics at an energy scale $\Lambda > M$, where $M$ generically denotes the mass a heavy RH Majorana neutrino, and work with the effective theory that is valid below $\Lambda$, obtained after integrating out all new heavy degrees of freedom. The EMDM interaction Lagrangian of interest is (rewritten here for convenience):
\begin{align}
\mathcal{L}_\text{EM}^\text{5D} &=
  -\lambda_{jk}\, 
   \overline{\nu}_{Lj}\,  
    \sigma^{\alpha\beta}\, P_R\, N_{k}\, F_{\alpha\beta}
    +\text{h.c.} \;,
 \label{eq4:BKL_toy5D_main_lag_A}\\
  &\equiv
   -\frac{1}{\Lambda}   
   (\lambda_0)_{jk}\, 
   \overline{\nu}_{Lj}\,  
    \sigma^{\alpha\beta}\, P_R\, N_{k}\, F_{\alpha\beta}
   +\text{h.c.} \;, 
   \label{eq4:BKL_toy5D_main_lag}
\end{align}
where $j=e,\mu,\tau$ and $k = 1,2,3$.  $F_{\alpha\beta} =
\partial_{\alpha}A_{\beta}-\partial_{\beta}A_{\alpha}$ is, as before, the electromagnetic field strength tensor, with $A_\alpha$ being the photon field. We have
defined $\lambda_0$ as a dimensionless $3\times 3$ matrix of complex coupling constants, and $\Lambda$ is the cut-off scale of our effective theory, which has dimensions of energy. 

An important observation is that the SM gauge symmetry, $SU(2)_L \times U(1)_Y$ is explicitly broken and the model is invariant only under the electromagnetic symmetry $U(1)_Q$. This does not seem to be an issue at first glance, for the SM symmetry will be broken down to $U(1)_Q$ at low energies. However, one major difficulty is that the theory demands (\ref{eq4:BKL_toy5D_main_lag}) to be valid up to the scale of $\Lambda$ (i.e above $M$), hence only $U(1)_Q$ is unbroken, while the SM implies that electroweak symmetry must be restored at that scale since $\Lambda ,M \gg \Lambda_\text{EW} \simeq 10^2$~GeV. Therefore as it stands, this toy model is incompatible with the framework of the SM. But we shall defer fixing this problem until the next section, because, to achieve the proof of principle that we seek, it is more transparent to work with such a toy model.

\begin{figure}[t]
\begin{center}
 \includegraphics[width=0.35\textwidth]{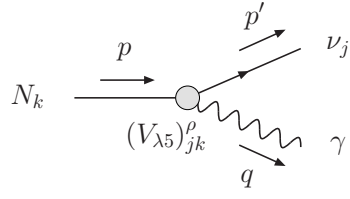}
\end{center} 
 \caption{The Feynman graph for the tree-level decay, $N_k\rightarrow \nu_j\,\gamma$ via the dimension-5 EMDM coupling of Eq.~(\ref{eq4:BKL_toy5D_main_lag}). Here $q=p-p'$ and $(V_{\lambda 5})_{jk}^\rho \equiv  2\lambda_{jk}\,P_R\,\sigma^{\alpha\rho}\,q_{\alpha}$ is the vertex factor.}
 \label{fig4:toy5D_tree}
\end{figure}

To ascertain whether electromagnetic leptogenesis in this model is possible, the key quantity of interest is the $CP$ asymmetry in the decay of $N_k$ (c.f. (\ref{eq2:LV_fcp1}) on page~\pageref{eq2:LV_fcp1}),
\begin{equation}
 \varepsilon_{k, j}^{(5)} = \frac{
 \Gamma_{(N_k\rightarrow \nu_j\,\gamma)}
 -\Gamma_{(N_k\rightarrow \overline{\nu}_j\,\gamma)}}
 {\Gamma_{(N_k\rightarrow \nu\,\gamma)}
 +\Gamma_{(N_k\rightarrow \overline{\nu}\,\gamma)}}\;,
 \label{eq4:BKL_toy5D_CP_defn}
\end{equation}
where $\Gamma_{(N_k\rightarrow \nu\,\gamma)} \equiv \sum_j
\Gamma_{(N_k\rightarrow \nu_j\,\gamma)}$ denotes the decay rate (summed over final state flavor~$j$). So with this in mind, we begin by calculating the lowest order contribution to the decay rate, $\Gamma_{(N_k\rightarrow \nu\,\gamma)}$. The tree-level diagram for this process is depicted in Fig.~\ref{fig4:toy5D_tree}. Using the Feynman rules developed in Secs.~\ref{subsec:BKL_simp_Maj-rules} and \ref{subsec:BKL_std_rules_results}, we can immediately write down the amplitude for the lowest order process as
\begin{align}
 -i\mmat &= \ubarj (2\lambda_{jk} P_R \sigma ^{\alpha\rho} q_\alpha)
  u_k^c \varepsilon_\rho^* \;,
  \\
\Rightarrow\qquad
 |\mmat|^2
  &= \ubarj (2 \lambda_{jk} P_R \sigma ^{\alpha\rho} q_\alpha)
  u_k^c \varepsilon_\rho^*
 \left[
  \ubarj (2 \lambda_{jk} P_R \sigma ^{\beta\sigma} q_\beta)
  u_k^c \varepsilon_\sigma^*\right]^\dagger
   \;, \nonumber \\
  &= 4(\lambda_{jk}^* \lambda_{jk}) \ubarj  P_R \sigma ^{\alpha\rho} q_\alpha
  u_k^c \varepsilon_\rho^* 
  \varepsilon_\sigma (-u_k^T C^\dagger \gammazero)\gammazero \sigma ^{\beta\sigma} q_\beta
  \gammazero P_R \gamma^0 u_j
   \;, \nonumber  \\
  &= -4(\lambda_{jk}^* \lambda_{jk}) \ubarj  P_R 
  \, \frac{i}{2}\left[\gamma^\alpha, \gamma^\rho\right]
   q_\alpha
  C \ubark^T  u_k^T C^\dagger 
  \, \frac{i}{2}\left[\gamma^\beta, \gamma^\sigma\right] q_\beta
  P_L u_j
  \,\varepsilon_\rho^* \varepsilon_\sigma
  \;.   
\end{align}
Averaging initial and summing final polarizations, we obtain
\begin{align}
 |\overline{\mmat}|^2
   &= (\lambda_{jk}^* \lambda_{jk}) \, P_R\, 
      (\qslash\gamma^\rho - \gamma^\rho\qslash)
      C\left[
       \frac{1}{2}\sum_s u_k \ubark\right]^T C^\dagger
      (\qslash\gamma^\sigma - \gamma^\sigma\qslash) 
       P_L\, \sum_{s'}u_j\ubarj 
       \sum_\text{pol} \varepsilon_\rho^* \varepsilon_\sigma \;,\nonumber
    \\
   &= \frac{1}{2}(\lambda_{jk}^* \lambda_{jk})\; 
   \text{Tr}\left[
       P_R\, 
      (\qslash\gamma^\rho - \gamma^\rho\qslash)
      (-\pslash +\mNk)
      (\qslash\gamma^\sigma - \gamma^\sigma\qslash) 
       P_L\, \ppslash\,
       (-g_{\rho\sigma})
       \right] \;, \nonumber\\
 &\hspace{140pt}\vdots\nonumber\\          
    &=(\lambda_{jk}^* \lambda_{jk})\left[
    16 (p \cdot q)(p' \cdot q)-4(p \cdot p')(q \cdot q)
   \right] 
       \;.\label{eq4:BKL_toy5D_treeM_final1}
\end{align}
where we have taken the masses of the light neutrino and photon to be zero.

Working in the centre-of-mass frame where
\begin{gather}
 p =(M_k\;,\;\vec 0)\;,\quad
 p' = (M_k/2\;,\;-\vec q)\;,\quad
 q = (M_k/2\;,\;\vec q)\;, \quad
 |\vec q| = M_k/2\;, \quad \text{ and }\nonumber\\
 p\cdot p' = p\cdot q= p' \cdot q=M_k^2/2\;,\quad
 p^2 \equiv (p \cdot p) = \mNk^2\;,\quad
 q^2 = (p')^2  =0\;,
  \label{eq4:BKL_toy5D_momenta}
\end{gather}
Eq.~(\ref{eq4:BKL_toy5D_treeM_final1}) becomes 
\begin{equation}\label{eq4:BKL_toy5D_treeM_final2}
|\overline{\mmat}|^2 
 = 4\,(\lambda_{jk}^*\lambda_{jk})\,\mNk^4\;.
\end{equation}
~
\noindent 
Hence, the tree-level decay rate for $N_k\rightarrow \nu\,\gamma$ is given by
\begin{align}
 \Gamma(N_k\rightarrow \nu\,\gamma)
  &=\frac{|\vec q|}{8\pi E_\text{cm}^2} |\overline{\mathcal{M}}|^2 \;,\nonumber\\
  &= \frac{1}{8\pi}\frac{M_{k}}{2}\frac{1}{M_{k}^2}\;
    4\,(\lambda^\dagger \lambda)_{kk}\,\mNk^4
     \;,\nonumber\\
  &= \frac{(\lambda^\dagger \lambda)_{kk}}{4\pi} \, M_{k}^3
  \equiv   \frac{(\lambda_0^\dagger \lambda_0)_{kk}}{4\pi} \, \frac{M_{k}^3}{\Lambda^2}
  \;,
   \label{eq4:BKL_toy5D_decay_rate}
\end{align}
where we have summed over $j$. Since we must necessarily have
$\Gamma(N_k\rightarrow \nu\,\gamma) \equiv \Gamma(N_k\rightarrow \overline{\nu}\,\gamma)$, the total decay rate is, $\Gamma_\text{tot} = 2\Gamma(N_k\rightarrow \nu\,\gamma)$, to first order.

\begin{figure}[h]
\begin{center}
\includegraphics[width=0.75\textwidth]{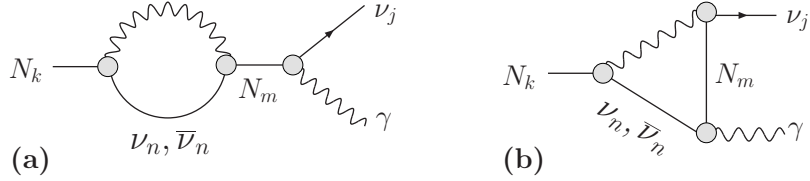}
\end{center} 
 \caption{{\bf (a)} Self-energy and {\bf (b)} vertex diagrams which contribute to
  the $CP$ asymmetry of $N_k$ decay via the dimension-5 EMDM coupling of Eq.~(\ref{eq4:BKL_toy5D_main_lag}). Note that since weak isospin is not conserved in
  this model, both $\nu_n$ and $\overline{\nu}_n$ are allowed in the
  loop of (b), unlike in standard leptogenesis. For simplicity, we have not drawn the arrow for the $\nu_n$ propagators as the two cases point in opposite directions.}
 \label{fig4:toy5D_1-loop_graphs}
\end{figure}

Next, we calculate the interference terms between the tree-level process of Fig.~\ref{fig4:toy5D_tree} and the one-loop diagrams with on-shell intermediate states depicted in Fig.~\ref{fig4:toy5D_1-loop_graphs}. Note that unlike in standard leptogenesis, 
both $\nu_n$ and $\overline{\nu}_n$ are allowed to propagate in the internal loop of the vertex correction graph (Fig.~\ref{fig4:toy5D_1-loop_graphs}b). This is because weak isospin is not conserved by the interaction Lagrangian (\ref{eq4:BKL_toy5D_main_lag}). As a result, one potentially has four interference terms (in the fully flavored regime where final flavor $j$ is not summed) contributing to the $CP$ asymmetry at leading order~\footnote{For standard leptogenesis, there are only three interference terms (before summing over final flavor $j$). One from the vertex correction shown in Sec.~\ref{subsec:BKL_maj_feyn_examples} and two from the self-energy correction discussed in Appendix~\ref{app:c_std_eg}.}. We shall discuss each of these in turn.

\begin{figure}[t]
\begin{center}
\includegraphics[width=0.80\textwidth]{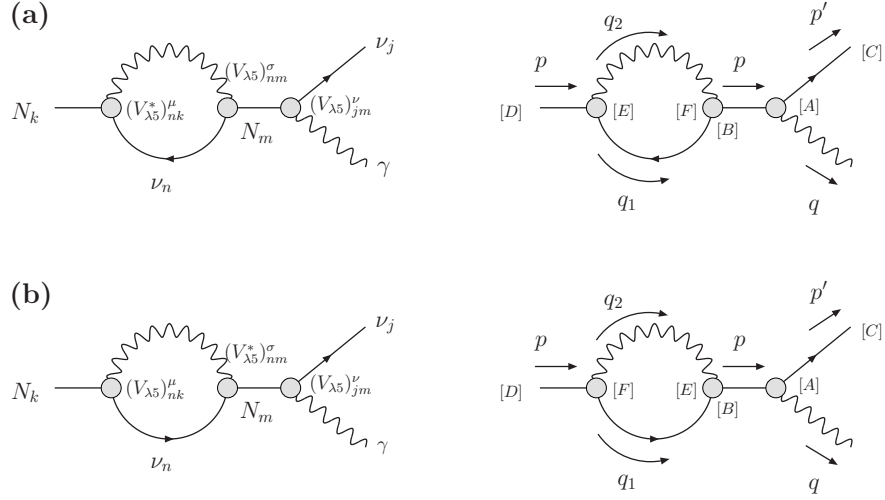}
\end{center} 
 \caption{Self-energy diagrams which contribute to the $CP$ asymmetry of $N_k$ decay via the dim-5 EMDM coupling with {\bf (a)} $\overline{\nu}_n$ and {\bf (b)} $\nu_n$ intermediate states respectively. LEFT: $(V_{\lambda 5})_{ab}^\rho \equiv  2\lambda_{ab}\,P_R\,\sigma^{\alpha\rho}\,q'_{\alpha} $ and $(V_{\lambda 5}^*)_{ab}^\rho \equiv  -2\lambda_{ab}^*\,C^\dagger \,\sigma^{\alpha\rho}\,q'_{\alpha}\,P_L $ denote the vertex factors. The corresponding diagram with momenta and spinor indices labeled for each case is displayed on the RIGHT.}
 \label{fig4:toy5D_sloop_diagrams}
\end{figure}

Firstly, let us consider the self-energy contributions. The two distinct graphs for these are shown in Fig.~\ref{fig4:toy5D_sloop_diagrams}a and Fig.~\ref{fig4:toy5D_sloop_diagrams}b respectively.  Applying the Feynman rules developed for the EMDM couplings, we can write down the interference term of Fig.~\ref{fig4:toy5D_sloop_diagrams}a and Fig.~\ref{fig4:toy5D_tree} as
\begin{align}
 I^\text{5D}_\text{self-(a)}
 &= \int \frac{d^4 q_1}{(2\pi)^4}\;
  (16 \lambda_{jk}^*  \lambda_{jm}  \lambda_{nm} \lambda_{nk}^*)
 \left[\ubarj\right]_{1C} 
 \left[P_R \sigma^{\alpha\nu} q_{\alpha}\right]_{CA} 
 \left[S_{N_m}(p)\right]_{AB}
 \left[P_R \sigma^{\beta\sigma} (-q_{2\beta})\right]_{FB}
 \nonumber\\
 & \;\;\times
 \left[S_\ell(-q_1)\right]_{EF} 
 \left[C^\dagger \sigma^{\delta\mu} q_{2\delta} P_L\right]_{DE} 
 \left[u_k^c\right]_{D1} 
 \left[D_{\sigma\mu}(q_2) \varepsilon^*_\nu\right]_{11} 
 \left[-u_k^T C^\dagger \sigma^{\eta\rho} q_{\eta} P_L u_j \varepsilon_\rho\right]_{11}
 \;,
 \nonumber
\end{align}
where the $-q_{2\beta}$ in $\left[\cdots\right]_{FB}$ comes from the fact that photon momentum,  $q_{2\beta}$ is flowing \emph{into} the vertex. Letting
$A_{\lambda}^{(5)} = \lambda_{jk}^* \lambda_{jm} \lambda_{nm} \lambda_{nk}^*$ and using matrix form, we then have
\begin{align}
  I^\text{5D}_\text{self-(a)}
 &=
  16 A_{\lambda}^{(5)} 
  \int\frac{d^4 q_1}{(2\pi)^4}
  \frac{
  \ubarj\, P_R \sigma^{\alpha\nu} q_{\alpha}
  (-i)(\pslash+\mNm)C\,
  (-\sigma^{\beta\sigma} q_{2\beta})^T P_R^T
  (i)(-\qoneslash)^T
   }
  {(p^2-M_m^2 +i\epsilon)(q_1^2+i\epsilon)(q_2^2+i\epsilon)} \nonumber\\
 &\qquad\qquad\qquad\times
  P_L^T (\sigma^{\delta\mu} q_{2\delta})^T C^* u_k^c
  (-i) g_{\sigma\mu}\, \varepsilon^*_\nu \varepsilon_\rho
  (-1)u_k^T C^\dagger \sigma^{\eta\rho} q_{\eta} P_L u_j \;, \nonumber\\
 &=
  8 i\,A_{\lambda}^{(5)}
  \int\frac{d^4 q_1}{(2\pi)^4}
  \frac{
  P_R \sigma^{\alpha\nu} q_{\alpha}
  (\pslash+M_m) 
  \sigma^{\beta\sigma} q_{2\beta}  P_R
  \qoneslash
  P_L   \sigma^{\delta\mu} q_{2\delta}
  C \left[\sum_s u_k \ubark\right]^T C^\dagger 
  }  
  {(p^2-M_m^2 +i\epsilon)(q_1^2+i\epsilon)(q_2^2+i\epsilon)}
  \nonumber\\
 &\qquad\qquad\qquad\times
 \sigma^{\eta\rho} q_{\eta} P_L g_{\sigma\mu}
  \sum_{s'} u_j\ubarj
  \, \sum_\text{pol}\varepsilon^*_\nu \varepsilon_\rho 
   \;,
 \nonumber\\
 &\hspace{140pt}\vdots\nonumber\\
 &=
  -\frac{i\,A_{\lambda}^{(5)}\mNm\mNk}{2}
  \int\frac{d^4 q_1}{(2\pi)^4}
  \frac{1}
  {(p^2-M_m^2 +i\epsilon)(q_1^2+i\epsilon)(q_2^2+i\epsilon)}
  \nonumber\\
 &\qquad\qquad\qquad\times
   \text{Tr}\left[
  P_R (\qslash\gamma^\nu - \gamma^\nu\qslash)
  (\qtwoslash\gamma^\sigma - \gamma^\sigma\qtwoslash)
   \qoneslash
   (\qtwoslash\gamma_\sigma - \gamma_\sigma\qtwoslash)
  (\qslash\gamma_\nu - \gamma_\nu\qslash) 
  \ppslash\right]
    \;,\nonumber\\
 &=
 i \,A_{\lambda}^{(5)} \mNm\mNk
  \int\frac{d^4 q_1}{(2\pi)^4}
  \frac{
  (p' \cdot q)\left[  
  -256 (q \cdot q_2)(q_1 \cdot q_2)
  +64 (q \cdot q_1) q_2^2
  \right]}
  {(p^2-M_m^2 +i\epsilon)(q_1^2+i\epsilon)(q_2^2+i\epsilon)}
   \;.\label{eq4:BKL_toy5D_self_a_used_in_6D}
\end{align}
The discontinuity of the integral
\begin{equation}
 I_\text{s-(a)}^\text{5D}
 \equiv
 i \, \mNm\mNk
  \int\frac{d^4 q_1}{(2\pi)^4}
  \frac{
  (p' \cdot q)\left[  
  -256 (q \cdot q_2)(q_1 \cdot q_2)
  +64 (q \cdot q_1) q_2^2
  \right]}
  {(p^2-M_m^2 +i\epsilon)(q_1^2+i\epsilon)(q_2^2+i\epsilon)}
   \;,
\end{equation}
may be determined by the cutting rules as described before, hence
\begin{align}
 \text{Disc}\left[I_\text{s-(a)}^\text{5D}\right]
 &=
 i\,\mNk\mNm \int\frac{d^4 q_1}{(2\pi)^4}
 \frac{
   (-2\pi i)^2
   \delta(q_1^2)
   \delta\left[(p-q_1)^2\right]\Theta(E_{1})\Theta(\mNk - E_{1})
   }
  {p^2-M_m^2} \nonumber\\
 &\qquad\qquad\qquad\qquad\times
 \mNk^2
 \left[  
 -128 (q \cdot q_2)(q_1 \cdot q_2)
  +32 (q \cdot q_1) q_2^2
 \right]\;, 
 \qquad (\epsilon \rightarrow 0)\;.\nonumber
\end{align}
Using $q_1 =(E_1,\vec q_1)$, $q_2=p-q_1$ and (\ref{eq4:BKL_toy5D_momenta}) to simplify, we eventually get
\begin{align}
 \text{Disc}\left[I_\text{s-(a)}^\text{5D}\right]
 &=
  \frac{-i\,\mNk^4\mNm}{4\pi^2 (\mNk^2-\mNm^2)} 
 \int d^3 q_1 dE_1
   \,\frac{1}{2\modq1}
   \delta(E_1-\modq1)
   \delta\left[(\mNk -E_1)^2-\modq1^2\right]\Theta(E_{1})
   \nonumber\\
 &\qquad
 \times
 \Theta(\mNk - E_{1})
 \left[  
 -64 \left(\mNk-E_1 + \modq1\costh\right)
      (\mNk E_1-E_1^2 +\modq1^2)
 \right.\nonumber\\
 &\qquad\qquad\qquad\qquad\qquad
 \left.  
  +16 \left(E_1 - \modq1\costh\right)
      \left((\mNk -E_1)^2-\modq1^2\right)
 \right]\;, \nonumber\\
\intertext{where $\theta$ is the smaller angle between $\vec q_1$ and $\vec q$. Performing the integrals using all the standard tricks, we obtain}
 \text{Disc}\left[I_\text{s-(a)}^\text{5D}\right]
 &=
 \frac{-i\,\mNk^4\mNm}{8\pi^2 (\mNk^2-M_m^2)} 
 \int \modq1^2 d\modq1 d\Omega \;
   \delta\left[\mNk^2-2\mNk\modq1\right]\Theta(\mNk - \modq1)
   \nonumber\\
 &\qquad
 \times
 \left[  
 -64 \left(\mNk-\modq1 + \modq1\costh\right)
      (\mNk)
  +16 \left(1 - \costh\right)
      \left(\mNk^2-2\mNk\modq1\right)
 \right]\;, \nonumber \\
 &\hspace{140pt}\vdots\nonumber\\
   &=
 \frac{-i\,\mNk^4\mNm}{16\pi^2 (\mNk^2-M_m^2)} 
 \int d\Omega \; \frac{\mNk^2}{4} 
 \left[  
 -64 \left(\halfmNk + \halfmNk\costh\right)
  +16 \left(1 - \costh\right)
      \right]\;, \nonumber  \\
  &=
 \frac{2i\,\mNk^7\mNm}{\pi (\mNk^2-M_m^2)}
  \;.
\end{align} 
The imaginary part of this interference term and its corresponding phase space, $V_\varphi$ are given by
\begin{equation}
 \text{Im}\left[I_\text{s-(a)}^\text{5D}\right]
 =
 \frac{\mNk^7\mNm}{\pi (\mNk^2-m_{Nm}^2)}
  \;,
  \qquad
  V_\varphi
   =  \frac{|\vec q|}{8\pi E_\text{cm}^2} 
   = \frac{1}{16\pi \mNk}\;.
\end{equation}
Note that unlike standard leptogenesis, there are no extra factors of 2 in the phase space for this diagram because only one intermediate (and final) state is possible. Putting everything together, the $CP$ asymmetry due to this interference term is
\begin{align}
 \varepsilon_{\text{self-(a)-}k,j}^\text{5D}
 &= 
 -\frac{4}{\Gamma_\text{tot}} \;
 \sum_{m\neq k}\sum_{n}
 \text{Im}\left[A_{\lambda}^{(5)}\right]\;
    \text{Im}\left[I_\text{s-(a)}^\text{5D} V_\varphi\right] \;, 
 \nonumber\\
 &=
  -\frac{\mNk^2}{2\pi(\lambda^\dagger \lambda)_{kk}}
  \sum_{m\neq k}
  \text{Im} \left[\lambda_{jk}^* \lambda_{jm} (\lambda^\dagger \lambda)_{km}\right]
      \frac{\sqrt{z}}{1-z}
\;,\label{eq4:BKL_toy5D_self_a_final}
\end{align}
where $z \equiv \mNm^2/\mNk^2$. 

Similarly, one can write down the amplitude due to Fig.~\ref{fig4:toy5D_sloop_diagrams}b~\footnote{Note that the direction of fermion-flow through $N_m$ is unambiguous in this case (c.f. Fig.~\ref{figC:std_cpsloop_cal_graphs_2} in Appendix~\ref{app:c_std_eg}).},
\begin{align}
 I^\text{5D}_\text{self-(b)}
 &= \int \frac{d^4 q_1}{(2\pi)^4}\;
  16\lambda_{jk}^* \lambda_{jm}\lambda_{nm}^*\lambda_{nk}
 \left[\ubarj\right]_{1C} 
 \left[P_R \sigma^{\alpha\nu} q_{\alpha}\right]_{CA} 
 \left[S_{N_m}(p)\right]_{AB}
 [C^\dagger \sigma^{\beta\sigma} (-q_{2\beta}) P_L]_{BE}
 \nonumber
 \\%
 & \;\;\times%
 \left[S_\ell (q_1)\right]_{EF} 
 \left[P_R\sigma^{\delta\mu} q_{2\delta}\right]_{FD} 
 \left[u_k^c\right]_{D1} 
 \left[D_{\sigma\mu}(q_2) \varepsilon^*_\nu\right]_{11}
 \left[-u_k^T C^\dagger \sigma^{\eta\rho} q_{\eta} P_L u_j\, \varepsilon_\rho\right]_{11}
 \;,\label{eq4:BKL_toy5D_self_b_start}
\end{align}
from which the following contribution to the $CP$ asymmetry is deduced (see Appendix~\ref{app:c_emdm_cal} for the full calculation):
\begin{equation}\label{eq4:BKL_toy5D_self_b_final}
 \varepsilon_{\text{self-(b)-}k,j}^\text{5D}
 =
  -\frac{\mNk^2}{2\pi(\lambda^\dagger \lambda)_{kk}}
  \sum_{m\neq k}
  \text{Im} \left[\lambda_{jk}^* \lambda_{jm} (\lambda^\dagger \lambda)_{mk}\right]
   \frac{1}{1-z}
  \;,
  \qquad z \equiv \frac{\mNm^2}{\mNk^2}
 \;.
\end{equation}
%

\begin{figure}[t]
\begin{center}
 \includegraphics[width=0.70\textwidth]{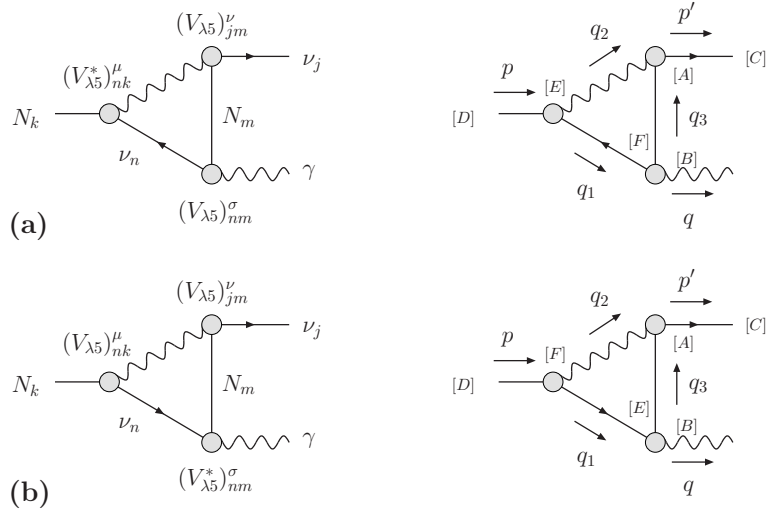}
\end{center} 
 \caption{Vertex corrections which contribute to the $CP$ asymmetry of $N_k$ decay via the dim-5 EMDM coupling with {\bf (a)} $\overline{\nu}_n$ and {\bf (b)} $\nu_n$ intermediate states respectively. LEFT: $(V_{\lambda 5})_{ab}^\rho \equiv  2\lambda_{ab}\,P_R\,\sigma^{\alpha\rho}\,q'_{\alpha} $ and $(V_{\lambda 5}^*)_{ab}^\rho \equiv  -2\lambda_{ab}^*\,C^\dagger \,\sigma^{\alpha\rho}\,q'_{\alpha}\,P_L $ denote the vertex factors. The corresponding diagram with momenta and spinor indices labeled for each case is displayed on the RIGHT.}
 \label{fig4:toy5D_vloop_diagrams}
\end{figure}

For the vertex interference, we study the graphs of Fig.~\ref{fig4:toy5D_vloop_diagrams}. Comparing them with the processes depicted in Fig.~\ref{fig4:BKL_std_cpvloop_cal_graphs}, it can be seen that the analogue of Fig.~\ref{fig4:toy5D_vloop_diagrams}b is noticeably absent in the standard scenario as we have alluded to earlier. However, the existence of such extra contribution is not expected to alter the overall $CP$ asymmetry in a significant way. This is because in a generic study of lepton asymmetry from heavy particle decays \cite{Liu:1993tg} where the analogous diagram to Fig.~\ref{fig4:toy5D_vloop_diagrams}b was included, the resulting loop function tends to zero when the squared-mass ratio, $z\gg 1$. Moreover, in the case of $z \simeq 1$, the self-energy contribution will overwhelm any vertex corrections due to resonant enhancement. In any case, upon summing over $j$ (the one-flavor approximation), such contribution will completely vanish. Therefore, one can expect the effect from Fig.~\ref{fig4:toy5D_vloop_diagrams}b to be sub-dominant.
Nevertheless, we calculate both terms here for completeness and to gain better insight, so that when it comes to the more realistic setup in Sec.~\ref{subsec:BKL_toy_model_6D}, we shall know what to expect from those inevitably more complicated diagrams even without explicitly computing all of them. 

To this end, we begin by turning to Fig.~\ref{fig4:toy5D_vloop_diagrams}a and write down the amplitude using our simplified Feynman rules:
\begin{align}
I^\text{5D}_\text{vert-(a)}
 &= \int \frac{d^4 q_1}{(2\pi)^4}\;
  16\lambda_{jk}^* \lambda_{jm} \lambda_{nm} \lambda_{nk}^*
 \left[\ubarj\right]_{1C} 
 \left[P_R \sigma^{\alpha\nu} (-q_{2\alpha})\right]_{CA} 
 \left[S_{N_m}(q_3)\right]_{AB}
      [P_R \sigma^{\beta\sigma} q_{\beta}]_{FB}
 \nonumber\\
 & \;\;\times
  \left[S_\ell(-q_1)\right]_{EF} 
 \left[C^\dagger \sigma^{\delta\mu} q_{2\delta} P_L\right]_{DE} 
 \left[u_k^c\right]_{D1} 
 \left[D_{\nu\mu}(q_2) \varepsilon^*_\sigma\right]_{11} 
 \left[[-u_k^T C^\dagger \sigma^{\eta\rho} q_{\eta} 
   P_L u_j \varepsilon_\rho\right]_{11}\;,
   \label{eq4:BKL_toy5D_vert_a_start}\\
  &=
  16 A_{\lambda}^{(5)} 
    \int\frac{d^4 q_1}{(2\pi)^4}
  \frac{
  \ubarj\, P_R \sigma^{\alpha\nu} (-q_{2\alpha})
  (-i)(\qthreeslash+\mNm)C\,
  (\sigma^{\beta\sigma} q_{\beta})^T P_R^T
  (i)(-\qoneslash)^T
  P_L^T 
  }
  {(q_3^2-\mNm^2 +i\epsilon)(q_1^2+i\epsilon)(q_2^2+i\epsilon)} \nonumber\\
 &\qquad\qquad\qquad\times
  (\sigma^{\delta\mu} q_{2\delta})^T 
  C^*  u_k^c
  (-i) g_{\nu\mu}\, \varepsilon^*_\sigma \varepsilon_\rho
  (-1)u_k^T C^\dagger \sigma^{\eta\rho} q_{\eta} P_L u_j \;, \nonumber\\
&\hspace{140pt}\vdots\nonumber\\
  &= \frac{-i A_{\lambda}^{(5)} \mNm\mNk}{2}
  \int\frac{d^4 q_1}{(2\pi)^4}
  \frac{1}
  {(q_3^2-\mNm^2 +i\epsilon)(q_1^2+i\epsilon)(q_2^2+i\epsilon)}
 \nonumber\\ 
  &\qquad\qquad\qquad \times
  \text{Tr}\left[
  P_R (\qtwoslash\gamma^\nu - \gamma^\nu\qtwoslash)
  (\qslash\gamma^\sigma - \gamma^\sigma\qslash)
   \qoneslash
   (\qtwoslash\gamma_\nu - \gamma_\nu\qtwoslash)
  (\qslash\gamma_\sigma - \gamma_\sigma\qslash) 
  \ppslash\right]
   \;,\nonumber\\
 &=
  i A_{\lambda}^{(5)}
  \int\frac{d^4 q_1}{(2\pi)^4}
  \frac{
  \mNm\mNk
  }
  {(q_3^2-\mNm^2 +i\epsilon)(q_1^2+i\epsilon)(q_2^2+i\epsilon)}
   \; 
    \left\{
    -32 \, \mNk^2 (q \cdot q_1) q_2^2
    \right.\nonumber\\
 &\left.\qquad\qquad
 +64 (q \cdot q_2)\left[
  -2 (p' \cdot q_1)(q \cdot q_2)
  +2 (p' \cdot q_2)(q \cdot q_1)
  + \mNk^2(q_1 \cdot q_2)
  \right]
  \right\}\;.\label{eq4:BKL_toy5D_vert_a_middle}
\end{align}
In deriving the above, we have used (\ref{eq4:BKL_toy5D_momenta}). Although (\ref{eq4:BKL_toy5D_vert_a_middle}) is considerably more complicated than any of the expressions we have tackled up to this point, the method (and tricks) required to evaluate it are the same (see the vertex example in Sec.~\ref{subsec:BKL_maj_feyn_examples}). Eventually, one obtains~\footnote{See Appendix~\ref{app:c_emdm_cal} for the full workings of this.}
\begin{equation}\label{eq4:BKL_toy5D_vert_a_final}
 \varepsilon_{\text{vert-(a)-}k,j}^\text{5D}
 =
  \frac{-\mNk^2}{2\pi(\lambda^\dagger \lambda)_{kk}}
  \sum_{m\neq k}
  \text{Im} \left[\lambda_{jk}^* \lambda_{jm} (\lambda^\dagger \lambda)_{km}\right]
  \sqrt{z}\left[1+2z\left(1-(z+1)\ln \left[\frac{z+1}{z}\right]\right)\right]
  \;,
\end{equation}
where $z \equiv \mNm^2/\mNk^2$ as usual.

Next, we repeat the procedure for the term corresponding to Fig.~\ref{fig4:toy5D_vloop_diagrams}b, which gives
\begin{align}
I^\text{5D}_\text{vert-(b)}
 &= 
  \int \frac{d^4 q_1}{(2\pi)^4}\;
  16\lambda_{jk}^* \lambda_{jm} \lambda_{nm}^* \lambda_{nk}
 [\ubarj]_{1C} 
 [P_R \sigma^{\alpha\nu} (-q_{2\alpha})]_{CA} 
 [S_{N_m}(q_3)]_{AB}
 [C^\dagger \sigma^{\beta\sigma} q_{\beta} P_L]_{BE}
\nonumber\\
 &\quad\times
 \left[S_\ell (q_1)\right]_{EF} 
 \left[P_R\sigma^{\delta\mu} q_{2\delta}\right]_{FD} 
 \left[u_k^c\right]_{D1} 
 \left[D_{\nu\mu}(q_2) \varepsilon^*_\sigma\right]_{11} 
 \left[-u_k^T C^\dagger \sigma^{\eta\rho} q_{\eta} P_L u_j\, \varepsilon_\rho\right]_{11}
 \;,
 \nonumber\\
 &\hspace{140pt}\vdots\nonumber\\
 &=
  B_{\lambda}^{(5)}
  \!\int\!\frac{d^4 q_1}{(2\pi)^4}
  \frac{  
  i \text{Tr}[
  P_R (\qtwoslash\gamma^\nu - \gamma^\nu\qtwoslash)
  \qthreeslash
  (\qslash\gamma^\sigma - \gamma^\sigma\qslash)
   \qoneslash
   (\qtwoslash\gamma_\nu - \gamma_\nu\qtwoslash)
   \pslash
  (\qslash\gamma_\sigma - \gamma_\sigma\qslash) 
  \ppslash ]}
  {2\,(q_3^2-\mNm^2 +i\epsilon)(q_1^2+i\epsilon)(q_2^2+i\epsilon)}
   ,\nonumber\\
 &= B_{\lambda}^{(5)}
     \int\frac{d^4 q_1}{(2\pi)^4}
  \frac{1}
  {(q_3^2-\mNm^2 +i\epsilon)(q_1^2+i\epsilon)(q_2^2+i\epsilon)}  \nonumber\\
 &\quad\times
 \left\{
  80i  (p \cdot q_3)(p' \cdot q_1)q^2 q_2^2
  -16i (p \cdot q_1)(p' \cdot q_3)q^2 q_2^2
  -16i (p \cdot p')(q_1 \cdot q_3)q^2 q_2^2
 \right.
  \nonumber\\
 &\quad\quad\; 
    +64 \LCtensor p'_{\mu}q_{\nu}q_{2\rho}q_{3\sigma}
    \left[(p \cdot q_2)(q \cdot q_1)-(p \cdot q)(q_1 \cdot q_2)\right]
 \nonumber\\
 &\quad\quad\;
     -64 \LCtensor p_{\mu}p'_{\nu}q_{1\rho}q_{3\sigma}(q \cdot q_2)^2
    +64 \LCtensor p'_{\mu}q_{\nu}q_{1\rho}q_{3\sigma}
    \left[(p \cdot q)q_2^2 -(p \cdot q_2)(q \cdot q_2)\right]
 \nonumber\\
 &\quad\quad\;
    +64 \LCtensor p'_{\mu}q_{1\nu}q_{2\rho}q_{3\sigma}(p \cdot q)(q \cdot q_2)
    +64 \LCtensor p_{\mu}p'_{\nu}q_{1\rho}q_{2\sigma}(q \cdot q_2)(q \cdot q_3)
 \nonumber\\
 &\quad\quad\;
    +64 \LCtensor p'_{\mu}q_{\nu}q_{1\rho}q_{2\sigma}
    \left[(p \cdot q_2)(q \cdot q_3) -(p \cdot q)(q_2 \cdot q_3)\right]
 \nonumber\\
 &\quad\quad\;
    +64 \LCtensor p_{\mu}p'_{\nu}q_{\rho}q_{2\sigma}
    \left[(q \cdot q_1)(q_2 \cdot q_3) -(q \cdot q_3)(q_1 \cdot q_2)\right]
 \nonumber\\
 &\quad\quad\;
    +64 \LCtensor p_{\mu}p'_{\nu}q_{\rho}q_{3\sigma}
    \left[(q \cdot q_2)(q_1 \cdot q_2) -(q \cdot q_1)q_2^2\right]
 \nonumber\\
 &\quad\quad\;
    +64 \LCtensor p_{\mu}p'_{\nu}q_{2\rho}q_{3\sigma}(q \cdot q_1)(q \cdot q_2)
 \nonumber\\
 &\quad\quad\;\left.
    +64 \LCtensor p_{\mu}p'_{\nu}q_{\rho}q_{1\sigma}
    \left[(q \cdot q_3)q_2^2 -(q \cdot q_2)(q_2 \cdot q_3)\right]
  \right\}
  \;,     \label{eq4:BKL_toy5D_vert_b_middle}
\end{align}
where $B_{\lambda}^{(5)} \equiv \lambda_{jk}^* \lambda_{jm} \lambda_{nm}^* \lambda_{nk}$ and $\LCtensor$ is the Levi-Civita tensor. An interesting observation from the final expression in (\ref{eq4:BKL_toy5D_vert_b_middle}) is that it is actually evaluate to \emph{zero}, thus
\begin{equation}\label{eq4:BKL_toy5D_vert_b_final}
 I^\text{5D}_\text{vert-(b)} =0\;.
\end{equation}
This comes about because, in the numerator, terms involving $\LCtensor$ are all contracted with 
four-momenta from the list, $\left\{p,p',q,q_1,q_2,q_3\right\}$ where only three are independent of each other, whilst the remaining terms are automatically zero for they are proportional to $q^2 \equiv 0$ (the photon is massless). As a result, even before other considerations, there cannot be any contribution to the final $CP$ asymmetry arising from the interference with the graph shown in Fig.~\ref{fig4:toy5D_vloop_diagrams}b.
So putting results (\ref{eq4:BKL_toy5D_self_a_final}), (\ref{eq4:BKL_toy5D_self_b_final}) and (\ref{eq4:BKL_toy5D_vert_a_final}) together, the total $CP$ asymmetry for the decay of $N_k\rightarrow \nu_j\,\gamma$ in this EMDM toy model is
\begin{equation}\label{eq4:BKL_toy5D_CP_total}
 \varepsilon_{k,j}^\text{5D}
  = \frac{-\mNk^2}{2\pi(\lambda^\dagger \lambda)_{kk}}
  \sum_{m\neq k}
  \text{Im} \!\!\left[\lambda_{jk}^* \lambda_{jm} 
  \left\{
  (\lambda^\dagger \lambda)_{km}
   \left[
   f_{Va}(z)
   +
   f_{Sa}(z)
   \right]
   +
   (\lambda^\dagger \lambda)_{mk}\;
   f_{Sb}(z)
   \right\}   
   \right]\!,
\end{equation}
or
\begin{equation}\label{eq4:BKL_toy5D_CP_total_2}
 \varepsilon_{k,j}^\text{5D}
  = \frac{-(\mNk/\Lambda)^2}{2\pi(\lambda_0^\dagger \lambda_0)_{kk}}
  \sum_{m\neq k}
  \text{Im}\!\! \left[(\lambda_0^*)_{jk} (\lambda_0)_{jm} 
  \left\{
  (\lambda_0^\dagger \lambda_0)_{km}
   \left[
   f_{Va}(z)
   +
   f_{Sa}(z)
   \right]
   +
   (\lambda_0^\dagger \lambda_0)_{mk}\;
   f_{Sb}(z)
   \right\}   
   \right]\!,
\end{equation}   
where $z = \mNm^2/\mNk^2$ and
\begin{equation}\label{eq4:BKL_toy5D_CP_loop_fns}
 f_{Va}(z) =\sqrt{z}\left[1+2z\left(1-(z+1)\ln 
               \left[\frac{z+1}{z}\right]\right)\right]\;,\;\;
 f_{Sa}(z) =\frac{\sqrt{z}}{1-z} \;,\;\;
 f_{Sb}(z) =\frac{1}{1-z}\;.
\end{equation}
In the limit of hierarchical heavy RH neutrinos and considering the $N_1$-dominated scenario (i.e. $k=1$), we have $z \gg 1$ and
\begin{equation}\label{eq4:BKL_toy5D_CP_loop_fns_approx}
 f_{Va}(z) \simeq \frac{1}{3\sqrt{z}}\;, 
 \qquad
 f_{Sa}(z) \simeq -\frac{1}{\sqrt{z}}\;, 
 \qquad
 f_{Sb}(z) \simeq -\frac{1}{z}\;.
\end{equation}
Therefore,
\begin{equation}\label{eq4:BKL_toy5D_CP_total_N1_approx}
 \varepsilon_{1,j}^\text{5D} 
   \simeq
    \frac{(M_1/\Lambda)^2}{2\pi(\lambda_0^\dagger \lambda_0)_{11}}
  \sum_{m\neq 1}
  \text{Im} \left[(\lambda_0^*)_{j1} (\lambda_0)_{jm} 
  \left\{
  \frac{2}{3\sqrt{z}}\;(\lambda_0^\dagger \lambda_0)_{1m}
   +
   \frac{1}{z}\;(\lambda_0^\dagger \lambda_0)_{m1}
   \right\}   
   \right]\;.
\end{equation}
This expression is almost identical  to the corresponding result from standard leptogenesis~\cite{Covi:1996wh, luty, CP_calculations_BP,CP_calculations_Flanz,Liu:1993tg}, and therefore we expect much of the subsequent discussion regarding the $CP$ asymmetry to be similar. In particular, we see that the dimensionless Yukawa coupling matrix $h$ which is central to the discussion on leptogenesis implications has simply been replaced by its EMDM counterpart, $\lambda_0$ which is again an arbitrary complex matrix (see Sec.~\ref{sec:BKL_EDMD_properties}).
 Also, the loop functions only differ by a multiplicative constant from before. Thus,  we can conclude that this type of EMDM interaction between light and heavy neutrinos can in general generate a lepton asymmetry in the early universe.

But before discussing the magnitude of this asymmetry and the parameter space in which  electromagnetic leptogenesis can be successful, we shall first look at a generalization of this EMDM scenario, one that will respect the SM gauge symmetries.

\subsection{A more realistic extension}\label{subsec:BKL_toy_model_6D}

As mentioned before, while the simplistic toy model in Sec.~\ref{subsec:BKL_toy_model_5D} can demonstrate the viability of lepton generation through EMDM operators, it is nonetheless unrealistic as it is incompatible with the SM.
We
now overcome this by considering only EMDM type operators that respect
the SM gauge group.  Again, we construct an effective theory by taking
the usual minimally extended SM Lagrangian with three generations of
heavy Majorana neutrinos, and augmenting it with EMDM operators.  The
most economical of such operators involving only (the minimally
extended) SM fields are of dimension six~\cite{Bell_mdm_PRL}, and the
interaction Lagrangian of interest is
\begin{align}
  \mathcal{L}_\text{EM} 
    &= 
 - \overline{\ell}_j
  \left[ \lambda'_{jk}\,\phi\,\sigma^{\alpha\beta}\,
    B_{\alpha\beta} 
+  \widetilde{\lambda}'_{jk}\,
\tau_i \,\phi\,\sigma^{\alpha\beta}\, W_{\alpha\beta}^i
  \right]P_R\, N_k
  +\text{h.c.} \;, 
  \label{eq4:BKL_6D_main_Lag_A}
  \\
  &\equiv 
- \frac{1}{\Lambda^2} \,\overline{\ell}_j
  \left[ (\lambda_0')_{jk}\,\phi\,\sigma^{\alpha\beta}\,
    B_{\alpha\beta} 
+  (\widetilde{\lambda}_0')_{jk}\,
\tau_i \,\phi\,\sigma^{\alpha\beta}\, W_{\alpha\beta}^i
  \right]P_R\, N_k
  +\text{h.c.} \;, 
\label{eq4:BKL_6D_main_Lag}
\end{align}
%
where the $\tau_i$ are the $SU(2)_L$ generators, $\ell_{j} = (\nu_{Lj},
e_{Lj})^T$ is the lepton doublet, and $\phi =
(\phi^{0}, \phi^-)^T$ is the SM Higgs doublet.  
The field strength tensors of $U(1)_Y$ and $SU(2)_L$ are given by $B_{\alpha\beta} =
\partial_{\alpha}B_{\beta}-\partial_{\beta}B_{\alpha}$ and
$W^i_{\alpha\beta} =
\partial_{\alpha}W^i_{\beta}-\partial_{\beta}W^i_{\alpha} - g\,\epsilon_{imn} W^m_{\alpha}W^n_{\beta}$, respectively, where $g'$ and
$g$ are the corresponding coupling constants.
As before, $\Lambda$ denotes the high energy cut-off of our effective theory, while the newly defined dimensionless EMDM coupling matrices, $\lambda_0'$ and $\widetilde{\lambda}_0'$, are in general complex. Note that $\lambda_0'$ and $\widetilde{\lambda}_0'$ play the exact same role as $\lambda_0$ in Lagrangian (\ref{eq4:BKL_toy5D_main_lag}).

The higher dimension (non-renormalizable) operators of
Eq.~(\ref{eq4:BKL_6D_main_Lag_A}) are assumed to be generated at the energy scale
$\Lambda$, beyond the electroweak scale.  Although the presence of these
operators would imply the existence of some new physics at high
energies, we shall not speculate on the nature of it here.  After spontaneous breaking of $SU(2)_L\otimes U(1)_Y$, these operators will then give rise to the usual transition moments between $N_k$ and $\nu_j$. But, for the purposes of leptogenesis, we are of course  interested in the regime above the electroweak symmetry breaking scale.

To proceed with the analysis, it is imperative to note that the decay of $N_k$ will now produce 3-body final states~\footnote{For simplicity, we shall ignore the 4-body final states such as $N_k\rightarrow \bar{\ell}_j\,\phi\, W^m_\alpha\,W^n_\beta$.} such as $N_k\rightarrow \bar{\ell}_j\,\phi\, B_\alpha$ and $N_k\rightarrow \bar{\ell}_j\,\phi\, W^i_\alpha$. As a result, the phase space calculation for all diagrams will be more involved than the toy model in Sec.~\ref{subsec:BKL_toy_model_5D}. Moreover, there are potentially several more inequivalent higher-order graphs that can interfere with each of the tree-level process.
Fortunately, we can expect that the results will be qualitatively similar to the previous model for two reasons. Firstly, having an extra scalar particle in the Lagrangian does not change the form of the vertex factors, which are now given by:
\begin{align}
 N_k \rightarrow \ell_{j}\,\bar{\phi}\, \bar{C}^{\rho}\;:\;\; 
\qquad 
 \SetScale{0.6}
   \begin{picture}(75,0)(30,-80)  
    \SetWidth{0.8}
    \SetColor{Black}
    \Line(90,-128)(125,-128)
    \ArrowLine(125,-128)(155,-102)
    \DashLine(125,-128)(160,-128){2.6}
    \SetWidth{0.5}
    \SetWidth{0.8}
    \Photon(125,-128)(155,-154){3}{3}
    \LongArrow(121,-139)(141,-157)
    \Vertex(125,-128){3}
    \Text(45,-77)[]{\small{\Black{$N$}}}
    \Text(100,-62)[]{\small{\Black{$\ell$}}}
    \Text(103,-76)[]{\small{\Black{$\bar{\phi}$}}}
    \Text(103,-92)[]{\small{\Black{$\bar{C}^\rho$}}}  
    \Text(71,-94)[]{\small{\Black{$q$}}}
    \end{picture}
\qquad  
  &= 2\, \zeta_{jk}\,P_R\,\sigma^{\alpha\rho} q_\alpha 
    \label{eq4:BKL_6D_vertex_rule1}\\
 N_k \rightarrow \bar{\ell}_{j}\,\phi\, C^{\rho}\;:\;\; 
\qquad 
 \SetScale{0.6}
   \begin{picture}(75,0)(30,-80)  
    \SetWidth{0.8}
    \SetColor{Black}
    \Line(90,-128)(125,-128)
    \ArrowLine(155,-102)(125,-128)
    \DashLine(125,-128)(160,-128){2.6}
    \SetWidth{0.5}
    \SetWidth{0.8}
    \Photon(125,-128)(155,-154){3}{3}
    \LongArrow(121,-139)(141,-157)
    \Vertex(125,-128){3}
    \Text(45,-77)[]{\small{\Black{$N$}}}
    \Text(100,-62)[]{\small{\Black{$\ell$}}}
    \Text(103,-76)[]{\small{\Black{$\phi$}}}
    \Text(103,-92)[]{\small{\Black{$C^\rho$}}} 
    \Text(71,-94)[]{\small{\Black{$q$}}}
    \end{picture}
\qquad  
  &= -2\, \zeta_{jk}^*\,C^\dagger\,\sigma^{\alpha\rho} q_\alpha\,P_L
   \rule{0pt}{40pt}
    \label{eq4:BKL_6D_vertex_rule2}
\end{align}
where the generic coupling matrix $\zeta$ represents $\lambda'$ when the vector boson $C_\rho \equiv B_\rho$, while it denotes $\widetilde{\lambda}'$ for $C_\rho \equiv W_\rho^i$. 
As we can see, these are basically the rules given in (\ref{eq4:vertex_rule_emdm1}) and (\ref{eq4:vertex_rule_emdm2}) with the replacement $\lambda \rightarrow \lambda'$ or $\widetilde{\lambda}'$.

Secondly, a scalar propagator (see (\ref{eq4:Feyn_scalar_prop})) does not contain any spinor structure nor can it affect the contractions between different momenta. Hence, once it is put on-shell, its inclusion in the diagrams cannot modify the form of the final amplitude of the interference terms. In the light of these observations, there is no need to (re-)compute all the possible Feynman graphs for this model. It would suffice to invoke the results of Sec.~\ref{subsec:BKL_toy_model_5D} in conjunction with a representative calculation for the new diagrams. The latter is done to demonstrate that the above claim (of a similar structure for the final equations) is indeed correct, as well as to elucidate the suppression factor expected from the additional $1/\Lambda$ (compare (\ref{eq4:BKL_6D_main_Lag}) with (\ref{eq4:BKL_toy5D_main_lag})), and the 3-body decay phase space. 

\begin{figure}[b]
\begin{center}
 \includegraphics[width=0.90\textwidth]{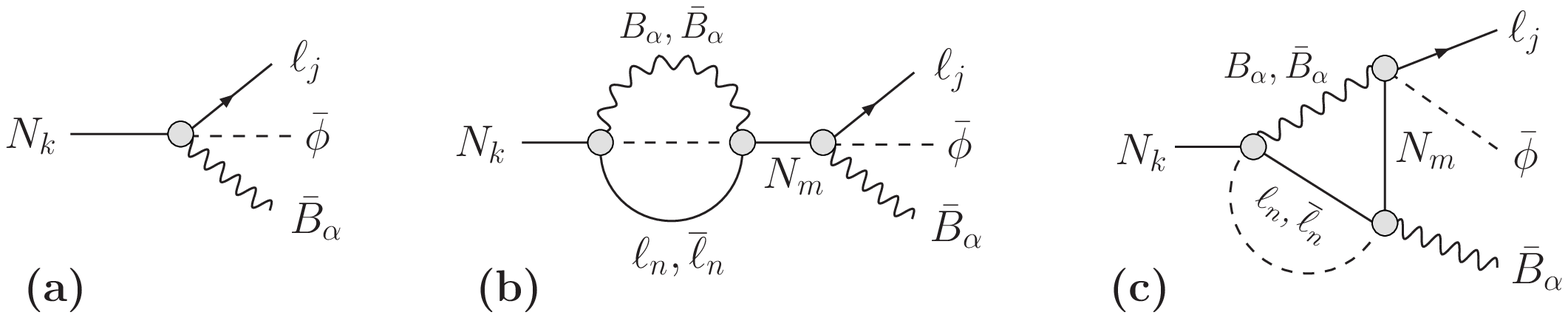}
\end{center} 
 \caption{{\bf (a)} The tree-level diagram for the 3-body decay: $N_k\rightarrow \ell_j\, \bar{\phi}\, \bar{B}_\alpha$ induced by the first term in Eq.~(\ref{eq4:BKL_6D_main_Lag_A}), and examples of the corresponding {\bf (b)} self-energy and {\bf (c)} vertex diagrams.}
 \label{fig4:BKL_6D_graphs_eg}
\end{figure}

With this aim in mind, we have drawn an example for each of the tree-level, self-energy and vertex processes for this model in Fig.~\ref{fig4:BKL_6D_graphs_eg}a, b and c respectively. Note that other graphs with the same structure as those shown in Fig.~\ref{fig4:BKL_6D_graphs_eg} exist since the mixing between bosons $B_\alpha$ and $W^i_\beta$ (for all $i$) will lead to different combinations of internal and final states. However, these extra diagrams are only expected to modify the overall self-energy and vertex contributions by an unimportant multiplicative constant, similar to the effect caused by having either $\ell_n$ or $\bar{\ell}_n$ running in the loops. Also, accidental cancellations amongst these are highly improbable based on what we have learned from the toy model in the previous section.  Hence, for simplicity, we shall only concentrate on the first term from (\ref{eq4:BKL_6D_main_Lag_A}) in the following discussion.

Firstly, by comparing the vertex factors arising from this model with those studied previously, it is clear that the tree-level process of Fig.~\ref{fig4:BKL_6D_graphs_eg}a has an amplitude given by
\begin{equation}\label{eq4:BKL_6D_tree_amp}
 |\overline{\mmat}|^2
  = 16 \,({\lambda'}_{jk}^* \, \lambda'_{jk})
     (p \cdot q)(p' \cdot q)
 \;,
\end{equation}
where $p, p'$ and $q$ are the four-momenta for $N_k, \ell_j$ and $\bar{B}_\alpha$ respectively, and we have used (\ref{eq4:BKL_toy5D_treeM_final1}) with $q^2=0$. In the centre-of-mass frame, we have the following definitions
\begin{equation}\label{eq4:BKL_6D_momenta}
 p = (\mNk\;,\;\vec 0)\;,\quad
 p' = (E'\;,\;\vec p~')\;,\quad
 p'' = (E''\;,\;\vec p~'')\;,\quad
 q = (E_q\;,\;\vec q)\;,
\end{equation}
where $|\vec p~'| = E'$, $|\vec p~''| = E''$, and  $|\vec q| = E_q$ as $m_{\ell, \phi, B} \rightarrow 0$. Here, $p''$ is the four-momentum of $\bar{\phi}$. The directions of the momenta are defined such that $p = p'+p''+q$ is satisfied. With (\ref{eq4:BKL_6D_momenta}) and the 3-body phase space of (\ref{eq4:freq_used_3-body-rate}), the differential decay rate for $N_k\rightarrow \ell \,\bar\phi\,\bar{B}_\alpha$ (i.e. summed over $j$) is
\begin{align}
 d\Gamma_{k}^\text{6D}
  &= 2\times 16\,({\lambda'}^\dagger {\lambda'})_{kk}(p\cdot q)(p'\cdot q)
  \nonumber\\
  &\qquad\times
  \frac{1}{2\mNk}\frac{d^3p'}{(2\pi)^3 2E'}\frac{d^3p''}{(2\pi)^3 2E''}
  \frac{d^3 q}{(2\pi)^3 2E_q} (2\pi)^4 \delta^{(4)}(p-p'-p''-q)\;,\\
\intertext{where the factor of 2 in front is to account for the two channels, $N_k\rightarrow \nu_L\,\phi^0\,\bar{B}$ and $N_k\rightarrow  e_L\,\phi^+\,\bar{B}$. Then, simplifying using the result $2(p'\cdot q) = (p-p'')^2$, we get}
 d\Gamma_{k}^\text{6D}
  &=
  \frac{8\,({\lambda'}^\dagger {\lambda'})_{kk}}{(2\pi)^5}
     \left[E_q (\mNk^2-2\mNk E'')\right]
  \frac{d^3p'}{2E'}\frac{d^3p''}{2E''}
  \frac{d^3 q}{2E_q} \,\delta^{(4)}(p-p'-p''-q)\;.\label{eq4:BKL_6D_diff_rate}
\end{align}
To obtain the full rate, $\Gamma_{k}^\text{6D}$, one has to perform the integrals over all possible values of $d^3p'$, $d^3p''$ and $d^3q$. The procedure is akin to the computation of the muon decay rate for massless final states ($\mu^- \rightarrow e^- \,\nu_e\, \nu_\mu$), where several standard tricks are needed. We shall leave the complete workings to Appendix~\ref{app:c_emdm_cal_3body} and only present the final result here, which is:
\begin{align}
 \Gamma_{k}^\text{6D}
  &= 
  \frac{({\lambda'}^\dagger {\lambda'})_{kk} \;\mNk^5}{256\pi^3}\;,
  \label{eq4:BKL_6D_decay_rate_final_A}
  \\
  &=
 \left(\frac{\mNk^2}{64\pi^2\Lambda^2}\right) 
 \frac{({\lambda'_0}^\dagger \lambda_0')_{kk}}{4\pi}\,  \frac{\mNk^3}{\Lambda^2}\;,\\
 &\equiv
 \left(\frac{\mNk}{8\pi\Lambda}\right)^2 \Gamma_{\text{5D},k}' \;,
 \label{eq4:BKL_6D_decay_rate_final}
\end{align}
where $\Gamma_{\text{5D},k}'$ denotes the decay rate from the toy model in Sec.~\ref{subsec:BKL_toy_model_5D} but with coupling matrix $\lambda_0$ replaced by $\lambda_0'$ (see Eq.~(\ref{eq4:BKL_toy5D_decay_rate})). Therefore, as expected, expression (\ref{eq4:BKL_6D_decay_rate_final}) indicates there is a suppression factor (due to the reasons alluded earlier) of $(\mNk/8\pi\Lambda)^2$.

\begin{figure}[t]
\begin{center}
 \includegraphics[width=0.85\textwidth]{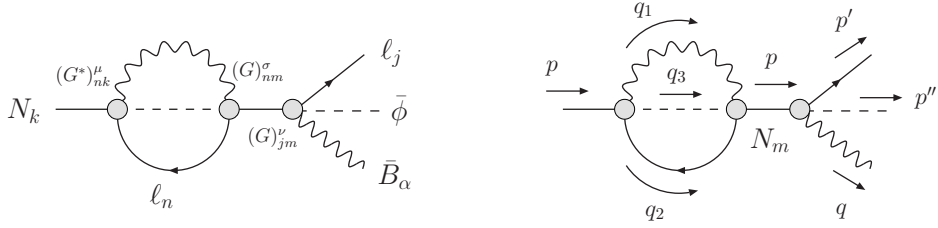}
\end{center} 
 \caption{LEFT: self-energy diagram of Fig.~\ref{fig4:BKL_6D_graphs_eg}b with $\bar{\ell}_n$ intermediate state, where $(G)_{ab}^\rho = 2\lambda'_{ab} P_R \sigma^{\beta\rho} q'_\beta$ and $(G^*)_{ab}^\rho = -2{\lambda'}_{ab}^* C^\dagger \sigma^{\beta\rho} q'_\beta P_L$ are the vertex factors. RIGHT: the corresponding diagram with momentum labels included.}
 \label{fig4:BKL_6D_self_for_calc}
\end{figure}

In order to demonstrate a similar modification for the $CP$ asymmetry, we compute the self-energy contribution coming from the interference between Fig.~\ref{fig4:BKL_6D_graphs_eg}a and the $\bar{\ell}_n$ intermediate case of  Fig.~\ref{fig4:BKL_6D_graphs_eg}b  as an illustrative example. To this end, we re-draw the  self-energy diagram with momentum labels and vertex factors inserted in Fig.~\ref{fig4:BKL_6D_self_for_calc}. By using the result of (\ref{eq4:BKL_toy5D_self_a_used_in_6D}) with the new momentum labels, while adding the scalar propagator, $i/(q_3^2+i\epsilon)$, and a new loop integral in $q_2$ to the expression, we can immediately deduce the amplitude for this interference term to be:
\begin{equation}\label{eq4:BKL_6D_self_amp}
 I_\text{self}^\text{6D}
 =
 \frac{64 \,A_{\lambda}^{(6)}}{(2\pi)^8}
  \int d^4 q_1\, d^4 q_2\,
  \frac{\mNm\mNk\,(p' \cdot q)
  \left[4(q \cdot q_1)(q_1 \cdot q_2)-(q \cdot q_2) q_1^2\right]}
   {(p^2-\mNm^2+i\epsilon)(q_2^2+i\epsilon)(q_3^2+i\epsilon)(q_1^2+i\epsilon)}\;,
\end{equation}
where $A_{\lambda}^{(6)} = {\lambda'}_{nk}^* {\lambda'}_{nm} {\lambda'}_{jm} {\lambda'}_{jk}^*$. After putting $\ell_n$, the internal scalar and vector boson propagators on-shell to pick out the imaginary part, we follow a similar procedure as before to integrate over the 3-body phase space, noting again that there are two possible final, as well as internal states ($\nu_L\,\phi^0\,\bar{B}$ and $e_L\,\phi^+\,\bar{B}$). Eventually, it can be shown that~\footnote{See Appendix~\ref{app:c_emdm_cal_3body} for the details.}
\begin{equation}\label{eq4:BKL_6D_self_cp_final}
 \varepsilon^\text{6D}_{\text{self-}k,j}
 =  \left(\frac{\mNk}{8\pi\Lambda}\right)^2
   \varepsilon_{\text{5D-self-}k,j}'\;,
\end{equation}
where $\varepsilon_{\text{5D-self-}k,j}'$ denotes the expression of the self-energy $CP$ asymmetry of (\ref{eq4:BKL_toy5D_self_a_final}) with coupling matrix $\lambda$ replaced by $\lambda'\Lambda$ everywhere, and $\Lambda$ is the cut-off scale defined in Lagrangian (\ref{eq4:BKL_6D_main_Lag}). 
Assuming that (\ref{eq4:BKL_6D_self_cp_final}) is representative of the various contributions to $\varepsilon^\text{EM}$, it follows that the
$CP$ asymmetry for this realistic EMDM leptogenesis model has the same dependence on $z$ ($\equiv \mNm^2/\mNk^2$) as the toy model in Sec.~\ref{subsec:BKL_toy_model_5D}. Hence, in the limit of a hierarchical heavy RH neutrino mass spectrum, the expression for the overall $CP$ asymmetry (summed over $j$) for $N_1$ decay via the dim-6 EMDM operators is approximately~\footnote{In writing this down, we have ignored the $\widetilde{\lambda}'$-terms in (\ref{eq4:BKL_6D_main_Lag}). Alternatively, one can get a similar structure by taking $\widetilde{\lambda}' \simeq \lambda'$.} 
\begin{equation}\label{eq4:BKL_6D_overall_CP_asym_N1}
 \varepsilon_1^\text{EM} \simeq 
 \frac{1}{2\pi}
 \sum_{m\neq 1}
 \frac{\textrm{Im}[({\lambda'_0}^\dagger \lambda_0')^2_{1m}]}
 {({\lambda_0'}^\dagger \lambda_0')_{11}}
 \frac{M_1}{M_m} 
 \left(\frac{M_1^2}{8\pi\Lambda^2}\right)^2
 \;.
\end{equation}
Since $\lambda'_0$ is in general arbitrary, it will contain complex phases that render $\varepsilon_1^\text{EM}$ nonzero. Consequently, the generation of a primordial lepton asymmetry via the transition electromagnetic dipole interactions between light and heavy neutrinos described here is a real possibility.\\

\section{A new scenario for leptogenesis}\label{sec:BKL_discussion}

Given the results of the previous section, it is important to explore some of the implications of demanding this electromagnetic leptogenesis scenario produce an asymmetry of the correct magnitude. Amongst many potential issues to address, the most relevant is, as always, the connection of leptogenesis parameters to neutrino properties. Indeed, as we have seen from our discussions of the standard Yukawa-mediated leptogenesis in previous chapters, neutrino masses and mixings play a critical role in determining the overall viability and predictions of the model. Therefore, it is imperative to gain some qualitative understanding on this front.

In the following, we first investigate the link between the EMDM operators in Sec.~\ref{subsec:BKL_toy_model_6D} and neutrino masses. Then, we put everything together to see whether a workable parameter space exists for this new leptogenesis scenario.

\subsection{Implications for neutrino mass}\label{subsec:BKL_numass}

As discussed in Sec.~\ref{subsec:numass_emdm}, massive neutrinos inevitably lead to nonzero neutrino dipole moments. Hence, a sensible question to ask is whether the existence of the EMDM operators of (\ref{eq4:BKL_6D_main_Lag_A}) can generate neutrino mass terms. Although it is well-known that via a careful choice for the new physics, one can eliminate the direct correspondence between large neutrino dipole moments and masses \cite{bell_fest_5_8, Fuk_Yan_book}, radiative corrections to the neutrino mass, induced by the dipole operators, generically link mass and dipole moments irrespective of the form of the new physics \cite{Bell_mdm_PRL, Bell_mdm_PLB, davidson_numass_mdm}. As a result, the new physics behind the origin of the effective EMDM operators in Lagrangian (\ref{eq4:BKL_6D_main_Lag_A}) is expected to give rise to neutrino mass terms at loop level. 

\begin{figure}[t]
\begin{center}
 \includegraphics[width=0.85\textwidth]{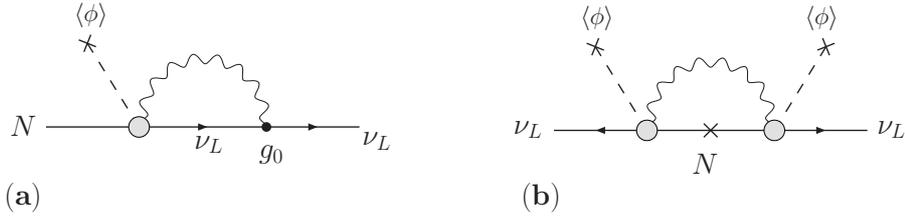}
\end{center} 
 \caption{Contribution to {\bf (a)} the neutrino Dirac mass and {\bf (b)} the neutrino Majorana mass induced by the dim-6 transition EMDM operator. Here $g_0$ and $\vev{\phi}$ denote the gauge coupling constant and Higgs VEV respectively.}
 \label{fig4:BKL_emdm_loop_mass}
\end{figure}

By examining the interactions in our model~\footnote{This includes all processes in the SM except maybe neutrino Yukawa terms which are not absolutely necessary in electromagnetic leptogenesis.}, it is not difficult to construct loop diagrams containing EMDM vertices that will contribute to the neutrino mass terms. Specifically, we note that a neutrino Dirac mass term would be induced by the one-loop graph depicted in Fig.~\ref{fig4:BKL_emdm_loop_mass}a. Moreover, there is a direct contribution to the light neutrino Majorana mass via the diagram in Fig.~\ref{fig4:BKL_emdm_loop_mass}b (c.f. Fig.~\ref{fig1:seesaw1} on page~\pageref{fig1:seesaw1}).

Unfortunately, the exact size of these contributions cannot be calculated in a model-independent way because one would require the precise relationship between the dipole moments and mass terms, which can only be resolved if the nature of the new physics at scale $\Lambda$ is specified. However, an estimate of these quantities can be obtained by applying simple dimensional analysis. To this end, we begin by writing down the approximate form of the amplitude for the Dirac mass diagram in Fig.~\ref{fig4:BKL_emdm_loop_mass}a, ignoring the external lines (i.e. setting their momentum to zero) and factors of $\pm i$:
\begin{align} 
 m_D &\sim
  \int^\Lambda \frac{d^4 k}{(2\pi)^4} 
  \left(g_0\gammamu\right)
  \frac{g_{\mu\nu}}{k^2 +i \epsilon}\,
  \frac{\slashed{k}+ m}{k^2-m^2 +i \epsilon}
  \left(\frac{2\,\lambda_0'\,\vev{\phi}}{\Lambda^2}\,
  P_R\,\sigma^{\alpha\nu}\,k_\alpha \right)
  \;, \nonumber\\
 &\simeq
 \frac{\lambda_0'\,g_0\,\vev{\phi}}{(2\pi)^4\Lambda^2}
  \int^\Lambda d^4 k \;
  \frac{1}{k^2}\,
  \frac{1}{k^2-m^2}
   \underbrace{\gammamu (\slashed{k}+ m)\, P_R\, 
  (\slashed{k}\gamma_\mu - \gamma_\mu\slashed{k})}_{\simeq P_R (k^2 +m \slashed{k})}
  \;,
\end{align}  
where $k$ is the loop momentum and $m$ is the mass of $\nu_L$. In the limit of $m\rightarrow 0$ and after dropping the projection operator, $P_R$ from the expression, we then have
\begin{align}
 m_D
  &\sim
   \frac{\lambda_0'\,g_0\,\vev{\phi}}{(2\pi)^4 \Lambda^2}
   \int^\Lambda d^4 k \; \frac{1}{k^2}
   \;, \nonumber\\
  &\simeq
   \frac{\lambda_0'\,g_0\,\vev{\phi}}{(2\pi)^4 \Lambda^2}
   \underbrace{\int^\Lambda (d k)^2 k^2 d\Omega \,\frac{1}{k^2}}_{\simeq\,\pi^2\Lambda^2}
   \;,\nonumber\\
  &\simeq
  \frac{\lambda_0'\,g_0\,\vev{\phi}}{16\pi^2}\;.
  \label{eq4:BKL_dirac_mass_est}
\end{align}
This Dirac mass term together with the heavy neutrino Majorana mass ($M$) can give rise to a light neutrino Majorana mass via the type~I seesaw mechanism (see Sec.~\ref{subsec:seesaw1}) with the form
\begin{equation}\label{eq4:BKL_emdm_seesaw_mass_1}
 m^{(1)}_\nu =  m_D\,  M^{-1}\, m_D^T\,
 \simeq\,
 \left(\frac{g_0\,\vev{\phi}}{16\pi^2} \right)^2  
 \lambda_0'\, M^{-1}\, (\lambda'_0)^T
  \;.
\end{equation}
Similarly, we can deduce the direct contribution to the light neutrino Majorana mass term coming from the loop diagram in Fig.~\ref{fig4:BKL_emdm_loop_mass}b via the following estimate of its amplitude:
\begin{align}
 m_\nu^{(2)} &\sim
  \int^\Lambda \frac{d^4 k}{(2\pi)^4}
  \left(
   \frac{2\,\lambda_0'\,\vev{\phi}}{\Lambda^2}\,P_R\,\sigma^{\alpha\nu} k_\alpha
  \right)
  \frac{g_{\mu\nu}}{k^2 +i \epsilon}\,\frac{\slashed{k}+ M}{k^2-M^2 +i \epsilon}
  \left(
   k_\beta \sigma^{\beta\mu}\, P_R \,\frac{2\,(\lambda_0')^T\,\vev{\phi}}{\Lambda^2}
  \right)
  \;,\label{eq4:BKL_emdm_direct_diagram_amp}
\end{align}
where we have omitted all the charge conjugation operators, $C$ in the definition of the Majorana propagator and from transposing for convenience. Note that this omission is inconsequential to our final outcome as the operator $C$ can always be absorbed into the external fields via relations such as $u=C\overline{v}^T$ and $v=C\overline{u}^T$ for any generic spinors. Simplifying (\ref{eq4:BKL_emdm_direct_diagram_amp}) using standard contraction identities for gamma matrices, we get
\begin{align}
 m_\nu^{(2)} &\sim
  \frac{\lambda_0'\,\vev{\phi}^2}{(2\pi)^4\Lambda^4}
  \int^\Lambda \frac{d^4 k}{k^2\, (k^2-M^2)}\,
  \underbrace{P_R\,(\slashed{k} \gamma_\mu - \gamma_\mu \slashed{k})\,
  (\slashed{k}+M)\,(\slashed{k}\gammamu- \gammamu\slashed{k})\,P_R}_{\simeq\, P_R M k^2}
  \;(\lambda_0')^T
  \;,\nonumber\\
 &\simeq
  \frac{\lambda_0'\,\vev{\phi}^2}{(2\pi)^4\Lambda^4}
   \underbrace{\int^\Lambda (d k)^2 k^2 d\Omega
    \;\frac{M}{k^2-M^2}}_{\simeq\,\pi^2 M \Lambda^2}
     \;(\lambda_0')^T
   \;,\nonumber\\  
 &\simeq
 \frac{\vev{\phi}^2}{16\pi^2 \Lambda^2}\;\lambda_0'\, M\,(\lambda_0')^T
 \;.\label{eq4:BKL_emdm_seesaw_mass_2}
\end{align}
Generically, one can expect that both (\ref{eq4:BKL_emdm_seesaw_mass_1}) and (\ref{eq4:BKL_emdm_seesaw_mass_2}) will add to the effective light neutrino Majorana mass although the contribution arising from (\ref{eq4:BKL_emdm_seesaw_mass_2}) is typically dominant (unless $\Lambda \gg M$) due to the extra $1/16\pi^2$ suppression factor in the definition of (\ref{eq4:BKL_emdm_seesaw_mass_1}). But in any case, this connection between the EMDM couplings, $\lambda_0'$, and neutrino masses presents an important constraint that can control the size of the $CP$ asymmetry arising from electromagnetic leptogenesis, as we shall discuss in the following.

\subsection{Discussion on the parameter space}\label{subsec:BKL_param}

As a result of the link between the EMDM couplings and light neutrino masses, we may derive a lower limit on the size of $M_1$, just like (\ref{eq1:lepto_std_lower_bound_M1}) on page~\pageref{eq1:lepto_std_lower_bound_M1} for standard $N_1$-leptogenesis. This is because there exists a relationship between $|\varepsilon_1^\text{EM}|_\text{max}$ and $M_1$, which is analogous to that of (\ref{eq1:lepto_std_eps_est_3}) on page~\pageref{eq1:lepto_std_eps_est_3}, with the role of the Yukawa coupling matrix, $h$ being replaced by $\lambda_0'$. Using Eqs.~(\ref{eq4:BKL_6D_overall_CP_asym_N1}) and (\ref{eq4:BKL_emdm_seesaw_mass_2}) with the assumptions $|(\lambda_0')_{jk}|\leq~\!|(\lambda_0')_{33}|$ and $M_1 \ll M_{2,3}$, we can retrace the steps leading to (\ref{eq1:lepto_std_eps_est_3}) and obtain
\begin{equation}\label{eq4:BKL_emdm_cp_M1_rel}
  |\varepsilon_1^\text{EM}|_\text{max}
   \sim
   10^{-6}\, \beta_\Lambda^4
    \left( \frac{M_1}{10^{10}~\text{GeV}} \right)
    \;,
\end{equation}
where we have taken $M\equiv M_3 \simeq \Lambda$ in (\ref{eq4:BKL_emdm_seesaw_mass_2}), and defined the quantity, $\beta_\Lambda \equiv M_1/\Lambda$ which indicates the level of hierarchy among the RH neutrinos. From (\ref{eq4:BKL_emdm_cp_M1_rel}), a couple of general observations for electromagnetic leptogenesis can be made. Firstly, it is clear that to obtain a reasonable size for the $CP$ asymmetry (e.g. $\order{10^{-6}}$), the scale for $M_1$ must be at least $\order{10^{10}}$~GeV, a result which is similar to that from standard $N_1$-leptogenesis~\footnote{It should be noted that although the scenario of TeV scale leptogenesis discussed in Ref.~\cite{Hambye:2001eu} involves 3-body decay processes, these are inherently different from the interactions we have investigated in the EMDM model. In the case of Ref.~\cite{Hambye:2001eu}, the 3-body decays studied are one-loop processes whereas ours involves two-loop graphs.}.
Secondly, the presence of the suppression factor, $\beta_\Lambda^4$ implies that a strong RH neutrino mass hierarchy is in general undesirable as $M_1 \ll M_3 \simeq \Lambda$ automatically means $\beta_\Lambda^4 \ll 1$. Hence, in order to acquire a sufficiently large raw $CP$ asymmetry together with experimentally acceptable light neutrino masses, some very mild fine-tunning between the scales of $M_i$ and $\Lambda$, as well as the absolute size of the coupling $\lambda_0'$ is required. 
Qualitatively speaking though, the conditions governing electromagnetic leptogenesis (modulo the suppression factor) are largely identical to the standard Yukawa-mediated case. To summarize the situation, we compare the various key quantities resulting from the EMDM model with the corresponding ones for the standard scenario in Table~\ref{table4:BKL_compare_exp}.


%
%
\begin{table}[t]
\begin{center}
\begin{tabular}{|c|c|}
\hline
Standard  & Electromagnetic  \\
\hline\hline
$\Gamma_1=\dfrac{1}{16\pi}\, (h^\dagger h)_{11} M_1$ &
 $\Gamma_1^\text{EM}=\dfrac{1}{4\pi}\, (\lambda_0'^\dagger \lambda_0')_{11} M_1
\left(\dfrac{M_1^2}{8\pi\Lambda^2}\right)^2$
  \rule{0pt}{2em}\\  [10pt]
\hline 
$\displaystyle{|\varepsilon_1| \simeq \frac{1}{\pi}\,
 \sum_{m\neq 1} \frac{\textrm{Im}\,(h^\dagger h)^2_{1m}]}{(h^\dagger h)_{11}}
 \frac{M_1}{M_m}}$
&
$\displaystyle{|\varepsilon_1^\text{EM}| \simeq \frac{1}{\pi}\,
 \sum_{m\neq 1} \frac{\textrm{Im}\,[(\lambda_0'^\dagger \lambda_0')^2_{1m}]}{(\lambda_0'^\dagger \lambda_0')_{11}}
 \frac{M_1}{M_m}} \left(\frac{M_1^2}{8\pi\Lambda^2}\right)^2$
  \rule{0pt}{2.5em}\\  [10pt]
\hline 
$\displaystyle{
 m_\nu \sim \vev{\phi}^2\,h\,M^{-1}\,h^T
}$
&
$\displaystyle{
 m_\nu^\text{EM}
  \sim 
 \frac{\vev{\phi}^2}{16\pi^2}
 \left[
 \frac{g_0^2}{16\pi^2}\,\lambda_0'\,M^{-1}\,(\lambda_0')^T
 +\frac{1}{\Lambda^2}\,\lambda_0'\,M\,(\lambda_0')^T
 \right]
}$
\rule{0pt}{2em}\\ [10pt]
\hline
\end{tabular}
\end{center}
\caption{Comparison of key quantities in standard and electromagnetic
  leptogenesis, where $h$ and $\lambda_0'$ denote the Yukawa and dimensionless EMDM coupling constants. We have assumed there is at least a mild hierarchy in the masses of the
  heavy neutrinos, such that the asymmetry is predominantly generated
  from the decay of the lightest state, $N_1$.}
\label{table4:BKL_compare_exp}
\end{table}

In a model where both the Yukawa and EMDM interactions are present in the Lagrangian, we have, in general, contributions to the decay rate, $CP$ asymmetry and neutrino mass coming from both the standard and electromagnetic sectors. As a result, either one mechanism will dominate, or there will be an interplay between the two, depending on the
relative size of $h$ and $\lambda_0'$ (and perhaps other factors). For our investigation here, we are particularly interested in examining if electromagnetic leptogenesis alone (i.e. when the Yukawa couplings are negligible) can give rise to the required asymmetry without contradicting any known experimental constraints. To this end, we select some representative input parameters for the EMDM model and study their implications.

Given that electromagnetic leptogenesis does not seem to favor a very strong hierarchy in the RH neutrino mass spectrum, as well as between $M$ and $\Lambda$, we shall adopt the values
\begin{equation}\label{eq4:BKL_param_spectrum}
 \Lambda \simeq 10 M_{2,3} \simeq 20 M_1
\;,
\end{equation}
as an illustrative example. In addition, we shall (for simplicity) make the crude assumption that all elements of the matrix $\lambda_0'$ are of similar magnitude, and ignoring any flavor structures it (as well as $m_\nu$ and $M$) may possess.
With these, it is straight forward to check using the expression for $|\varepsilon_1^\text{EM}|$ in Table~\ref{table4:BKL_compare_exp} that a value of 
\begin{equation}\label{eq4:BKL_param_lambda0}
 \lambda_0' \simeq 35\;,
\end{equation}
for the EMDM coupling is sufficient to the produce an asymmetry of $|\varepsilon_1^\text{EM}| \sim 10^{-6}$. To understand the general behavior of washout, we can follow the standard leptogenesis analysis and define the decay parameter (see (\ref{eq2:LV_decayK}) on page~\pageref{eq2:LV_decayK}):
\begin{equation}\label{eq4:BKL_param_decayK}
 K_1 = \frac{\Gamma_1^\text{EM}}{H(T=M_1)} \;,
\end{equation}
where $H$ is the Hubble parameter as defined in (\ref{eq2:LV_hubble}) on page~\pageref{eq2:LV_hubble}. Recall that $K_1$ controls whether the $N_1$ decays are in equilibrium, and is also a measure of washout effects via inverse decay. For definiteness, if we set 
\begin{equation}\label{eq4:BKL_param_M1}
 M_1 \simeq 5 \times 10^{12}~\text{GeV} \;,
\end{equation}
then the parameters in (\ref{eq4:BKL_param_spectrum}) and (\ref{eq4:BKL_param_lambda0}) imply that $K_1 \simeq 0.3$. As we know from Chapter~\ref{ch_work_LV}, $K_1\ll 1$ ($K_1 \gg 1$) corresponds to the weak (strong) washout regime, hence this parameter choice for the EMDM model should lead to a moderate washout of the asymmetry~\footnote{A detailed study of washout effects is beyond the scope of this work.}. So, qualitatively speaking, we expect that successful electromagnetic leptogenesis is achievable with these parameter choices.

The implications for the light neutrino mass can be readily obtained using the two terms for $m_\nu^\text{EM}$ in Table~\ref{table4:BKL_compare_exp}, and one finds
\begin{equation}\label{eq4:BKL_param_numass}
 m_\nu^{(1)} \simeq 4 \times 10^{-2}~\text{eV}\;,\qquad 
 m_\nu^{(2)} \simeq 1 \times 10^{-1}~\text{eV} \;,
\end{equation}
where we have set $g_0 \equiv g'$. In evaluating the numerical values for $g'$, we have used $M_W = g \vev{\phi}/\sqrt{2} \simeq 80$~GeV and $M_Z= \vev{\phi}\sqrt{(g^2+g'^2)/2} \simeq 91$~GeV with $\vev{\phi} = 174$~GeV. It is worth pointing out that, as expected, the direct contribution to the Majorana mass term, $ m_\nu^{(2)}$ (Fig.~\ref{fig4:BKL_emdm_loop_mass}b) dominates over $m_\nu^{(1)}$ (Fig.~\ref{fig4:BKL_emdm_loop_mass}a).

\begin{figure}[b]
\begin{center}
 \includegraphics[width=0.45\textwidth]{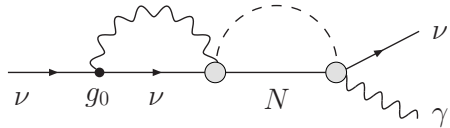}
\end{center} 
 \caption{An example of a two-loop digram that contributes to the effective light neutrino dipole moments.}
 \label{fig4:BKL_2-loop_dm}
\end{figure}

Finally, we note that the presence of the EMDM operators in the model will not only give rise to transition moments between a light and a heavy neutrino, they will also induce effective dipole moment interactions between \emph{two light} neutrino states via two-loop diagrams such as the one depicted in Fig.~\ref{fig4:BKL_2-loop_dm}. Therefore, it is imperative to check that this contribution to the light neutrino dipole moments is not in conflict with the current experimental upper limits, when we take the particular input values assumed above.

Applying dimensional analysis and assuming the worst case scenario where all quantities in the numerator with unit of mass are replaced by the cut-off scale $\Lambda$, we estimate the amplitude of Fig.~\ref{fig4:BKL_2-loop_dm} as
\begin{align}
  \mu_{\nu}^{\text{eff}} &\sim
  \frac{g_0}{(16\pi^2)^2}\,\frac{\lambda_0'}{\Lambda^2}\,
  \frac{\lambda_0'}{\Lambda^2}
  \times \Lambda^3
  \;,\nonumber\\
  &=
  g_0 \left(\frac{\lambda_0'}{16\pi^2}\right)^2\, \frac{1}{\Lambda}
  \;.\label{eq4:BKL_light_nu_dm_est}
\end{align}
Substituting in the value for $\Lambda$ assumed previously, and expressing in units of the Bohr magneton, $\mu_B = e/2m_e \simeq 3\times 10^{-7} \text{ eV}^{-1}$, we have $\mu_{\nu}^{\text{eff}} \sim 5 \times 10^{-19}\,\mu_B$. So, it is well within the current experimental limits which are of $\order{10^{-11}\,\mu_B}$~\cite{borexino,mdm_bound_exp1, mdm_bound_exp2}.

It is worth mentioning that when the Yukawa couplings are switched on, there will be additional contributions to these effective light neutrino dipole moments originating from light and heavy neutrino mixings. However, such contributions are hugely suppressed by a factor of $\vev{\phi}^2\Lambda^{-2} M^{-1}$ (instead of $\Lambda^{-1}$ as in (\ref{eq4:BKL_light_nu_dm_est})), and so their effects will be sub-dominant.

In summary, we have demonstrated that there is a region of parameter space where electromagnetic leptogenesis is viable, thus providing a new alternative that can stand alone or complement the standard scenario.\\

\section{Summary and outlook}\label{sec:BKL_summary}

The inclusion of the heavy RH neutrinos to the SM has not only provided an elegant way to generate small neutrino masses via the type~I seesaw mechanism, it has also opened up many new possibilities for model builders. In this chapter, we have seized upon this and investigated the natural extension of introducing electromagnetic dipole moment interactions between the heavy RH neutrinos and their light counterparts. One immediate result is that lepton violating decays of the heavy neutrinos are no longer solely mediated by the Yukawa term, but also by the newly included EMDM operators. Consequently, it is of great interest to ascertain whether such radiative decay is capable of creating a cosmic baryon asymmetry in a similar way to that of standard leptogenesis.

By carefully studying the properties of the EMDM couplings and explicitly calculating the decay amplitudes for the tree-level and interference terms, we have shown that, in general, the $CP$ asymmetry induced by the EMDM operators is nonzero. As a result, we have proved that, in principle at least, electromagnetic leptogenesis can be a viable mechanism for generating a lepton asymmetry in the early universe. 

Furthermore, in order to embed this model into the SM, a realistic generalisation to the basic dimension-5 EMDM operator has also been investigated. The end result is that this new scenario of leptogenesis is very much akin to the standard case, except that the role of the Yukawa couplings is now played by their EMDM counterparts, and there is an additional suppression factor ($\sim M^2/\Lambda^2$) since the EMDM operator is of higher dimension. Because of the presence of this suppression, the region of applicability for leptogenesis is slightly different in the EMDM scenario, and it disfavors a strong hierarchy in the RH neutrino mass spectrum. Moreover, washout effects in this model will differ from the standard case since the principle $L$ violating processes all have a 3-body initial or final state and involve a vector boson, hence the set of interactions participating in washout will be drastically different.

Other constraints on the model come from the light neutrino mass and effective dipole moment terms induced by loop diagrams involving the EMDM operators. These contributions directly restrict the magnitude of the EMDM coupling allowed, although the latter effects are normally not as significant due to the heaviness of $\Lambda$. In fact, we have discovered that the constraints from the neutrino mass terms demand that the scale of $M$, hence $\Lambda$, must be at least of $\order{10^{10}}$~GeV to produce a $m_\nu$ at the correct scale while maintaining a reasonably sized $|\varepsilon^\text{EM}|$. Thus, the conditions imposed by the effective light neutrino dipole contribution are usually unimportant for most viable input parameters.

In addition, this result implies a typical leptogenesis temperature of $T \sim 10^{10}$~GeV, which is comparable to the usual case. Consequently, it opens up an intriguing (but yet to be explored) possibility of Yukawa-EMDM interplay in the generation of lepton number. Indeed, there is a plethora of new ideas and issues one may further study on this topic. For instance, in the specific analysis, the flavor structure of the EMDM couplings, $\lambda_0'$ has not been included. Therefore, an investigation of $\lambda_0'$ similar to that for the Yukawas, $h$ (or the seesaw orthogonal matrix, $\Omega$ discussed in Sec.~\ref{subsec:lepto_N2} and in Chapter~\ref{ch_work_LV}) would be worthwhile. Also, owing to the fact that a strongly hierarchical RH neutrino spectrum is undesirable for electromagnetic leptogenesis, it would be quite natural to study the scenario where the RH neutrinos are quasi-degenerate, where resonant effects (see Sec.~\ref{subsec:lepto_resonant}) in the radiative decays may become important. Finally, there is a possibility that the newly included EMDM operators can give rise to non-trivial 
interactions between all neutrino species and the primordial magnetic field in the thermal plasma, hence affecting the evolution of lepton number.

But even without venturing into these complications, our work here has demonstrated that a new scenario beyond the standard leptogenesis framework (due to transition neutrino dipole moments) is viable, and it provides yet another example of the important link between neutrino properties and cosmology.


%
%
%
%
%



\chapter{Conclusion}\label{ch_conc}

\ECapC{O}{n~e}{7}{0}
of the most fundamental concepts in the study of physics is the idea of symmetry. Yet, Nature as we know it does not always seem to be perfectly symmetrical. Indeed, the principal theme for this current work is motivated by none other than the apparent asymmetry between matter and antimatter in the universe. Therefore, along with the appeal of symmetry, a major topic of interest is the mechanism of symmetry breaking or asymmetry creation. 

For the cosmic baryon asymmetry, this presents a unique challenge as the study of baryogenesis inevitably brings together the two giants of particle physics and Big Bang cosmology. To date though the solution to this puzzle is still highly speculative, and in the absence of more experimental inputs, it may remain unresolved for some time to come. Owing to this, the choice of the model employed is often based entirely on its theoretical appeal, as well as the prejudices of the investigator. It is fair to say that our approach of taking leptogenesis as the correct answer to the baryogenesis problem throughout this work is no exception, and it is strongly influenced by our interest in neutrino physics.

Without a doubt, it is quite fascinating that two seemingly unrelated problems--- the tiny masses of light neutrino and the matter-antimatter asymmetry--- may be explained in a consistent manner by the mere introduction of heavy RH neutrinos to the SM. As we have seen in our earlier discussions, the former may be explained by the type~I seesaw mechanism while thermal leptogenesis provides an attractive solution to the later. This naturally means that an intricate link between neutrino properties and the baryon asymmetry can be established. Consequently, it has been the purpose of this work to explore the implications of several different neutrino models in the leptogenesis context.

Taking the type~I seesaw framework (with three heavy RH neutrinos) as the starting point, we investigated three distinct possibilities in extending the typical setup for the neutrino sector, and studied their connection to the generation of lepton asymmetry. In the first part of our work, we performed a comprehensive analysis on the leptogenesis implications of the neutrino models proposed in Ref.~\cite{Low:2005yc}. These models have the advantage of requiring less free parameters to define the theory than the default seesaw setup, a unique feature which is made possible by the imposition of an abelian family symmetry and the inclusion of a new Higgs singlet. As a corollary, $\theta_{13} =0$ in the $\Upmns$ matrix, as well as a fully hierarchical light neutrino spectrum are predicted. Since it is the leptogenesis credentials of these models that interest us, we have carefully identified all the key ingredients which can modify the overall scenario for lepton production, and compared how each element may differ in its predictions from those in the standard seesaw. The general conclusion we obtained is that having an extra abelian symmetry and singlet Higgs do not hinder the possibility of successful leptogenesis in these setups. In fact, their predictions are largely identical to the default case for both unflavored and flavored $N_1$-leptogenesis. Consequently, this shows another important feature of these economical seesaw models. 

In the second part of our work, we noted that conventional type~I seesaw inevitably gives rise to a largely unconstrained heavy RH seesaw sector, hence making it difficult to experimentally test this mass generation mechanism. Moreover, given the close relationship between neutrino models and leptogenesis, one would like to ascertain more information from the seesaw sector. Therefore, it is well-motivated to explore the possibility of constraining the heavy seesaw sector so that it is completely determined by low-energy observables. The strategy we employed in tackling this involves combining the powers of an expanded gauge group and an inter-generational flavor symmetry such that the former can enforce the desired mass relations while the latter leads to predictable diagonalization matrices. Subsequently, the simplest models of this type lead to a RH neutrino mass matrix that can be written in terms of only low-energy fermion masses, mixing angles and $CP$ phases:
\begin{equation}
 M_R \simeq \widehat{m}_f\,\Upmns^*\,\widehat{m}_\nu^{-1}\,\Upmns^\dagger\,\widehat{m}_f\;,
 \qquad \text{where } f=e,d,u \;.
\nonumber
\end{equation}
In the representative models that we have constructed to achieve the above relation, it has been found that successful leptogenesis is only possible in a very fine-tuned region of the parameter space. Specifically, one must select the $f=u$ case, as well as certain combinations of Dirac and Majorana phases in $\Upmns$ such that a lepton asymmetry can be generated via either resonant or flavored $N_2$-leptogenesis. Furthermore, it has been shown that although the $f=e$ case can yield a TeV scale RH neutrino, the probability of detecting it at the LHC or a next-generation collider such as the ILC is far too small.

But it is enlightening to note that this part of our work has demonstrated the ability of symmetries to bridge the missing link between the heavy seesaw sector and experimentally accessible parameters. In addition, it is highly probable that, with a different choice of the underlying symmetry and Higgs structure as shown here, interesting phenomenology can result and potentially provide new hints on building realistic neutrino seesaw models that are experimentally verifiable.

Finally, in the last part of our work, we investigated the effects of introducing an effective transition electromagnetic dipole moment operator between the LH light and RH heavy neutrinos. Since nonzero neutrino dipole moments are a direct consequence of having massive neutrinos, the inclusion of RH neutrinos to the SM via type~I seesaw immediately opens up the possibility of active-sterile neutrino transition moments. Such interactions are potentially important in cosmology because the radiative decay of the heavy RH neutrino ($N\rightarrow \nu\,\gamma$) violates lepton number, and in general, $CP$ . As a result, a new scenario for leptogenesis whereby the lepton asymmetry is solely generated by the EMDM-type (instead of the usual Yukawa-mediated) interactions is possible. By exploring the key ingredients leading to $CP$ violation, we have shown by explicit computations of the relevant diagrams in a toy model that, in principle, electromagnetic leptogenesis is a viable alternative for creating a lepton asymmetry. 

Eventually, to build a realistic theory that can be embedded into the SM, the simplistic setup was generalized such that it became compatible with the SM gauge group. Although in this case all relevant processes involved consisted of 3-body initial or final states, it has been found that a nonzero $CP$ asymmetry can still be generated, and that the effect of the suppression factor due to 3-body phase space can be reduced by the appropriate choice of the RH neutrino mass hierarchy. In fact, this scenario disfavors the usual strongly hierarchical RH mass spectrum normally assumed for standard leptogenesis. However, we have demonstrated that for a mild hierarchy, where $N_1$-leptogenesis is still a good approximation, there exists a workable parameter space where electromagnetic leptogenesis alone can produce the necessary lepton asymmetry to solve the baryogenesis problem. In addition, foregoing the hierarchical requirement completely and having quasi-degenerate RH neutrinos is perfectly acceptable, and perhaps even more desirable in some situations as resonant effects can open up a larger region for successful electromagnetic leptogenesis. So, there is no doubt that transition EMDM interactions between light and heavy neutrinos can have far-reaching consequences in the early universe.

In conclusion, this work has further highlighted the already well-known fact of the inseparability between neutrino physics and cosmology. Even though we have only explored a tiny region of these two vast topics, it is nonetheless enlightening to witness the richness of ideas allowed and motivated by them through our meticulous investigations. It is true however that many important questions remain to be answered, and therefore we await the new data which will come with future technological advances to inform us of the model that really represents how Nature works.





{\singlespacing
\chapter*{List of Publications}\label{listofpub}
  \addcontentsline{toc}{chapter}{\numberline{}List of Publications} }

\vspace{30pt}

\begin{enumerate}
 \item   S.~S.~C.~Law and R.~R.~Volkas,\\
  {\it Leptogenesis implications in models with Abelian family symmetry and one
  extra real Higgs singlet},\\
  Phys.\ Rev.\  D {\bf 75}, 043510 (2007)
  [arXiv:hep-ph/0701189].\\
 \item   N.~F.~Bell, B.~J.~Kayser and S.~S.~C.~Law,\\
  {\it Electromagnetic Leptogenesis},\\
  Phys.\ Rev.\  D {\bf 78}, 085024 (2008)
  [arXiv:0806.3307 [hep-ph]]\\
 \item   X.~G.~He, S.~S.~C.~Law and R.~R.~Volkas,\\
  {\it Determining the heavy seesaw neutrino mass matrix from low-energy parameters},\\
  Phys.\ Rev.\  D {\bf 78}, 113001 (2008)
  [arXiv:0810.1104 [hep-ph]].
\end{enumerate}


{\singlespacing  

\renewcommand{\rightmark}{\cmarkfont Bibliography}
\renewcommand{\leftmark}{\cmarkfont Bibliography}

\addcontentsline{toc}{chapter}{\numberline{}Bibliography}

%
%
%
%
%



}


\appendix

\renewcommand{\chaptermark}[1]{\markboth{\cmarkfont \appendixname~\thechapter. #1}{}}

%
%
%
%
%



\chapter{Supplement materials for chapter~1}

\section{Boltzmann equation for \textit{N}$_\text{1}$}\label{app:a_BE_N1}

In this section, we include all the steps leading to the derivation of Eq.~(\ref{eq1:lepto_std_BE_N1_raw}) on page~\pageref{eq1:lepto_std_BE_N1_raw}. Starting with the Boltzmann equation of (\ref{eq1:lepto_std_BE_final_form}) and  writing down all the relevant interactions on the RHS, we obtain

\begin{align}
 \frac{d n_{N_1}}{d t} +3H n_{N_1} =&
  - \frac{n_{N_1}}{n_{N_1}^\text{eq}}\gamma(N_1 \rightarrow \ell\bar{\phi})
  + \frac{n_{\bar{\ell}}}{n_{\ell}^\text{eq}}\gamma(\bar{\ell}\phi \rightarrow N_1)
  - \frac{n_{N_1}}{n_{N_1}^\text{eq}}\gamma(N_1 \rightarrow \bar{\ell}\phi)
  \nonumber\\
  &\; 
  + \frac{n_{\ell}}{n_{\ell}^\text{eq}}\gamma(\ell\bar{\phi} \rightarrow N_1)
  - \frac{n_{N_1}n_{\ell}}{n_{N_1}^\text{eq}n_{\ell}^\text{eq}}
        \gamma(\ell N_1 \rightarrow \bar{t}_R q_L)_s
  + \gamma(\bar{t}_R q_L \rightarrow \ell N_1)_s
  \nonumber\\
  &\; 
  - \frac{n_{N_1}n_{\bar{\ell}}}{n_{N_1}^\text{eq}n_{\ell}^\text{eq}}
        \gamma(\bar{\ell} N_1 \rightarrow t_R \bar{q}_L)_s
  + \gamma(t_R \bar{q}_L \rightarrow \bar{\ell} N_1)_s
  - \frac{n_{N_1}}{n_{N_1}^\text{eq}}\gamma(N_1 q_L \rightarrow \ell t_R)_t
  \nonumber\\
  &\; 
  - \frac{n_{N_1}}{n_{N_1}^\text{eq}}\gamma(N_1 \bar{q}_L \rightarrow \bar{t}_R\bar{\ell})_t
  + \frac{n_{\ell}}{n_{\ell}^\text{eq}}\gamma(\ell t_R \rightarrow N_1 q_L)_t
  + \frac{n_{\ell}}{n_{\ell}^\text{eq}}\gamma(\ell\bar{q}_L \rightarrow N_1 \bar{t}_R)_t
  \nonumber\\
  &\; 
  - \frac{n_{N_1}}{n_{N_1}^\text{eq}}\gamma(N_1 t_R\rightarrow q_L\bar{\ell} )_t
  - \frac{n_{N_1}}{n_{N_1}^\text{eq}}\gamma(N_1 \bar{t}_R\rightarrow \bar{q}_L\ell )_t
  + \frac{n_{\bar{\ell}}}{n_{\ell}^\text{eq}}
      \gamma(\bar{\ell} \bar{t}_R\rightarrow N_1 \bar{q}_L)_t
  \nonumber\\
  &\; 
  + \frac{n_{\bar{\ell}}}{n_{\ell}^\text{eq}}
      \gamma(\bar{\ell} q_L \rightarrow N_1 t_R)_t \;,
      \label{eqa1:BE_N1_first}
\end{align}

\noindent
where the subscripts $s$ and $t$ denote $s$- and $t$-channel processes respectively. In writing down (\ref{eqa1:BE_N1_first}), we have assumed that $n_\ell^\text{eq}\equiv n_{\bar{\ell}}^\text{eq}$ and $n_{\phi,t_R, q_L} \equiv n_{\phi,t_R, q_L}^\text{eq}$. Using 
(\ref{eq1:lepto_std_BE_new_rate_D1}) and (\ref{eq1:lepto_std_BE_new_rate_D2}), and the assumption that $\gamma(aX\rightarrow Y) \equiv \gamma(Y\rightarrow aX)$ for all processes, (\ref{eqa1:BE_N1_first}) becomes
\begin{align}
 \frac{d n_{N_1}}{d t} +3H n_{N_1} =&
  - \frac{n_{N_1}}{n_{N_1}^\text{eq}}(1+\varepsilon_1)\gamma_D
  + \frac{n_{\bar{\ell}}}{n_{\ell}^\text{eq}}(1+\varepsilon_1)\gamma_D
  - \frac{n_{N_1}}{n_{N_1}^\text{eq}}(1-\varepsilon_1)\gamma_D
  + \frac{n_{\ell}}{n_{\ell}^\text{eq}}(1-\varepsilon_1)\gamma_D
    \nonumber\\
  &\; 
  + 2 \gamma_{\phi,s}
  - \frac{n_{N_1}(n_{\ell}+n_{\bar{\ell}})}{n_{N_1}^\text{eq}n_{\ell}^\text{eq}}
        \gamma_{\phi,s}
  -  \frac{4 n_{N_1}}{n_{N_1}^\text{eq}}\gamma_{\phi,t}
  +  \frac{2(n_{\ell}+n_{\bar{\ell}})}{n_{\ell}^\text{eq}}\gamma_{\phi,t} \;,
  \label{eqa1:BE_N1_second}
\end{align}
where 
\begin{align}
\gamma_{\phi,s} 
  &=  \gamma(\ell N_1 \rightarrow \bar{t}_R q_L)_s 
   = \gamma(\bar{t}_R q_L \rightarrow \ell N_1)_s 
  = \gamma(t_R \bar{q}_L \rightarrow \bar{\ell} N_1)_s 
   =  \gamma(\bar{\ell} N_1 \rightarrow t_R \bar{q}_L)_s\;,
\end{align}
and
\begin{align}
 \gamma_{\phi,t} 
 &= \gamma(N_1 q_L \rightarrow \ell t_R)_t
 = \gamma(N_1 \bar{q}_L \rightarrow \bar{t}_R\bar{\ell})_t
 = \gamma(\ell t_R \rightarrow N_1 q_L)_t 
 = \gamma(\ell\bar{q}_L \rightarrow N_1 \bar{t}_R)_t \;,\nonumber\\
 &= \gamma(N_1 t_R\rightarrow q_L\bar{\ell} )_t 
 = \gamma(N_1 \bar{t}_R\rightarrow \bar{q}_L\ell )_t\;,
 = \gamma(\bar{\ell} \bar{t}_R\rightarrow N_1 \bar{q}_L)_t 
 = \gamma(\bar{\ell} q_L \rightarrow N_1 t_R)_t\;.
\end{align}
Simplifying (\ref{eqa1:BE_N1_second}), we get
\begin{align}
 \frac{d n_{N_1}}{d t} +3H n_{N_1} =&
  - \frac{2 n_{N_1}}{n_{N_1}^\text{eq}}\gamma_D
  + \left(\frac{n_{\ell}+n_{\bar{\ell}}}{n_{\ell}^\text{eq}}\right)\gamma_D
  + \frac{n_{\bar{\ell}}-n_{\ell}}{n_{\ell}^\text{eq}}\varepsilon_1\gamma_D
  + 2 \gamma_{\phi,s}  
     \nonumber\\
  &\; 
  - \left(\frac{n_{\ell}+n_{\bar{\ell}}}{n_{\ell}^\text{eq}}\right)
    \frac{n_{N_1}}{n_{N_1}^\text{eq}}\gamma_{\phi,s}
  -  \frac{4 n_{N_1}}{n_{N_1}^\text{eq}}\gamma_{\phi,t}
  + 2 \left(\frac{n_{\ell}+n_{\bar{\ell}}}{n_{\ell}^\text{eq}}\right)
  \gamma_{\phi,t} \;,  \\
   =&  
  - \frac{2 n_{N_1}}{n_{N_1}^\text{eq}}\gamma_D
  + 2\gamma_D
  + 2 \gamma_{\phi,s}  
  -  \frac{2n_{N_1}}{n_{N_1}^\text{eq}}\gamma_{\phi,s}
  -  \frac{4 n_{N_1}}{n_{N_1}^\text{eq}}\gamma_{\phi,t}
  + 4 \gamma_{\phi,t} 
  +\order{\varepsilon_1, \frac{\mu_\ell}{T}}\;, \label{eqa1:BE_N1_third} \\
  =&
 -2 \left[
 \frac{n_{N_1}}{n_{N_1}^\text{eq}} -1
   \right] (\gamma_D + \gamma_{\phi,s} + 2 \gamma_{\phi,t})
   + \order{\varepsilon_1, \frac{\mu_\ell}{T}}\;,
\end{align}
which is the result of (\ref{eq1:lepto_std_BE_N1_raw}). In (\ref{eqa1:BE_N1_third}), we have used definition (\ref{eq1:lepto_std_BE_n_density}) to write
\begin{align}
 \frac{n_{\ell}+n_{\bar{\ell}}}{n_{\ell}^\text{eq}} 
 &= 
  \left(
  \frac{g_\ell}{2\pi^2} \int f_\ell^\text{eq}(E)\,E^2\, dE\right)^{-1}
  \left(
   \frac{g_\ell}{2\pi^2} \int \left[f_\ell(E) + f_{\bar{\ell}}(E)\right]
    \,E^2\, dE
  \right)
  \;, \quad m_\ell \ll T\;,\nonumber\\
 &=
  \left(
  \int e^{-E/T}\,E^2\, dE
  \right)^{-1}
 \left(
   \int \left[e^{-(E -\mu_\ell)/T} +
   e^{-(E +\mu_{\ell})/T} \right]
    \,E^2\, dE
  \right)
  \;,\nonumber\\
  &= \left(
  \int e^{-E/T}\,E^2\, dE
  \right)^{-1} 
  2 \cosh \left(\frac{\mu_\ell}{T}\right)  \int e^{-E/T}\,E^2\, dE 
  \;,\nonumber\\
   &= 2 + \order{\frac{\mu_\ell}{T}}\;. \label{eqa1:BE_nl+nlb=2}
\end{align}  
where we have used Maxwell-Boltzmann distribution for the phase space densities and imposed the condition for kinetic equilibrium $\mu_\ell \equiv -\mu_{\bar{\ell}}$.

\section{Boltzmann equation for \textit{B}$-$\textit{L}}\label{app:a_BE_B-L}

To obtain the result of (\ref{eq1:lepto_std_BE_B-L_raw}) on page~\pageref{eq1:lepto_std_BE_B-L_raw}, we begin by writing down the evolution equation for particle density $n_{\bar\ell}$: 
\begin{align}
 \frac{d n_{\bar\ell}}{d t} +3H n_{\bar\ell} =&\;
   \frac{n_{N_1}}{n_{N_1}^\text{eq}}\gamma(N_1 \rightarrow \bar{\ell}\phi)
  - \frac{n_{\bar{\ell}}}{n_{\ell}^\text{eq}}\gamma(\bar{\ell}\phi \rightarrow N_1)
  + \frac{n_{\ell}}{n_{\ell}^\text{eq}}
     \widetilde{\gamma}(\ell\bar{\phi} \rightarrow \bar{\ell}\phi)_s
  \nonumber\\
  &\; 
  - \frac{n_{\bar\ell}}{n_{\ell}^\text{eq}}
     \widetilde{\gamma}(\bar{\ell}\phi \rightarrow\ell\bar{\phi} )_s
  + \frac{n_{\ell}}{n_{\ell}^\text{eq}}
     \gamma(\ell\bar{\phi} \rightarrow \bar{\ell}\phi)_t
  - \frac{n_{\bar\ell}}{n_{\ell}^\text{eq}}
     \gamma(\bar{\ell}\phi \rightarrow\ell\bar{\phi} )_t
  \nonumber\\
  &\; 
  + \gamma(t_R \bar{q}_L \rightarrow \bar{\ell} N_1)_s
  - \frac{n_{N_1}n_{\bar{\ell}}}{n_{N_1}^\text{eq}n_{\ell}^\text{eq}}
        \gamma(\bar{\ell} N_1 \rightarrow t_R \bar{q}_L)_s
  + \frac{n_{N_1}}{n_{N_1}^\text{eq}}\gamma(N_1 t_R\rightarrow q_L\bar{\ell} )_t
  \nonumber\\
  &\; 
  -\frac{n_{\bar{\ell}}}{n_{\ell}^\text{eq}}
      \gamma(\bar{\ell} q_L \rightarrow N_1 t_R)_t
  + \frac{n_{N_1}}{n_{N_1}^\text{eq}}\gamma(N_1 \bar{q}_L \rightarrow \bar{t}_R\bar{\ell})_t
  - \frac{n_{\bar{\ell}}}{n_{\ell}^\text{eq}}
      \gamma(\bar{\ell} \bar{t}_R\rightarrow N_1 \bar{q}_L)_t
  \;,
        \label{eqa1:BE_B-L_ell1}\\
 =&\;
   \frac{n_{N_1}}{n_{N_1}^\text{eq}} (1-\varepsilon_1)\gamma_D
  - \frac{n_{\bar{\ell}}}{n_{\ell}^\text{eq}}(1+\varepsilon_1)\gamma_D
  + \frac{n_{\ell}}{n_{\ell}^\text{eq}}
     (\gamma_{N,s}+\varepsilon_1\gamma_D)
  - \frac{n_{\bar\ell}}{n_{\ell}^\text{eq}}
     (\gamma_{N,s}-\varepsilon_1\gamma_D)  
   \nonumber\\
  &\; 
  + \frac{n_{\ell}}{n_{\ell}^\text{eq}}
     \gamma_{N,t}
  - \frac{n_{\bar\ell}}{n_{\ell}^\text{eq}}
     \gamma_{N,t}
  + \gamma_{\phi,s}
  - \frac{n_{N_1}n_{\bar{\ell}}}{n_{N_1}^\text{eq}n_{\ell}^\text{eq}}
        \gamma_{\phi,s}
  + \frac{n_{N_1}}{n_{N_1}^\text{eq}}\gamma_{\phi,t}
  -\frac{n_{\bar{\ell}}}{n_{\ell}^\text{eq}}
      \gamma_{\phi,t}
  + \frac{n_{N_1}}{n_{N_1}^\text{eq}}\gamma_{\phi,t}
   \nonumber\\
  &\; 
  - \frac{n_{\bar{\ell}}}{n_{\ell}^\text{eq}}
      \gamma_{\phi,t}
  \;,\\
 \frac{d n_{\bar\ell}}{d t} +3H n_{\bar\ell}=&\;
   \frac{n_{N_1}}{n_{N_1}^\text{eq}} (1-\varepsilon_1)\gamma_D
  - \frac{n_{\bar{\ell}}- n_{\ell}\,\varepsilon_1}{n_{\ell}^\text{eq}} \gamma_D
  - \frac{n_{\bar{\ell}}- n_{\ell}}{n_{\ell}^\text{eq}}\gamma_{N,s}
  - \frac{n_{\bar\ell}-n_{\ell}}{n_{\ell}^\text{eq}}
     \gamma_{N,t}
  + \gamma_{\phi,s}
 \nonumber\\
 &\;
  - \frac{n_{N_1}n_{\bar{\ell}}}{n_{N_1}^\text{eq}n_{\ell}^\text{eq}}
        \gamma_{\phi,s}
  + \frac{2 n_{N_1}}{n_{N_1}^\text{eq}}\gamma_{\phi,t}
  -\frac{2 n_{\bar{\ell}}}{n_{\ell}^\text{eq}}
      \gamma_{\phi,t}
  \;. \label{eqa1:BE_B-L_ell1_final}
\end{align}
Similarly, we can write down the equation for $n_\ell$ as
\begin{align}
 \frac{d n_{\ell}}{d t} +3H n_{\ell}=&\;
   \frac{n_{N_1}}{n_{N_1}^\text{eq}} (1+\varepsilon_1)\gamma_D
  - \frac{n_{\ell}+ n_{\bar{\ell}}\,\varepsilon_1}{n_{\ell}^\text{eq}} \gamma_D
  - \frac{n_{\ell}- n_{\bar{\ell}}}{n_{\ell}^\text{eq}}\gamma_{N,s}
  - \frac{n_{\ell}-n_{\bar\ell}}{n_{\ell}^\text{eq}}
     \gamma_{N,t}
  + \gamma_{\phi,s}
 \nonumber\\
 &\;
  - \frac{n_{N_1}n_{\ell}}{n_{N_1}^\text{eq}n_{\ell}^\text{eq}}
        \gamma_{\phi,s}
  + \frac{2 n_{N_1}}{n_{N_1}^\text{eq}}\gamma_{\phi,t}
  -\frac{2 n_{\ell}}{n_{\ell}^\text{eq}}
      \gamma_{\phi,t}
  \;.
 \label{eqa1:BE_B-L_ell2_final}
\end{align}
Subtracting (\ref{eqa1:BE_B-L_ell2_final}) from (\ref{eqa1:BE_B-L_ell1_final}), we have
\begin{align}
 \frac{d n_{B-L}}{d t} +3H n_{B-L}=&\;
  -2\varepsilon_1 \gamma_D \frac{n_{N_1}}{n_{N_1}^\text{eq}} 
  +\left(\frac{n_{\ell}+ n_{\bar{\ell}}}{n_{\ell}^\text{eq}}\right)\varepsilon_1 \gamma_D
  - \frac{n_{B-L}}{n_{\ell}^\text{eq}} \gamma_D
  - \frac{2n_{B-L}}{n_{\ell}^\text{eq}}\gamma_{N,s}
 \nonumber\\
 &\;
  - \frac{2 n_{B-L}}{n_{\ell}^\text{eq}} \gamma_{N,t}
  - \frac{n_{N_1}n_{B-L}}{n_{N_1}^\text{eq}n_{\ell}^\text{eq}}
        \gamma_{\phi,s}
  -\frac{2 n_{B-L}}{n_{\ell}^\text{eq}}
      \gamma_{\phi,t}
  \;,
   \label{eqa1:BE_B-L_final_0}
\end{align}
where we have defined $n_{\bar\ell} - n_\ell \equiv n_{B-L}$. Using (\ref{eqa1:BE_nl+nlb=2}) to simplify, we then get
\begin{align}
 \frac{d n_{B-L}}{d t} +3H n_{B-L} &=
  -2\varepsilon_1  \left[\frac{n_{N_1}}{n_{N_1}^\text{eq}}-1\right]\gamma_D
  - \frac{n_{B-L}}{n_{\ell}^\text{eq}} 
  \left[
  \gamma_D
  - 2\gamma_{N,s}
  - 2\gamma_{N,t}
  - \frac{n_{N_1}}{n_{N_1}^\text{eq}}\gamma_{\phi,s}
  -2\gamma_{\phi,t}
  \right]
  \;,\nonumber
\end{align}
\begin{equation}
  =
  -2\varepsilon_1  \left[\frac{n_{N_1}}{n_{N_1}^\text{eq}}-1\right]\gamma_D
  - \frac{n_{B-L}}{n_{\ell}^\text{eq}} 
  \gamma_W + \order{\varepsilon_1^2, \frac{\mu_\ell}{T}}
  \;,
\end{equation}
which is the result of (\ref{eq1:lepto_std_BE_B-L_raw}).


%
%
%
%
%



\chapter{Supplement materials for chapter~3}

\section[Properties of the \textit{A}$_\text{4}$ group]
{Properties of the \textit{A}$_\text{4}$ group}\label{app:b_A4_def}

$A_4$ is the alternating group of order 4. It is isomorphic to the group representing the proper rotational symmetries of a regular tetrahedron. It has 12 elements and 4 conjugacy classes. One set contains the identity: $\left\{I\right\}$, two sets contain four 3-fold rotations each: 
$\left\{R_1,R_2,R_3,R_4\right\},\left\{R_1^2,R_2^2,R_3^2,R_4^2\right\}$,
and one set contains three 2-fold rotations: $\left\{r_1,r_2,r_3\right\}$. 
The character table for this group is shown in Table~\ref{tableB:a4_character}.
By the dimensionality theorem, we know that $A_4$ must have four irreducible representations: $\rep, \repp,\reppp$ and $\rept$, where $\rep$ is the trivial representation, $\repp$ and $\reppp$ are non-trivial one-dimensional representations that are complex conjugates of each other, while $\rept$ is a real three-dimensional representation.

All elements of the group may be generated by two generators: $S$ and $T$ with the properties~\cite{group_theory_books}:
\begin{equation}
 S^2=T^3=(ST)^3=I\;.
\end{equation}
So, for example, one can make the identification: $S\equiv r_1$ and $T\equiv R_1$. The one-dimensional representations can be generated by:
\begin{align}
 &\rep:\quad S=1\;, \quad T=1\;, \\
 &\repp:\quad S=1\;, \quad T=e^{2\pi i/3}\equiv \omega\;, \\
 &\reppp:\quad S=1\;, \quad T=e^{-2\pi i/3}\equiv \omega^2\;,
\end{align}
while for the real three-dimensional representation, we may choose:
\begin{equation}
 S = \Matdiag{1}{-1}{-1} 
 \;,\qquad
T = \ththMat
 {0}{1}{0}
 {0}{0}{1}
 {1}{0}{0} 
 \;. 
\end{equation}
Using this, we can obtain the rest of the $3\times 3$ representation matrices for $A_4$ ($I$ not shown):
\begin{align}
 & ST = \ththMat{0}{1}{0}
                {0}{0}{-1}
                {-1}{0}{0}\;,\;\; 
  TS = \ththMat{0}{-1}{0}
                {0}{0}{-1}
                {1}{0}{0}\;, \;\;               
  STS =\ththMat{0}{-1}{0}
                {0}{0}{1}
                {-1}{0}{0}\;,\\
 & (ST)^2 = \ththMat{0}{0}{-1}
                    {1}{0}{0}
                    {0}{-1}{0}\;,\;\; 
  (TS)^2 = \ththMat{0}{0}{1}
                   {-1}{0}{0}
                   {0}{-1}{0}\;, \;\;               
  TST =\ththMat{0}{0}{-1}
                {-1}{0}{0}
                {0}{1}{0}\;,\\
 & T^2 =\ththMat{0}{0}{1}
                {1}{0}{0}
                {0}{1}{0}\;,\;\;
  TST^2S =\ththMat{-1}{0}{0}
                  {0}{1}{0}
                  {0}{0}{-1}\;,\;\;                        
  ST^2ST =\ththMat{-1}{0}{0}
                  {0}{-1}{0}
                  {0}{0}{1}\;.\;\;             
\end{align}

%
\begin{table}[t]
\begin{center}
\begin{tabular}{|c||c|c|c|c|c|c|}
\hline
class  &$\chi_1$ &$\chi_2$ &$\chi_3$ &$\chi_4$ &$n_{\mathcal{C}_i}$&$h_{\mathcal{C}_i}$ \\
\hline
$\mathcal{C}_1$  &1 &1 &1 &3 &1 &1\\
$\mathcal{C}_2$  &1 &$\omega$ &$\omega^2$ &0 &4 &3\\
$\mathcal{C}_3$  &1 &$\omega^2$ &$\omega$ &0 &4 &3\\
$\mathcal{C}_4$  &1 &1 &1 &$-1$ &3 &2\\
\hline
\end{tabular}\caption{Character table for the $A_4$ group. In the table, $\omega$ denotes $e^{2\pi i/3}$, while $n_{\mathcal{C}_i}$ and $h_{\mathcal{C}_i}$ are the number of distinct elements and the order of the elements in $i$th conjugacy class respectively.}
\label{tableB:a4_character}
\end{center}
\end{table}

%
%
%
The corresponding multiplication table for this group is given in Table~\ref{tableB:a4_multiply}. In calculations, it is useful to know the basic tensor product rules, which are given by:
\begin{align}
 &\rep \otimes \rep = \rep \;,\\
 &\repp \otimes \reppp = \rep \;,\\
 &\repp \otimes \repp = \reppp \;,\\
 &\rept \otimes \rept = \rep \oplus \repp \oplus \reppp \oplus \rept_a \oplus \rept_s\;,\label{eqb1:3x3}
\end{align}
where subscripts $a$ and $s$ denote ``asymmetric'' and ``symmetric'' respectively. For the multiplication of two triplets in Eq.~(\ref{eqb1:3x3}), we use the following convention: suppose $x_{\rept}=(x_1,x_2,x_3)$ and $y_{\rept}=(y_1,y_2,y_3)$ are triplets in $A_4$, then the three singlets and two triplets contained in the product $(x_{\rept}\otimes y_{\rept})$ are given by
\begin{align}
 (x_{\rept}\; y_{\rept})_{\rep} &= x_1 y_1 + x_2 y_2 + x_3 y_3\;,\\
 (x_{\rept}\; y_{\rept})_{\repp} &= x_1 y_1 + \omega x_2 y_2 + \omega^2 x_3 y_3\;,\\
 (x_{\rept}\; y_{\rept})_{\reppp} &= x_1 y_1 + \omega^2 x_2 y_2 + \omega x_3 y_3\;,\\
 (x_{\rept}\; y_{\rept})_{\rept_a} &= (x_2 y_3 - x_3 y_2\;, x_3 y_1 - x_1 y_3\;, x_1 y_2 - x_2 y_1)\;,\\
 (x_{\rept}\; y_{\rept})_{\rept_s} &= (x_2 y_3 + x_3 y_2\;, x_3 y_1 + x_1 y_3\;, x_1 y_2 + x_2 y_1)\;,
\end{align}
where $\omega = e^{2\pi i/3}$ and we have abbreviated $(x_{\rept}\otimes y_{\rept})$ with $(x_{\rept}\; y_{\rept})$.\\

%
\begin{table}[t]
\begin{center}
\begin{tabular}{|>{\columncolor[gray]{0.8}[\tabcolsep]}c|c|c|c|c|c|c|c|c|c|c|c|c|}
\hline
\rowcolor[gray]{0.8}
\cellcolor[gray]{1}$*$
&$I$ &$R_1$ &$R_2$ &$R_3$ &$R_4$ &$R_1^2$ &$R_2^2$ &$R_3^2$ &$R_4^2$ &$r_1$ &$r_2$ &$r_3$\\
\hline
$I$&$I$&$R_1$ &$R_2$ &$R_3$ &$R_4$ &$R_1^2$ &$R_2^2$ &$R_3^2$ &$R_4^2$ &$r_1$ &$r_2$ &$r_3$\\
\hline
$R_1$ &$R_1$ &$R_1^2$ &$R_4^2$ &$R_2^2$ &$R_3^2$& $I$  &$r_3$ &$r_1$ &$r_2$ &$R_4$&$R_2$ &$R_3$ \\
\hline
$R_2$ &$R_2$ &$R_3^2$ &$R_2^2$ &$R_4^2$ &$R_1^2$ &$r_3$ &$I$ &$r_2$ &$r_1$ &$R_3$ &$R_1$&$R_4$ \\
\hline
$R_3$ &$R_3$ &$R_4^2$ &$R_1^2$ &$R_3^2$ &$R_2^2$ &$r_1$ &$r_2$ &$I$ &$r_3$ &$R_2$ &$R_4$&$R_1$ \\
\hline
$R_4$ &$R_4$ &$R_2^2$ &$R_3^2$ &$R_1^2$ &$R_4^2$ &$r_2$ &$r_1$ &$r_3$ &$I$ &$R_1$ &$R_3$&$R_2$ \\
\hline
$R_1^2$ &$R_1^2$ &$I$  &$r_2$ &$r_3$ &$r_1$ &$R_1$ &$R_3$ &$R_4$ &$R_2$ &$R_3^2$ &$R_4^2$ &$R_2^2$ \\
\hline
$R_2^2$ &$R_2^2$ &$r_2$ &$I$ &$r_1$ &$r_3$ &$R_4$ &$R_2$ &$R_1$ &$R_3$ &$R_4^2$ &$R_3^2$ &$R_1^2$ \\
\hline
$R_3^2$ &$R_3^2$ &$r_3$ &$r_1$ &$I$ &$r_2$ &$R_2$ &$R_4$ &$R_3$ &$R_1$ &$R_1^2$ &$R_2^2$ &$R_4^2$ \\
\hline
$R_4^2$ &$R_4^2$ &$r_1$ &$r_3$ &$r_2$ &$I$ &$R_3$ &$R_1$ &$R_2$ &$R_4$ &$R_2^2$ &$R_1^2$ &$R_3^2$ \\
\hline
$r_1$ &$r_1$ &$R_3$ &$R_4$ &$R_1$ &$R_2$ &$R_4^2$ &$R_3^2$ &$R_2^2$ &$R_1^2$ &$I$ &$r_3$ &$r_2$ \\
\hline
$r_2$ &$r_2$ &$R_4$ &$R_3$ &$R_2$ &$R_1$ &$R_2^2$ &$R_1^2$ &$R_4^2$ &$R_3^2$ &$r_3$ &$I$  &$r_1$ \\
\hline
$r_3$ &$r_3$ &$R_2$ &$R_1$ &$R_4$ &$R_3$ &$R_3^2$ &$R_4^2$ &$R_1^2$ &$R_2^2$ &$r_2$ &$r_1$ &$I$ \\
\hline
\end{tabular}\caption{Multiplication table for the $A_4$ group. Note that multiplication is performed in following order: $G_\text{col} * G_\text{row}$, where $G_\text{col}$ is an element from the grayed column whilst $G_\text{row}$ is an element from the grayed row in the table.}
\label{tableB:a4_multiply}
\end{center}
\end{table}

\newpage

{\bf\noindent
The following is the explicit calculation for each of the mass matrices omitted in Sec.~\ref{subsec:HLV_QL} and Sec.~\ref{subsec:HLV_LR}.
}

\section{Mass matrices in the quark-lepton model}
\label{app:b_QL_details}

\noindent 
$\boxed{m_e}\;\;$ \textbf{charged lepton mass matrix} 
\begin{align}
&g_1\, (\overline{e}_L\,\vev{\phi_{2}^0})_{\underline{1}}\,e_R
 + g_1'\,(\overline{e}_L\,\vev{\phi_{2}^0})_{\underline{1}'}\,e_R''
 + g_1''\,(\overline{e}_L\,\vev{\phi_{2}^0})_{\underline{1}''}\,e_R' \nonumber\\
&\quad
 -g_2\, (\overline{e}_L\,\vev{\phi_{1}^{0*}})_{\underline{1}}\,e_R
 - g_2'\,(\overline{e}_L\,\vev{\phi_{1}^{0*}})_{\underline{1}'}\,e_R''
 - g_2''\,(\overline{e}_L\,\vev{\phi_{1}^{0*}})_{\underline{1}''}\,e_R'
 + \text{h.c.}\;,
\end{align} 
\begin{align}
 =& \threerMat{\overline{e}_{L1}}{\overline{e}_{L2}}{\overline{e}_{L3}}
 \left[
  \ththMat
  {g_1\, v_{2} }{g_1'\, v_{2} }{g_1''\, v_{2} }
  {g_1\, v_{2} }{\omega g_1'\, v_{2} }{\omega^2 g_1''\, v_{2} }
  {g_1\, v_{2} }{\omega^2 g_1'\, v_{2} }{\omega g_1''\, v_{2} }
  -
  \ththMat
  {g_2\, v_1^* }{g_2'\, v_1^* }{g_2''\, v_1^* }
  {g_2\, v_1^* }{\omega g_2'\, v_1^* }{\omega^2 g_2''\, v_1^* }
  {g_2\, v_1^* }{\omega^2 g_2'\, v_1^* }{\omega g_2''\, v_1^* }
 \right]
  \threecMat{e_R}{e_R''}{e_R'}
  \nonumber
  \\ 
  &~\hspace{11cm}
   + \text{h.c.}\;,
\intertext{where subscripts 1,2,3 of $e_L$ are the flavor indices and we have substituted in the vacuum values $\vev{\phi_{2}^0} = (v_{2},v_{2},v_{2})$ and
$\vev{\phi_1^{0*}} = (v_1^*, v_1^*, v_1^*)$,}
 =&
  \threerMat{\overline{e}_{L1}}{\overline{e}_{L2}}{\overline{e}_{L3}}
  \Uw
  \underbrace{\sqrt{3}
  \Matdiag
  {g_1 v_2 - g_2 v_1^*}{g_1' v_2 - g_2' v_1^*}{g_1'' v_2 - g_2'' v_1^*}
  }_{\widehat{m}_e}
  \threecMat{e_R}{e_R''}{e_R'}+ \text{h.c.}\;,
\end{align}
where
\begin{equation}
 \Uw = \MatU\;.
\end{equation}
From this we can see that \fbox{$V_{eL}^\dagger = \Uw$ and $V_{eR} = I$} .\\

\noindent 
$\boxed{m_u}\;\;$ \textbf{up-type quark mass matrix} 
\begin{align}
  &g_1\, (\overline{u}_L\,\vev{\phi_{1}^0})_{\underline{1}}\,u_R
 + g_1'\,(\overline{u}_L\,\vev{\phi_{1}^0})_{\underline{1}'}\,u_R''
 + g_1''\,(\overline{u}_L\,\vev{\phi_{1}^0})_{\underline{1}''}\,u_R' \nonumber\\
&\quad
 +g_2\, (\overline{u}_L\,\vev{\phi_{2}^{0*}})_{\underline{1}}\,u_R
 + g_2'\,(\overline{u}_L\,\vev{\phi_{2}^{0*}})_{\underline{1}'}\,u_R''
 + g_2''\,(\overline{u}_L\,\vev{\phi_{2}^{0*}})_{\underline{1}''}\,u_R'
 + \text{h.c.}\;,
\end{align}
\begin{align}
 =& \threerMat{\overline{u}_{L1}}{\overline{u}_{L2}}{\overline{u}_{L3}}
 \left[
  \ththMat
  {g_1\, v_1 }{g_1'\, v_1 }{g_1''\, v_1 }
  {g_1\, v_1 }{\omega g_1'\, v_1 }{\omega^2 g_1''\, v_1 }
  {g_1\, v_1 }{\omega^2 g_1'\, v_1 }{\omega g_1''\, v_1 }
  +
  \ththMat
  {g_2\, v_2^* }{g_2'\, v_2^* }{g_2''\, v_2^* }
  {g_2\, v_2^* }{\omega g_2'\, v_2^* }{\omega^2 g_2''\, v_2^* }
  {g_2\, v_2^* }{\omega^2 g_2'\, v_2^* }{\omega g_2''\, v_2^* }
 \right]
  \threecMat{u_R}{u_R''}{u_R'}
    \nonumber
  \\ 
  &~\hspace{11cm}
   + \text{h.c.}\;,
\intertext{where we have substituted in the vacuum values $\vev{\phi_{1}^0} = (v_1,v_1,v_1)$ and
$\vev{\phi_2^{0*}} = (v_2^*, v_2^*, v_2^*)$,}
 =& 
 \threerMat{\overline{u}_{L1}}{\overline{u}_{L2}}{\overline{u}_{L3}}
  \Uw
  \underbrace{\sqrt{3}
  \Matdiag
  {g_1 v_1 + g_2 v_2^*}{g_1' v_1 + g_2' v_2^*}{g_1'' v_1 + g_2'' v_2^*}
  }_{\widehat{m}_u}
  \threecMat{u_R}{u_R''}{u_R'}+ \text{h.c.}\;.  
\end{align}
Hence, \fbox{$V_{uL} = \Uw^\dagger$ and $V_{uR} = I$} .\\

\noindent 
$\boxed{m_d}\;\;$ \textbf{down-type quark mass matrix} 
\begin{align}
  &g_3\, (\overline{d}_L\,\vev{\phi_{d}^0})_{\underline{1}}\,d_R
 + g_3'\,(\overline{d}_L\,\vev{\phi_{d}^0})_{\underline{1}'}\,d_R''
 + g_3''\,(\overline{d}_L\,\vev{\phi_{d}^0})_{\underline{1}''}\,d_R' + \text{h.c.}\;,\\
 =& \threerMat{\overline{d}_{L1}}{\overline{d}_{L2}}{\overline{d}_{L3}}
  \ththMat
  {g_3\, v_d }{g_3'\, v_d }{g_3''\, v_d }
  {g_3\, v_d }{\omega g_3'\, v_d }{\omega^2 g_3''\, v_d }
  {g_3\, v_d }{\omega^2 g_3'\, v_d }{\omega g_3''\, v_d }
  \threecMat{d_R}{d_R''}{d_R'}+ \text{h.c.}\;,
\intertext{where we have set the vacuum to be $\vev{\phi_{d}^0} =  (v_{d},v_{d},v_{d})$ with $v_{d} \in \mathbb{R}$,}
 =&
  \threerMat{\overline{d}_{L1}}{\overline{d}_{L2}}{\overline{d}_{L3}}
  \Uw
  \underbrace{
  \Matdiag
  {\sqrt{3}\,g_3\,v_d}{\sqrt{3}\,g_3'\,v_d}{\sqrt{3}\,g_3''\,v_d}
  }_{\widehat{m}_d}
  \threecMat{d_R}{d_R''}{d_R'}+ \text{h.c.}\;,\\
 \Rightarrow \qquad &
  \boxed{V_{dL}^\dagger = \Uw \quad\text{and}\quad V_{dR} = I}\;.
\end{align}

\noindent 
$\boxed{\mnud}\;\;$ \textbf{neutrino Dirac mass matrix} 
\begin{align}
 &g_3\, (\overline{\nu}_L\,\vev{\phi_{d}^{0*}})_{\underline{1}}\,\nu_R
 + g_3'\,(\overline{\nu}_L\,\vev{\phi_{d}^{0*}})_{\underline{1}'}\,\nu_R''
 + g_3''\,(\overline{\nu}_L\,\vev{\phi_{d}^{0*}})_{\underline{1}''}\,\nu_R'
 + \text{h.c.}\;,
\end{align}
\begin{align}
=& \threerMat{\overline{\nu}_{L1}}{\overline{\nu}_{L2}}{\overline{\nu}_{L3}}
  \ththMat
  {g_3\, v_d }{g_3'\, v_d }{g_3''\, v_d }
  {g_3\, v_d }{\omega g_3'\, v_d }{\omega^2 g_3''\, v_d }
  {g_3\, v_d }{\omega^2 g_3'\, v_d }{\omega g_3''\, v_d }
  \threecMat{\nu_R}{\nu_R''}{\nu_R'}+ \text{h.c.}\;,
\intertext{where we have set the vacuum to be $\vev{\phi_{d}^{0*}} =  (v_{d},v_{d},v_{d})$ with $v_{d} \in \mathbb{R}$,}
 =&
  \threerMat{\overline{\nu}_{L1}}{\overline{\nu}_{L2}}{\overline{\nu}_{L3}}
  \Uw
  \underbrace{
  \Matdiag
  {\sqrt{3}\,g_3\,v_d}{\sqrt{3}\,g_3'\,v_d}{\sqrt{3}\,g_3''\,v_d}
  }_{\widehat{m}_d}
  \threecMat{\nu_R}{\nu_R''}{\nu_R'}+ \text{h.c.}\;,\\
 \Rightarrow \qquad &
  \boxed{\mnud = \Uw\,\widehat{m}_d = V_{eL}^\dagger\,\widehat{m}_d}\;.
\end{align}

\noindent 
$\boxed{M_R}\;\;$ \textbf{neutrino Majorana mass matrix}
\begin{align}
  &
   h_1^{(11)} \overline{\nu_R^c} \nu_R \vev{\Delta^0_{1}}
 +h_1^{(22)} \overline{\nu_R^c}'' \nu_R'' \vev{{\Delta^{0}_{1}}''}
 +h_1^{(33)} \overline{\nu_R^c}' \nu_R' \vev{{\Delta^{0}_{1}}'}
 +h_1^{(12)}
    \left[\overline{\nu_R^c} \nu_R''
         +\overline{\nu_R^c}'' \nu_R\right] \vev{{\Delta^{0}_{1}}'}\nonumber\\
 &\quad
 +h_1^{(13)}
    \left[\overline{\nu_R^c} \nu_R'
         +\overline{\nu_R^c}' \nu_R\right] \vev{{\Delta^{0}_{1}}''}
 +h_1^{(23)}
    \left[\overline{\nu_R^c}'' \nu_R'
         +\overline{\nu_R^c}' \nu_R''\right] \vev{\Delta^0_{1}}+\text{h.c.}\;.
\end{align}
Letting $\vev{\Delta^{0}_{1}} = v_{\delta 1}, \vev{{\Delta^{0}_{1}}'} = v_{\delta 1}', \vev{{\Delta^{0}_{1}}''} = v_{\delta 1}''$, this becomes the following in matrix form:
\begin{align}
 =& \threerMat{\overline{\nu^c_R}}{\overline{\nu^c_R}''}{\overline{\nu^c_R}'}
  \underbrace{
  \ththMat
   {h_1^{(11)}v_{\delta 1}}
   {h_1^{(12)}v_{\delta 1}'}
   {h_1^{(13)}v_{\delta 1}''}
   {h_1^{(12)}v_{\delta 1}'}
   {h_1^{(22)}v_{\delta 1}''}
   {h_1^{(23)}v_{\delta 1}}
   {h_1^{(13)}v_{\delta 1}''}
   {h_1^{(23)}v_{\delta 1}}
   {h_1^{(33)}v_{\delta 1}'}
  }_{M_R \text{ is complex symmetric}}
  \threecMat{\nu_R}{\nu_R''}{\nu_R'}+ \text{h.c.}\;.
\end{align}

\section{Mass matrices in the left-right model}
\label{app:b_LR_details}

\noindent 
$\boxed{m_e}\;\;$ \textbf{charged lepton mass matrix}
\begin{align}
 &-\left[
    y_1 \,(\overline{e}_L\,\phi^{0*})_{\underline{1}}\,e_R
  + y_1' \,(\overline{e}_L\,\phi^{0*})_{\underline{1}'}\,e_R''
  + y_1'' \,(\overline{e}_L\,\phi^{0*})_{\underline{1}''}\,e_R'
 \right]\nonumber\\
 &\quad
  + \widetilde{y}_1 \,(\overline{e}_L\,\phi^{0*})_{\underline{1}}\,e_R
  + \widetilde{y}_1' \,(\overline{e}_L\,\phi^{0*})_{\underline{1}'}\,e_R''
  + \widetilde{y}_1'' \,(\overline{e}_L\,\phi^{0*})_{\underline{1}''}\,e_R'
  +\text{h.c.}\;,\\
  =\;&
  - y_1 v_\ell (\overline{e}_{L1}+\overline{e}_{L2}+\overline{e}_{L3}) e_R
  - y'_1 v_\ell (\overline{e}_{L1}+ \omega\overline{e}_{L2}+\omega^2\overline{e}_{L3}) e''_R
  \nonumber\\
  &\quad
  - y''_1 v_\ell (\overline{e}_{L1}+\omega^2\overline{e}_{L2}+\omega\overline{e}_{L3}) e'_R
  + \widetilde{y}_1 v_\ell (\overline{e}_{L1}+\overline{e}_{L2}+\overline{e}_{L3}) e_R
  \nonumber\\
  &\qquad
  + \widetilde{y}'_1 v_\ell (\overline{e}_{L1} + 
   \omega\overline{e}_{L2}+\omega^2\overline{e}_{L3}) e''_R
  + \widetilde{y}''_1 v_\ell (\overline{e}_{L1}+\omega^2\overline{e}_{L2}+\omega\overline{e}_{L3}) e'_R
 +\text{h.c.}\;,
\end{align}
where we have used $\vev{\phi^{0*}} \equiv \vev{\phi^0}
  = (v_\ell, v_\ell, v_\ell)$ with $v_\ell = v_\ell^*$. In matrix form, we obtain
\begin{align}
  =&
   \threerMat{\overline{e}_{L1}}{\overline{e}_{L2}}{\overline{e}_{L3}}
 \left[
  -\ththMat
  {y_1\, v_{\ell} }{y_1'\, v_{\ell} }{y_1''\, v_{\ell} }
  {y_1\, v_{\ell} }{\omega y_1'\, v_{\ell} }{\omega^2 y_1''\, v_{\ell} }
  {y_1\, v_{\ell} }{\omega^2 y_1'\, v_{\ell} }{\omega y_1''\, v_{\ell} } 
  +
  \ththMat
  {\widetilde{y}_1\, v_\ell }{\widetilde{y}_1'\, v_\ell }{\widetilde{y}_1''\, v_\ell }
  {\widetilde{y}_1\, v_\ell }{\omega \widetilde{y}_1'\, v_\ell }{\omega^2 \widetilde{y}_1''\, v_\ell }
  {\widetilde{y}_1\, v_\ell }{\omega^2 \widetilde{y}_1'\, v_\ell }{\omega \widetilde{y}_1''\, v_\ell }
 \right]
  \threecMat{e_R}{e_R''}{e_R'}
 \nonumber\\
 &~\hspace{11.5cm}
  + \text{h.c.}\;,\\
  =&
  \threerMat{\overline{e}_{L1}}{\overline{e}_{L2}}{\overline{e}_{L3}}
  \Uw\;
  \underbrace{\sqrt{3}\;v_\ell
  \Matdiag
  {-y_1 +\widetilde{y}_1}{-y_1' +\widetilde{y}_1'}{-y_1''  +\widetilde{y}_1''}
  }_{\widehat{m}_e}
  \threecMat{e_R}{e_R''}{e_R'}
  + \text{h.c.}\;,
\end{align}
So, \fbox{$V_{eL}^\dagger = \Uw$ and $V_{eR} = I$} .

\noindent 
$\boxed{\mnud}\;\;$ \textbf{neutrino Dirac mass matrix} 
\begin{align}
 &
    y_1 \,(\overline{\nu}_L\,\phi^{0})_{\underline{1}}\,\nu_R
  + y_1' \,(\overline{\nu}_L\,\phi^{0})_{\underline{1}'}\,\nu_R''
  + y_1'' \,(\overline{\nu}_L\,\phi^{0})_{\underline{1}''}\,\nu_R'
 \nonumber\\
 &\quad
  -\left[ \widetilde{y}_1 \,(\overline{\nu}_L\,\phi^{0})_{\underline{1}}\,\nu_R
  + \widetilde{y}_1' \,(\overline{\nu}_L\,\phi^{0})_{\underline{1}'}\,\nu_R''
  + \widetilde{y}_1'' \,(\overline{\nu}_L\,\phi^{0})_{\underline{1}''}\,\nu_R'
  \right]
  +\text{h.c.}\;,\\
 =&
  \threerMat{\overline{\nu}_{L1}}{\overline{\nu}_{L2}}{\overline{\nu}_{L3}}
 \left[
  \ththMat
  {y_1\, v_{\ell} }{y_1'\, v_{\ell} }{y_1''\, v_{\ell} }
  {y_1\, v_{\ell} }{\omega y_1'\, v_{\ell} }{\omega^2 y_1''\, v_{\ell} }
  {y_1\, v_{\ell} }{\omega^2 y_1'\, v_{\ell} }{\omega y_1''\, v_{\ell} }
  -
  \ththMat
  {\widetilde{y}_1\, v_\ell }{\widetilde{y}_1'\, v_\ell }{\widetilde{y}_1''\, v_\ell }
  {\widetilde{y}_1\, v_\ell }{\omega \widetilde{y}_1'\, v_\ell }{\omega^2 \widetilde{y}_1''\, v_\ell }
  {\widetilde{y}_1\, v_\ell }{\omega^2 \widetilde{y}_1'\, v_\ell }{\omega \widetilde{y}_1''\, v_\ell }
 \right]
  \threecMat{\nu_R}{\nu_R''}{\nu_R'}
 \nonumber\\
 &~\hspace{11.5cm}
  + \text{h.c.}\;,\\
  =&
  \threerMat{\overline{\nu}_{L1}}{\overline{\nu}_{L2}}{\overline{\nu}_{L3}}
  \Uw\;
  \underbrace{\sqrt{3}\;v_\ell
  \Matdiag
  {y_1 -\widetilde{y}_1}{y_1' -\widetilde{y}_1'}{y_1''  -\widetilde{y}_1''}
  }_{-\widehat{m}_e}
  \threecMat{\nu_R}{\nu_R''}{\nu_R'}\;\;+ \text{h.c.}\;,  
  \\
  \therefore &\;\;
  \boxed{\mnud = \Uw\,(-\widehat{m}_e) = -V_{eL}^\dagger\,\widehat{m}_e}\;.
\end{align}

\noindent 
$\boxed{M_R}\;\;$ \textbf{neutrino Majorana mass matrix}
\begin{align}
  &\hat{y}_4\,\overline{\ell_R^c}\, i \tau_2\, \vev{\Delta_R}_{(1+1'+1'')}\, \ell_R  
  +\text{h.c.}\;,\nonumber\\
 =\;&\hat{y}_4\,\begin{pmatrix}
                \overline{\nu_R^c}\; & \overline{e_R^c}
              \end{pmatrix}
              \twtwMat{0}{1}{-1}{0}
              \twtwMat{0}{0}{v_\delta}{0}_{(1+1'+1'')}
              \twocMat{\nu_R}{e_R}+\text{h.c.}\;,\nonumber\\
 =\;&
     \hat{y}_4\, \overline{\nu_R^c}\, \nu_R\, {v_\delta}_{(1+1'+1'')}+\text{h.c.}\;.
\end{align}
Letting ${v_\delta}_{(1)} = v_\delta,\; {v_\delta}_{(1')} = v_\delta',\; {v_\delta}_{(1'')} = v_\delta''$ and expanding out the $A_4$ components to get
\begin{align}
  &
   y_4^{(11)} \overline{\nu_R^c} \nu_R \,v_\delta
 +y_4^{(22)} \overline{\nu_R^c}'' \nu_R'' \,v_\delta''
 +y_4^{(33)} \overline{\nu_R^c}' \nu_R' \,v_\delta'
 +y_4^{(12)}
    \left[\overline{\nu_R^c} \nu_R''
         +\overline{\nu_R^c}'' \nu_R\right] \,v_\delta'\nonumber\\
 &\quad
 +y_4^{(13)}
    \left[\overline{\nu_R^c} \nu_R'
         +\overline{\nu_R^c}' \nu_R\right] \,v_\delta''
 +y_4^{(23)}
    \left[\overline{\nu_R^c}'' \nu_R'
         +\overline{\nu_R^c}' \nu_R''\right] \,v_\delta\;\;+\text{h.c.}\;,\\
=& \threerMat{\overline{\nu^c_R}}{\overline{\nu^c_R}''}{\overline{\nu^c_R}'}
  \underbrace{
  \ththMat
   {y_4^{(11)}v_{\delta }}
   {y_4^{(12)}v_{\delta }'}
   {y_4^{(13)}v_{\delta }''}
   {y_4^{(12)}v_{\delta }'}
   {y_4^{(22)}v_{\delta }''}
   {y_4^{(23)}v_{\delta }}
   {y_4^{(13)}v_{\delta }''}
   {y_4^{(23)}v_{\delta }}
   {y_4^{(33)}v_{\delta }'}
  }_{M_R \text{ is complex symmetric}}
  \threecMat{\nu_R}{\nu_R''}{\nu_R'}\;\;+ \text{h.c.}\;.
\end{align}

\noindent 
$\boxed{m_u}\;\;$ \textbf{up-type quark mass matrix} 
\begin{align}
  &
   y_{2s}\,(\overline{u}_L\,u_R)_{\rept s} \,\phi^{0}
  +y_{2a}\,(\overline{u}_L\,u_R)_{\rept a} \,\phi^{0}
  -\widetilde{y}_{2s}\,(\overline{u}_L\,u_R)_{\rept s} \,\phi^{0}
  -\widetilde{y}_{2a}\,(\overline{u}_L\,u_R)_{\rept a} \,\phi^{0} \nonumber\\
  &\quad
  +y_3\,(\overline{u}_L\,u_R)_{\rep} \,\phi^{0}_A
  +y_3'\,(\overline{u}_L\,u_R)_{\repp} \,{\phi^{0}_A}''
  +y_3''\,(\overline{u}_L\,u_R)_{\reppp} \,{\phi^{0}_A}' \nonumber\\
  &\quad\quad
  +\widetilde{y}_3\,(\overline{u}_L\,u_R)_{\rep} \,\phi^{0*}_B
  +\widetilde{y}_3'\,(\overline{u}_L\,u_R)_{\repp} \,{\phi^{0*}_B}''
  +\widetilde{y}_3''\,(\overline{u}_L\,u_R)_{\reppp} \,{\phi^{0*}_B}'+\text{h.c.}\;.
\end{align}
Letting $\vev{\phi^0} = (v_\ell , v_\ell , v_\ell),\;
\vev{\phi^0_A} = v_A,\;
\vev{{\phi^{0}_A}'} = v_A',\;
\vev{{\phi^{0}_A}'} = v_A''$ and
$\vev{\phi^{0*}_B} = v_B^*,\;
\vev{{\phi^{0*}_B}'} = {v_B^{*}}',\;
\vev{{\phi^{0*}_B}''} = {v_B^{*}}''$
and expanding (h.c. omitted for brevity):
\begin{align}
  =\;&\label{LR:001}
  y_{2s}\,v_\ell
  (\overline{u}_{L2}\,u_{R3} +\overline{u}_{L3}\,u_{R2}
   +\overline{u}_{L3}\,u_{R1} +\overline{u}_{L1}\,u_{R3}
   +\overline{u}_{L1}\,u_{R2} +\overline{u}_{L2}\,u_{R1})\\
  &\;\label{LR:002}
 +y_{2a}\,v_\ell
   (\overline{u}_{L2}\,u_{R3} -\overline{u}_{L3}\,u_{R2}
   +\overline{u}_{L3}\,u_{R1} -\overline{u}_{L1}\,u_{R3}
   +\overline{u}_{L1}\,u_{R2} -\overline{u}_{L2}\,u_{R1})\\
  &\;\;\label{LR:003}
 -\widetilde{y}_{2s}\,v_\ell
   (\overline{u}_{L2}\,u_{R3} +\overline{u}_{L3}\,u_{R2}
   +\overline{u}_{L3}\,u_{R1} +\overline{u}_{L1}\,u_{R3}
   +\overline{u}_{L1}\,u_{R2} +\overline{u}_{L2}\,u_{R1})\\
  &\;\;\;\label{LR:004}
 -\widetilde{y}_{2a}\,v_\ell
   (\overline{u}_{L2}\,u_{R3} -\overline{u}_{L3}\,u_{R2}
   +\overline{u}_{L3}\,u_{R1} -\overline{u}_{L1}\,u_{R3}
   +\overline{u}_{L1}\,u_{R2} -\overline{u}_{L2}\,u_{R1})\\
  &\;\;\;\;\label{LR:005}
 +y_3\,{v_A}
  (\overline{u}_{L1}\,u_{R1} +\overline{u}_{L2}\,u_{R2} +\overline{u}_{L3}\,u_{R3})\\
  &\;\;\;\;\;\label{LR:006}
 +y_3'\,{v_A}''
  (\overline{u}_{L1}\,u_{R1} +\omega\,\overline{u}_{L2}\,u_{R2}
  +\omega^2\,\overline{u}_{L3}\,u_{R3})\\
  &\;\;\;\;\;\;\label{LR:007}
 +y_3''\,{v_A}'
  (\overline{u}_{L1}\,u_{R1} +\omega^2\,\overline{u}_{L2}\,u_{R2}
  +\omega\,\overline{u}_{L3}\,u_{R3})\\
  &\;\;\;\;\;\;\;\label{LR:008}
 +\widetilde{y}_3\,{v_B^*}
  (\overline{u}_{L1}\,u_{R1} +\overline{u}_{L2}\,u_{R2} +\overline{u}_{L3}\,u_{R3})\\
  &\;\;\;\;\;\;\;\;\;\label{LR:009}
 +\widetilde{y}_3'\,{v_B^*}''
  (\overline{u}_{L1}\,u_{R1} +\omega\,\overline{u}_{L2}\,u_{R2}
  +\omega^2\,\overline{u}_{L3}\,u_{R3})\\
  &\;\;\;\;\;\;\;\;\;\;\;\label{LR:010}
 +\widetilde{y}_3''\,{v_B^*}'
  (\overline{u}_{L1}\,u_{R1} +\omega^2\,\overline{u}_{L2}\,u_{R2}
  +\omega\,\overline{u}_{L3}\,u_{R3})
\end{align}
%
%
%
\begin{align}
 \text{(\ref{LR:001}) and (\ref{LR:003}): }\;
 &
 \underbrace{(y_{2s} - \widetilde{y}_{2s})\,v_\ell}_{Y_{2s}}\,
 \threerMat{\overline{u}_{L1}}{\overline{u}_{L2}}{\overline{u}_{L3}}
  \ththMat
  {0}{1}{1}
  {1}{0}{1}
  {1}{1}{0}
  \threecMat{u_{R1}}{u_{R2}}{u_{R3}}\;,\label{LR:I}
\end{align}  
\begin{align}
 \text{(\ref{LR:002}) and (\ref{LR:004}): }\;
&
 \underbrace{(y_{2a} - \widetilde{y}_{2a})\,v_\ell}_{Y_{2a}}\,
 \threerMat{\overline{u}_{L1}}{\overline{u}_{L2}}{\overline{u}_{L3}}
  \ththMat
  {0}{1}{-1}
  {-1}{0}{1}
  {1}{-1}{0}
  \threecMat{u_{R1}}{u_{R2}}{u_{R3}}\;,\label{LR:II}\\
 \text{(\ref{LR:005}) and (\ref{LR:008}): }\;
&
 \underbrace{(y_{3}\,v_A + \widetilde{y}_{3}\,{v_B^*})}_{Y_{A}}\,
 \threerMat{\overline{u}_{L1}}{\overline{u}_{L2}}{\overline{u}_{L3}}
  \Matdiag{1}{1}{1}
  \threecMat{u_{R1}}{u_{R2}}{u_{R3}}\;,\label{LR:III}\\
 \text{(\ref{LR:006}) and (\ref{LR:009}): }\;
&
 \underbrace{(y_{3}'\,v_A'' + \widetilde{y}_{3}'\,{v_B^*}'')}_{Y_{A}'}\,
 \threerMat{\overline{u}_{L1}}{\overline{u}_{L2}}{\overline{u}_{L3}}
  \Matdiag{1}{\omega}{\omega^2}
  \threecMat{u_{R1}}{u_{R2}}{u_{R3}}\;,\label{LR:IV}\\
 \text{(\ref{LR:007}) and (\ref{LR:010}): }\;
&
 \underbrace{(y_{3}''\,v_A' + \widetilde{y}_{3}''\,{v_B^*}')}_{Y_{A}''}\,
 \threerMat{\overline{u}_{L1}}{\overline{u}_{L2}}{\overline{u}_{L3}}
  \Matdiag{1}{\omega^2}{\omega}
  \threecMat{u_{R1}}{u_{R2}}{u_{R3}}\;.\label{LR:V}
\end{align}
Putting (\ref{LR:I}), (\ref{LR:II}), (\ref{LR:III}), (\ref{LR:IV}) and (\ref{LR:V}) together, we obtain
\begin{align}
 &
 \threerMat{\overline{u}_{L1}}{\overline{u}_{L2}}{\overline{u}_{L3}}
  \ththMat
  {Y_A+Y_A'+Y_A''}{Y_{2s}+Y_{2a}}{Y_{2s}-Y_{2a}}
  {Y_{2s}-Y_{2a}}{Y_A+\omega Y_A'+\omega^2 Y_A''}{Y_{2s}+Y_{2a}}
  {Y_{2s}+Y_{2a}}{Y_{2s}-Y_{2a}}{Y_A+\omega^2 Y_A'+\omega Y_A''}
  \threecMat{u_{R1}}{u_{R2}}{u_{R3}} 
  \nonumber\;,\\
 \equiv &
  \threerMat{\overline{u}_{L1}}{\overline{u}_{L2}}{\overline{u}_{L3}}
  \underbrace{
  \ththMat
  {Y_A^{(1)}}{Y_2^{+}}{Y_2^{-}}
  {Y_2^{-}}{Y_A^{(1')}}{Y_2^{+}}
  {Y_2^{+}}{Y_2^{-}}{Y_A^{(1'')}}
  }_{m_u}
  \threecMat{u_{R1}}{u_{R2}}{u_{R3}}
  \;.
\end{align}
The mass matrix $m_u$ contains 10 real parameters only, and so we expect that $V_{uL}$ and $V_{uR}$ are not completely arbitrary and may have a special form. However, there should be enough degrees of freedom to produce the observed quark mixings.\\

\noindent 
$\boxed{m_d}\;\;$ \textbf{down-type quark mass matrix} 
\begin{align}
  &
  -y_{2s}\,(\overline{d}_L\,d_R)_{\rept s} \,\phi^{0*}
  -y_{2a}\,(\overline{d}_L\,d_R)_{\rept a} \,\phi^{0*}
  +\widetilde{y}_{2s}\,(\overline{d}_L\,d_R)_{\rept s} \,\phi^{0*}
  +\widetilde{y}_{2a}\,(\overline{d}_L\,d_R)_{\rept a} \,\phi^{0*} \nonumber\\
  &\quad
  +y_3\,(\overline{d}_L\,d_R)_{\rep} \,\phi^{0}_B
  +y_3'\,(\overline{d}_L\,d_R)_{\repp} \,{\phi^{0}_B}''
  +y_3''\,(\overline{d}_L\,d_R)_{\reppp} \,{\phi^{0}_B}' \nonumber\\
  &\quad\quad
  +\widetilde{y}_3\,(\overline{d}_L\,d_R)_{\rep} \,\phi^{0*}_A
  +\widetilde{y}_3'\,(\overline{d}_L\,d_R)_{\repp} \,{\phi^{0*}_A}''
  +\widetilde{y}_3''\,(\overline{d}_L\,d_R)_{\reppp} \,{\phi^{0*}_A}'
  +\text{h.c.}\;,
  \\ \nonumber
\end{align}
\begin{align}
 =&\label{LR:001d}
  \,-y_{2s}\,v_\ell
  (\overline{d}_{L2}\,d_{R3} +\overline{d}_{L3}\,d_{R2}
   +\overline{d}_{L3}\,d_{R1} +\overline{d}_{L1}\,d_{R3}
   +\overline{d}_{L1}\,d_{R2} +\overline{d}_{L2}\,d_{R1})\\
  &\;\label{LR:002d}
 -y_{2a}\,v_\ell
   (\overline{d}_{L2}\,d_{R3} -\overline{d}_{L3}\,d_{R2}
   +\overline{d}_{L3}\,d_{R1} -\overline{d}_{L1}\,d_{R3}
   +\overline{d}_{L1}\,d_{R2} -\overline{d}_{L2}\,d_{R1})\\
  &\;\;\label{LR:003d}
 +\widetilde{y}_{2s}\,v_\ell
   (\overline{d}_{L2}\,d_{R3} +\overline{d}_{L3}\,d_{R2}
   +\overline{d}_{L3}\,d_{R1} +\overline{d}_{L1}\,d_{R3}
   +\overline{d}_{L1}\,d_{R2} +\overline{d}_{L2}\,d_{R1})\\
  &\;\;\;\label{LR:004d}
 +\widetilde{y}_{2a}\,v_\ell
   (\overline{d}_{L2}\,d_{R3} -\overline{d}_{L3}\,d_{R2}
   +\overline{d}_{L3}\,d_{R1} -\overline{d}_{L1}\,d_{R3}
   +\overline{d}_{L1}\,d_{R2} -\overline{d}_{L2}\,d_{R1})\\
  &\;\;\;\;\label{LR:005d}
 +y_3\,{v_B}
  (\overline{d}_{L1}\,d_{R1} +\overline{d}_{L2}\,d_{R2} +\overline{d}_{L3}\,d_{R3})\\
  &\;\;\;\;\;\label{LR:006d}
 +y_3'\,{v_B}''
  (\overline{d}_{L1}\,d_{R1} +\omega\,\overline{d}_{L2}\,d_{R2}
  +\omega^2\,\overline{d}_{L3}\,d_{R3})\\
  &\;\;\;\;\;\;\label{LR:007d}
 +y_3''\,{v_B}'
  (\overline{d}_{L1}\,d_{R1} +\omega^2\,\overline{d}_{L2}\,d_{R2}
  +\omega\,\overline{d}_{L3}\,d_{R3})\\
  &\;\;\;\;\;\;\;\label{LR:008d}
 +\widetilde{y}_3\,{v_A^*}
  (\overline{d}_{L1}\,d_{R1} +\overline{d}_{L2}\,d_{R2} +\overline{d}_{L3}\,d_{R3})\\
  &\;\;\;\;\;\;\;\;\;\label{LR:009d}
 +\widetilde{y}_3'\,{v_A^*}''
  (\overline{d}_{L1}\,d_{R1} +\omega\,\overline{d}_{L2}\,d_{R2}
  +\omega^2\,\overline{d}_{L3}\,d_{R3})\\
  &\;\;\;\;\;\;\;\;\;\;\;\label{LR:010d}
 +\widetilde{y}_3''\,{v_A^*}'
  (\overline{d}_{L1}\,d_{R1} +\omega^2\,\overline{d}_{L2}\,d_{R2}
  +\omega\,\overline{d}_{L3}\,d_{R3})\;,
\end{align}
where we have omitted the h.c. in the above. Following a similar procedure as in the $m_u$ case, we find:
\begin{align}
 &
 \threerMat{\overline{d}_{L1}}{\overline{d}_{L2}}{\overline{d}_{L3}}
  \ththMat
  {Y_B+Y_B'+Y_B''}{-Y_{2s}-Y_{2a}}{-Y_{2s}+Y_{2a}}
  {-Y_{2s}+Y_{2a}}{Y_B+\omega Y_B'+\omega^2 Y_B''}{-Y_{2s}-Y_{2a}}
  {-Y_{2s}-Y_{2a}}{-Y_{2s}+Y_{2a}}{Y_B+\omega^2 Y_B'+\omega Y_B''}
  \threecMat{d_{R1}}{d_{R2}}{d_{R3}}
  \;,\nonumber \\
 \equiv &
  \threerMat{\overline{d}_{L1}}{\overline{d}_{L2}}{\overline{d}_{L3}}
  \underbrace{
  \ththMat
  {Y_B^{(1)}}{-Y_2^{+}}{-Y_2^{-}}
  {-Y_2^{-}}{Y_B^{(1')}}{-Y_2^{+}}
  {-Y_2^{+}}{-Y_2^{-}}{Y_B^{(1'')}}
  }_{m_d}
  \threecMat{d_{R1}}{d_{R2}}{d_{R3}}
  \;,
\end{align}
where $Y_{2s,2a} = -(y_{2s,2a}-\widetilde{y}_{2s,2a}) v_\ell$,
$Y_B =y_{3} v_B+\widetilde{y}_{3} v_A^*$, $Y_B' =y_{3}' v_B''+\widetilde{y}_{3}' {v_A^*}''$
and  $Y_B'' =y_{3}'' v_B''+\widetilde{y}_{3}'' {v_A^*}'$. Like before, we expect that $V_{dL}$ and $V_{dR}$ are not completely arbitrary.


%
%
%
%
%



\chapter{Supplement materials for chapter~4}\label{app:c_chapter}

\section{More illustrative examples from standard leptogenesis}\label{app:c_std_eg}

In this section, we present the computations of the self-energy contributions to the $CP$ asymmetry in standard leptogenesis with the aim to further illustrate the usage of the simplified Majorana Feynman rules, as well as to confirm that the known results can be obtained this way. Note that there are actually \emph{two} separate self-energy graphs that contribution to the interference term when final state flavor $j$ is not summed over. So for completeness, we shall calculate them both here~\footnote{For the vertex correction, there is only one graph because weak isospin conservation forbids the existence of a second diagram.}.

\subsubsection{Self-energy contribution to the $CP$ asymmetry (1)}

\begin{figure}[t]
\begin{center}
 \includegraphics[width=0.90\textwidth]{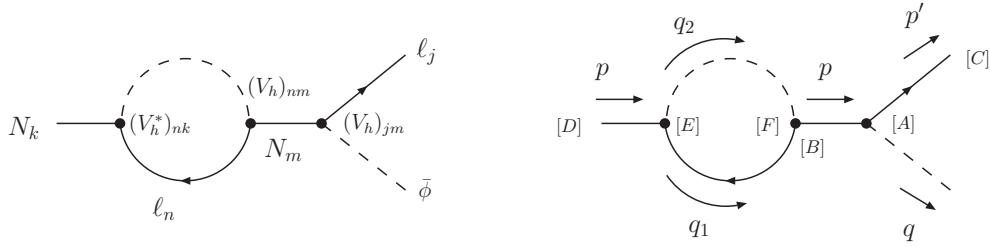}
\end{center} 
 \caption{One-loop self-energy correction graph (1) for the process $N_k\rightarrow \ell_j \bar\phi$. LEFT: $(V_h)_{ab} \equiv -i\,h_{ab}\,P_R$ and $(V_h^*)_{ab} \equiv i\,h_{ab}^*\,C^\dagger P_L$ are the vertex factors; RIGHT: we have included the momentum flows and spinor indices~$[X]$, where $q=p-p'$ and $q_2=p-q_1$.}
 \label{figC:std_cpsloop_cal_graphs}
\end{figure}

The first self-energy contribution is given by interference between one-loop graph in Fig.~\ref{figC:std_cpsloop_cal_graphs} and the tree-level diagram in Fig.~\ref{fig4:BKL_std_tree_cal_graphs} on page~\pageref{fig4:BKL_std_tree_cal_graphs}. Following the procedure discussed in Sec.~\ref{subsec:BKL_maj_feyn_examples}, we have for the interference term:
\begin{align}
 I_\text{self-(1)}
 &=\int \frac{d^4 q_1}{(2\pi)^4}
 (i  h_{jk}^*)(-i  h_{jm})
 (-i  h_{nm})(i  h_{nk}^*)
 \left[\ubarj\right]_{1C} \left[P_R\right]_{CA} \left[S_{N_m}(p)\right]_{AB}
   \nonumber\\
 &\qquad\times  
 \left[P_R\right]_{FB}
 \left[S_\ell(-q_1)\right]_{EF} [C^\dagger P_L]_{DE} \left[u_k^c\right]_{D1} 
 \left[D(q_2)\right]_{11}
 \left[-u_k^T C^\dagger P_L u_j\right]_{11}\;,
\end{align}
where we have shown all spinor indices explicitly. Letting $A_h = h_{jk}^* h_{jm} h_{nm} h_{nk}^*$, this then becomes ($m_\phi, m_{\ell_j} \approx 0$)
\begin{align}
 I_\text{self-(1)}
 &=
  A_h 
  \int\frac{d^4 q_1}{(2\pi)^4}
  \frac{\ubarj\, P_R (-i)(\pslash+M_m)C\,P_R^T \, i(-\qoneslash)^T P_L^T C^* u_k^c (i)
  (-1)u_k^T C^\dagger P_L u_j}
  {(p^2-M_m^2 +i\epsilon)(q_1^2+i\epsilon)(q_2^2+i\epsilon)}\;, \nonumber\\
 &=
  -i A_h\int\frac{d^4 q_1}{(2\pi)^4}
  \frac{\ubarj\, P_R (\pslash+M_m)C\,C^\dagger P_R C \, 
  C^\dagger\qoneslash C C^\dagger P_L C C^*\; C\ubark^T 
  u_k^T C^\dagger P_L u_j}
  {(p^2-M_m^2 +i\epsilon)(q_1^2+i\epsilon)(q_2^2+i\epsilon)}\;, \nonumber\\
 &=
  \frac{i A_h}{2}\int\frac{d^4 q_1}{(2\pi)^4}
  \frac{P_R (\pslash+M_m)\,P_R \qoneslash P_L \; 
  C\left(\sum_s u_k\ubark\right)^T C^\dagger P_L \sum_{s'}u_j\ubarj}
  {(p^2-M_m^2 +i\epsilon)(q_1^2+i\epsilon)(q_2^2+i\epsilon)}\;, 
  \nonumber\\
 &=
  \frac{i A_h}{2}\int\frac{d^4 q_1}{(2\pi)^4}
  \frac{\text{Tr}\left[
  P_R (\pslash+M_m)\,P_R \qoneslash P_L\,
  C\left(\pslash^T +M_k\right)C^\dagger P_L \ppslash
  \right]}
  {(p^2-M_m^2 +i\epsilon)(q_1^2+i\epsilon)(q_2^2+i\epsilon)}\;,\nonumber\\
  &\hspace{140pt}\vdots \nonumber\\
   &=
  i A_h\,M_k\, M_m\int\frac{d^4 q_1}{(2\pi)^4}
  \frac{q_1\cdot p'}
  {(p^2-M_m^2 +i\epsilon)(q_1^2+i\epsilon)(q_2^2+i\epsilon)}\;.
\end{align}
To pick out the discontinuity of the integral
\begin{equation}
 I_{(1)} =
  i\,M_k M_m \int\frac{d^4 q_1}{(2\pi)^4}
  \frac{q_1\cdot p'}
  {(p^2-M_{m}^2 +i\epsilon)(q_1^2+i\epsilon)(q_2^2+i\epsilon)}\;,
\end{equation}
we note that there is only one sensible way to cut the diagram, namely, through the propagators associated with $q_1$ and $q_2$. So, with the replacement:
\begin{equation}
 \frac{1}{q_1^2 + i\epsilon} \;\rightarrow\; -2\pi i \delta(q_1^2)\Theta(E_{1})\;,
 \;\;\text{ and }\;\;\;
 \frac{1}{q_2^2 + i\epsilon} \;\rightarrow\; -2\pi i \delta(q_2^2)\Theta(E_{2})
  \;,
\end{equation}
we have (using $q_2 = p-q_1$)
\begin{align}
 \text{Disc}(I_{(1)})
 &=
 i\,M_kM_m \int\frac{d^4 q_1}{(2\pi)^4}
 \frac{
   (-2\pi i)^2
   (q_1 \cdot p')
   \delta(q_1^2)
   \delta((p-q_1)^2)\Theta(E_{1})\Theta(M_k - E_{1})
   }
  {p^2-M_m^2}\;.
\end{align}  
%
Simplifying this using all the tricks and relations mentioned in Sec.~\ref{subsec:BKL_maj_feyn_examples}, we obtain
\begin{align}
 \text{Disc}(I_{(1)})
  &=
 \frac{-i\,\mNk\mNm}{4\pi^2(\mNk^2-\mNm^2)}
 \int d^4 q_1
   \left[E_1 \halfmNk -(-\vec q \cdot \vec q_1)\right]  
   \delta(E_1^2-\modq1^2)
   \nonumber\\
 &\qquad\qquad\qquad\qquad\qquad \times
   \delta\left[(\mNk -E_1)^2-\modq1^2\right]
   \Theta(E_{1})\Theta(\mNk - E_{1})
   \;, \nonumber\\
 &=
 \frac{-i\,\mNk^2\mNm}{8\pi^2(\mNk^2-\mNm^2)}
 \int dE_1 d^3 q_1
   (E_1 +\modq1\costh)  
  \frac{1}{2\modq1}
   \delta(E_1-\modq1)
  \nonumber\\
 &\qquad\qquad\qquad\qquad\qquad \times
    \delta\left[(\mNk -E_1)^2-\modq1^2\right]
   \Theta(E_{1})\Theta(\mNk - E_{1})
   \;, \nonumber\\
&=
 \frac{-i\,\mNk^2\mNm}{16\pi^2(\mNk^2-\mNm^2)}
 \int \modq1^2 d\modq1 d\Omega\; 
   \modq1(1 +\costh)  
  \frac{1}{\modq1}
  \nonumber\\
 &\qquad\qquad\qquad\qquad\qquad \times  
  \delta\left[(\mNk -\modq1)^2-\modq1^2\right] \Theta(\mNk - \modq1)
   \;, \nonumber\\
&=
 \frac{-i\,\mNk^2\mNm}{16\pi^2(\mNk^2-\mNm^2)}
 \int \modq1^2 d\modq1 d\Omega\; 
   (1 +\costh)\;  
   \delta\left[\mNk^2-2\mNk\modq1\right] \Theta(\mNk - \modq1)
   \;, \nonumber\\
&=
 \frac{-i\,\mNk^2\mNm}{16\pi^2(\mNk^2-\mNm^2)}
 \int \modq1^2 d\modq1 d\Omega\; 
   \;  
   \frac{(1 +\costh)}{|-2\mNk|}\;
   \delta\left[\modq1-\halfmNk\right] \Theta(\mNk - \modq1)
   \;, \nonumber\\
&=
 \frac{-i\,\mNk\mNm}{32\pi^2(\mNk^2-\mNm^2)}
 \int d\Omega\;
  \frac{\mNk^2}{4} 
   (1 +\costh)\;  
   \;, \nonumber\\
&=
 \frac{-i\,\mNk^3\mNm}{32\pi^2(\mNk^2-\mNm^2)}  \frac{2\pi}{4}
 \underbrace{\int_{-1}^{1} dx\;  (1 +x)}_{= 2}  
    \;, \nonumber\\
&=
 \frac{-i\,\mNk^3\mNm}{32\pi(\mNk^2-\mNm^2)}
    \;.   
\end{align}
From this the imaginary part is given by
\begin{align}
 \text{Im}\left[I_{(1)}\right]
 &= \frac{1}{2i}\;
 \text{Disc}\left[I_{(1)}\right] =
 -\frac{\mNk^2}{64\pi}\;\left(\frac{\mNk\mNm}{\mNk^2-\mNm^2}\right)
  \;.
\end{align}
The 2-body decay phase space for this case is given by
\begin{equation}
 V_\varphi' = 2\times 2\times \frac{|\vec q|}{8\pi E_\text{cm}^2} 
   = 2\times 2\times \frac{1}{8\pi}\;\frac{\mNk}{2\,\mNk^2} 
   = \frac{1}{4\pi\,\mNk} \;,
\end{equation}
where one of the factor of 2 is to account for the two channels of final decay products while the other is to account for the two types of intermediate state ($\nu\phi^0$ or $e^-\phi^+$) inside the self-energy loop. Putting all these together and summing over all heavy Majorana neutrino species $m\neq k$, as well as the internal lepton species $n$, we get a contribution to the asymmetry due to this interference as
\begin{equation}
 \varepsilon_\text{self-(1)}
 = -\frac{4}{\Gamma_\text{tot}} \;
 \sum_{m\neq k}\sum_{n}
 \text{Im} (A_h)\;
    \text{Im} (I_{(1)}V_\varphi') \;,
\end{equation}
\begin{align}
 \varepsilon_\text{self-(1)}
 &=
  4\times \frac{8\pi}{(h^\dagger h)_{kk}\;\mNk}
  \sum_{m\neq k}\sum_{n}
  \text{Im} (h_{jk}^* h_{jm} h_{nm} h_{nk}^*)
  \,\frac{\mNk^2}{64\pi}\; 
  \left(\frac{\mNk\mNm}{\mNk^2-\mNm^2}\right)
  \frac{1}{4\pi\,\mNk}\;,\nonumber\\
 &=
  \frac{1}{8\pi(h^\dagger h)_{kk}}
  \sum_{m\neq k}
  \text{Im} \left[h_{jk}^* h_{jm} (h^\dagger h)_{km}\right]
  \left(\frac{\mNk\mNm}{\mNk^2-\mNm^2}\right)
  \;,\nonumber\\
 &=
  \frac{1}{8\pi(h^\dagger h)_{kk}}
  \sum_{m\neq k}
  \text{Im} \left[h_{jk}^* h_{jm} (h^\dagger h)_{km}\right]
  \;\frac{\sqrt{z}}{1-z}\;,
  \qquad\qquad z \equiv \frac{\mNm^2}{\mNk^2}
  \;.
\end{align}
Observe that this is the same result as presented in \cite{Covi:1996wh}. If we sum over $j$, then we recover the expression used in Chapters~\ref{ch_intro} and \ref{ch_work_LV}.

\subsubsection{Self-energy contribution to the $CP$ asymmetry (2)}

\begin{figure}[t]
\begin{center}
 \includegraphics[width=0.90\textwidth]{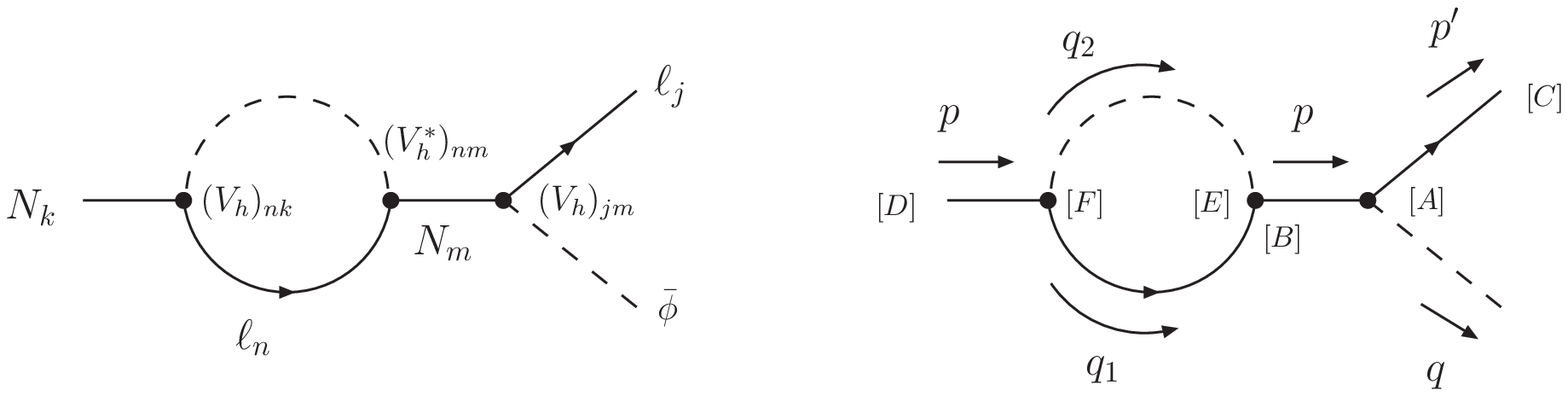}
\end{center} 
\caption{One-loop self-energy correction graph (2) for the process $N_k\rightarrow \ell_j \bar\phi$. LEFT: $(V_h)_{ab} \equiv -i\,h_{ab}\,P_R$ and $(V_h^*)_{ab} \equiv i\,h_{ab}^*\,C^\dagger P_L$ are the vertex factors; RIGHT: we have included the momentum flows and spinor indices~$[X]$, where $q=p-p'$ and $q_2=p-q_1$.}
 \label{figC:std_cpsloop_cal_graphs_2}
\end{figure}

The second self-energy contribution is given by interference between one-loop graph in Fig.~\ref{figC:std_cpsloop_cal_graphs_2} and the tree-level diagram in Fig.~\ref{fig4:BKL_std_tree_cal_graphs} on page~\pageref{fig4:BKL_std_tree_cal_graphs}. Note that for this self-energy diagram, there is \emph{no} ambiguity in the direction of fermion-flow through the Majorana propagator, hence things are more straight forward than before. The interference term can be readily  written down as
\begin{align}
I_\text{self-(2)}
 &= \int \frac{d^4 q_1}{(2\pi)^4}
 (i h_{jk}^*)(-i h_{jm})
 (i h_{nm}^*)(-i h_{nk})
 \left[\ubarj\right]_{1C} \left[P_R\right]_{CA} \left[S_{N_m}(p)\right]_{AB}
 \nonumber\\
 &\qquad\times
  [C^\dagger P_L]_{BE}
 \left[S_\ell(q_1)\right]_{EF} \left[P_R\right]_{FD} \left[u_k^c\right]_{D1} 
 \left[D(q_2)\right]_{11}  
 \left[-u_k^T C^\dagger P_L u_j\right]_{11}\;, \label{eqC:self-result_1}
\end{align}
where we have again shown all spinor indices explicitly. Letting $B_h = h_{jk}^* h_{jm} h_{nm}^* h_{nk}$, and putting into matrix form, we get
\begin{align}
I_\text{self-(2)}
 &=
  B_h 
  \int\frac{d^4 q_1}{(2\pi)^4}
  \frac{\ubarj\, P_R (-i)(\pslash+\mNm)C C^\dagger P_L \, i(\qoneslash) P_R u_k^c (i)
  (-1)u_k^T C^\dagger P_L u_j}
  {(p^2-M_m^2 +i\epsilon)(q_1^2+i\epsilon)(q_2^2+i\epsilon)}\;, \nonumber\\
 &=
  -i B_h\int\frac{d^4 q_1}{(2\pi)^4}
  \frac{\ubarj\, P_R (\pslash+\mNm)P_L \,(\qoneslash) P_R\, 
  C \ubark^T u_k^T C^\dagger \, 
  P_L u_j}
  {(p^2-M_m^2 +i\epsilon)(q_1^2+i\epsilon)(q_2^2+i\epsilon)}\;, \nonumber\\
 &=
  \frac{-i B_h}{2} \int\frac{d^4 q_1}{(2\pi)^4}
  \frac{P_R (\pslash+\mNm)P_L \,(\qoneslash)  
  C \left[\sum_s u_k\ubark\right]^T C^\dagger \, 
  P_L \sum_{s'}u_j\ubarj}
  {(p^2-M_m^2 +i\epsilon)(q_1^2+i\epsilon)(q_2^2+i\epsilon)}\;,
  \quad\text{(index form)} \nonumber\\  
 &=
  \frac{-i B_h}{2} \int\frac{d^4 q_1}{(2\pi)^4}
  \frac{
  \text{Tr}\left[
  P_R (\pslash+\mNm)P_L \,(\qoneslash)  
  (-\pslash+\mNk) \, 
  P_L \ppslash
  \right]}
  {(p^2-M_m^2 +i\epsilon)(q_1^2+i\epsilon)(q_2^2+i\epsilon)}\;,
  \nonumber\\    
 &=
  \frac{i B_h}{2} \int\frac{d^4 q_1}{(2\pi)^4}
  \frac{
  \text{Tr}\left[
  P_R \pslash \,\qoneslash
  \pslash\,\ppslash
  \right]}
  {(p^2-M_m^2 +i\epsilon)(q_1^2+i\epsilon)(q_2^2+i\epsilon)}\;,
  \nonumber\\     
 &=
  i B_h \int\frac{d^4 q_1}{(2\pi)^4}
  \frac{
  2(p \cdot p')(p \cdot q_1)
  - p^2 (p' \cdot q_1)
  }
  {(p^2-M_m^2 +i\epsilon)(q_1^2+i\epsilon)(q_2^2+i\epsilon)}\;.     
\end{align}
We now concentrate on the integral
\begin{equation}
 I_{(2)} =
  i \int\frac{d^4 q_1}{(2\pi)^4}
  \frac{
  2(p \cdot p')(p \cdot q_1)
  - p^2 (p' \cdot q_1)
  }
  {(p^2-M_m^2 +i\epsilon)(q_1^2+i\epsilon)(q_2^2+i\epsilon)}\;.     
\end{equation}
Like before, there is only one way to cut the diagram (through $q_{1}$ and $q_{2}$), and the discontinuity is given by ($\epsilon \rightarrow 0$)
\begin{align}
 \text{Disc}(I_{(2)})
 &=
  i \int\frac{d^4 q_1}{(2\pi)^4}
  (-2\pi i)^2 \delta(q_1^2)
   \delta((p-q_1)^2)\Theta(E_{1})\Theta(\mNk - E_{1})
  \nonumber\\
  &\qquad\qquad\qquad\qquad\qquad
  \times 
   \frac{
  2(p \cdot p')(p \cdot q_1)
  - p^2 (p' \cdot q_1)
  }
  {p^2-M_m^2 +i\epsilon}
  \;,\nonumber\\
 &=
  \frac{-i}{4\pi^2 (\mNk^2-\mNm^2)} 
  \int dE_1 d^3 q_1
  \delta(q_1^2)
   \delta((p-q_1)^2)\Theta(E_{1})\Theta(\mNk - E_{1}) 
 \nonumber\\
 &\qquad\qquad\qquad\qquad\qquad
  \times \left[2\times \frac{\mNk^2}{2} \;\mNk E_1
  - \mNk^2 \left(\halfmNk E_1 +\modq1 |\vec q| \costh\right)
  \right]
   \;,\nonumber\\  
  &=
  \frac{-i}{4\pi^2 (\mNk^2-\mNm^2)} 
  \int dE_1 d^3 q_1
  \delta(E_1^2-\modq1^2)
   \delta\left((\mNk-E_1)^2-\modq1^2\right)\Theta(E_{1})
   \nonumber\\
 &\qquad\qquad\qquad\qquad\qquad
  \times 
  \Theta(\mNk - E_{1})
   \left[\mNk^3 E_1
  - \frac{\mNk^3}{2} \left(E_1 +\modq1\costh \right)
  \right]
  \;,\nonumber  \\
&\hspace{140pt}\vdots \nonumber\\
    %
  &=
  \frac{-i\,\mNk^3}{16\pi^2 (\mNk^2-\mNm^2)} 
  \int \modq1^2 d\modq1 d\Omega \;
  (1-\costh)
  \,\frac{1}{|-2\mNk|}\,
   \delta\left(\modq1-\halfmNk\right)
   \nonumber\\
  &\qquad\qquad\qquad\qquad\qquad 
   \times \Theta(\mNk - \modq1)
  \;,\nonumber \\
  &=
  \frac{-i\,\mNk^4}{32\pi^2 (\mNk^2-\mNm^2)} 
  \;\frac{2\pi}{4}
  \underbrace{\int_{-1}^{1} d(\costh)\,(1-\costh)}_{=2}
  \;,\nonumber\\
  &=
  \frac{-i\,\mNk^4}{32\pi (\mNk^2-\mNm^2)} 
  \;.             
\end{align}
%

\noindent
Hence, the imaginary part is given by
\begin{align}
 \text{Im}\left[I_{(2)}\right]
 &= \frac{1}{2i}\;
 \text{Disc}\left[I_{(2)}\right]
  =
 -\frac{\mNk^2}{64\pi}\;\left(\frac{\mNk^2}{\mNk^2-\mNm^2}\right)
  \;.
\end{align}
Therefore, asymmetry due to this interference term is given by
\begin{align}
 \varepsilon_\text{self-(2)}
 &= -\frac{4}{\Gamma_\text{tot}} \;
 \sum_{m\neq k}\sum_{n}
 \text{Im} (B_h)\;
    \text{Im} (I_{(2)} V_\varphi') \;,\nonumber\\
 &=
  4\times \frac{8\pi}{(h^\dagger h)_{kk}\;\mNk}
  \sum_{m\neq k}\sum_{n}
  \text{Im} (h_{jk}^* h_{jm} h_{nm}^* h_{nk})
  \,\frac{\mNk^2}{64\pi} 
 \left(\frac{\mNk^2}{\mNk^2-\mNm^2}\right)
  \frac{1}{4\pi\,\mNk}\;,\nonumber\\
 &=
  \frac{1}{8\pi(h^\dagger h)_{kk}}
  \sum_{m\neq k}
  \text{Im} \left[h_{jk}^* h_{jm} (h^\dagger h)_{mk}\right]
 \left(\frac{\mNk^2}{\mNk^2-\mNm^2}\right)
 \;.  \nonumber\\
  &=
  \frac{1}{8\pi(h^\dagger h)_{kk}}
  \sum_{m\neq k}
  \text{Im} \left[h_{jk}^* h_{jm} (h^\dagger h)_{mk}\right]
 \left(\frac{1}{1-z}\right)\;,
 \qquad z \equiv \frac{\mNm^2}{\mNk^2}\;.
  \label{eqC:self-result_2}
\end{align}
It should be noted that upon summing over $j$, the above expression vanishes (because the argument of Im$[\cdots]$ is real), thus in the one-flavor approximation, this term is absent. Nonetheless,
combining this with result (\ref{eqC:self-result_1}), we get the full contribution due to the self-energy correction graphs:
\begin{align}
 \varepsilon_\text{self}
 &=
  \frac{1}{8\pi(h^\dagger h)_{kk}}
  \sum_{m\neq k}
  \text{Im} \left[h_{jk}^* h_{jm}
  \left\{
  (h^\dagger h)_{km}\,
  \frac{\sqrt{z}}{1-z}
   +
   (h^\dagger h)_{mk}\,
 \frac{1}{1-z}
 \right\}
 \right]
 \;,\\
 &=
  \frac{1}{8\pi(h^\dagger h)_{kk}}
  \sum_{m\neq k}
  \frac{\mNk}{\mNk^2-\mNm^2}\, 
  \text{Im}\left[h_{jk}^* h_{jm}
  \left\{
  (h^\dagger h)_{km}\,
  \mNm
   +
   (h^\dagger h)_{mk}\,
 \mNk
 \right\}
 \right]\;.
\end{align}
This is basically Eq.~6 of \cite{Covi:1996wh}, hence we have reproduced the standard result given in the literature.

\newpage

\section{Full workings of selected items from Sec.~\ref{subsec:BKL_toy_model_5D}} \label{app:c_emdm_cal}

\subsubsection{Interference term involving the self-energy correction of Fig.~\ref{fig4:toy5D_sloop_diagrams}b}

The following is related to Eq.~(\ref{eq4:BKL_toy5D_self_b_start}) on page~\pageref{eq4:BKL_toy5D_self_b_start}. It shows all the steps leading to the final result of (\ref{eq4:BKL_toy5D_self_b_final}). Unless otherwise stated, all symbols have the same meaning as in Eqs.~(\ref{eq4:BKL_toy5D_self_b_start}) and (\ref{eq4:BKL_toy5D_self_b_final}).
\begin{align}
 I^\text{5D}_\text{self-(b)}
 &= \int \frac{d^4 q_1}{(2\pi)^4}\;
  16\,B_{\lambda}^{(5)}
 \left[\ubarj\right]_{1C} 
 \left[P_R \sigma^{\alpha\nu} q_{\alpha}\right]_{CA} 
 \left[S_{N_m}(p)\right]_{AB}
 \left[C^\dagger \sigma^{\beta\sigma} (-q_{2\beta}) P_L \right]_{BE}
 \nonumber
 \\%
 & \qquad\times%
 \left[S_\ell (q_1)\right]_{EF} 
 \left[P_R\sigma^{\delta\mu} q_{2\delta}\right]_{FD} 
 \left[u_k^c\right]_{D1} 
 \left[D_{\sigma\mu}(q_2) \varepsilon^*_\nu\right]_{11}
 \left[-u_k^T C^\dagger \sigma^{\eta\rho} q_{\eta} P_L u_j\, \varepsilon_\rho\right]_{11}
 \;.\nonumber
 \\
 \intertext{where $B_{\lambda}^{(5)} = \lambda_{jk}^* \lambda_{jm}\lambda_{nm}^*\lambda_{nk}$,}
  &=
  16 B_{\lambda}^{(5)}
  \int\frac{d^4 q_1}{(2\pi)^4}
  \frac{
  \ubarj\, P_R \sigma^{\alpha\nu} q_{\alpha}
  (-i\Lambda_m)(\pslash+\mNm)CC^\dagger\,
  \sigma^{\beta\sigma} (-q_{2\beta}) P_L
  (i)\qoneslash
  P_R \sigma^{\delta\mu} q_{2\delta}
  u_k^c
  }
  {(p^2-M_m^2 +i\epsilon)(q_1^2+i\epsilon)(q_2^2+i\epsilon)} \nonumber\\
 &\qquad\qquad\qquad\times
  (-i) g_{\sigma\mu}\, \varepsilon^*_\nu \varepsilon_\rho
  (-1)u_k^T C^\dagger \sigma^{\eta\rho} q_{\eta} P_L u_j \;, \nonumber\\
  &=
  -8 i B_{\lambda}^{(5)}
  \int\frac{d^4 q_1}{(2\pi)^4}
  \frac{
  P_R \sigma^{\alpha\nu} q_{\alpha}
  (\pslash+\mNm) 
  \sigma^{\beta\sigma} q_{2\beta}  P_L
  \qoneslash
  P_R   \sigma^{\delta\mu} q_{2\delta}
  C \left[\sum_s u_k \ubark\right]^T C^\dagger 
  }  
  {(p^2-M_m^2 +i\epsilon)(q_1^2+i\epsilon)(q_2^2+i\epsilon)}
  \nonumber\\
 &\qquad\qquad\qquad\times
  \sigma^{\eta\rho} q_{\eta} P_L g_{\sigma\mu}
  \sum_{s'} u_j\ubarj
  \, \sum_\text{pol}\varepsilon^*_\nu \varepsilon_\rho 
   \;,
   \nonumber\\
 &=
  B_{\lambda}^{(5)}\!
  \int\frac{d^4 q_1}{(2\pi)^4}
  \frac{-i  
  \text{Tr}[
  P_R (\qslash\gamma^\nu - \gamma^\nu\qslash)
  \pslash
  (\qtwoslash\gamma^\sigma - \gamma^\sigma\qtwoslash)
   \qoneslash
   (\qtwoslash\gamma_\sigma - \gamma_\sigma\qtwoslash)
   \pslash
  (\qslash\gamma_\nu - \gamma_\nu\qslash) 
  \ppslash]}
  {2\,(p^2-M_m^2 +i\epsilon)(q_1^2+i\epsilon)(q_2^2+i\epsilon)}
  ,\nonumber\\
  &=
   B_{\lambda}^{(5)}
  \int\frac{d^4 q_1}{(2\pi)^4}
  \frac{32i \mNk^4
  \left[
  4(q \cdot q_2)(q_1 \cdot q_2)-(q \cdot q_1)q_2^2
  -4(p \cdot q_2)(q_1 \cdot q_2)+(p \cdot q_1)q_2^2
  \right]
  }
  {(p^2-M_m^2 +i\epsilon)(q_1^2+i\epsilon)(q_2^2+i\epsilon)}
   ,
\end{align}
Focusing on the integral:
\begin{equation}
 I^\text{5D}_\text{s-(b)}
  \equiv
 32i\,\mNk^4
  \int\frac{d^4 q_1}{(2\pi)^4}\;
  \frac{
  4(q \cdot q_2)(q_1 \cdot q_2)
  -4(p \cdot q_2)(q_1 \cdot q_2)
  -(q \cdot q_1)q_2^2
  +(p \cdot q_1)q_2^2
  }
  {(p^2-M_m^2 +i\epsilon)(q_1^2+i\epsilon)(q_2^2+i\epsilon)}
   \;.
\end{equation}
The discontinuity of this integral is determined by cutting through the propagators with momenta $q_1$ and $q_2$, which then results in ($q_2=p-q_1, q_1\equiv (E_1,\vec{q}_1)$):
\begin{align}
 \text{Disc}(I^\text{5D}_\text{s-(b)})
 &=
  \frac{32i\,\mNk^4}{\mNk^2-\mNm^2}
  \int\frac{d^4 q_1}{(2\pi)^4}\;
  (-2\pi i)^2
   \delta(q_1^2)
   \delta\left[(p-q_1)^2\right]\Theta(E_{1})\Theta(\mNk - E_{1})
   \nonumber\\
 &\qquad\qquad\qquad\qquad\times
  \left[  
  4(q \cdot q_2)(q_1 \cdot q_2)
  -4(p \cdot q_2)(q_1 \cdot q_2)
  -(q \cdot q_1)q_2^2
  +(p \cdot q_1)q_2^2
  \right]
  \;,\nonumber
\end{align}
\begin{align}
&=
  \frac{32i\,\mNk^4}{\mNk^2-M_m^2}
  \int\frac{d^4 q_1}{(2\pi)^4}\;
  (-2\pi i)^2
   \delta(q_1^2)
   \delta\left[(p-q_1)^2\right]\Theta(E_{1})\Theta(\mNk - E_{1})
   \nonumber\\
 &\qquad\qquad\qquad\qquad\times
  \left[  
  4(q \cdot q_2)(q_1 \cdot q_2)
  -4(p \cdot q_2)(q_1 \cdot q_2)
  -(q \cdot q_1)q_2^2
  +(p \cdot q_1)q_2^2
  \right]
  \;,\nonumber\\
&=
  \frac{-8i\,\mNk^4}{\pi^2(\mNk^2-M_m^2)}
  \int d^3 q_1 dE_1\;
   \delta(E_1^2-\modq1^2)
   \delta\left[(\mNk -E_1)^2-\modq1^2\right]\Theta(E_{1})\Theta(\mNk - E_{1})
  \nonumber\\
 &\qquad\qquad\qquad\qquad\times
  \left[  
  4(\mNk E_1- E_1^2+\modq1^2)
  \left(
  \frac{\mNk^2}{2} - \halfmNk (E_1-\modq1\costh)
  -\mNk^2+\mNk E_1
  \right)
  \right.
    \nonumber\\
 &\qquad\qquad\qquad\qquad\qquad
  \left. 
   +
   \left(
   \mNk E_1-\halfmNk (E_1-\modq1\costh)
   \right)
   ((\mNk -E_1)^2-\modq1^2)
  \right]
  \;,\nonumber
\end{align}  
where $\theta$ is the smaller angle between $\vec q$ and $\vec{q}_1$,  
\begin{align}
 \text{Disc}(I^\text{5D}_\text{s-(b)})
 &=
  \frac{-2i\,\mNk^5}{\pi^2(\mNk^2-M_m^2)}
  \int \modq1^2 d\modq1 d\Omega \;
  \frac{1}{\modq1}\,
   \delta\left[(\mNk -\modq1)^2-\modq1^2\right]\Theta(\mNk - \modq1)
  \nonumber\\
 &\quad\times
  \left[  
  4\mNk \modq1
  \left(
  -\mNk + \modq1+\modq1\costh
  \right)
   +
   \modq1\left(
   1+\costh
   \right)
   ((\mNk -\modq1)^2-\modq1^2)
  \right]
   ,\nonumber    \\
 &=
  \frac{-2i\,\mNk^5}{\pi^2(\mNk^2-M_m^2)}
  \int \modq1^2 d\modq1 d\Omega \;
  \frac{1}{|-2\mNk|}\,
   \delta\left[\modq1-\halfmNk\right]\Theta(\mNk - \modq1)
  \nonumber\\
 &\qquad\quad\times
  \left[  
  4\mNk 
  \left(
  -\mNk + \modq1+\modq1\costh
  \right)
   +
   \mNk
   \left(
   1+\costh
   \right)
   (\mNk-2\modq1)
  \right]
  \;,\nonumber \\
 &=
  \frac{-i\,\mNk^5}{\pi^2(\mNk^2-M_m^2)}
  \int d\phi \int d(\costh) \; \frac{\mNk^2}{4} \times
  4 \times -\frac{\mNk}{2}
  (1-\costh)
  \;,\nonumber \\  
 &=
  \frac{2i\,\mNk^8}{\pi(\mNk^2-M_m^2)}
  \;. 
\end{align}
So, the the imaginary part is given by
\begin{align}
 \text{Im}\left[I^\text{5D}_\text{s-(b)}\right]
 = \frac{1}{2i}\;
 \text{Disc}\left[I^\text{5D}_\text{s-(b)}\right]
 =
 \frac{\mNk^8}{\pi (\mNk^2-M_m^2)}
  \;.
\end{align}
The total decay rate is given by the twice of (\ref{eq4:BKL_toy5D_decay_rate}) and phase space is same as for Fig.~\ref{fig4:toy5D_sloop_diagrams}a with
$V_\varphi
   = 1/16\pi \mNk$. Therefore,
\begin{align}
  \varepsilon_{\text{self-(b)-}k,j}^\text{5D}
 &= -\frac{4}{\Gamma_\text{tot}} \;
 \sum_{m\neq k}\sum_{n}
 \text{Im} (B_{\lambda}^{(5)})\;
    \text{Im} (I_\text{s-(b)}^\text{5D} V_\varphi)\;,\nonumber
\\    
 &=
  -\frac{\mNk^2}{2\pi(\lambda^\dagger \lambda)_{kk}}
  \sum_{m\neq k}
  \text{Im} \left[\lambda^*_{jk} \lambda_{jm} 
  (\lambda^\dagger \lambda)_{mk}\right]
   \frac{1}{1-z}
  \;,
\nonumber
\\   
 &\equiv
  -\frac{(\mNk/\Lambda)^2}{2\pi(\lambda_0^\dagger \lambda_0)_{kk}}
  \sum_{m\neq k}
  \text{Im} \left[(\lambda_0^*)_{jk} (\lambda_0)_{jm} 
  (\lambda_0^\dagger \lambda_0)_{mk}\right]
   \frac{1}{1-z}
  \;,  
 \qquad z \equiv \frac{\mNm^2}{\mNk^2}
 \;,
\end{align}
where we have summed over all heavy Majorana neutrino species $m\neq k$, as well as internal lepton species $n$.

\subsubsection{Interference term involving the vertex correction of Fig.~\ref{fig4:toy5D_vloop_diagrams}a}

This subsection includes all the workings in the computation of the integrals in Eq.~(\ref{eq4:BKL_toy5D_vert_a_middle}) on page~\pageref{eq4:BKL_toy5D_vert_a_middle}. Unless otherwise stated, all symbols have the same meaning as was first introduced in Eqs.~(\ref{eq4:BKL_toy5D_vert_a_middle}) and (\ref{eq4:BKL_toy5D_vert_a_final}).

We would like to evaluate the discontinuity of the integral
\begin{align}
 I_\text{vert-(a)}^\text{5D}  &=
  i
  \int\frac{d^4 q_1}{(2\pi)^4}
  \frac{
  \mNm\mNk
  }
  {(q_3^2-\mNm^2 +i\epsilon)(q_1^2+i\epsilon)(q_2^2+i\epsilon)}
   \; 
    \left\{
    -32 \, \mNk^2 (q \cdot q_1) q_2^2
    \right.\nonumber\\
 &\left.\qquad\qquad
 +64 (q \cdot q_2)\left[
  -2 (p' \cdot q_1)(q \cdot q_2)
  +2 (p' \cdot q_2)(q \cdot q_1)
  + \mNk^2(q_1 \cdot q_2)
  \right]
  \right\}\;.
\end{align}
To this end, we note that there is only one way to cut Fig.~\ref{fig4:toy5D_vloop_diagrams}a which is through $q_1$ and $q_2$. Therefore, we make the replacement
\begin{align}
 \frac{1}{q_1^2 + i\epsilon} \;&\rightarrow\; -2\pi i \delta(q_1^2)\Theta(E_{1})\;,\nonumber\\
 \frac{1}{q_2^2 + i\epsilon} \;&\rightarrow\; -2\pi i \delta(q_2^2)\Theta(E_{2})
  = -2\pi i \delta((p-q_1)^2)\Theta(\mNk - E_{1})\;,\nonumber
\end{align}
where $q_1 = (E_{1},\vec q_1),\, q_2 = (E_{2},\vec q_2)\equiv (\mNk - E_{1}, -\vec q_1)$. 
Letting $\epsilon \rightarrow 0$, and substituting $q_3=q_1 -q$ and $q_2= p-q_1$, we get
\begin{align}
  \text{Disc}\left[I_\text{vert-(a)}^\text{5D}\right]
  &= 
     i\, \mNm\mNk
  \int\frac{d^4 q_1}{(2\pi)^4}
  \frac{ (-2\pi i)^2 
  \delta(q_1^2)
  \delta\left((p-q_1)^2\right) \Theta(E_{1})\Theta(\mNk - E_{1})
  }
  {(q_1-q)^2-M_m^2}
 \nonumber\\
 &\qquad\qquad\times
 \left\{
    64 \left(\frac{\mNk^2}{2}-q \cdot q_1\right)
  \left[
  -\mNk (E_{1} +|\vec q_1| \costh)\left(\frac{\mNk^2}{2}-q \cdot q_1\right)
  \right.\right.
  \nonumber\\
 &\qquad\qquad\qquad \left.
  +2 \left(\frac{\mNk^2}{2} - \halfmNk (E_{1} +|\vec q_1| \costh)\right)(q \cdot q_1)
  + \mNk^2(\mNk E_1 - q_1^2)
  \right]
  \nonumber\\
 &\qquad\qquad\qquad \left.
  -32 \, \mNk^2 (q \cdot q_1) ((\mNk-E_1)^2-\modq1^2)\frac{}{}
  \right\}
  \;,
\end{align}
where we have used 
\begin{align}
 p\cdot q_1 &= \mNk E_1 \;,\\
 p'\cdot q_1 &= E_{1} \frac{\mNk}{2} - (-\vec q) \cdot \vec q_1
  = E_{1} \frac{\mNk}{2} + |\vec q_1| |\vec q| \cos \theta
  = \frac{\mNk}{2} (E_{1} +|\vec q_1| \cos \theta)\;,
\end{align}
where $\theta$ is the smaller angle between $\vec q_1$ and $\vec q$. In addition, we have
\begin{align}
 q \cdot q_1 &= E_{1} \frac{\mNk}{2} - \vec q \cdot\vec q_1
  = E_{1} \frac{\mNk}{2} - |\vec q_1| |\vec q| \cos \theta
  = \frac{\mNk}{2} (E_{1} -|\vec q_1| \cos \theta)\;.
\end{align}
Thus,
\begin{align}
  \text{Disc}\left[I_\text{vert-(a)}^\text{5D}\right]
  &= 
     \frac{-i\, \mNm\mNk}{(2\pi)^2}
  \int d^3 q_1 dE_1
  \frac{
  \delta(E_{1}^2-|\vec q_1|^2)
  \delta\left((\mNk-E_{1})^2-|\vec q_1|^2\right) \Theta(E_{1})
  }
  {(E_1-\halfmNk)^2 -|\vec q_1-\vec q|^2-M_m^2}
 \nonumber\\
 &\;\;\times
 \Theta(\mNk - E_{1})
 \left\{
    32 \mNk
    \left(\mNk- E_{1} +|\vec q_1| \cos \theta)\right)
  \left[
  -\frac{\mNk^2}{2} (E_{1} +|\vec q_1| \costh)
  \right.\right.
  \nonumber\\
 &\;\; 
 \times
 (\mNk- E_{1} +|\vec q_1| \cos \theta)
  +\frac{\mNk^2}{2} 
  (E_{1} -|\vec q_1| \cos \theta)
  \left(\mNk - E_{1} -|\vec q_1| \costh\right)
  \nonumber\\
 &\quad \left.
     + \mNk^2(\mNk E_1 - E_{1}^2+|\vec q_1|^2)
  \phantom{\frac{1}{1}}\!\!\! 
  \right]
\nonumber\\
 &\qquad \left.
  -32 \, \mNk^2 
  \left( \frac{\mNk}{2} (E_{1} -|\vec q_1| \cos \theta)\right)
   ((\mNk-E_1)^2-\modq1^2)
   \phantom{\frac{1}{1}}\!\!\!
  \right\}
  \;,\\
\intertext{and simplifying using 
$\delta(E_{1}^2-|\vec q_1|^2) = \left[\delta(E_1-\modq1)
  +\delta(E_1+\modq1)\right]/2\modq1$, to get}  
 \text{Disc}\left[I_\text{vert-(a)}^\text{5D}\right]
  &= 
     \frac{-i\, \mNm\mNk}{(2\pi)^2}
  \int d^3 q_1 \frac{1}{2\modq1}
  \frac{
  \delta\left((\mNk-\modq1)^2-|\vec q_1|^2\right) \Theta(\mNk - \modq1)
  }
  {( \modq1-\halfmNk)^2 -|\vec q_1-\vec q|^2-M_m^2}
 \nonumber\\
 &\times
 \left\{
    32 \mNk
    \left(\mNk  -  \modq1(1 - \cos \theta))\right)
  \left[
  \frac{-\mNk^2 \modq1}{2} ( 1 + \costh)(\mNk-  \modq1(1 - \cos \theta))
  \right.\right.
  \nonumber\\
 &\;\;\qquad \left.
  +\frac{\mNk^2\modq1}{2} 
  (1 - \cos \theta)
  \left(\mNk -  \modq1 -|\vec q_1| \costh\right)
    + \mNk^3  \modq1
  \right]
  \nonumber\\
 &\;\;\qquad \left.
  - 16\mNk^3\modq1 (1 - \cos \theta)
   \left((\mNk- \modq1)^2-\modq1^2\right)\frac{}{}
  \right\}
  \;,   \nonumber\\  
  \rule{0pt}{35pt}
  &= 
     \frac{-i\, \mNm\mNk}{2(2\pi)^2}
  \int \modq1^2 d\modq1 d\Omega 
  \frac{
  \delta\left(\mNk^2 - 2\mNk|\vec q_1|\right) \Theta(\mNk - \modq1)
  }
  {( \modq1-\halfmNk)^2 -|\vec q_1-\vec q|^2-M_m^2}
 \nonumber\\
 &\;\;\times
 \left\{
    32 \mNk
    \left(\mNk-  \modq1(1 - \cos \theta))\right)
  \left[
  -\frac{\mNk^2}{2} ( 1 + \costh)(\mNk-  \modq1(1 - \cos \theta))
  \right.\right.
  \nonumber\\
 &\qquad\qquad\qquad \left.
  +\frac{\mNk^2}{2} 
  (1 - \cos \theta)
  \left(\mNk -  \modq1 -|\vec q_1| \costh\right)
    + \mNk^3 
  \right]
  \nonumber\\
 &\qquad\qquad\qquad \left.
  - 16\mNk^3(1 - \cos \theta)
   \left(\mNk^2 - 2\mNk|\vec q_1|\right)\frac{}{}
  \right\}
  \;, \nonumber  \\
  \rule{0pt}{35pt}
  &= 
     \frac{-i}{8\pi^2}
  \int 
  \frac{\modq1^2 d\modq1 d\Omega}{|-2\mNk|}\,
  \frac{
  \mNm\mNk\;\delta\left[|\vec q_1|-\halfmNk\right] \Theta(\mNk - \modq1)
  }
  {(\modq1-\halfmNk)^2 -\modq1^2 -\frac{\mNk^2}{4}+ \modq1 \mNk \costh - M_m^2}
 \nonumber\\
 &\;\times
 \left\{
    32 \mNk
    \left(\mNk-  \modq1 +|\vec q_1| \cos \theta)\right)
  \left[
  -\frac{\mNk^2}{2} ( 1 + \costh)(\mNk-  \modq1 +|\vec q_1| \cos \theta)
  \right.\right.
  \nonumber\\
 &\qquad\qquad\qquad \left.
  +\frac{\mNk^2}{2} 
  (1 - \cos \theta)
  \left(\mNk -  \modq1 -|\vec q_1| \costh\right)
    + \mNk^3 
  \right]
  \nonumber\\
 &\qquad\qquad\qquad\qquad\qquad\quad \left.
  - 16\mNk^3(1 - \cos \theta)
   \left(\mNk^2 - 2\mNk|\vec q_1|\right)\phantom{\frac{1}{1}}\!\!\!
  \right\}
  \;, 
\end{align}
where in the last step we have used the identity, $\delta(ax)=\delta(x)/|a|$ and the fact that
\begin{align}
 -|\vec q_1-\vec q|^2
 &= -\left(
   |\vec q_1|^2 +|\vec q|^2 - 2|\vec q_1||\vec q|\cos (\theta)
   \right),\nonumber\\
 &= -|\vec q_1|^2 -\frac{\mNk^2}{4} + |\vec q_1|\mNk \cos \theta,
\end{align}
where $\theta$ is again the smaller angle between $\vec q_1$ and $\vec q$. Performing the $d\modq1$ integral to get
\begin{align}
 \text{Disc}\left[I_\text{vert-(a)}^\text{5D}\right]
  &= 
     \frac{-i\, \mNm}{16\pi^2}
  \int d\Omega \;\frac{\mNk^2}{4}\;
  \frac{4\mNk^5 (1+\costh)
  \left[
  -(1+\costh)^2 +(1-\costh)^2 +4
  \right]
  }
  {-\frac{\mNk^2}{2}+ \frac{\mNk^2}{2} \costh - M_m^2}
 \;,\nonumber\\
  &= 
     \frac{i\, \mNm\mNk^5}{4\pi^2}
  \int d\Omega \;
  \frac{(1+\costh)
  \left[
  -(1+\costh)^2 +(1-\costh)^2 +4
  \right]
  }
  {2(1- \costh) + 4z}\;,
 \;,\nonumber\\
  &= 
     \frac{i\, \mNm\mNk^5}{4\pi^2}
  \int d\phi \int_{-1}^{1} dx \;
  \frac{(1+x)
  \left[
  -(1+x)^2 +(1-x)^2 +4
  \right]
  }
  {2(1- x) + 4z}
 \;,\nonumber   \\
  &= 
     \frac{i\, \mNm\mNk^5}{\pi}
  \int_{-1}^{1} dx \;
  \frac{
  1-x^2
  }
  {1- x + 2z}
 \;,\nonumber     \\
  &= 
   \frac{i\, \mNm\mNk^5}{\pi}\;
   \left(2+4z-4z(z+1)\ln \left[-2(z+1)\right]+ 4z(z+1)\ln\left[-2z\right]\right)
   \;,\nonumber   \\ 
 \text{Disc}\left[I_\text{vert-(a)}^\text{5D}\right]
  &= 
   \frac{i\,\mNk^6}{\pi}\;
   \sqrt{z}\left(2+4z-4z(z+1)\ln \left[\frac{z+1}{z}\right]\right)
   \;,  \qquad \text{ with } z \equiv \frac{\mNm^2}{\mNk^2}\;.      
\end{align} 
So,
\begin{align}
 \text{Im}\left[I_\text{vert-(a)}^\text{5D}\right]
 = \frac{1}{2i}\;
 \text{Disc}\left[I_\text{vert-(a)}^\text{5D}\right] 
 =
  \frac{\mNk^6}{\pi}\; 
  \sqrt{z}\left[1+2z\left(1-(z+1)\ln \left[\frac{z+1}{z}\right]\right)\right]\;.  
\end{align}
Total decay rate and phase space factor are as in the previous case, hence the contribution to $CP$ asymmetry for this vertex correction graph is
\begin{align}
\varepsilon_{\text{vert-(a)-}k,j}^\text{5D}
 &=
  \frac{-\mNk^2}{2\pi(\lambda^\dagger \lambda)_{kk}}
  \sum_{m\neq k}
  \text{Im} \left[\lambda_{jk}^* \lambda_{jm} (\lambda^\dagger \lambda)_{km}\right]
  f_{Va}(z)
 \;,
 \nonumber\\
 &\equiv
  \frac{-(\mNk/\Lambda)^2}{2\pi(\lambda_0^\dagger \lambda_0)_{kk}}
  \sum_{m\neq k}
  \text{Im} \left[
  (\lambda_0^*)_{jk} (\lambda_0)_{jm} (\lambda_0^\dagger \lambda_0)_{km}\right]
  f_{Va}(z)
%
 \;,\\
\intertext{where}
&
 f_{Va}(z) \equiv
 \sqrt{z}\left[1+2z\left(1-(z+1)\ln \left[\frac{z+1}{z}\right]\right)\right]
 \;.
\end{align}


\newpage
\section{Full workings of selected items from Sec.~\ref{subsec:BKL_toy_model_6D}} \label{app:c_emdm_cal_3body}

The following are the intermediate steps between Eqs.~(\ref{eq4:BKL_6D_diff_rate}) and (\ref{eq4:BKL_6D_decay_rate_final_A}) on page~\pageref{eq4:BKL_6D_diff_rate}, which show how to evaluate the 3-body phase space integrals. Starting from (\ref{eq4:BKL_6D_diff_rate}):
\begin{equation}
  d\Gamma_{k}^\text{6D}
  =
  \frac{8\,({\lambda'}^\dagger {\lambda'})_{kk}}{(2\pi)^5}
     \left[E_q (\mNk^2-2\mNk E'')\right]
  \frac{d^3p'}{2E'}\frac{d^3p''}{2E''}
  \frac{d^3 q}{2E_q} \,\delta^{(4)}(p-p'-p''-q)\;,
  \label{eqC:6d_tree_starting}
\end{equation}
we integrate over all possible values in $d^3p'$, $d^3p''$ and $d^3q$ to get the decay rate:
\begin{equation}
 \Gamma_{k}^\text{6D} = \int_{p',p'',q} (\text{\ref{eqC:6d_tree_starting}})\;.
\end{equation}
Using the relation for massless particles with four-momentum $k$:
\begin{equation}\label{eqC:general_intg_trick}
  \int \, \frac{d^3 k}{2E_k}\, f(k) = \int \, d^4 k\,\Theta(k_0)\,\delta(k^2)\,f(k)\;,
\end{equation}
where $f(k)$ is any function of $k$,  $\Theta(x)$ is the unit step function, and 
$\delta(x)$ is the Dirac-delta function, we can simlify the $d^3p'$ integral and get
\begin{align}
 \Gamma_{k}^\text{6D} &= 
 \int
  \frac{8\,(\lambda'^\dagger \lambda')_{kk}}{(2\pi)^5}
  \left[E_q (\mNk^2-2\mNk E'')\right]
  d^4 p'\,\Theta(p'_0)\,\delta(p'^2)\, 
  \frac{d^3p''}{2E''}
  \frac{d^3 q}{2E_q} \,\delta^{(4)}(p-p'-p''-q)\;,\nonumber
  \\
  &= \frac{8\,(\lambda'^\dagger \lambda')_{kk}}{(2\pi)^5}
  \left[E_q (\mNk^2-2\mNk E'')\right]
  dp_0'd^3p'\Theta(p'_0)\delta(p_0'^2 - |\vec p~'|^2)
  \nonumber\\
  &\hspace{90pt} 
  \times  
  \frac{d^3p''}{2E''}
  \frac{d^3 q}{2E_q} \,
  \delta(\mNk-E'-E''-E_q)\,
  \delta^{(3)}(\vec p~' + \vec p~'' + \vec q)\;.  
\end{align}
Performing the $dp_0'$ integral (NB: $dp_0'\equiv dE'$) using delta function: $\delta(\mNk-E'-E''-E_q)$, we then have
\begin{align}
  \Gamma_{k}^\text{6D}
 &=
 \int
 \frac{8\,(\lambda'^\dagger \lambda')_{kk}}{(2\pi)^5}
  \left[E_q (\mNk^2-2\mNk E'')\right]
 \Theta(\mNk-E''-E_q)\delta\left[(\mNk-E''-E_q)^2 - |\vec p~'|^2\right]\, 
  \nonumber\\
 &\hspace{90pt} 
  \times   d^3 p'\,  
  \frac{d^3p''}{2E''}
  \frac{d^3 q}{2E_q} \,
  \delta^{(3)}(\vec p~' + \vec p~'' + \vec q)\;.
\end{align}
Next we do the $d^3 p'$ integral using delta function: $\delta^{(3)}(\vec p~' + \vec p~'' + \vec q)$ to obtain
\begin{align}
   \Gamma_{k}^\text{6D}
 &=
 \int
 \frac{8\,(\lambda'^\dagger \lambda')_{kk}}{(2\pi)^5}
  \left[E_q (\mNk^2-2\mNk E'')\right]
  \frac{d^3p''}{2E''}
  \frac{d^3 q}{2E_q}  \nonumber\\
 &\hspace{90pt}
 \times  
  \underbrace{\Theta(\mNk-E''-E_q)}_{=\,1}
  \underbrace{\delta\left[(\mNk-E''-E_q)^2 - |\vec p~'' + \vec q|^2\right]}_{=\,
  \delta\left[(p-p''-q)^2\right]} \;,
\end{align}
\begin{equation}
=
 \int
 \frac{2(\lambda'^\dagger \lambda')_{kk}}{(2\pi)^5}
  \left[\mNk^2-2\mNk E''\right]
  \delta\left[\mNk^2 - 2\mNk E'' - 2 \mNk E_q + 2E''E_q(1-\cos\beta)\right]
  \frac{d^3p''\,d^3 q}{E''}\;,\label{eqC:6d_tree_mark1} 
\end{equation}
where we have used
\begin{align}
  |\vec p~'' + \vec q|^2 
  &= |\vec p~''|^2 + |\vec q|^2- 2|\vec p~''||\vec q|\cos(\pi-\beta)\;,\nonumber\\
  &= E''^2+E_q^2 + 2E''E_q \cos\beta\;.
\end{align}
Here, $\beta$ is the smaller angle between $\vec p~''$ and $\vec q$. In polar coordinates, the measure $d^3p''\,d^3q$ can be rewritten as
\begin{align}
  d^3p''\,d^3q 
  &= |\vec p~''|^2\, d|\vec p~''|\, d\Omega''\, |\vec q|^2\, d|\vec q|\,d\Omega_q \;,\\
  &= E''^2\, dE''\,(4\pi)\, E_q^2\, dE_q\, (2\pi)\, d(\cos\beta)\;,\\
  &= 2\,(2\pi)^2\, E''^2\, dE''\, E_q^2\, dE_q\, d(\cos\beta)\;.
\end{align}
Using this in (\ref{eqC:6d_tree_mark1}), we have
\begin{align}
   \Gamma_{k}^\text{6D}
 &=
 \frac{4(\lambda'^\dagger \lambda')_{kk}}{(2\pi)^3}
  \int E''^2\, dE''\, E_q^2\, dE_q\, d(\cos\beta)\,
  \frac{\mNk^2-2\mNk E''}{E''}
  \nonumber\\
 &\hspace{90pt}
 \times  
  \delta\left[\mNk^2 - 2\mNk E'' - 2 \mNk E_q + 2E''E_q(1-\cos\beta)\right]
  \;,  \nonumber\\
 &=
 \frac{4(\lambda'^\dagger \lambda')_{kk}}{(2\pi)^3}
  \int E''^2\, dE''\, E_q^2\, dE_q\, d(\cos\beta)\,
  \frac{\mNk^2-2\mNk E''}{E''}
  \nonumber\\
 &\hspace{90pt}
 \times  
  \frac{1}{2E''E_q}\;
  \delta\left[\frac{\mNk^2}{2E''E_q} 
  - \frac{\mNk}{E''} - \frac{\mNk}{E_q} + 1-\cos\beta\right]
  \;,\nonumber\\
 &=
 \frac{2(\lambda'^\dagger \lambda')_{kk}}{(2\pi)^3}
   \int E_q\, dE''\, dE_q\, d(\cos\beta)\,
  (\mNk^2-2\mNk E'')\;
  \nonumber\\
 &\hspace{90pt}
 \times   
  \delta\left[\frac{\mNk^2}{2E''E_q} 
  - \frac{\mNk}{E''} - \frac{\mNk}{E_q} + 1-\cos\beta\right]
  \;,
\end{align}  
and after performing the $d(\cos\beta)$ integral, we obtain
\begin{align}
  \Gamma_{k}^\text{6D}
 &=
  \frac{2(\lambda'^\dagger \lambda')_{kk}}{(2\pi)^3}
   \int dE''\, dE_q\,
  E_q\, (\mNk^2-2\mNk E'')\;. \label{eqC:6d_tree_markA}
\end{align}
In order to evaluate the $dE_q$ and $dE''$ integrals, we must first ascertain the respective integration limits for $E_q$ and $E''$. To this end, we first note that we have $E'', E_q \geq 0$ by our definitions~\footnote{This can be seen from the way we have written down the argument in the step function $\Theta(\mNk - E''-E_q)$ and compare it with the direction of momentum-flows in the Feynman diagram.}. Next, we observe that in getting to (\ref{eqC:6d_tree_markA}), the integration over $(\cos\beta)$ necessarily picks out an appropriate value such that the argument of the $\delta$-function vanishes. In other words, the expression of the argument,
\begin{equation}
 \frac{\mNk^2}{2E''E_q}-\frac{\mNk}{E''} - \frac{\mNk}{E_q} + 1-\cos\beta\;,
\end{equation}
must equal to zero somewhere between its maximum (when $\cos\beta = -1$) and minimum (when $\cos\beta = 1$) since $-1\leq \cos\beta \leq 1$. This requirement then gives rise to a set of limits for $E''$ and $E_q$. Specifically, we can write
\begin{align}
 \mNk^2 - 2\mNk E'' - 2\mNk E_q &\leq 0\;,\qquad 
 (\cos\beta \;\;\text{at minimum})\;,\\
 \mNk - 2E'' - 2 E_q  &\leq 0 \;,\\
 \Rightarrow\qquad\qquad
 \halfmNk - E''   &\leq  E_q \;, \label{eqC:6d_tree_mark2}
\end{align}
which implies $E''\leq \mNk/2$ as $E_q \geq 0$. Furthermore, conservation of energy implies that $\mNk \geq E'' + E_q$ (in the centre-of-mass frame of the decaying particle), so for a given \mNk, $E_q$ is maximum when $E''$ is at its smallest possible value. Therefore, to find the upper bound for $E_q$, we observe that
\begin{align}
 \mNk^2 - 2\mNk E'' - 2\mNk E_q &\leq 0\;,\qquad 
 (\cos\beta \;\;\text{at minimum})\;,\label{eqC:6d_tree_mark4}\\
  \mNk^2 - 2\mNk E'' - 2\mNk E_q &\geq 0\;,\qquad 
 (\cos\beta \;\;\text{at maximum})\;.\label{eqC:6d_tree_mark5}
\end{align} 
Eqs.~(\ref{eqC:6d_tree_mark4}) and (\ref{eqC:6d_tree_mark5}) are true for all allowed values of $E''$ and $E_q$, and in particular, if we set $E''=0$ (so to maximise $E_q$), we get
\begin{align}
 \mNk^2 - 2\mNk E_q &\leq 0\;,\qquad 
 \Rightarrow\quad \halfmNk \leq E_q\;,\\
  \mNk^2 - 2\mNk E_q &\geq 0\;,\qquad 
 \Rightarrow\quad \halfmNk \geq E_q\;.
\end{align} 
These two relations must be true simultaneously and so the upper bound for $E_q$ is $\mNk/2$. For the lower bound, we simply take (\ref{eqC:6d_tree_mark2}). In summary, the limits for $E''$ and $E_q$ are~\footnote{Note that we could have expressed $E''$ in terms of $E_q$ instead (and their roles will interchange in the formulas). The choice itself is not important, but once the limits are set, it will dictate the order in which we integrate these variables.}
\begin{align}
 \halfmNk - E'' &\leq E_q \leq \halfmNk\;,\label{eqC:6d_tree_Epp_limits}
 \\
 0 &\leq E'' \leq \halfmNk\;.\label{eqC:6d_tree_Eq_limits}
\end{align}
Returning to (\ref{eqC:6d_tree_markA}) and performing the $dE_q$ and $dE''$ integrals to get
\begin{align}
 \Gamma^\text{6D}_k 
 &=
 \frac{2(\lambda'^\dagger \lambda')_{kk}}{(2\pi)^3}
 \int_{0}^{\halfmNk} dE''\int_{\halfmNk - E''}^{\halfmNk}  dE_q\,  
  E_q (\mNk^2-2\mNk E'')\;,\nonumber\\
 &=
 \frac{2(\lambda'^\dagger \lambda')_{kk}}{(2\pi)^3}
 \int_{0}^{\halfmNk} dE''\;  
  \frac{\mNk E''}{2} \left(2E''^2 - 3 \mNk E'' + \mNk^2 \right)\;,\nonumber\\
 &=
 \frac{(\lambda'^\dagger \lambda')_{kk}}{(2\pi)^3}\,
 \frac{\mNk^5}{32}\;,\nonumber\\ 
 &=
 \frac{(\lambda'^\dagger \lambda')_{kk} \;\mNk^5}{256\pi^3}
 \;,
\end{align}
which is the result of (\ref{eq4:BKL_6D_decay_rate_final_A}) on page~\pageref{eq4:BKL_6D_decay_rate_final_A}.

~\\

Next, we present the calculations leading to Eq.~(\ref{eq4:BKL_6D_self_cp_final}) on page~\pageref{eq4:BKL_6D_self_cp_final}. Beginning with the expression for the amplitude given in (\ref{eq4:BKL_6D_self_amp}), we apply the cutting rules and put the propagators associated with $q_1$, $q_2$ and $q_3~(= p - q_1 - q_2)$ on-shell~\footnote{Note that this is the only possible cut for this self-energy diagram.}. In other words, we make the following replacements:
\begin{gather}
 \frac{1}{q_1^2+i\epsilon} \;\rightarrow\; -2\pi i\; \delta(q_1^2)\Theta(E_1)\;;
 \qquad
 \frac{1}{q_2^2+i\epsilon} \;\rightarrow\; -2\pi i\; \delta(q_2^2)\Theta(E_2)\;;\\
 \frac{1}{q_3^2+i\epsilon} \;\rightarrow\; -2\pi i\; \delta(q_3^2)\Theta(E_3)
 =-2\pi i\; \delta((p-q_1-q_2)^2)\Theta(\mNk-E_1-E_2) \;,
\end{gather}
where we have used (\ref{eq4:BKL_6D_momenta}) in writing these out. So the discontinuity of (\ref{eq4:BKL_6D_self_amp}) is
\begin{align}
 \text{Disc}\left[I_\text{self}^\text{6D}\right]
 &=
 \frac{64 i\,A_{\lambda}^{(6)}\mNm\mNk}{(2\pi)^5}
  \int d^4 q_1\, d^4 q_2\,
  \frac{(p' \cdot q)
  \left[4(q \cdot q_1)(q_1 \cdot q_2)-(q \cdot q_2) q_1^2\right]}
   {\mNk^2-\mNm^2}
  \; \delta(q_1^2)
  \,\delta(q_2^2) 
   \nonumber\\
 &\hspace{80pt}  
   \times
   \delta[(p-q_1-q_2)^2]\,
   \Theta(E_1)\, \Theta(E_2)\, \Theta(\mNk-E_1-E_2)
  \;,
\end{align}
where we have substituted $p^2 = \mNk^2$ and taken $\epsilon \rightarrow 0$. Using (\ref{eqC:general_intg_trick}) and the fact that $q_1^2 = E_1^2 -|\vec q_1|^2 = 0$ (i.e. massless on-shell particle) to rewrite the integral while expanding out the scalar products, we get
\begin{align}
   \text{Disc}\left[I_\text{self}^\text{6D}\right]
 &=
 \underbrace{
 \frac{64 i\,A_{\lambda}^{(6)}\mNm\mNk}{(2\pi)^5 (\mNk^2-\mNm^2)}}_{\widetilde{\mathcal{A}}}
  \int \frac{d^3 q_1}{2E_1}\,\frac{d^3 q_2}{2E_2}\,
  (p' \cdot q)
  \left[4(q \cdot q_1)(q_1 \cdot q_2)\right]
   \,  \delta[(p-q_1-q_2)^2]
   \nonumber\\
 &\hspace{110pt}  
   \times
   \Theta(\mNk-E_1-E_2)
  \;,
\end{align}
\begin{align}
 &=
   \widetilde{\mathcal{A}}\, (p' \cdot q)
  \int \frac{d^3 q_1}{2E_1}\,\frac{d^3 q_2}{2E_2}\,
  \left[E_q E_1(1-\cos\alpha_{q1})\, E_1 E_2 (1-\cos\alpha_{21})\right]
 \nonumber\\
 &\hspace{90pt}
 \times  
  \delta\left[\mNk^2 - 2E_1\mNk - 2E_2\mNk + 2 E_1 E_2 (1-\cos\alpha_{21})\right]
  \;,
\end{align}
where $\alpha_{21}$ and $\alpha_{q1}$ are the smaller angles between 
$\vec q_2,\vec q_1$ and $\vec q, \vec q_1$ respectively. Note that we have used 
$E_1=|\vec q_1|, E_2=|\vec q_2|$ and $E_q=|\vec q|$ because the on-shell particles are assumed to be massless. For convenience, we have also dropped the $\Theta(\mNk-E_1-E_2)$ since it equates to one. Next we replace
\begin{align}
 d^3 q_1 d^3 q_2
  &= E_1^2\, dE_1\,E_2^2\, dE_2\,d\Omega_1\, d\Omega_2\;,\nonumber\\
  &= E_1^2\, dE_1\,E_2^2\, dE_2\,(2\pi)\,d(\cos\alpha_{q1})\,(2\pi)\,d(\cos\alpha_{21})\;,
\end{align}     
and get
\begin{align}
   \text{Disc}\left[I_\text{self}^\text{6D}\right]
&=
  \widetilde{\mathcal{A}}\,(2\pi)^2(p' \cdot q) 
  \int E_1^2 E_2^2 dE_1 dE_2 d(\cos\alpha_{q1})d(\cos\alpha_{21})
  \,E_q E_1\,(1-\cos\alpha_{q1})\nonumber\\
&\hspace{50pt}   
   \times
   (1-\cos\alpha_{21})\;
  \delta\left[\mNk^2 - 2E_1\mNk - 2E_2\mNk + 2 E_1 E_2 (1-\cos\alpha_{21})\right]
 \;, \nonumber\\
 &=
   2\widetilde{\mathcal{A}}\,(2\pi)^2(p' \cdot q) 
  \int E_1^2 E_2^2\, dE_1\, dE_2\, d(\cos\alpha_{21})
  \left[E_q E_1 (1-\cos\alpha_{21})\right]\nonumber\\
&\hspace{90pt}   
   \times
   \frac{1}{2E_1E_2}\,
  \delta\left[\frac{\mNk^2}{2E_1E_2} - \frac{\mNk}{E_1} - \frac{\mNk}{E_2} +
   1-\cos\alpha_{21}\right]
   \;, \nonumber\\
  &=
   \widetilde{\mathcal{A}}\,(2\pi)^2(p' \cdot q) 
  \int E_1\, E_2\, dE_1\, dE_2
  \,E_q\, E_1
  \left[\frac{\mNk}{E_1} +\frac{\mNk}{E_2}-\frac{\mNk^2}{2E_1E_2}\right]
     \;, \nonumber\\
  &=
  2\pi^2 \widetilde{\mathcal{A}}\,(p' \cdot q)
 \int dE_1\, dE_2\, E_q\,E_1
 \left(2E_1\mNk + 2E_2\mNk - \mNk^2 \right)\;.\\
 \Rightarrow\qquad
  &= 
\end{align}
Similar to the tree-level 3-body phase space case, we see that after integrating over those delta functions in the above, the limits for $E_1$ and $E_2$ are fixed and are given by $E_1$ and $E_2$ are $0\leq E_1\leq \mNk/2$ and $\mNk/2-E_1\leq E_2\leq \mNk/2$. So we have
\begin{align}
    \text{Disc}\left[I_\text{self}^\text{6D}\right]
 &=
 2\pi^2 \widetilde{\mathcal{A}}\,(p' \cdot q)
 \int_{0}^{\halfmNk} dE_1 \int_{\halfmNk-E_1}^{\halfmNk} dE_2\, E_q\,E_1
 \left(2E_1\mNk + 2E_2\mNk - \mNk^2 \right)
 \;,\nonumber
 \\
  &= 
  2\pi^2 \widetilde{\mathcal{A}}\,(p' \cdot q)\,
  E_q \int_{0}^{\halfmNk} dE_1 \, E_1^3\,\mNk
 \;,\nonumber
 \\
  &= 
  2\pi^2 \widetilde{\mathcal{A}}\,(p' \cdot q)\,
  E_q \,\frac{\mNk^5}{64}\;,\nonumber
 \\
  &=
  \frac{A_{\lambda}^{(6)} \mNk^5}{16\pi^3}\,
  \frac{i\,\mNm\mNk}{\mNk^2-\mNm^2}\,
   (p' \cdot q)\, E_q
   \;.
\end{align}
Hence,
\begin{equation}
  \text{Im}\left[I_\text{self}^\text{6D}\right]
    =   
    \frac{A_{\lambda}^{(6)} \mNk^5}{32\pi^3}\,
  \frac{\mNm\mNk}{\mNk^2-\mNm^2}\,
   (p' \cdot q)\, E_q
   \;.
\end{equation}
To deduce the relevant (imaginary part of the) kinematic factor that enters into the $CP$ asymmetry formula (\ref{eq1:CP_aysm_final_form}) on page~\pageref{eq1:CP_aysm_final_form}, we must integrate over the 3-body phase space and account for all the different decay channels. To this end, we need to evaluate
\begin{align}
 \text{Im}\left[I_\text{self}'\right]
 &=
 4   \int 
 \underbrace{
 \frac{\mNk^5}{32\pi^3}\,\frac{\mNm\mNk}{\mNk^2-\mNm^2}}_{\widetilde{\mathcal{B}}}
  \,(p' \cdot q)\, E_q \,
   \frac{1}{2\mNk}\frac{d^3 p'}{(2\pi)^3 2E'}\frac{d^3 p''}
   {(2\pi)^3 2E''}\frac{d^3 q}{(2\pi)^3 2E_q}
   \nonumber\\
  &\hspace{90pt}
  \times
   (2\pi)^4\,\delta^{(4)}(p-p'-p''-q)\;,
\end{align}
where the factor of 4 in front accounts for the two possible decay channels  and two possible internal states ($\nu\,\phi^0 \bar{B}$ and $e^-\phi^+ \bar{B}$). The procedure in computing this is almost identical to the tree-level case where we begin by rewriting the $d^3p''$ integral using (\ref{eqC:general_intg_trick}) to get
\begin{align}
  \text{Im}\left[I_\text{self}'\right]
 &=
 4\widetilde{\mathcal{B}}
  \int
   (p' \cdot q) E_q \;
   \frac{1}{2(2\pi)^5\mNk}
   \frac{d^3 p'}{2E'}\frac{d^3 q}{2E_q}\, 
   \underbrace{\Theta(\mNk- E'-E_q)}_{=\,1}
   \,\delta \left[(p-p'-q)^2\right]\;,\nonumber
   \\
  &= 
  4\widetilde{\mathcal{B}}\int
   \frac{(p' \cdot q)}{8(2\pi)^5\mNk}
   \frac{d^3 p'd^3 q}{E'}
   \,\delta \left[(\mNk-E'-E_q)^2-|\vec p~'+\vec q|^2\right]\;.   
\end{align}
Using
\begin{align}
  |\vec p~' + \vec q|^2 
  &= |\vec p~'|^2 + |\vec q|^2- 2|\vec p~'||\vec q|\cos(\pi-\theta)\;,\nonumber\\
  &= E'^2+E_q^2 + 2E'E_q \cos\theta\;,
\end{align}
where $\theta$ is the smaller angle between $\vec p~'$ and $\vec q$, to obtain
\begin{align}
   \text{Im}\left[I_\text{self}'\right]
 &=
   4 \widetilde{\mathcal{B}}\int
   \frac{(E'E_q - |\vec p~'||\vec q|\costh)}{8(2\pi)^5\mNk}
   \frac{d^3 p'd^3 q}{E'}
  \nonumber\\
  &\hspace{90pt}
  \times
  \delta \left[\mNk^2 - 2E'\mNk - 2E_q\mNk + 2 E' E_q (1-\costh)\right]\;,\nonumber
  \\
   &=
   4 \widetilde{\mathcal{B}}\int
   |\vec p~'|^2 d|\vec p~'|\,|\vec q|^2 d|\vec q|\,d\Omega' d\Omega_q
   \frac{E_q (1- \costh)}{8(2\pi)^5\mNk}
   \nonumber\\
    &\hspace{90pt}
  \times
  \delta \left[\mNk^2 - 2E'\mNk - 2E_q\mNk + 2 E' E_q (1-\costh)\right]\;,\nonumber
  \\
   &=
   4 \widetilde{\mathcal{B}}\int
   E'^2 dE'\,E_q^2 dE_q\,(4\pi)(2\pi)\,d(\costh)
   \frac{E_q (1- \costh)}{8(2\pi)^5\mNk}
   \nonumber\\
    &\hspace{90pt}
  \times
  \delta \left[\frac{\mNk^2}{2 E' E_q} 
  - \frac{\mNk}{E'} -\frac{\mNk}{E_q} + 1-\costh\right]\;, \nonumber
\end{align}
\begin{align}
  &=
   4 \widetilde{\mathcal{B}}\int  
     2E' dE'\,E_q^2 dE_q\,d(\costh)
   \frac{1- \costh}{16(2\pi)^3\mNk}
   \;
   \delta \left[\frac{\mNk^2}{2 E' E_q} - \frac{\mNk}{E'} -\frac{\mNk}{E_q} + 1-\costh\right]\;,\nonumber  \\
  &=
   4 \widetilde{\mathcal{B}}\int 
     \frac{2E' dE'\,E_q^2 dE_q}{16(2\pi)^3\mNk}
   \left[\frac{\mNk}{E'} +\frac{\mNk}{E_q}-\frac{\mNk^2}{2 E' E_q}\right]\;, 
    \nonumber  \\
  &=
   4 \widetilde{\mathcal{B}}\int 
     \frac{dE'\, dE_q}{16(2\pi)^3\mNk}
   E_q\left[2E'\mNk + 2E_q\mNk - \mNk^2\right]\;.
\end{align}
Inserting the limits for $E_q$ and $E'$ (see the analogous workings for the tree-level case), to obtain
\begin{align}
   \text{Im}\left[I_\text{self}'\right]
 &=
   \frac{\widetilde{\mathcal{B}}}{4(2\pi)^3\mNk}
   \int_{0}^{\halfmNk} dE_q \int_{\halfmNk-E_q}^{\halfmNk} dE'\,
   E_q\,(2E'\mNk + 2E_q\mNk - \mNk^2)\;,\nonumber\\
  &=
       \frac{\widetilde{\mathcal{B}}}{4(2\pi)^3\mNk}
   \int_{0}^{\halfmNk} dE_q\,
   E_q^3\mNk\;,\nonumber\\
  &=
       \frac{\widetilde{\mathcal{B}}}{4(2\pi)^3\mNk}\,
       \frac{\mNk^5}{64}\;,\nonumber\\
  &= \frac{\mNk^9}{65536\pi^6}\,\frac{\mNm\mNk}{\mNk^2-\mNm^2}
 \;,\\
  &\equiv
      \frac{\mNk^9}{65536\pi^6}\,\frac{\sqrt{z}}{1-z}\;,
      \qquad\qquad \text{ with }\quad z = \frac{\mNm^2}{\mNk^2}\;.
\end{align}
Thus, the $CP$ asymmetry due to the self-energy interference term for $N_k$ decay is given by
\begin{align}
 \varepsilon_{\text{self-}k,j}^\text{6D}
  &= -\frac{4}{2\times\Gamma^\text{6D}_k}
  \sum_{m\neq k}
  \sum_n
   \text{Im}\left[A_{\lambda}^{(6)}\right]\text{Im}\left[I_\text{self}'\right]
  \;,\nonumber\\
  &=
  -\frac{512\pi^3}{(\lambda'^\dagger \lambda')_{kk} \mNk^5}  
  \sum_{m\neq k} \sum_n
  \text{Im}\left[{\lambda'}_{nk}^* {\lambda'}_{nm} {\lambda'}_{jm} {\lambda'}_{jk}^*\right]
  \frac{\mNk^9}{65536\pi^6}\,\frac{\sqrt{z}}{1-z}
  \;,
\nonumber\\
  &=   
   -\frac{\mNk^4}{128 \pi^3 (\lambda'^\dagger \lambda')_{kk}}  
  \sum_{m\neq k}
  \text{Im}\left[{\lambda'}^*_{jk} {\lambda'}_{jm}
      ({\lambda'}^\dagger {\lambda'})_{km}\right] \frac{\sqrt{z}}{1-z}
      \;,
\nonumber\\
   &=   
   -\left(\frac{\mNk^2}{64\pi^2 \Lambda^2}\right)
   \frac{\mNk^2}{2 \pi (\Lambda^2\lambda'^\dagger \lambda')_{kk}}
  \sum_{m\neq k}   
   \text{Im}\left[{\lambda'}^*_{jk} {\lambda'}_{jm}
      ({\lambda'}^\dagger {\lambda'})_{km} \Lambda^4 \right] \frac{\sqrt{z}}{1-z}      
      \;,
\nonumber\\
    &\equiv
  \left(\frac{\mNk}{8\pi \Lambda}\right)^2
  \varepsilon_{\text{5D-self-}k,j}'      
      \;,
\end{align}
where we have defined (c.f. Eq.~(\ref{eq4:BKL_toy5D_self_a_final}))
\begin{equation}
  \varepsilon_{\text{5D-self-}k,j}'
  \equiv
  -\frac{\mNk^2}{2 \pi (\zeta^\dagger \zeta)_{kk}}
  \sum_{m\neq k}   
   \text{Im}\left[\zeta^*_{jk} \zeta_{jm}
      (\zeta^\dagger \zeta)_{km} \right] \frac{\sqrt{z}}{1-z} 
      \;,
\end{equation}
with $\zeta = \lambda'\Lambda$.

%


\end{document}